%% file: volume-detectors.tex
\documentclass[final]{cdr}
\pdfoutput=1            % must be in first 5 lines so arXiv finds it
\graphicspath{ {figures/}{volume-detectors/figures/}{volume-detectors/generated/} }

\input{common/preamble}

\renewcommand\thedoctitle{\voldunetitle}
\def\titleextra{\includegraphics[width=0.8\textwidth]{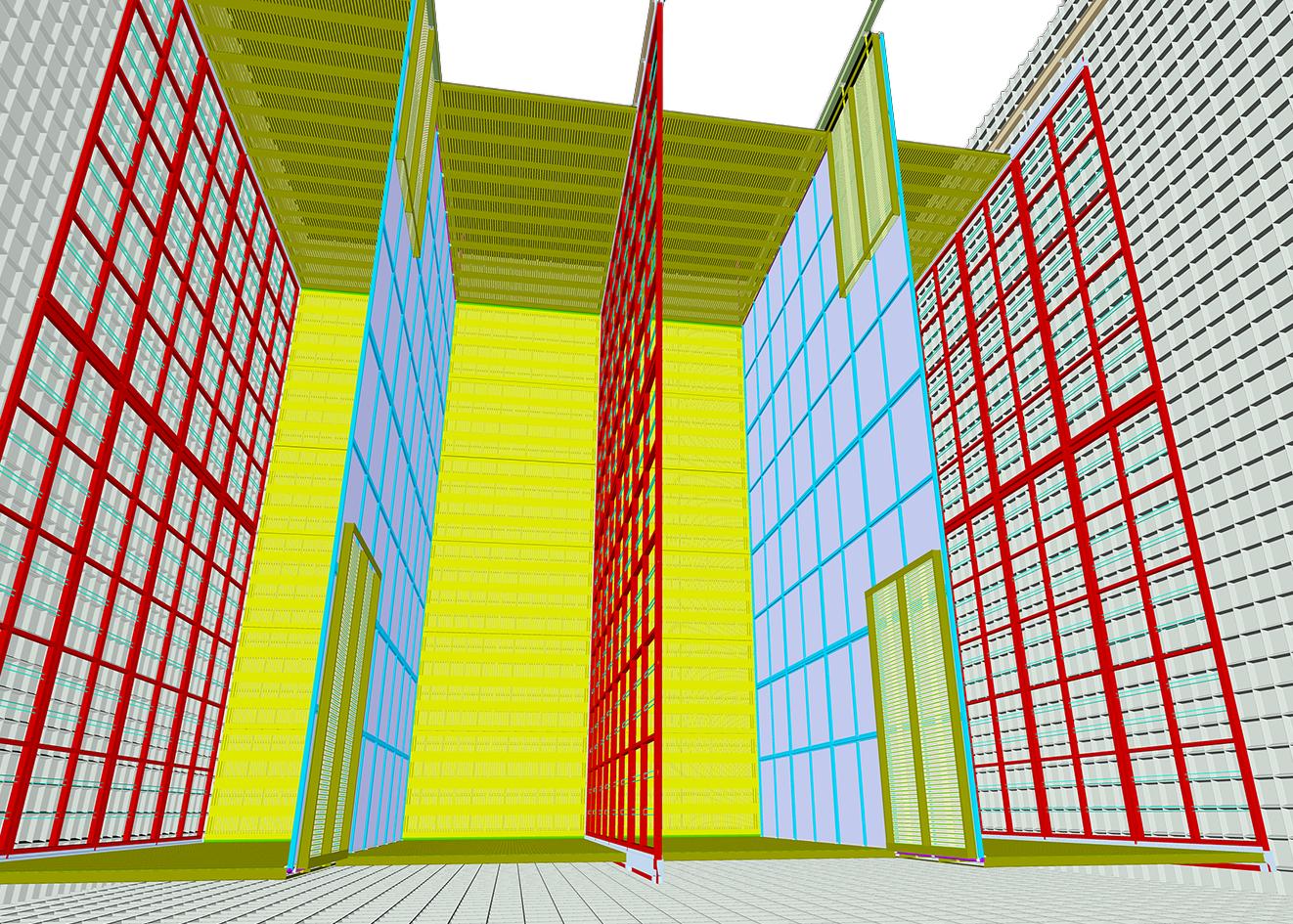}}

\begin{document}

\input{common/init}

\input{common/acronyms-shared-vol1-2-4}
\input{common/acronyms-shared-vol1-4}
\input{common/acronyms-shared-vol2-4}
\input{common/acronyms-vol4}

\input{volume-detectors/chapter-overview}
\cleardoublepage

\input{volume-detectors/chapter-strategy}
\cleardoublepage

%\input{volume-detectors/chapter-pm}  <--- put this info in Project Volume (1)
%\cleardoublepage

\input{volume-detectors/chapter-mgmt-struc}  % <--- new 5/7/15
\cleardoublepage

\input{volume-detectors/chapter-fd-ref}

\cleardoublepage

\input{volume-detectors/chapter-fd-alt}

\cleardoublepage

\input{volume-detectors/chapter-synergy}

\cleardoublepage

\input{volume-detectors/chapter-nd-ref}

\cleardoublepage

\input{volume-detectors/chapter-sc}
\cleardoublepage

\input{volume-detectors/chapter-proto}

\cleardoublepage

\input{volume-detectors/chapter-summary}

\cleardoublepage

\input{common/final}
\end{document}

%% file: common/preamble.tex
\input{common/defs.tex}
\input{common/units.tex}

\input{generated/parameters.tex}

\hypersetup{
    pdftitle={\expshort CDR \thedocsubtitle},
    pdfauthor={\expshort Collaboration},
    final=true,
    colorlinks=false,
    linktocpage=true,
    linkbordercolor=blue,
    citebordercolor=green,
    urlbordercolor=magenta,
    filecolor=black,
    pdfpagemode=UseOutlines,
    pdfborderstyle={/S/U},  
}

%% file: common/defs.tex
\def\expshort{DUNE\xspace}
\def\explong{The Deep Underground Neutrino Experiment\xspace}

\def\thedocsubtitle{Long-Baseline Neutrino Facility (LBNF) and 
Deep Underground Neutrino Experiment (DUNE)}% 
\def\cdrtitle{Conceptual Design Report}

\def\voldunetitle{Volume 4: The DUNE Detectors at LBNF\xspace}

\def\volintro{Volume 1: \textit{The LBNF and DUNE Projects\xspace}}
\def\volphys{Volume 2: \textit{The Physics Program for DUNE at LBNF\xspace}}
\def\vollbnf{Volume 3: \textit{The Long-Baseline Neutrino Facility for DUNE\xspace}}
\def\voldune{Volume 4: \textit{The DUNE Detectors at LBNF\xspace}}

\def\anxlbnefd{Annex 4A: \textit{The LBNE Design for a Deep Underground Single-Phase Liquid Argon TPC\xspace}}
\def\anxrates{Annex 4B: \textit{Expected Data Rates for the DUNE Detectors\xspace}}
\def\anxreco{Annex 4C: \textit{Simulation and Reconstruction\xspace}}

\def\anxlbnob{Annex 4E: \textit{LAGUNA/LBNO Part 2\xspace}}

\def\anxndref{Annex 4G: \textit{DUNE Near Detector Reference Design\xspace}}

\def\anxdualtdr{Annex 4I: \textit{WA105 TDR\xspace}}
\def\anxcernproto{Annex 4J: \textit{CERN Single-phase Prototype Detector Proposal\xspace}}

\def\cernsingleproto{\textit{Full-Scale Detector Engineering Test and Test Beam Calibration of a Single-Phase LArTPC\xspace}}
\def\cerndualproto{\textit{Long Baseline Neutrino Observatory Demonstration (WA105)\xspace}}

\def\Ar39{$^{39}$Ar}

%% file: common/units.tex
\DeclareSIUnit \kton {\kilo\tonne}
\DeclareSIUnit \kt {\kilo\tonne}
\DeclareSIUnit \Mt {\mega\tonne}
\DeclareSIUnit \eV {\electronvolt}
\DeclareSIUnit \keV {\kilo\electronvolt}
\DeclareSIUnit \MeV {\mega\electronvolt}
\DeclareSIUnit \GeV {\giga\electronvolt}
\DeclareSIUnit \km {\kilo\meter}
\DeclareSIUnit \kW {\kilo\watt}
\DeclareSIUnit \MW {\mega\watt}
\DeclareSIUnit \MHz {\mega\hertz}
\DeclareSIUnit \mrad {\milli\radian}
\DeclareSIUnit \year {year}
\DeclareSIUnit \POT {POT}
\DeclareSIUnit \sig {$\sigma$}
\DeclareSIUnit\parsec{pc}
\DeclareSIUnit\lightyear{ly}
\DeclareSIUnit\foot{ft}
\DeclareSIUnit\ft{ft}
\def\ktyr{\si[inter-unit-product=\ensuremath{{}\cdot{}}]{\kt\year}\xspace}

\def\ktMWyr{\si[inter-unit-product=\ensuremath{{}\cdot{}}]{\kt\MW\year}\xspace}

\newcommand{\SIadj}[2]{\SI[number-unit-product = -]{#1}{#2}}
\newcommand{\ktadj}[1]{\SIadj{#1}{\kt}}
\newcommand{\MeVadj}[1]{\SIadj{#1}{\MeV}}
\newcommand{\GeVadj}[1]{\SIadj{#1}{\GeV}}
\newcommand{\MWadj}[1]{\SIadj{#1}{\MW}}

%% file: generated/parameters.tex
\newcommand{\daqsamplerate}{\SI[round-mode=places,round-precision=1]{2.0}{\mega\hertz}\xspace}

\newcommand{\beamspillcycle}{\SI[round-mode=places,round-precision=1]{1.2}{\second}\xspace}

\newcommand{\dunefsreadoutsize}{\SI[round-mode=places,round-precision=1]{24.8832}{\giga\byte}\xspace}

\newcommand{\tpcmodulemass}{\SI[round-mode=places,round-precision=0]{10.0}{\kilo\tonne}\xspace}

\newcommand{\dunenumberchannels}{\num[round-mode=places,round-precision=0]{1536000.0}\xspace}

\newcommand{\nddatarate}{\SI[round-mode=places,round-precision=1]{1.0048}{\mega\byte\per\second}\xspace}

\newcommand{\beamrunfraction}{\num[round-mode=places,round-precision=3]{0.667}\xspace}

\newcommand{\ndmuidchannels}{\num[round-mode=places,round-precision=0]{165888.0}\xspace}

\newcommand{\tpcdrifttime}{\SI[round-mode=places,round-precision=2]{2.25}{\milli\second}\xspace}

\newcommand{\beameventsizecompressed}{\SI[round-mode=places,round-precision=1]{0.1}{\mega\byte}\xspace}

\newcommand{\ndecalchannels}{\num[round-mode=places,round-precision=0]{52224.0}\xspace}

\newcommand{\beameventsize}{\SI[round-mode=places,round-precision=1]{2.5}{\mega\byte}\xspace}

\newcommand{\daqdriftsperreadout}{\num[round-mode=places,round-precision=1]{2.4}\xspace}

\newcommand{\ndsstchannels}{\num[round-mode=places,round-precision=0]{215040.0}\xspace}

\newcommand{\daqchannelspermodule}{\num[round-mode=places,round-precision=0]{384000.0}\xspace}

\newcommand{\dunedetectormass}{\SI[round-mode=places,round-precision=0]{40.0}{\kilo\tonne}\xspace}

%% file: common/init.tex
% This should be \input first thing after \begin{document}

\pagestyle{titlepage}

\begin{center}
   {\Huge  \thedocsubtitle}  %Yes, I know title and subtitle are reversed!

  \vspace{5mm}

  {\Huge  \cdrtitle}  

  \vspace{10mm}

 {\LARGE \thedoctitle}

  \vspace{15mm}

\titleextra

  \vspace{10mm}
  \today
    \vspace{15mm}
    
\end{center}

\cleardoublepage



# The DUNE Collaboration

R. Acciarri[43], M. A. Acero[10], M. Adamowski[43], C. Adams[144], P. Adamson[43], S. Adhikari[77], Z. Ahmad[135],
C. H. Albright[43], T. Alion[120], E. Amador[131], J. Anderson[8], K. Anderson[43], C. Andreopoulos[76], M. Andrews[43],
R. Andrews[43], I. Anghel[62], J. d. Anjos[99], A. Ankowski[37], M. Antonello[72], A. Aranda Fernandez[31], A. Ariga[13],
T. Ariga[13], D. Aristizabal[75], E. Arrieta-Diaz[87], K. Aryal[131], J. Asaadi[129], D. Asner[103], M. S. Athar[6], M. Auger[13],
A. Aurisano[29], V. Aushev[68], D. Autiero[63], M. Avila[131], J. J. Back[139], X. Bai[122], B. Baibussinov[106], M. Baird[61],
B. Balantekin[142], B. Baller[43], P. Ballett[41], B. Bambah[52], M. Bansal[107], S. Bansal[107], G. J. Barker[139],
W. A. Barletta[83], G. Barr[102], N. Barros[109], L. Bartoszek[78], A. Bashyal[131], M. Bass[102], F. Bay[133], J. Beacom[100],
B. R. Behera[107], G. Bellettini[110], V. Bellini[26], O. Beltramello[16], P. A. Benetti[108], A. Bercellie[114], M. Bergevin[19],
E. Berman[43], H. Berns[20], R. Bernstein[43], S. Bertolucci[16], B. Bhandari[49], V. Bhatnagar[107], B. Bhuyan[60], J. Bian[90],
K. Biery[43], M. Bishai[15], T. Blackburn[128], A. Blake[73], F. d. M. Blaszczyk[14], E. Blaufuss[82], B. Bleakley[124],
E. Blucher[28], V. Bocean[43], F. Boffelli[108], J. Boissevain[78], S. Bolognesi[117], T. Bolton[65], M. Bonesini[89], T. Boone[33],
C. Booth[118], S. Bordoni[53], M. Borysova[68], B. Bourguille[53], S. B. Boyd[139], D. Brailsford[73], A. Brandt[131],
J. Bremer[16], S. Brice[43], C. Bromberg[87], G. Brooijmans[34], G. Brown[131], R. Brown[15], G. Brunetti[43], X. Bu[43],
N. Buchanan[33], H. Budd[114], B. Bugg[130], P. Calafiura[74], E. Calligarich[108], E. Calvo[17], L. Camilleri[34],
M. Campanelli[77], C. Cantini[147], B. Carls[43], R. Carr[34], M. Cascella[77], C. Castromonte[44], E. Catano Mur[62],
F. Cavanna[43], S. Centro[106], A. Cervera Villanueva[136], V. B. Chandratre[12], A. Chatterjee[131], S. Chattopadhyay[43],
S. Chattopadhyay[135], L. Chaussard[63], S. Chembra[52], H. Chen[15], K. Chen[15], M. Chen[21], D. Cherdack[33], C. Chi[34],
S. Childress[43], S. Choubey[47], B. C. Choudhary[38], G. Christodoulou[76], C. Christofferson[122], E. Church[103],
D. Cianci[81], D. Cline[21], T. Coan[125], A. Cocco[93], J. Coelho[134], P. Cole[58], G. Collin[83], J. M. Conrad[83],
M. Convery[115], R. Corey[122], L. Corwin[122], J. Cranshaw[8], P. Crivelli[147], D. Cronin-Hennessy[90], A. Curioni[89],
J. Cushing[28], D. L. Adams[15], D. Dale[58], S. R. Das[60], T. Davenne[116], G. S. Davies[61], J. Davies[128], J. Dawson[2],
K. De[131], A. de Gouvea[97], J. K. de Jong[102], P. de Jong[92], P. De Lurgio[8], M. Decowski[92], A. Delbart[117],
C. Densham[116], R. Dharmapalan[8], N. Dhingra[107], S. Di Luise[147], M. Diamantopoulou[9], J. S. Diaz[61],
G. Diaz Bautista[114], M. Diwan[15], Z. Djurcic[8], J. Dolph[15], Q. Drake[8], D. Duchesneau[71], M. Duvernois[142],
H. Duyang[120], D. A. Dwyer[74], S. Dye[48], S. Dytman[111], B. Eberly[115], R. Edgecock[50], D. Edmunds[87], S. Elliott[78],
M. Elnimr[21], S. Emery[117], E. Enedretos[105], S. Eno[82], A. Ereditato[13], C. O. Escobar[43], J. Evans[81], A. Falcone[108],
L. Falk[128], A. Farbin[131], C. Farnese[106], Y. Farzan[57], A. Fava[106], L. Favilli[18], J. Felde[82], J. Felix[46], S. Fernandes[4],
L. Fields[43], A. Finch[73], M. Fitton[116], B. Fleming[144], T. Forest[58], J. Fowler[40], W. Fox[61], J. Fried[15], A. Friedland[78],
S. Fuess[43], B. Fujikawa[74], A. Gago[105], H. Gallagher[134], S. Galymov[63], T. Gamble[118], R. Gandhi[47],
D. Garcia-Gamez[81], S. Gardiner[20], G. Garvey[78], V. M. Gehman[74], A. Gendotti[147], G. d. Geronimo[15], C. Ghag[77],
P. Ghoshal[47], D. Gibin[106], I. Gil-Botella[17], R. Gill[15], D. Girardelli[25], A. Giri[55], S. Glavin[109], D. Goeldi[13],
S. Golapinni[65], M. Gold[96], R. A. Gomes[44], J. J. Gomez Cadenas[136], M. C. Goodman[8], D. Gorbunov[56],
S. Goswami[47], N. Graf[111], N. Graf[115], M. Graham[115], E. Gramelini[144], R. Gran[91], C. Grant[20], N. Grant[139],
V. Greco[26], H. Greenlee[43], L. Greenler[142], C. Greenley[79], M. Groh[29], S. Grullon[109], T. Grundy[73], K. Grzelak[138],
E. Guardincerri[78], V. Guarino[8], E. Guarnaccia[137], G. P. Guedes[42], R. Guenette[102], A. Guglielmi[106], A. T. Habig[91],
R. W. Hackenburg[15], A. Hackenburg[144], H. Hadavand[131], R. Haenni[13], a. Hahn[43], M. D. Haigh[139], T. Haines[78],
T. Hamernik[43], T. Handler[130], S. Hans[15], D. Harris[43], J. Hartnell[128], T. Hasegawa[67], R. Hatcher[43],
A. Hatzikoutelis[130], S. Hays[43], E. Hazen[14], M. Headley[123], A. Heavey[43], K. Heeger[144], J. Heise[123], K. Hennessy[76],
J. Hewes[81], A. Higuera[49], T. Hill[58], A. Himmel[43], M. Hogan[33], P. Holanda[25], A. Holin[77], W. Honey[73],
S. Horikawa[147], G. Horton-Smith[65], B. Howard[61], J. Howell[43], P. Hurh[43], J. Huston[87], J. Hylen[43], R. Imlay[79],
J. Insler[79], G. Introzzi[108], D. Ioanisyan[145], A. Ioannisian[145], K. Iwamoto[114], A. Izmaylov[136], C. Jackson[131],
D. E. Jaffe[15], C. James[43], E. James[43], F. Jediny[36], C. Jen[137], A. Jhingan[107], S. Jiménez[17], J. H. Jo[27],
M. Johnson[43], R. Johnson[29], J. Johnstone[43], B. J. Jones[83], J. Joshi[15], H. Jostlein[43], C. K. Jung[127], T. Junk[43],
A. Kaboth[59], R. Kadel[74], T. Kafka[134], L. Kalousis[137], Y. Kamyshkov[130], G. Karagiorgi[81], D. Karasavvas[9],
Y. Karyotakis[71], A. Kaur[107], P. Kaur[107], B. Kayser[43], N. Kazaryan[145], E. Kearns[14], P. Keener[109], S. Kemboi[131],
E. Kemp[25], S. H. Kettell[15], M. Khabibullin[56], M. Khandaker[58], A. Khotjantsev[56], B. Kirby[15], M. Kirby[43],
J. Klein[109], T. Kobilarcik[43], S. Kohn[19], G. Koizumi[43], A. Kopylov[56], M. Kordosky[141], L. Kormos[73], U. Kose[16],
A. Kostelecky[61], M. Kramer[19], I. Kreslo[13], K. Kriske[90], W. Kropp[21], Y. Kudenko[56], V. A. Kudryavtsev[118],
S. Kulagin[56], A. Kumar[107], G. Kumar[69], J. Kumar[48], A. Kumar[107], T. Kutter[79], A. Laminack[4], K. Lande[109],
C. Lane[39], K. Lang[132], F. Lanni[15], J. Learned[48], P. Lebrun[43], D. Lee[78], H. Lee[114], K. Lee[21], W. M. Lee[43],
M. A. Leigui de Oliveira[1], Q. Li[43], S. Li[15], S. Li[100], X. Li[127], Y. Li[15], Z. Li[40], J. Libo[120], C. S. Lin[74], S. Lin[33],
J. Ling[15], J. Link[137], Z. Liptak[32], D. Lissauer[15], L. Littenberg[15], B. Littlejohn[54], Q. Liu[78], T. Liu[125], S. Lockwitz[43],
N. Lockyer[43], T. Loew[74], M. Lokajicek[3], K. Long[59], M. D. L. Lopes[43], J. P. Lopez[32], J. Losecco[98], W. Louis[78],

J. Lowery[61], M. Luethi[13], K. Luk[19], B. Lundberg[43], T. Lundin[43], X. Luo[144], T. Lux[53], J. Lykken[43],
A. A. Machado[72], J. R. Macier[43], S. Magill[8], G. Mahler[15], K. Mahn[87], M. Malek[59], S. Malhotra[12], D. Malon[8],
F. Mammoliti[26], S. Mancina[8], S. K. Mandal[38], S. Mandodi[52], S. L. Manly[114], A. Mann[134], A. Marchionni[43],
W. Marciano[15], C. Mariani[137], J. Maricic[48], A. Marino[32], M. Marshak[90], C. Marshall[114], J. Marshall[24],
J. Marteau[63], J. Martin-Albo[102], D. Martinez[54], S. Matsuno[48], J. Matthews[79], C. Mauger[78], K. Mavrokoridis[76],
D. Mayilyan[145], E. Mazzucato[117], N. McCauley[76], K. E. McCluskey[43], N. McConkey[118], K. McDonald[112],
K. S. McFarland[114], A. M. McGowan[114], C. McGrew[127], R. McKeown[141], D. McNulty[58], R. McTaggart[124],
A. Mefodiev[56], M. Mehrian[33], P. Mehta[95], D. Mei[121], O. Mena[136], S. Menary[146], H. Mendez[85], A. Menegolli[108],
G. Meng[106], Y. Meng[21], H. Merritt[59], D. Mertins[4], M. Messier[61], W. Metcalf[79], M. Mewes[61], H. Meyer[140],
T. Miao[43], R. Milincic[48], W. Miller[90], G. Mills[78], O. Mineev[56], O. Miranda[30], C. S. Mishra[43], S. R. Mishra[120],
B. Mitrica[51], D. Mladenov[16], I. Mociu[104], R. Mohanta[52], N. Mokhov[43], C. Montanari[16], D. Montanari[43],
J. Moon[43], M. Mooney[15], C. Moore[43], J. Morfin[43], B. Morgan[139], C. Morris[49], W. Morse[15], Z. Moss[83], C. Mossey[43],
C. A. Moura[1], J. Mousseau[86], L. Mualem[23], M. Muether[140], S. Mufson[61], S. Murphy[147], J. Musser[61], R. Musser[131],
Y. Nakajima[74], J. Naples[111], J. Navarro[10], D. Navas[17], J. Nelson[141], M. Nessi[16], M. Newcomer[109], Y. Ng[131],
R. Nichol[77], T. C. Nicholls[116], K. Nikolics[147], E. Niner[61], B. Norris[43], F. Noto[16], P. Novakova[15], P. Novella[136],
J. Nowak[73], M. S. Nunes[25], H. O'Keeffe[73], R. Oldeman[24], R. Oliveira[25], T. Olson[134], V. Onishchuk[68], J. Osta[43],
T. Ovsjannikova[56], B. Page[87], S. Pakvasa[48], S. Pal[118], O. Palamara[43], A. Palazzo[84], J. Paley[43], C. Palomares[17],
E. Pantic[20], V. Paolone[111], V. Papadimitriou[43], J. Park[137], S. Parke[43], Z. Parsa[15], S. Pascoli[41], R. Patterson[23],
S. Patton[74], T. Patzak[2], B. Paulos[142], L. Paulucci[1], Z. Pavlovic[43], G. Pawloski[90], S. Peeters[128], E. Pennacchio[63],
A. Perch[77], G. N. Perdue[43], L. Periale[147], J. D. Perkin[118], H. Pessard[7], G. Petrillo[43], R. Petti[120], A. Petukhov[122],
F. Pietropaolo[106], R. Plunkett[43], S. Pordes[43], M. Potekhin[15], R. Potenza[26], B. Potukuchi[64], N. Poudyal[121],
O. Prokofiev[43], N. Pruthi[107], P. Przewlocki[94], D. Pushka[43], X. Qian[15], J. L. Raaf[43], R. Raboanary[7], V. Radeka[15],
A. Radovic[141], G. Raffelt[84], I. Rakhno[43], H. T. Rakotondramanana[7], L. Rakotondravohitra[7], Y. A. Ramachers[139],
R. Rameika[43], J. Ramsey[78], A. Rappoldi[108], G. Raselli[108], P. Ratoff[73], B. Rebel[43], C. Regenfus[147],
J. Reichenbacher[122], D. Reitzner[43], A. Remoto[71], A. Renshaw[49], S. Rescia[15], M. Richardson[118], K. Rielage[78],
K. Riesselmann[43], M. Robinson[118], L. Rochester[115], O. B. Rodrigues[25], P. Rodrigues[114], B. Roe[86], M. Rosen[48],
M. Roser[43], M. Ross-Lonergan[41], M. Rossella[108], A. Rubbia[147], C. Rubbia[45], R. Rucinski[43],
C. Rudolph von Rohr[13], B. Russell[144], D. Ruterbories[114], R. Saakyan[77], N. Sahu[55], P. Sala[88], N. Samios[15],
F. Sanchez[53], M. Sanchez[62], B. Sands[112], S. Santana[25], R. Santorelli[17], G. Santucci[127], N. Saoulidou[9],
A. Scaramelli[88], H. Schellman[101], P. Schlabach[43], R. Schmitt[43], D. Schmitz[28], J. Schneps[134], K. Scholberg[40],
A. Schukraft[43], J. Schwehr[33], E. Segreto[72], S. Seibert[109], J. A. Sepulveda-Quiroz[62], F. Sergiampietri[147],
L. Sexton-Kennedy[43], D. Sgalaberna[147], M. Shaevitz[34], J. Shahi[107], S. Shahsavarani[131], P. Shanahan[43],
S. U. Shankar[47], R. Sharma[15], R. K. Sharma[113], T. Shaw[43], R. Shrock[127], I. Shyrma[68], N. Simos[15], G. Sinev[40],
I. Singh[107], J. Singh[107], J. Singh[80], V. Singh[11], G. Sinnis[78], W. Sippach[34], D. Smargianaki[16], M. Smy[21],
E. Snider[43], P. Snopok[54], J. Sobczyk[143], H. Sobel[21], M. Soderberg[129], N. Solomey[140], W. Sondheim[78], M. Sorel[136],
A. Sousa[29], K. Soustruznik[27], J. Spitz[83], J. Spitz[86], N. J. Spooner[118], M. Stancari[43], I. Stancu[4], D. Stefan[16],
H. M. Steiner[74], J. Stewart[15], J. Stock[122], S. Stoica[51], J. Stone[14], J. Strait[43], M. Strait[28], T. Strauss[43],
S. Striganov[43], R. Sulej[94], G. Sullivan[82], Y. Sun[48], L. Suter[8], C. M. Sutera[26], R. Svoboda[20], B. Szczerbinska[37],
A. Szelc[81], S. Söldner-Rembold[81], R. Talaga[8], M. Tamsett[128], S. Tariq[43], E. Tatar[58], R. Tayloe[61], C. Taylor[78],
D. Taylor[123], K. Terao[34], M. Thiesse[118], J. Thomas[77], L. F. Thompson[118], M. Thomson[24], C. Thorn[15],
M. Thorpe[116], X. Tian[120], D. Tiedt[122], S. C. Timm[43], A. Tonazzo[2], T. Tope[43], A. Topkar[12], F. R. Torres[25],
M. Torti[108], M. Tortola[136], F. Tortorici[26], M. Toups[43], C. Touramanis[76], M. Tripathi[20], I. Tropin[43], Y. T. Tsai[115],
K. V. Tsang[74], R. Tsenov[119], S. Tufanli[144], C. Tull[74], J. Turner[41], M. Tzanov[79], E. Tziaferi[9], Y. Uchida[59],
J. Urheim[61], T. Usher[115], M. Vagins[66], P. Vahle[141], G. A. Valdiviesso[5], L. Valerio[43], Z. Vallari[127], J. Valle[136],
R. Van Berg[109], R. Van de Water[78], P. Van Gemmeren[8], F. Varanini[106], G. Varner[48], G. Vasseur[117], K. Vaziri[43],
G. Velev[43], S. Ventura[106], A. Verdugo[17], T. Viant[147], T. V. Vieira[25], C. Vignoli[72], C. Vilela[127], B. Viren[15],
T. Vrba[36], T. Wachala[70], D. Wahl[142], M. Wallbank[118], N. Walsh[20], B. Wang[125], H. Wang[22], L. Wang[121],
T. Wang[41], T.K. Warburton[118], D. Warner[33], M. Wascko[59], D. Waters[77], T. B. Watson[131], A. Weber[116],
M. Weber[13], W. Wei[121], A. Weinstein[62], D. Wells[122], D. Wenman[142], M. Wetstein[62], A. White[131], L. Whitehead[49],
D. Whittington[61], M. Wilking[127], J. Willhite[43], P. Wilson[43], R. J. Wilson[33], L. Winslow[83], P. Wittich[35],
S. Wojcicki[126], H. H. Wong[19], K. Wood[120], E. Worcester[15], M. Worcester[15], S. Wu[147], T. Xin[62], C. Yanagisawa[127],
S. Yang[29], T. Yang[43], K. Yarritu[78], J. Ye[125], M. Yeh[15], N. Yershov[56], K. Yonehara[43], B. Yu[15], J. Yu[131],
J. Zalesak[3], L. Zambelli[7], B. Zamorano[128], L. Zang[118], A. Zani[108], A. Zani[16], G. Zavala[46], G. Zeller[43],
C. Zhang[15], C. Zhang[121], E. D. Zimmerman[32], M. Zito[117], and R. Zwaska[43]

[1] *ABC Federal University, Santo André –SP, 09210-580, Brazil*
[2] *APC-Paris, Batiment Condorcet; 10, rue Alice Domon et L'eonie Duquet; F-75205 Paris CEDEX 13, France*
[3] *Institute of Physics ASCR, v. v. i. , Na Slovance 2; 182 21 Praha 8, Czech Republic*
[4] *Univ. of Alabama (Tuscaloosa), Tuscaloosa, AL 35487-0324, USA*

[5] *Univ. Federal de Alfenas em Poços de Caldas, 11999, CEP 37715-900 Poços de Caldas-MG, Brazil*

[6] *Aligarh Muslim University, Department of Physics Aligarh-202002, India*

[7] *Antananarivo, Présidence de l'Université d'Antananarivo: BP 566, Antananarivo 101, Madagascar*

[8] *Argonne National Lab., Argonne, IL 60439, USA*

[9] *University of Athens, University Campus, Zografou GR 157 84 Greece, Greece*

[10] *Universidad del Atlantico, Km 7 antigua vi a Puerto Colombia, Barranquilla, Colombia*

[11] *Banaras Hindu University, Dept. of Physics, Varanasi UP 221005, India*

[12] *Bhabha Atmoic Research Center, Trombay, Mumbai, India*

[13] *Univ. of Bern, Lab for High Energy Physics; Sidlerstrasse 5; CH-3012 Bern, Switzerland*

[14] *Boston University, Boston, MA 02215, USA*

[15] *Brookhaven National Lab., Upton, NY 11973-5000, USA*

[16] *CERN, European Organization for Nuclear Research European Laboratory for Particle Physics; 1211 Gen'eve 23, Switzerland*

[17] *CIEMAT, División de Física de Partículas; Avenida Complutense 40; E-28040 Madrid, Spain*

[18] *CNR Pisa, 1, 56124 Pisa PI, Italy*

[19] *University of California at Berkeley, #7300; Berkeley, CA 94720-7300, USA*

[20] *Univ. of California (Davis), Davis, CA 95616, USA*

[21] *Univ. of California (Irvine), Irvine, CA 92697-4575, USA*

[22] *Univ. of California (Los Angeles), Los Angeles, CA 90095-1547, USA*

[23] *California Inst. of Tech., MC 356-48; Pasadena, CA 91125, USA*

[24] *Univ. of Cambridge, JJ Thomson Avenue, Cambridge CB3 0HE, UK*

[25] *Univ. de Campinas, Av. Sérgio Buarque de Holanda, 777 CEP 13083-859 Campinas-SP, Brazil*

[26] *Univ. di Catania, Via Santa Sofia, I-95123 Catania, Italy*

[27] *Institute of Particle and Nuclear Physics of the Faculty of Mathematics and Physics of the Charles University in Prague, V Holešovičkách 747/2; 180 00 Praha 8-Libeň, Czech Republic*

[28] *Univ. of Chicago, Chicago, IL 60637-1434, USA*

[29] *Univ. of Cincinnati, Cincinnati, OH 45221-0011, USA*

[30] *Cinvestav, 07360 Ciudad de México, D.F., Mexico*

[31] *Universidad de Colima, Facultad de Ciencias Bernal Diaz del Castillo 340 Colonia Villa San Sebastian Colima, Colima, Mexico*

[32] *Univ. of Colorado, Boulder, CO 80309, USA*

[33] *Colorado State University, Fort Collins, CO 80523, USA*

[34] *Columbia University, New York, NY 10027, USA*

[35] *Cornell Univ., Laboratory for Elementary-Particle Physics; Newman Lab; Ithaca, NY 14853–5001, USA*

[36] *Czech Technical University in Prague, Brehova 7; 115 19 Praha 1, Czech Republic*

[37] *Dakota State University, Madison, SD 57042, USA*

[38] *Univ. of Delhi, Department of Physics and Astrophysics, Delhi 110007, India*

[39] *Drexel University, Philadelphia, PA 19104, USA*

[40] *Duke University, Durham, NC 27706, USA*

[41] *Univ. of Durham, Institute for Particle Physics Phenomenology; Dept. of Physics, Ogden Centre for Fund. Physics; South Road; Durham DH1 3LE, UK*

[42] *Univ. Estadual de Feira de Santana, S/N;44036-900, Feira de Santana-BA, Brazil*

[43] *Fermi National Accelerator Lab, Batavia, IL 60510-0500, USA*

[44] *Univ. Federal de Goias, Goiania, GO, Brazil*

[45] *Gran Sasso Science Institute, 7, L'Aquila, Italy*

[46] *Universidad de Guanajuato, Gto., C.P.37000, Mexico*

[47] *Harish-Chandra Research Institute, Jhunsi, Allahabad 211 019, India*

[48] *Univ. of Hawaii, Honolulu, HI 96822-2219, USA*

[49] *Univ. of Houston, Houston, TX 77204, USA*

[50] *Huddersfield, Huddersfield, West Yorkshire HD1 3DH, UK*

[51] *Horia Hulubei National Institute of Physiscs and Nuclear Engineering, Horia Hulubei National Institute of Physiscs and Nuclear Engineering, Bucharest-Magurele, Romania*

[52] *University of Hyderabad, Gachibowli, Hyderabad - 500 046, India*

[53] *Institut de Física d'Altes Energies (IFAE); Campus Universitat Autonoma de Barcelona , E-08193 Cerdanyola del Vallès (Barcelona), Spain*

[54] *Illinois Institute of Technology , Room 182LS; Chicago, IL 60616, USA*

[55] *IIT Hyderabad, Kandi, Hyderabad - 502205, India*

[56] *Institute for Nuclear Research of the Russian Academy of Sciences (INR RAS), 60th October Anniversary Prosp. 7a Moscow, Russia 117312, Russia*

[57] *Institute for Research in Fundamental Sciences (IPM), Tehran, Iran. Postal code: 19538-33511 PO Box: 19395-5531, Iran*

[58] *Idaho State University, Department of Physics 921 South 8th Ave. Stop 8106 Pocatello, ID 83209-8106, USA*

[59] *Imperial College of Science Tech. & Medicine, Blackett Lab.; Prince Consort Road; London SW7 2BZ, UK*

[60] *Indian Institute of Technology Guwahati, Guwahati, 781 039, India*

[61] *Indiana University, Bloomington, IN 47405-7105, USA*

[62] *Iowa State University, Ames, IA 50011, USA*

[63] *Institut de Physique Nucleaire de Lyon (IPNL), Rue E. Fermi 4 69622 Villeurbanne, France*

[64] *University of Jammu, Physics Department, JAMMU-180006, India*

[65] *Kansas State University, Manhattan, KS 66506, USA*

[66] *Kavli IPMU, Univ. of Tokyo, Kashiwa Shi, Chiba 277-8568, Japan*

[67] *KEK, High Energy Accelerator Research Organization 1–1 Oho, Tsukuba-shi; Ibaraki-ken 305-0801, Japan*

[68] *KYIV National University, Department of Nuclear Physics, 64, 01601 Kyiv, Ukraine*

[69] *Koneru Lakshmaiah, , India*

[70] *Krakow, , Poland*

[71] *Lab. d'Annecy-le-Vieux de Phys. des Particules, BP 110; F-74941 Annecy-le-Vieux CEDEX, France*

[72] *Laboratori Nazionali del Gran Sasso, I-67010 Assergi, AQ, Italy*

[73] *Lancaster University, Bailrigg, Lancaster LA1 4YB, UK*

[74] *Lawrence Berkeley National Lab., Berkeley, CA 94720-8153, USA*

[75] *Université de Liège, Bat B5, Sart Tilman B-4000, Belgium*

[76] *Univ. of Liverpool, L69 7ZE, Liverpool, UK*

[77] *University College London, London, WC1E 6BT, UK*

[78] *Los Alamos National Laboratory, Los Alamos, NM 87545, USA*

[79] *Louisiana State University, Baton Rouge, LA 70803-4001, USA*

[80] *University of Lucknow, Department of Physics Lucknow 226007 Uttar Pradesh, India*

[81] *Univ. of Manchester, Oxford Road, Manchester M13 9PL, UK*

[82] *Univ. of Maryland, College Park, MD 20742-4111, USA*

[83] *Massachusetts Institute of Technology, Cambridge, MA 02139-4307, USA*

[84] *Max Planck MPP, Max-Planck-Institut fuer Physik (Werner-Heisenberg-Institut) Foehringer Ring 6 80805 Muenchen, Germany*

[85] *Univ. of Puerto Rico, Box 9016; Mayaguez, PR 00681-9000, USA*

[86] *University of Michigan, 450 Church Street Ann Arbor, Michigan 48109-1040, USA*

[87] *Michigan State University, East Lansing, MI 48824, USA*

[88] *Univ. di Milano, INFN Sezione di Milano, I-20133 Milano, Italy*

[89] *INFN Sezione di Milano Bicocca, Piazza della Scienza 3, 20126 Milano, Italy*

[90] *Univ. of Minnesota, Minneapolis, MN 55455, USA*

[91] *Univ. of Minnesota (Duluth), Duluth, MN 55812, USA*

[92] *NIKHEF, Science Park, Amsterdam, Netherlands*

[93] *Istituto Nazionale di Fisica Nucleare - Sezione di Napoli, Complesso Universitario di Monte S. Angelo, I-80126 Napoli, Italy*

[94] *National Centre for Nuclear Research, A. Soltana 7, 05 400 Otwock, Poland*

[95] *Jawaharlal Nehru University, New Delhi 110067, INDIA, India*

[96] *Univ. of New Mexico, MSC07 4220; Albuquerque, NM 87131, USA*

[97] *Northwestern University, Evanston, Il 60208, USA*

[98] *Univ. of Notre Dame, Notre Dame, IN 46556-5670, USA*

[99] *Observatorio Nacional, , Brazil*

[100] *Ohio State Univ., Dept. of Physics; 191 W. Woodruff Ave.; Columbus, OH 43210, USA*

[101] *Oregon State University, Dept. of Physics; 301 Weniger Hall; Corvallis, OR 97331–6507, USA*

[102] *Univ. of Oxford, Oxford, OX1 3RH, UK*

[103] *Pacific Northwest National Lab, , USA*

[104] *Pennsylvania State University, PMB 264; University Park, PA 16802-6300, USA*

[105] *PUCP, Av. Universitaria 1801, Lima, Peru*

[106] *Univ. of Padova, Dip. Fisica e Astronomia G. Galilei and INFN Sezione di Padova, I-35131 Padova, Italy*

[107] *Panjab University, Chandigarh, 160014 U.T., India*

[108] *Univ. of Pavia, INFN Sezione di Pavia, I-27100 Pavia, Italy*

[109] *Univ. of Pennsylvania, Philadelphia, PA 19104-6396, USA*

[110] *Univ. di Pisa, Theor. Division; Largo B. Pontecorvo 3, Ed. B-C; I-56127 Pisa, Italy*

[111] *Univ. of Pittsburgh, Pittsburgh, PA 15260, USA*

[112] *Princeton University, Princeton, New Jersey 08544-0708, USA*

[113] *Punjab Agri. University, Centre for High Energy Physics; Lahore - 54590, India*

[114] *Univ. of Rochester, Rochester, NY 14627-0171, USA*

[115] *SLAC National Accelleratory Laboratory, Menlo Park, CA 94025, USA*

[116] *STFC Rutherford Appleton Laboratory, Harwell Oxford, Didcot OX11 0QX, UK*

[117] *CEA/Saclay, IPhT; Inst. de Physique Theorique; Orme des Merisiers, Point Courrier 136; F-91191 Gif-sur-Yvette CEDE, France*

[118] *Univ. of Sheffield, Sheffield, S3 7RH, UK*

[119] *Univ. of Sofia, Atomic Physics Dept., Faculty of Physics; 5 James Bourchier Blvd.; BG-1164 Sofia, Bulgaria*

[120] *Univ. of South Carolina, Columbia, SC 29208, USA*

[121] *Univ. of South Dakota, Vermillion, SD 57069, USA*

[122] *South Dakota School of Mines and Technology, Rapid City, SD 57701, USA*

[123] *South Dakota Science And Technology Authority, Lead, SD 57754, USA*

[124] *South Dakota State University, Brookings, SD 57007, USA*

[125] *Southern Methodist University, Dallas, TX 75275, USA*

[126] *Stanford University, Varian Physics Bldg.; 382 Via Pueblo Mall; Stanford, CA 94305–4060, USA*

[127] *Stony Brook University, Nucleon Decay and Neutrino Physics Group Department of Physics and Astronomy Stony Brook University Stony Brook, NY 11794-3800, USA*

[128] *Univ. of Sussex, Brighton, BN1 9RH, UK*

[129] *Syracuse University, Syracuse, NY 13244-1130, USA*

[130] *Univ. of Tennessee at Knoxville, Knoxville, TN, USA*

[131] *Univ. of Texas (Arlington), Arlington, TX 76019, USA*

[132] *Univ. of Texas (Austin), Austin, TX 78712-0264, USA*

[133] *TUBITAK Space Technologies Research Institute , TR-06800, Ankara, Turkey*

[134] *Tufts University, Medford, MA 02155, USA*

[135] *Variable Energy Cyclotron Centr, 1/AF, Bidhannagar Kolkata - 700 064 West Bengal, India*

[136] *Instituto de Fisica Corpuscular, C/Catedratico Jose Beltran, 2 E-46980 Paterna (Valencia), Spain*

[137] *Virginia Tech., Blacksburg, VA 24061-0435, USA*

[138] *Univ. of Warsaw, Faculty of Physics ul. Pasteura 5 02-093 Warsaw, Poland*

[139] *Univ. of Warwick, Coventry CV4 7AL, UK*

[140] *Wichita State University, Physics Division 1845 Fairmount St. Wichita, KS 67260-0032, USA*

[141] *College of William and Mary, Williamsburg, VA 23187-8795, USA*

[142] *Univ. of Wisconsin, Madison, WI 53706, USA*

[143] *Wroclaw University, Plac Maxa Borna 9, 50-204 Wroclaw, Poland*

[144] *Yale University, New Haven, CT 06520, USA*

[145] *Yerevan Institute for Theoretical Physics and Modeling, Halabian Str. 34; Yerevan 0036, Armenia*

[146] *York, Physics and Astronomy Dept.; 4700 Keele St.; Toronto M3J 1P3, Canada*

[147] *ETH Zurich, HPK F 23 Schafmattstr. 20 8093 Zürich, Switzerland*

2015-11-25



\renewcommand{\familydefault}{\sfdefault}
\renewcommand{\thepage}{\roman{page}}
\setcounter{page}{0}

\pagestyle{plain} 

%\clearpage

%\setcounter{tocdepth}{3}
\setcounter{tocdepth}{2}
\textsf{\tableofcontents}
%\clearpage

\textsf{\listoffigures}
%\clearpage

\textsf{\listoftables}
%\clearpage

%For acronym list to appear just after TOC, TOF, TOT
\printnomenclature
%\clearpage

\iffinal\else
\textsf{\listoftodos}
\clearpage
\fi

\renewcommand{\thepage}{\arabic{page}}
\setcounter{page}{1}

\pagestyle{fancy}

% Set how header/footers look
\renewcommand{\chaptermark}[1]{%
\markboth{Chapter \thechapter:\ #1}{}}
\fancyhead{}
\fancyhead[RO,LE]{\textsf{\footnotesize \thechapter--\thepage}}
\fancyhead[LO,RE]{\textsf{\footnotesize \leftmark}}

\fancyfoot{}
\fancyfoot[RO]{\textsf{\footnotesize LBNF/DUNE Conceptual Design Report}}
\fancyfoot[LO]{\textsf{\footnotesize \thedoctitle}}
\fancypagestyle{plain}{}

\renewcommand{\headrule}{\vspace{-4mm}\color[gray]{0.5}{\rule{\headwidth}{0.5pt}}}

%% file: common/acronyms-shared-vol1-2-4.tex
%%%%%%%%%%%%%%%%%%%%%%%%% COMMON list for acronyms below %%%%%%%%%%%%%%%
\nomenclature{$\mathcal{O}(n)$}{of order $n$}
\nomenclature{3D}{3 dimensional (also 1D, 2D, etc.)} % not phys
\nomenclature{CDR}{Conceptual Design Report}
\nomenclature{CF}{Conventional Facilities}
\nomenclature{CP}{product of charge and parity transformations}
\nomenclature{CPT}{product of charge, parity and time-reversal transformations}
\nomenclature{CPV}{violation of charge and parity symmetry}
\nomenclature{DAQ}{data acquisition}
\nomenclature{DOE}{U.S. Department of Energy}
\nomenclature{DUNE}{Deep Underground Neutrino Experiment}
\nomenclature{ESH}{Environment, Safety and Health}
\nomenclature{eV}{electron volt, unit of energy (also keV, MeV, GeV, etc.)}
\nomenclature{FD}{far detector}
\nomenclature{FGT}{Fine-Grained Tracker}
\nomenclature{FSCF}{far site conventional facilities}
\nomenclature{NSCF}{near site conventional facilities}
\nomenclature{GUT}{grand unified theory}
\nomenclature{\ktyr}{exposure (without beam), expressed in kilotonnes times years}
\nomenclature{\ktMWyr}{exposure, expressed in kilotonnes $\times$ megawatts $\times$ years, based on 56\% beam uptime and efficiency} 
\nomenclature{L}{level, indicates depth in feet underground at the far site, e.g., 4850L}
\nomenclature{LAr}{liquid argon}
\nomenclature{LArTPC}{liquid argon time-projection chamber}
\nomenclature{LBL}{long-baseline (physics)}
\nomenclature{LBNF}{Long-Baseline Neutrino Facility}
\nomenclature{MH}{mass hierarchy}
\nomenclature{MI}{Main Injector (at Fermilab)}
\nomenclature{ND}{near neutrino detector}
\nomenclature{NDS}{Near Detector Systems; refers to the collection of detector systems at the near site }
\nomenclature{near detector}{except in Volume 4 Chapter 7, \textit{near detector} refers to the \textit{neutrino} detector system in the NDS}
\nomenclature{POT}{protons on target}
\nomenclature{QA}{quality assurance}
\nomenclature{SM}{Standard Model of particle physics}
\nomenclature{t}{metric ton, written \textit{tonne} (also kt)}
\nomenclature{tonne}{metric ton}
\nomenclature{TPC}{time-projection chamber (not used as `total project cost' in the CDR)}

%% file: common/acronyms-shared-vol1-4.tex
%%%%%%%%%%%%% PROJECT AND DETECTORS VOLUME list for acronyms below %%%%%%%%%%%%

\nomenclature{APA}{anode plane assembly} 
\nomenclature{BLM}{(in Volume 4) beamline measurement (system); (in Volume 3) beam loss monitor}
\nomenclature{CPA}{cathode plane assembly}
\nomenclature{ECAL}{electromagnetic calorimeter}
\nomenclature{GAr}{gaseous argon}
\nomenclature{HV}{high voltage}

%% file: common/acronyms-shared-vol2-4.tex
%%%%%%%%%%%%% PHYSICS AND DETECTORS VOLUME list for acronyms below %%%%%%%%%%%%
\nomenclature{CC}{charged current (interaction)}
\nomenclature{DIS}{deep inelastic scattering}
\nomenclature{FSI}{final-state interactions}
\nomenclature{GEANT4}{GEometry ANd Tracking, a platform for the simulation of the passage of particles through matter using Monte Carlo methods} 
\nomenclature{GENIE}{Generates Events for Neutrino Interaction Experiments (an object-oriented neutrino Monte Carlo generator)} 
\nomenclature{MC}{Monte Carlo (detector simulation methods)}
\nomenclature{QE}{quasi-elastic (interaction)}

%% file: common/acronyms-vol4.tex
%%%%%%%%%%%%%%%%%%%%%%%%% DETECTORS VOLUME list for acronyms below %%%%%%%%%%%%%%%

\nomenclature{CE}{Cold Electronics}
\nomenclature{COB}{cluster on-board (motherboards)} %?
\nomenclature{CRP}{Charge-Readout Planes }
\nomenclature{DRAM}{dynamic random access memory}
\nomenclature{FE}{front end (electronics)}
%\nomenclature{Fermilab (also FNAL)}{Fermi National Accelerator Laboratory (in Batavia, IL, the Near Site)}
%\nomenclature{FNAL}{see Fermilab}
\nomenclature{FPGA}{field programmable gate array} %?
\nomenclature{FGT}{Fine-Grained Tracker}
\nomenclature{FS}{full stream (data volumes)} %?
\nomenclature{LEM}{Large Electron Multiplier}
\nomenclature{LNG}{liquefied natural gas}
%\nomenclature{LNGS}{Laboratori Nazionali (National Laboratory) del Gran Sasso (in L'Aquila, Italy)}
%\nomenclature{MaVaNs}{mass varying neutrinos}
\nomenclature{MIP}{minimum ionizing particle}
\nomenclature{MTS}{Materials Test Stand}
\nomenclature{MuID}{muon identifier (detector)}
%\nomenclature{OPERA}{Oscillation Project with Emulsion-Racking Apparatus (experiment at LNGS)}
\nomenclature{NND}{(used only in Volume 4 Chapter 7) near neutrino detector, same as ND}
%\nomenclature{OD}{outer diameter}
\nomenclature{PD}{photon detection (system)}
\nomenclature{PMT}{photomultiplier tube}
\nomenclature{PPM/PPB/PPT}{parts per million/billion/trillion}
\nomenclature{RCE}{reconfigurable computing element}
\nomenclature{RIO}{reconfigurable input output}
\nomenclature{RPC}{resistive plate chamber}
\nomenclature{SiPM}{silicon photomultiplier}
\nomenclature{S/N}{signal-to-noise (ratio)}
\nomenclature{SSP}{SiPM signal processor}
\nomenclature{SBN}{Short-Baseline Neutrino program (at Fermilab)}
\nomenclature{STT}{straw tube tracker}
%\nomenclature{SURF (also Sanford Lab)}{Sanford Underground Research Facility (in Lead, SD, the Far Site)}
\nomenclature{TR}{transition radiation}
%\nomenclature{W}{Watt (also mW, kW, MW) }
%\nomenclature{WA105}{Single-Phase LArTPC and the Long Baseline Neutrino Observatory Demonstration}
\nomenclature{WLS}{wavelength shifting}
\nomenclature{ZS}{zero suppression}

%% file: volume-detectors/chapter-overview.tex
\chapter{Overview}
\label{ch:detectors-overview}

\input{common/intro}

%%%%%%%%%%%%%%%%%%%%%%%%%%%%%%%
\subsection{About this Volume}

The first part of \voldune{} of the CDR describes the strategies for
implementing the near and far detectors
(Chapter~\ref{ch:detectors-strategy}) and outlines the DUNE management
structure (Chapter~\ref{ch:detectors-pm}). The next part describes the
technical designs: the reference and alternative designs for the far
detector and the synergies between them
(Chapters~\ref{ch:detectors-fd-ref},~\ref{ch:detectors-fd-alt}
and~\ref{ch:detectors-synergy}), and the near detector systems design
(Chapter~\ref{ch:detectors-nd-ref}).  Following this,
Chapter~\ref{ch:detectors-sc} describes the designs for the computing
infrastructure and physics software, and
Chapter~\ref{ch:detectors-proto} provides an overview of the ongoing
and planned prototyping effort.  The software and computing efforts,
as well as some of the prototyping activities are
off-project. Chapter~\ref{ch:detectors-summary} summarizes and
concludes the volume.
 
%%%%%%%%%%%%%%%%%%%%%%%%%%%%%%%%%%%%%%%%%%%%%%%%%%%%%%%%%%%%%%
\section{Introduction to the DUNE Detectors}
\label{sec:intro-dune-det}

%%%%%%%%%%%%%%%%%%%%%%%%%%%%%%%
\subsection{Far Detector}
\label{sec:intro-dune-far-det}

The proposed far detector (FD) will be located deep underground at the
SURF 4850L with a fiducial mass of 40~kt. It consists of four
cryostats instrumented with Liquid Argon Time Projection Chambers
(LArTPCs).  It is assumed that all four detector modules will be
similar but not necessarily identical, allowing for evolution of the
LArTPC technology to be implemented.

LArTPC technology provides excellent tracking and calorimetry
performance. It is ideal for massive neutrino detectors that require
high signal efficiency, effective background discrimination,
capability to identify and precisely measure neutrino events over a
wide range of energies and high resolution reconstruction of kinematic
properties. The full imaging of events in the DUNE detector will allow
study of neutrino interactions and other rare events with
unprecedented detail. The detector's huge mass will result in data
sets large enough to enable precision studies and the search for CP
violation.

The mature LArTPC technology, pioneered by ICARUS, is the result of
several decades of worldwide R\&D.  Nonetheless, the size of a single
\ktadj{10} DUNE detector module represents an extrapolation by over
one order of magnitude relative to the ICARUS~T600, which is the
largest detector of this kind operated to date. To address this
challenge, DUNE is developing both a reference and an alternative
design (see Figure~\ref{fig:FarDet-overview-SPDP}), and is engaged in
a comprehensive prototyping effort.  A list of synergies between the
reference and alternative designs has been identified and is
summarized in Chapter~\ref{ch:detectors-synergy}. Common solutions for
DAQ, electronics, HV feedthroughs, and so on, will be pursued and
implemented, independent of the details of the TPC design choice. The
development of the two detector module designs is a considerable
advantage, and it is made possible by the convergence of previously
separate international neutrino efforts into the DUNE collaboration.
\begin{cdrfigure}[3D models of the DUNE far detector designs]{FarDet-overview-SPDP}
{3D models of two 10-kt detectors using the single-phase reference design (left) 
and the dual-phase alternate design (right) for the DUNE far detector to be 
located at 4850L.}
\centering
\begin{minipage}[b]{1.0\textwidth}
\begin{center}
\includegraphics[width=.5\textwidth]{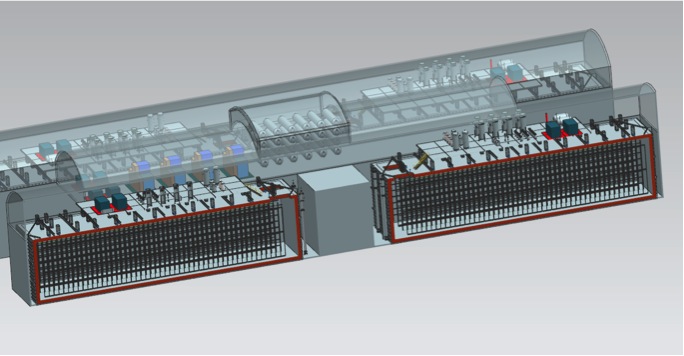}
\includegraphics[width=0.46\textwidth]{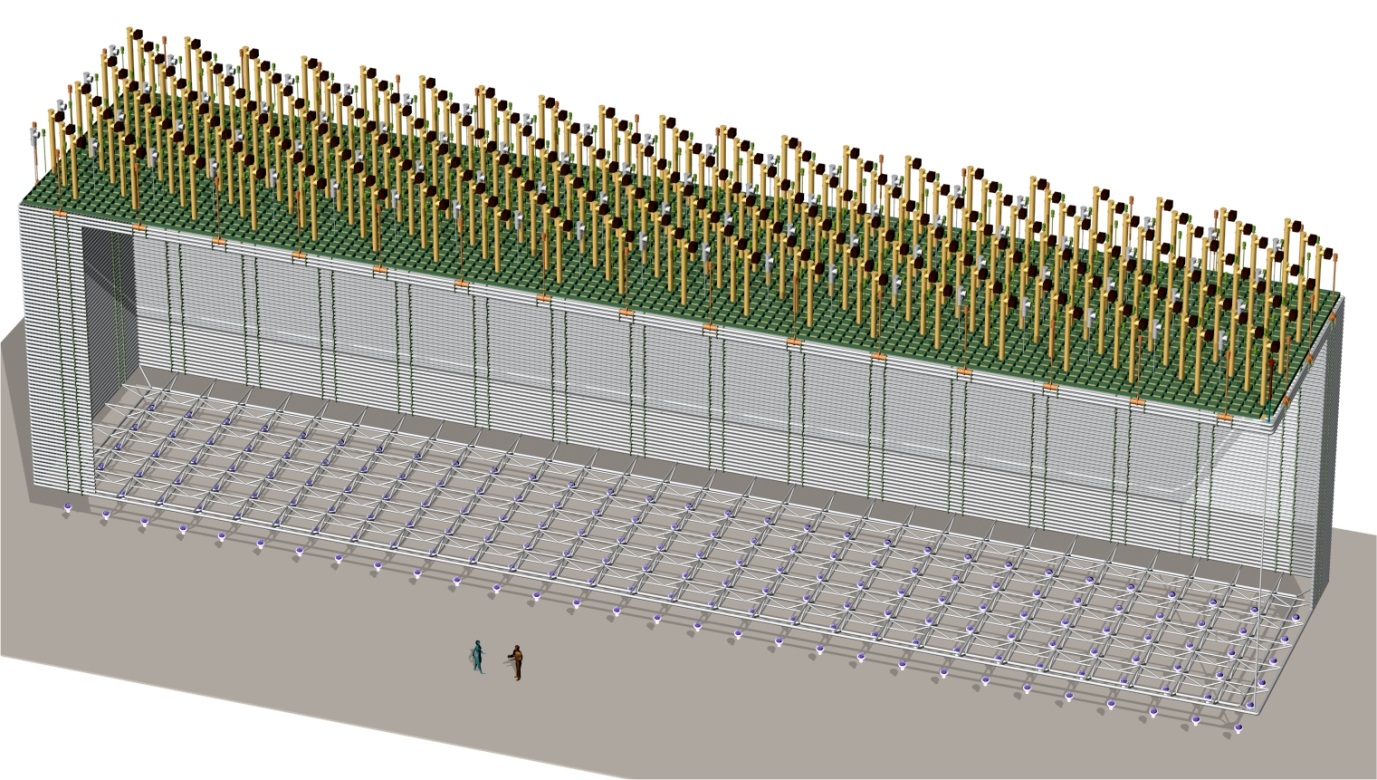}
\end{center}
\end{minipage}
\end{cdrfigure}

Interactions in liquid argon (LAr) produce ionization charge and
scintillation light.  The electrons drift in a constant electric field
away from the cathode plane towards the segmented anode plane.  The
prompt scintillation light is observed by photodetectors that provide
the absolute time of the event.  The reference design, described in
Chapter~\ref{ch:detectors-fd-ref}, adopts a single-phase readout, in
which the readout anode is composed of wire planes in the LAr volume.
The alternate design, discussed in Chapter~\ref{ch:detectors-fd-alt},
considers the dual-phase approach, where ionization charge is
extracted, amplified and detected in gaseous argon above the liquid
surface.  The dual-phase design allows a finer readout pitch (3~mm), a
lower detection-energy threshold, and better pattern reconstruction of
the events.  Both the reference and alternate designs include systems
to collect the scintillation light.

A comprehensive prototyping strategy for both designs is being
actively pursued, as described in Chapter~\ref{ch:detectors-proto}.
The reference design, closer to the original ICARUS design, is
currently being validated in the 35-t prototype LAr detector at
Fermilab (see Section~\ref{sec:proto-35t}).  The novel alternative
design approach has been proven on several small-scale prototypes, and
a 20-t dual-phase prototype is being constructed at CERN, intended for
operation in 2016.  Full-scale engineering prototypes will be
assembled and commissioned at the CERN neutrino platform; 
%~\footnote{See  CERN Bulletin article at  \href{http://cds.cern.ch/journal/CERNBulletin/2014/51/News\%20Articles/1975980?ln=en}{http://cds.cern.ch/journal/CERNBulletin/2014/51/News\%20Articles/1975980?ln=en}.};
they are expected to provide the ultimate validation of the engineered
solutions for both far detector designs around the year 2018.

A test-beam data campaign will be executed at the CERN Neutrino Platform in the following years to
collect a large sample of charged particle interactions to study the
detector response with high precision.

The deployment of the four 10-kt modules at SURF will take several
years and be guided by principles detailed in
Chapter~\ref{ch:detectors-strategy}. According to this strategy, DUNE
adopts the lowest-risk design that satisfies the physics and detector
requirements and allows installation of the first 10-kt detector
module as early as possible.  Accordingly, the first 10-kt module will
implement the reference design.  

A clear and transparent decision process will be adopted for
determining the design of the second and subsequent modules.  The
decision will be based on physics performance, technical and schedule
risks, costs and funding opportunities.  Besides taking advantage of
technological developments, a flexible approach to the far detector
design acknowledges the diversity of DUNE and offers the potential to
attract additional interest and resources into the collaboration. A
staged approach provides access to an early science program while
allowing for new developments to be implemented over the relatively
long installation period of the experiment.

%%%%%%%%%%%%%%%%%%%%%%%%%%%%%%%
\subsection{Near Detector Systems}
\label{sec:intro-dune-near-det}

DUNE will install a near neutrino detector (NND) $\sim$0.5~km
downstream of the target and a Beamline Measurement System (BLM)
$\sim$300~m upstream of the NND. These are collectively called the
Near Detector Systems (NDS).  The NDS will allow DUNE to reduce
systematic errors to match the high-statistics phase precision sensitivity
for the long-baseline neutrino oscillation studies.  The primary role
of the neutrino detector is to measure the spectrum and flavor
composition of the beam to high precision. This detector will be
magnetized so that it can charge-discriminate electrons and muons
produced in the neutrino charged current interactions; it will
therefore be capable of making separate measurements of the neutrino
and antineutrino fluxes.
%
%In order to reach the ultimate
%sensitivity for the long-baseline neutrino oscillation studies, the neutrino detector will
%measure the spectrum and flavor composition of the (neutrino beam to high precision.  
%Separate measurements of the fluxes of neutrinos and antineutrinos requires a
%magnetized neutrino detector to charge-discriminate electrons and
%muons produced in the neutrino charged current interactions.  This is
%the primary role of the DUNE near detector; % system; 
%however, 

In addition, exposure to the intense neutrino flux provides the
opportunity to collect neutrino interaction data sets of unprecedented
size, enabling an extended science program.  The near detector
therefore provides an opportunity for a wealth of fundamental neutrino
interaction measurements which are an important part of the ancillary
scientific goals of the DUNE collaboration.

The reference design for the neutrino near detector (NND) design is
the NOMAD-inspired fine-grained tracker (FGT) and is described in
Chapter~\ref{ch:detectors-nd-ref}. The NND subsystems include a
central straw-tube tracker and an electromagnetic calorimeter embedded
in a 0.4-T dipole field. The magnet yoke steel will be instrumented
with muon identifiers.

The Beamline Measurement System (BLM), designed to measure the muon
flux from hadron decay, is located in the region of the beam absorber
at the downstream end of the decay region. It is intended to monitor
the beam profile on a spill-by-spill basis and will operate for the
life of the experiment.

%% file: common/intro.tex
% Intro shared by all subsections

\section{An International Physics Program}

The global neutrino physics community is developing a multi-decade
physics program to measure unknown parameters of the Standard Model of
particle physics and search for new phenomena.  The program will be carried out as an international,
leading-edge, dual-site experiment for neutrino science and proton decay studies, which 
is known as the Deep Underground Neutrino Experiment (DUNE).
The detectors for this experiment will be designed, built, commissioned and operated by the international DUNE Collaboration. The facility required to support this experiment, the Long-Baseline Neutrino Facility (LBNF), is hosted by Fermilab and its design and construction is organized as a DOE/Fermilab project incorporating international partners. Together LBNF and DUNE will comprise the world's highest-intensity neutrino beam at Fermilab, in Batavia, IL, a high-precision near detector on the Fermilab site, a massive liquid argon time-projection chamber (LArTPC) far detector installed deep underground at the Sanford Underground Research Facility (SURF) \SI{1300}{\km} away in Lead, SD, and all of the conventional and technical facilities necessary to support the beamline and detector systems.

The strategy for executing the experimental program presented in this Conceptual 
Design Report (CDR) has been developed to meet the requirements 
set out in the P5 report~\cite{p5report} and takes into account the recommendations of the European Strategy for Particle Physics~\cite{ESPP-2012}. It adopts a model where U.S. and international funding agencies 
share costs on the DUNE detectors, and CERN and other participants provide in-kind contributions 
to the supporting infrastructure of LBNF. LBNF and DUNE will be tightly coordinated as DUNE collaborators 
design the detectors and infrastructure that will carry out the scientific program.
  
The scope of LBNF is
\begin{itemize}
\item an intense neutrino beam aimed at the far site,
\item conventional facilities at both the near and far sites, and
\item cryogenics infrastructure to support the DUNE detector at the far site.
\end{itemize}

The DUNE detectors include
\begin{itemize}
\item a high-performance neutrino detector and beamline  
measurement system
located a few hundred meters downstream of the neutrino source, and
\item a massive liquid argon time-projection chamber (LArTPC) neutrino detector located deep underground at the far site.
\end{itemize}

With the facilities provided by LBNF and the detectors provided by
DUNE, the DUNE Collaboration proposes to mount a focused attack on the
puzzle of neutrinos with broad sensitivity to neutrino oscillation
parameters in a single experiment.  The focus of the scientific
program is the determination of the neutrino mass hierarchy and the
explicit demonstration of leptonic CP violation, if it exists, by
precisely measuring differences between the oscillations of muon-type
neutrinos and antineutrinos into electron-type neutrinos and
antineutrinos, respectively. Siting the far detector deep underground
will provide exciting additional research opportunities in nucleon
decay, studies utilizing atmospheric neutrinos, and neutrino
astrophysics, including measurements of neutrinos from a core-collapse
supernova should such an event occur in our galaxy during the
experiment's lifetime.

%%%%%%%%%%%%%%%%%%%%%%%%%%%%%%%%%%%%%%%%%%%%%%%%%%%%%%%%%%%%%%%
\section{The LBNF/DUNE Conceptual Design Report Volumes}

%%%%%%%%%%%%%%%%%%%%%%%%%%%%%%%%%%%
\subsection{A Roadmap of the CDR}

The LBNF/DUNE CDR describes the proposed physics program and 
technical designs at the conceptual design stage.  At this stage, the design is
still undergoing development and the CDR therefore presents a \textit{reference design} 
for each element as well as \textit{alternative designs} that are under consideration.

The CDR is composed of four volumes and is supplemented by several annexes that 
provide details on the physics program and technical designs. The volumes are as follows

\begin{itemize}
\item \volintro{}\cite{cdr-vol-1} provides an executive summary of and strategy for the experimental 
program and introduces the CDR.
\item \volphys{}\cite{cdr-vol-2} outlines the scientific objectives and describes the physics studies that 
the DUNE Collaboration will undertake to address them.
\item \vollbnf{}\cite{cdr-vol-3} describes the LBNF Project, which includes design and construction of the 
beamline at Fermilab, the conventional facilities at both Fermilab and SURF, and the cryostat
 and cryogenics infrastructure required for the DUNE far detector.
\item \voldune{} describes the DUNE Project, which includes the design, construction and 
commissioning of the near and far detectors. 
\end{itemize}

More detailed information for each of these volumes is provided in a set of annexes listed on LBNF and DUNE's shared  \href{https://web.fnal.gov/project/LBNF/SitePages/Proposals%20and%20Design%20Reports.aspx}{\textit{Proposals and Design Reports} page}. 

%%%%%%%%%%%%%%%%%%%%%%%%%%%%%%%%%%%
%\subsection{About this Volume}  <----- follows in overview chapter file of indiv volume

%% file: volume-detectors/chapter-strategy.tex
\chapter{Implementation Strategy}
\label{ch:detectors-strategy}

\section{Overview}

Recommendation 12 of the Report of the Particle Physics Prioritization
Panel (P5) states that for a Long-Baseline Neutrino Oscillation
Experiment to proceed ``The minimum requirements to proceed are the
identified capability to reach an exposure of
120~\ktMWyr{} by the 2035 timeframe, the far detector
situated underground with cavern space for expansion to at least \ktadj{40}
LAr fiducial volume, and \MWadj{1.2} beam power upgradable to
multi-megawatt power. The experiment should have the demonstrated
capability to search for supernova bursts and for proton decay,
providing a significant improvement in discovery sensitivity over
current searches for the proton lifetime.''  The strategy presented
here meets these criteria.  The P5 recommendation is in line with the CERN
European Strategy for Particle Physics (ESPP) of 2013, which
classified the long-baseline neutrino program as one of the four
scientific objectives with required international infrastructure.

\section{Strategy for Implementing the DUNE Far Detector}
\label{sec:detectors-strategy-FD}
The LBNF Project will provide four cryostats at the 4850L of the Sanford
Underground Research Facility (SURF) in which the DUNE Collaboration
will deploy four \ktadj{10} (fiducial) mass far detector LArTPCs. 
DUNE contemplates two options for the
read out of the ionization signals: single-phase readout, where the
ionization is detected using wire planes in the liquid argon volume;
and dual-phase readout, where the ionization signals are amplified and
detected in gaseous argon above the liquid surface.  An active
development program for both technologies is being pursued in the
context of the CERN neutrino platform, as well as the
the Fermilab SBN program.

The viability of the LArTPC technology has been proven by the ICARUS
experiment with single-phase wire plane LArTPC readout, where data was
successfully accumulated over a period of three years.  An
extrapolation of the observed performance and implementation of
improvements in the design (e.g., cold electronics) will allow the
single-phase approach (see
Chapter~\ref{ch:detectors-fd-ref}) to meet DUNE requirements, and
is hence adopted as the \textit{reference design}. 

The reference design is already relatively advanced for  the 
conceptual design
stage. Modifications of the reference design will
be approved by the DUNE technical board. A preliminary design
review will take place as early as possible, utilizing the experience
from the DUNE 35-t prototype; the design review will define the
baseline design that will form the basis of the TDR (CD-2).  Once
defined, changes to the baseline will fall under a formal
change-control process. An engineering prototype consisting of
six full-sized drift cells will be validated at the CERN neutrino
platform.  This engineering
prototype at CERN is a central part of the risk mitigation strategy
for the first \ktadj{10} FD module. Following experience at the CERN
neutrino platform, the DUNE technical coordinator will organize a
final design review. The CERN single-phase prototype provides the opportunity for
production sites to validate manufacturing procedures ahead of
large-scale production for the far detector. Three major operational
milestones are defined for this single-phase prototype: 1) engineering
validation -- successful cool-down; 2) operational validation --
successful TPC readout with cosmic-ray muons; and 3) physics
validation with test beam data. Reaching milestone 2 will allow the
retirement of a number of technical risks for the construction of the
first \ktadj{10} detector module.

In parallel with preparation for construction of the first \ktadj{10}
detector module, the DUNE Collaboration recognizes the potential of
the dual-phase technology and strongly endorses the already
approved WA105 experiment
 at the CERN neutrino platform, which
includes the operation of the 20-ton prototype and the
6$\times$6$\times$6~m$^3$ WA105 demonstrator. 
Many DUNE collaborators are participants in the WA105
experiment. A concept for the dual-phase implementation of a far detector module
is presented in detail as an \textit{alternative design} in
Chapter~\ref{ch:detectors-fd-alt}. This alternative design, if
demonstrated, could form the basis of the second or subsequent \ktadj{10}
far detector modules, to achieve improved detector performances in a
cost-effective way.

The DUNE program at the CERN neutrino platform will be coordinated by
a single manager. Common technical solutions will be adopted
wherever possible.  The charged-particle test-beam data will provide
essential calibration samples for both technologies and will enable a
direct comparison of the relative physics benefits.

For the purposes of cost and schedule, the reference design for the
first far detector module is adopted as the reference design for the subsequent
three modules. However, the experience with the first \ktadj{10}
module and the development activities at the CERN platform are likely
to lead to the evolution of the TPC technology, both in terms of
refinements to single-phase design and the validation of the operation
of the dual-phase design.  The technology choice for the second and
subsequent LArTPCs will be based on risk, cost (including the
potential benefits of additional non-DOE funding) and physics
performance (as established in the CERN charged-particle test beam).

As already stated, this strategy allows flexibility with respect to international
contributions and provides the possibility of attracting interest and
resources from a broader community with space for flexibility to
respond to the funding constraints from different sources. 

\section{Strategy for Implementing the DUNE Near Detector Systems}

%The LBNF project will provide the civil (conventional) facilities for the DUNE near
%detector systems (muon monitors and near neutrino detector). 

%The
%primary scientific motivation for the DUNE near detector system is to
%constrain the beam spectrum for the long-baseline neutrino oscillation
%studies. It also provides large data samples for precision studies of
%neutrino-argon interactions. The near detector, which is exposed to an
%intense flux of neutrinos, also provides an opportunity for a wealth
%of fundamental neutrino interaction measurements, which are an
%important part of the secondary scientific goals of the DUNE
%collaboration. 

Within the former LBNE collaboration the neutrino near
detector (NND) design was the NOMAD-inspired fine-grained tracker
(FGT), which was developed through a strong collaboration of Indian and
U.S. institutions. DUNE adopts the FGT concept as the 
\textit{reference design} for the NND according to the following guidelines:
\begin{itemize}
%\item Recognition of the central importance of the reference design for NND;
\item The primary design consideration of the DUNE NND
is the ability to adequately constrain the systematic
  errors in the DUNE long-baseline oscillation analysis.
\item The secondary design consideration for the DUNE NND is the
  self-contained non-oscillation neutrino physics program.
\item It is recognized that a detailed cost-benefit study of potential
  NND options has yet to take place and such a study is of high
  priority to DUNE. 
\end{itemize}
The cost and resource-loaded schedule are based on the design presented in
Chapter~\ref{ch:detectors-nd-ref}.

The contribution of Indian institutions to the design and construction
of the DUNE FGT near detector is a vital part of the strategy for the
construction of the experiment. The reference design will provide a
rich, self-contained physics program. From the perspective of the
high-statistics phase of the long-baseline oscillation program, there
may be benefits of augmenting the FGT with a relatively small LArTPC
that would allow for a direct comparison with the far detector, or
adding a high-pressure gaseous argon TPC. At this stage, the benefits
of such options have not been studied, nor are alternative designs for
the NND presented in the CDR; they will be the subject of detailed
studies in the coming months.

A full end-to-end study of the impact of the reference NND 
design on the
oscillation systematics has yet to be performed. Many of the elements
of such a study are in development, for example the Monte Carlo
simulation of the FGT and the adoption of the T2K framework for
implementing near detector measurements as constraints in the propagation of
systematic uncertainties to the far detector.  After the CD-1-R review, the DUNE
Collaboration will initiate a detailed study of the optimization of
the NND. 
To this end a task force will be set up with the
charge to
\begin{itemize}
\item Deliver the simulation of the reference design of the NND and
  possible alternatives;
\item Undertake an end-to-end study to provide a quantitative
  understanding of the power of the NND designs to constrain the
  systematic uncertainties on the long-baseline oscillation measurements; and
\item Quantify the benefits of augmenting the reference design with
  a LArTPC or high-pressure gaseous argon TPC.
\end{itemize}
High priority will be placed on this work and the intention is to
engage a broad cross section of the collaboration in this process. The
task force will be charged to deliver a report by July 2016. Based on
this report and input from the DUNE technical board,
the DUNE executive board will refine the DUNE strategy for the near
detector. 

%% file: volume-detectors/chapter-mgmt-struc.tex
\chapter{DUNE Project Management}
\label{ch:detectors-pm}

\section{Overview}

The international DUNE Project is responsible for managing all
contributions to the design, construction, installation and
commissioning of the DUNE near and far detectors.

As described in CDR \volintro, the DUNE Project is integrated within the
DUNE Collaboration (\textit{the collaboration}).  The collaboration, in consultation with the
Fermilab director, is responsible for forming the international
project team.  The leaders of this team are the Technical Coordinator
(TC) and Resource Coordinator (RC), who are jointly appointed by the
DUNE Co-Spokespersons and the Fermilab Director.  The project receives
appropriate oversight from stakeholders including the Fermilab
Directorate and DOE.  A detailed description of the collaboration
structure along with its advisory and coordinating structures is
contained within Chapter 4  of CDR \volintro.

The DUNE Project is responsible for coordinating the work
packages assigned to each of the international partners contributing
to the effort.  These individual work packages, which are supported
through independent funding agencies, have internal management
structures that are responsible for satisfying the tracking and
reporting requirements of the supporting agencies.  However, the
entire project scope (including non-DOE partner contributions) will be
subject to the DOE critical decision process.

The international DUNE Project Office maintains a schedule for the
entire project and tracks individual contributions through detailed
sets of milestones embedded within the schedule.  It is also responsible
for ensuring that the interfaces between the different work package
deliverables are well defined and that all of these deliverables meet
safety and operational readiness requirements for installation at
Fermilab and the Sanford Underground Research Facility.  Project
Office members including the Project Manager are appointed by the TC.
The managers of the collaboration detector and prototyping
organizations report to the Project Manager and provide the required
interface between the DUNE project and the other members of the
collaboration contributing to these efforts.  As members of these
organizations, they participate in all discussions related to the
design, construction, installation and commissioning of individual
detector elements.  Managers have the primary responsibility for
implementing collaboration plans developed within their organizations.
 
Following this model, the DUNE-US project will have its own project
office and management structure.  All normal DOE project management
requirements will apply to the DUNE-US project.  In its role as host,
the DOE will provide financial support for both the international DUNE
Project Office and the DUNE-US Project Office. The DUNE TC, who acts
as Project Director in the context of the international DUNE project,
also serves as the DUNE-US Project Director.  However, each project
office will have its own project manager.  Other equivalent positions
within the project offices may be filled by the same individuals in
cases where the TC believes this sharing of resources to be most
efficient. Some project office staff may also overlap with the LBNF
project office as appropriate.

\section[Work Breakdown Structure (WBS)]{WBS}

The DUNE Project will manage all contributions to the design,
construction, installation and commissioning of the DUNE near and far
detectors through an international Work Breakdown Structure (WBS).
The WBS organizes the Project's tasks and deliverables into convenient
components and is used to organize the cost and schedule for the DUNE
project. The DUNE Project consists of two major subsystems: the Near Detector
(WBS 130.03), shown to WBS level 4 in Figure~\ref{fig:ND_WBS}, and the
Far Detector (WBS 130.05), shown to WBS level 3 in
Figure~\ref{fig:FD_WBS}.
\begin{cdrfigure}[Near detector WBS]{ND_WBS}{Near detector Work Breakdown Structure.}
\centering
\begin{center}
\includegraphics[width=0.85\textwidth]{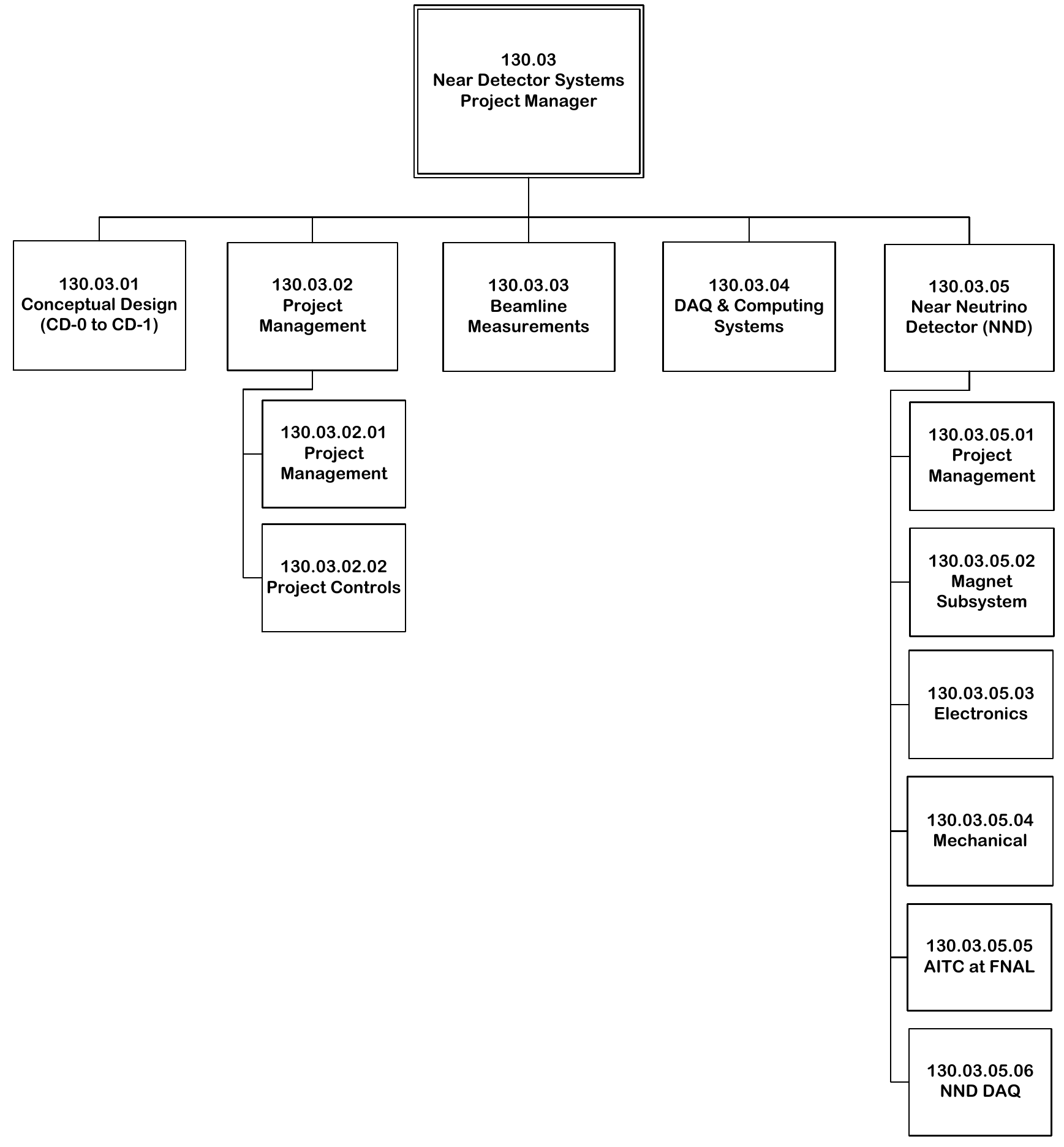}
\end{center}
\end{cdrfigure}
\begin{cdrfigure}[Far detector WBS]{FD_WBS}{Far detector Work Breakdown Structure.}
\centering
\begin{center}
\includegraphics[width=0.9\textwidth]{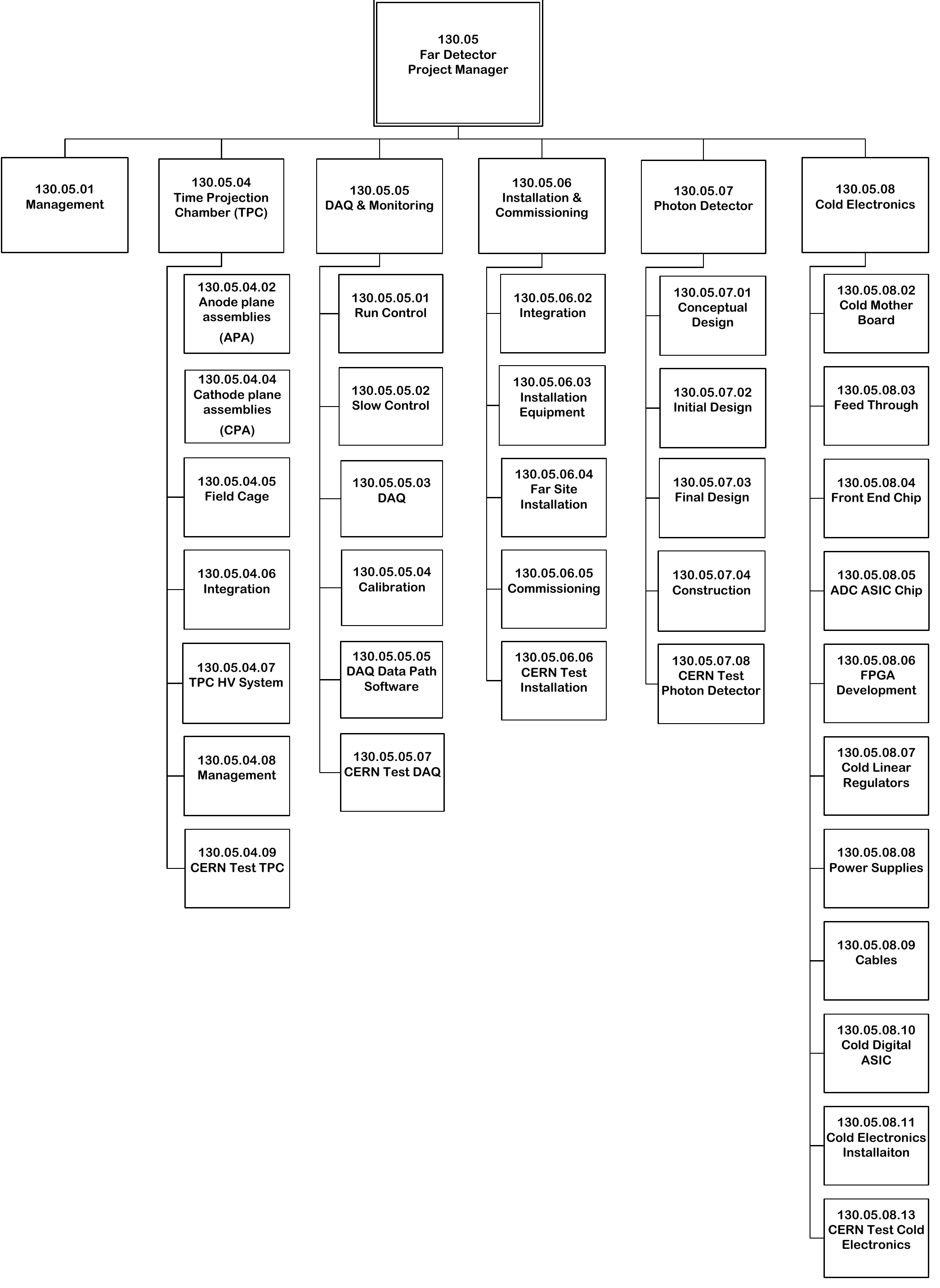}
\end{center}
\end{cdrfigure}
The DUNE Project organization and structure will evolve as the project
becomes more fully internationalized.

%% file: volume-detectors/chapter-fd-ref.tex
\chapter{Far Detector Reference Design: Single-Phase LArTPC}
\label{ch:detectors-fd-ref}

%%%%%%%%%%%%%%%%%%%%%%%%%%%%%%%%
\input{volume-detectors/fd-ref-sections/fd-ref-overview}

\input{volume-detectors/fd-ref-sections/fd-ref-performance}

\input{volume-detectors/fd-ref-sections/fd-ref-optimization}

\input{volume-detectors/fd-ref-sections/fd-ref-tpc}
\input{volume-detectors/fd-ref-sections/fd-ref-daq}
\input{volume-detectors/fd-ref-sections/fd-ref-ce}
\input{volume-detectors/fd-ref-sections/fd-ref-pd}
\input{volume-detectors/fd-ref-sections/fd-ref-install}

%% file: volume-detectors/fd-ref-sections/fd-ref-overview.tex
%%%%%%%%%%%%%%%%%%%%%%%%%%%%%%%%
\section{Overview}
\label{sec:detectors-fd-ref-ov}

This chapter describes the reference design of the DUNE far detector.
The reference design consists of four nominal \ktadj{10} fiducial
mass, single-phase LArTPC modules, augmented with photon detection
systems.  A ``single-phase'' detector is one in which the charge
generation, drift and collection all occur in liquid argon (LAr). The
scope of the far detector includes the design, procurement,
fabrication, testing, delivery, installation and commissioning of the
detector components:
\begin{itemize}
\item Time Projection Chamber (TPC)
\item Data Acquisition System (DAQ)  
\item Cold Electronics (CE)
\item Photon Detector System (PD)
\end{itemize}
The LArTPCs will be housed in cryostats provided by LBNF, described in
\vollbnf. The reference design is based largely on the LBNE far
detector design as of January 2015, documented in \anxlbnefd~\cite{cdr-annex-lbne-design}. This
annex provides the detailed descriptions of the systems and components
that the DUNE reference design incorporates; the differences between
the DUNE and LBNE designs are clearly indicated in this
chapter. Differences include detector size, APA and CPA
placement, and small changes to the APA dimensions.

The detector modules will be constructed sequentially
with the first module coming online as soon as possible and the rest
at a regular pace thereafter. A model of the underground experimental area with
the four \ktadj{10} LArTPCs is shown in
Figure~\ref{fig:FarDet-overview-SP}. 
\begin{cdrfigure}[FD reference design]{FarDet-overview-SP}{Left: 3D model of the reference design for the DUNE far detector to be located at the 4850L. Right: Schematic view of the active detector elements showing the plane ordering of the TPC inside the detector.}
\centering
\begin{minipage}[b]{1.0\textwidth}
\begin{center}
\includegraphics[width=.58\textwidth]{FarDet-3D-SP.jpg}
\includegraphics[width=0.38\textwidth]{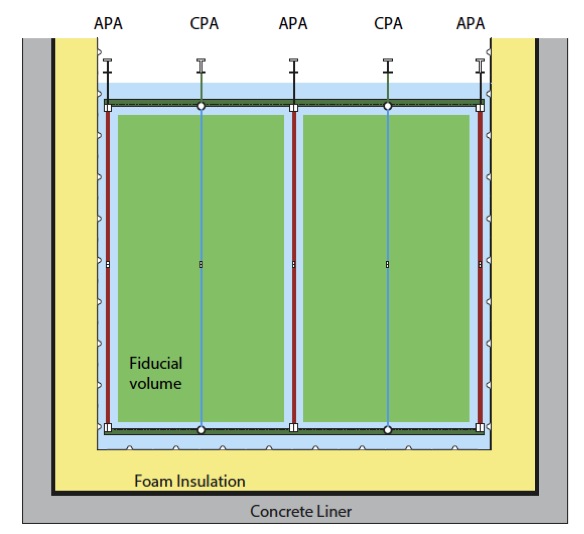}
\end{center}
\end{minipage}
\end{cdrfigure}
Planning for the conventional facilities calls for construction of the
second cryostat to be completed prior to filling the first so that it
may serve initially as a liquid storage facility.  The detector
technology is expected to improve in the coming years with MicroBooNE,
the SBN program and the CERN neutrino platform.  DUNE's staged program
allows selection of optimal designs for each module as the technology
evolves.

The reference design presented in this chapter and documented in the
project cost and schedule is patterned after the successful ICARUS
experiment, but adapted to the local site requirements at SURF and the
scaled up detector size.  The TPC configuration is shown on the right
in Figure~\ref{fig:FarDet-overview-SP}.  The TPC, described in
Section~\ref{sec:detectors-fd-ref-tpc}, is constructed by placing
alternating high-voltage cathode planes and anode readout planes in a
bath of ultra-pure liquid argon. Particles interacting in the argon
generate electron-ion pairs and scintillation light.

The single-phase design offers the advantage that the charge is
collected directly without gain, enabling precision charge
calibration. However, signal levels are low, requiring the use of cold
electronics (Section~\ref{sec:detectors-fd-ref-ce}). The readout is
based on stereo induction and collection planes, requiring a
deconvolution of the induced signal. A photon detection system
(Section~\ref{sec:detectors-fd-ref-pd}) provides the $t_0$ (event
time) for physics processes that are uncorrelated with the LBNF
neutrino beam.

%% file: volume-detectors/fd-ref-sections/fd-ref-performance.tex
\section{Reference Design Expected Performance}
\label{sec:detectors-fd-ref-perf}

The physics requirements are described in \volphys,
for the long-baseline oscillation, atmospheric, supernova
and nucleon decay physics programs.  This section outlines the
numerical detector performance parameters needed to meet the
requirements and the ability of the far detector reference design
to achieve these performance parameters.  

The expected performance of the far detector reference design is based
on the measured performance of the ICARUS\cite{Amerio:2004ze} and
ArgoNeuT\cite{Anderson:2012vc} detectors, scanned Monte Carlo
events\cite{docdb-6954} and newer studies with automated
reconstruction, which are described in
Sections~\ref{sec:detectors-sc-physics-software-simulation-fd},~\ref{sec:detectors-sc-physics-software-reconstruction-fd}
and in \anxreco~\cite{cdr-annex-reco}.  Simulation and reconstruction studies are ongoing.
While many components are in place, a full end-to-end simulation,
reconstruction and analysis chain does not yet exist. Many of the
numerical detector performance requirements are estimates; some of
them correspond to achievements by ICARUS and ArgoNeuT, although these
detectors differ somewhat from the DUNE far detector.  Additional
parameters will be calibrated using the data from LArIAT and the two
CERN prototypes, the \cernsingleproto{} and the \cerndualproto.

Table~\ref{tab:TPC-metric} lists the required performance values,
achieved values (if any) and the values expected from DUNE. The rest
of this section describes each parameter and its connection to the
detector design and physics goals.
\begin{cdrtable}[Preliminary far detector performance expectations]{llll}{TPC-metric}{Preliminary summary of the most 
important performance parameters of the DUNE reference design far
detector.  For each parameter, the table lists the 
performance requirement, performance achieved by other detectors and
projected performance for DUNE. References are given.  Notes: $^1$For
a MIP at the CPA, minimum in all three views, for any track angle;
$^2$Achieved for the collection view; $^3$In order for the fiducial
volume to be known to $\pm 1\%$, the resolution performances are
reported separately in the $x$, $y$, and $z$ directions, where $z$
points along the neutrino beam axis; $^4$For a sample of stopping
muons; $^5$For short electron tracks (stubs) with $E>5$\,MeV.  }
%The third argument (reads {cc}) can use c, l, r or p{some length}  e.g. {clll} or {llp{3cm}}, for instance.
Parameter & Requirement & Achieved Elsewhere & Expected Performance \\ \toprowrule
Signal/Noise Ratio$^1$ & $9:1$ & $10:1$~\cite{Antonello:2015zea,Antonello:2014eha}$^2$ & $9:1$ \\ \colhline
Electron Lifetime & $3\,$ms & $>15$\,ms~\cite{Antonello:2014eha} & $>3$\,ms \\ \colhline
Uncertainty on Charge & & & \\
Loss due to Lifetime  &   $<5\%$  & $<1\%$~\cite{Antonello:2014eha} & $<1\%$ \\ \colhline
Dynamic Range of Hit & & & \\
Charge Measurement & $15$\,MIP & & $15$\,MIP \\ \colhline
% two-hit resolution needs more study.  And likely a different resolution along the drift axis than in the other two directions
% Two-Hit Resolution & 2~mm & & 2~mm \\ \colhline
% Track-finding efficiency is expected to be high.  Needs study to connect it to physics performance
%Track-Finding Efficiency\footnote{For tracks with $L>5$~cm} & $>98\%$ & & $>98\%$ \\ \colhline
Vertex Position Resolution$^3$ & (2.5,2.5,2.5)\,cm & & (1.1,1.4,1.7)\,cm~\cite{Marshall:2013bda,Marshall:2012hh}\\ \colhline
$e-\gamma$ separation $\epsilon_e$ & $>0.9$ & & 0.9 \\ \colhline
$e-\gamma$ separation $\gamma$ rejection & $>0.9$ & & 0.99 \\ \colhline
Multiple Scattering Resolution & & & \\
on muon momentum$^4$ & $\sim18\%$ & $\sim18\%$~\cite{gibinmuon,Ankowski:2006ts} & $\sim18\%$ \\ \colhline
% electron energy scale uncertainty requirement from LBNE DocDB 8741
Electron Energy Scale & & & From LArIAT \\
Uncertainty & $\sim5\%$ & $\sim2.2\%$\cite{ICARUS-pizero} &  and CERN Prototype \\ \colhline
Electron Energy Resolution & $0.15/\sqrt{E{\rm (MeV)}}$ &$0.33/\sqrt{E{\rm (MeV)}}$  \cite{ICARUS-pizero} & From LArIAT \\
 & $\oplus 1\%$ &  +1\% & and CERN Prototype \\ \colhline
Energy Resolution for & & & From LArIAT\\
Stopping Hadrons & $<10\%$ & & and CERN Prototype \\ \colhline
Stub-Finding Efficiency$^5$ & $>90\%$ & & $>90\%$ \\ 
%Stub Arrival Time Resolution & 0.1~ms & & \\
%Efficiency for & & & \\
%finding $t_0$ for & & & \\
%a contained & & & \\
%100 MeV $K^\pm$ & 99\% & & 99\% \\
\end{cdrtable}

The signal-to-noise ratio requirement is motivated by the need to
detect small signals in a large detector that has a low signal rate,
while limiting the required output data volume.  It is set at 9:1 for
a minimum-ionizing particle (MIP) in all three views, for any
orientation of the track.  This ratio is required for all particles in
the detector, specifically those ionizing the liquid argon close to
the CPA, where the reattachment effects are greatest.  Since the
strategy is to zero-suppress the data, it is important that the data
volume from random excursions of the noise over the zero-suppression
threshold compose a vanishingly small fraction of all ADC samples, and
that it preserve the detector's ability to detect sub-MIP signals
(e.g., signals from nuclear de-excitation photons or isolated hits on
the edges of electromagnetic showers).  Since the noise in the
detector may vary by channel and by time, and in addition to thermal
noise from the wires and the electronics, may include coherent noise
sources from electromagnetic pickup and acoustical vibrations of the
wires, among other sources, sufficient contingency on the
signal-to-noise ratio is necessary to ensure that the detector meets
the physics requirements.  A value of 10:1 was achieved by
ICARUS\cite{Antonello:2015zea,Antonello:2014eha} and even higher
values were achieved by Long Bo\cite{Bromberg:2015uia}. Similar to the
DUNE design, Long Bo used cold electronics, although its wires were
much shorter than those planned for DUNE.

The electron lifetime requirement is set at 3\,ms to preserve the
signal-to-noise ratio across the entire detector volume in the
presence of noise sources that are not yet foreseen.  A shorter
lifetime also places demands on the dynamic range of the ADCs: the
gain will need to be large enough to detect weak signals at the CPA,
but small enough to record strong signals near the APAs without
saturation, if possible.  The calorimetric energy resolution of
low-energy electrons is highly sensitive to the lifetime for electrons
that do not record flashes in the photon-detection system.  The energy
resolution is approximately 20\% for electrons of energy below 50\,MeV
(see Section~\ref{sec:detectors-fd-ref-perf-lowe}) without
corresponding photon flashes in a detector with a 2.5-m maximum drift
length, assuming only an average correction is applied for the
lifetime.  This resolution rapidly degrades for shorter lifetimes and
longer drift lengths.  For a maximum drift length of 3.6\,m and an
electron lifetime of 1.5\,ms, the energy resolution is estimated to
degrade to 44\%.  A lifetime of 3\,ms is consistent with that achieved
by the 35-t prototype.  The ICARUS Collaboration has reported a much
longer lifetime, $>$15\,ms\cite{Antonello:2014eha}.

The charge loss due to lifetime effects is expected to be well
measured in the DUNE far detector.  In addition to the cosmic-ray
muons which accumulate at a rate of 0.26\,Hz (see \anxrates~\cite{cdr-annex-rates}), the
laser calibration system and purity monitors will provide detailed
time-dependent measurements of the electron lifetime.  This
limit is placed at 1\% in order to meet the energy-scale and resolution
requirements for electrons, and to a lesser extent, to meet the requirements
of $dE/dx$-based particle identification algorithms.

The dynamic range requirement is placed at 15 MIP in order to detect,
without ADC saturation, particle ionization densities from one MIP up
to the last hit on a track before a particle stops.  The typical
application is for protons, where data from ArgoNeuT show roughly a
factor of 15 between the lowest-charge hit and the highest.
Nonetheless, particles also travel along wires and dense showers may
require even more dynamic range before saturation.  The desire to
measure sub-MIP signals also expands the desired dynamic range.
MicroBooNE set a requirement of 50 on the signal dynamic
range\cite{microboonetdr}.  The dynamic range requirement is
effectively a compound requirement on the noise level, electron
lifetime and number of bits in the ADC.

The primary vertex position resolution requirement is intended 
to keep it from being a significant source of uncertainty on the
fiducial volume determination, though this effect is mitigated if the
resolution is well known.  The current resolution from PANDORA, 
a pattern recognition program developed at Cambridge, easily
meets this requirement.  The axis along which the resolution is the
weakest is that of the neutrino beam direction and the asymmetry in the
achieved resolution is not a result of detector anisotropy.   Tighter
demands on the primary vertex position resolution will be made by
topological selection of $\pi^0\rightarrow\gamma\gamma$ decays, which
require pointing of the photon-induced showers back to the primary
vertex.

In order to reduce the neutral-current background to $\nu_e$CC events
by a factor of roughly 100, information from the $dE/dx$ of the
initial $\sim$2.5\,cm of an electromagnetic shower must be combined
with the topological $\pi^0\rightarrow\gamma\gamma$
selection\cite{docdb-6954}.  Current Monte Carlo studies indicate
that, for showers with enough hits in the initial part to measure
$dE/dx$, the performance of the ionization method is roughly 90\%
electron efficiency with a 90\% rejection factor for single photons.
A topological hand-scan indicates that a $\nu_e$CC signal
efficiency of $\sim$80\% with a 95\% rejection of neutral-current
background can be obtained.  With optimizations to the $dE/dx$
analysis and automating the pattern-recognition identification of
$\pi^0$ decays by topology, it is anticipated that the requested level
of 99\% $\pi^0$ rejection can be obtained at 90\% signal efficiency.

The momentum of muons in $\nu_\mu$CC events is an important ingredient
in measuring the $\nu_\mu$ energy spectrum in the far detector, which
is one of the inputs to the oscillation parameter fits.  Muons that
stop in the detector volume will be well measured using their range.
For those that are not contained the distribution of deviations of the
muon track from a straight line is a function of the muon momentum.
The expected performance of $\pm$18\% on the muon momentum was
achieved by ICARUS for a sample of stopping muons, where the momentum
measured by multiple scattering was compared against that obtained
from the range.  It is anticipated that the resolution will
deteriorate for higher-energy muons because they scatter less.

The requirements on the electron energy-scale uncertainty and the
resolution are driven by the need to analyze the reconstructed $\nu_e$
energy spectrum to extract oscillation parameters in the
high-statistics phase of DUNE.  A fraction of the energy of an
electromagnetic shower escapes in undetected low-energy photons that
can be simulated, but this fraction must be calibrated in data in order to give
confidence in the uncertainty.  An absolute energy scale will need to
come from test-beam data --- LArIAT and the CERN prototypes, the
\cernsingleproto{} and the \cerndualproto.  Analyzing
$\pi^0\rightarrow\gamma\gamma$ decays in ICARUS\cite{ICARUS-pizero}
gives an achieved $\pm 2.2$\% uncertainty on the electromagnetic
energy scale.  The same data also constrain the energy resolution.
The proposed data sets from the test-beam experiments will 
measure these, though the results will need to be extrapolated to the
DUNE far detector geometry and readout details using a full
simulation.  Similarly, the detector response to hadrons --- protons,
charged pions and kaons --- will be calibrated to the necessary precision
by LArIAT and the CERN prototypes.

Absent from the list is a requirement on the two-hit resolution.  In the
direction parallel to the drift field, this resolution is expected to
be very good, of the order of 2\,mm, given existing ArgoNeuT data and
simulations.  The resolution in the other two dimensions is governed
by the wire spacing.  Separation of hits is important for pattern
recognition, for counting tracks near the primary vertex (which is
important for classifying neutrino scatters as quasi-elastic,
resonant, or DIS), and for associating dense groups of tracks in
showers between views.  More study is required to determine
the required two-hit resolution.

It is expected that as the software tools improve and as measurements
from MicroBooNE and the dedicated test-beam programs become
available, the uncertainties on the projected performance will become
smaller.

\subsection{Expected Performance for Low-Energy Events}\label{sec:detectors-fd-ref-perf-lowe}

Low-energy (5--50\,MeV) events require special consideration.
Electron-type neutrino interactions appearing close together in time
constitute the signature for a supernova burst event.  A \MeVadj{5} electron
is expected to hit four wires in the DUNE far detector, and given the
signal-to-noise requirement above, it is anticipated that this signal
will be easy to separate from noise with the required 90\% efficiency.
(Similarly, the photon-detection system is expected to detect the energy
from a proton decay event resulting in a \MeVadj{100} kaon with high
efficiency.)

\begin{figure}[!htb] %  figure placement: here, top, bottom, or page
 \centering
\includegraphics[width=0.45\textwidth]{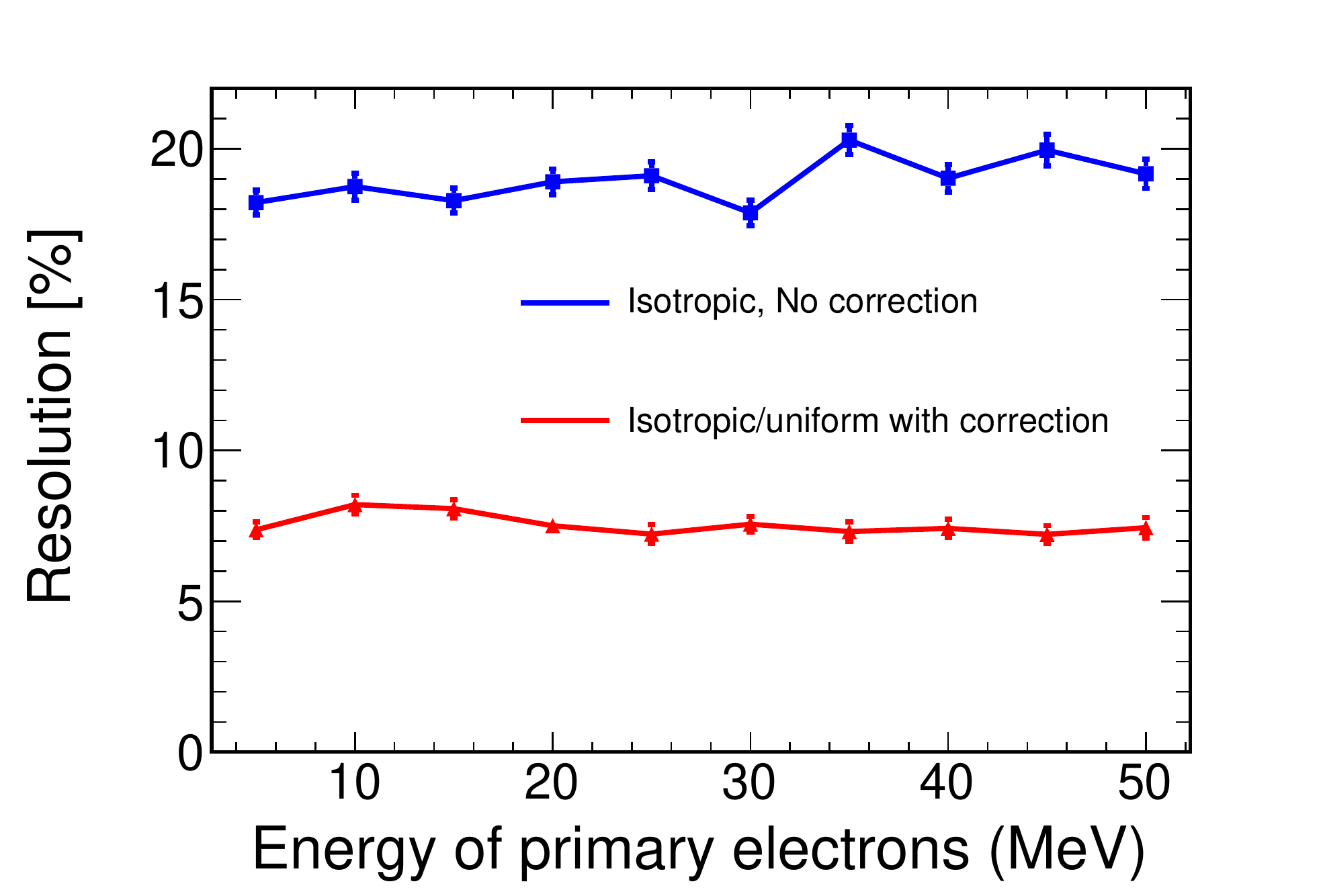} 
\includegraphics[width=0.45\textwidth]{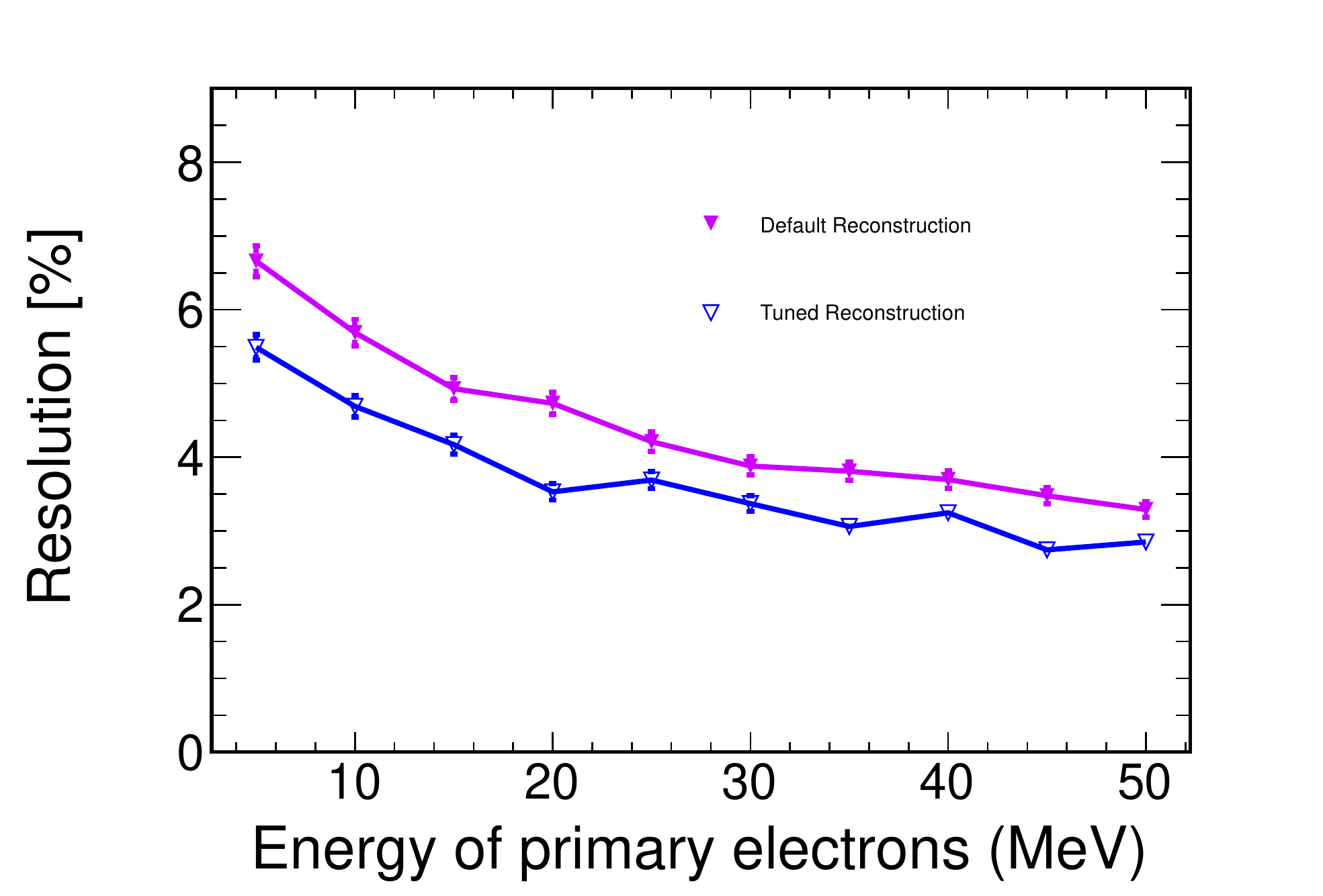} 
 \caption[Comparisons of energy resolution]{Left: Comparison of energy
   resolution (defined as $\sigma/E$, where $\sigma$ is the spread of
   the collection-plane-charge-based event energy $E$ for a
   monoenergetic electron), with and without electron-lifetime
   correction, as a function of electron energy (assumptions: 3\,ms
   drift time, 1.63\,mm/$\mu$s drift velocity, and 2.5\,m maximum drift
   length). The blue curve is the energy resolution of isotropic and
   uniform electrons without electron-lifetime correction. The red
   curve is the energy resolution with electron-lifetime correction
   based on MC truth.  Right: Comparison of energy resolution before
   and after tuning the reconstruction algorithm (for fixed
   position/direction electron events).}\label{fig:lowe_res}
\end{figure}

Work is currently underway using the LArSoft simulation package to
characterize low-energy response for realistic DUNE detector
configurations.
%Figure~\ref{fig:evdisplays} shows a sample \MeVadj{20} event in the DUNE
%\SIadj{35}{t} prototype
%\ktadj{10} 
%geometry simulated with LArSoft. 
So far, most studies have been done with the MicroBooNE geometry, with
the results expected to be generally applicable to the larger DUNE
detector.  For a preliminary understanding of achievable energy
resolution, isotropic and uniform monoenergetic electrons with
energies of 5--50\,MeV (which should approximate the $\nu_e$CC
electron products) were simulated and reconstructed with the LArSoft
package.  The charge of reconstructed hits on the collection plane was
used to reconstruct the energy of the primary electrons. (Compared to induction-plane 
and track-length-based reconstruction in
 this preliminary study, collection-plane hit charge gave the best resolution results;
 however, improved reconstruction based on
  broader information should be possible.)
Figure~\ref{fig:lowe_res} shows the results of a resolution study.  
A correction to compensate for loss of electrons during
drift, $Q_{\rm collection}=Q_{\rm production}\times e^{-T_{\rm drift} / T_{\rm
    electron}}$ (where $T_{\rm drift}$ is the drift time of the ionization
electrons, and $T_{\rm electron}$ is the electron lifetime), using Monte
Carlo truth to evaluate $T_{\rm drift}$, improved resolution
significantly.  This study indicated that photon-time information will
be valuable for low-energy event reconstruction.  Some of the
resolution was determined to be due to imperfect hit-finding by the
nominal reconstruction software.  A tuned hit-finding algorithm did
somewhat better (Figure~\ref{fig:lowe_res}), and further improvements
for reconstruction algorithms optimized for low-energy events are
expected.

Also under study is the potential for tagging $\nu_e$CC absorption
events ($\nu_e +{}^{40}{\rm Ar} \rightarrow e^- +{}^{40}{\rm K^*}$)
using the cascade of de-excitation $\gamma$ rays (that Compton-scatter
in the detector); this should serve the dual purposes of rejecting
background and isolating the CC component of the supernova burst
signal.  Reconstructing these gammas also improves the neutrino energy
measurement.

%% file: volume-detectors/fd-ref-sections/fd-ref-optimization.tex
\subsection{Reference Design Optimization}
\label{sec:detectors-fd-ref-optimization}

Considerations of physics reach as well as cost, schedule and risk
enter the optimization of the detector design parameters.  Ideally one
would like to estimate the asymptotic far-future performance of a full
simulation, reconstruction and analysis chain, replicated for each
design parameter choice and then choose the values that maximize
sensitivity while minimizing the cost and producing timely physics
results. While the GEANT4 simulation is fairly mature, it needs to be
tuned to data from the 35-t prototype and MicroBooNE so that realistic
signal and noise modeling, which are inputs to the optimization
procedures, can improve the performance modelling.  The reconstruction
tools are under development (see
Section~\ref{sec:detectors-sc-physics-software}) and thus, physics
sensitivity is currently optimized using estimates of detector
performance that are input to the Fast Monte Carlo. The DUNE
collaboration plans to establish a detector performance optimization
task force to review various possible detector optimizations in light
of the new collaboration and project organizations.  This section
briefly outlines the considerations and procedures that have been and
will be used to optimize these parameters: the wire pitch, wire angle,
wire length and maximum drift length.  The wire length, angle and
pitch are directly related to the APA dimensions, as discussed in
Section~\ref{sec:detectors-fd-ref-tpc} and the APA dimensions are
constrained by the needs of manufacturing, storage, transport and
assembly.

\subsubsection{Wire Pitch}

The spacing between neighboring sense wires in the APAs is an
optimizable parameter.  In principle it is freely adjustable for all
three wire planes, though to minimize the anisotropy of the detector
response, similar wire pitch should be chosen in all three planes.
The choice of a $\sim$5~mm pitch is documented in\cite{docdb-3407}.
The pitch of the grid plane wires is less important as they are not
instrumented, though the grid wires do shadow the optical detectors
and therefore should not be made with too fine a pitch.

The signal-to-noise ratio is expected to be proportional to the wire
spacing, assuming that the noise on a channel is not impacted by the
presence of nearby wires, and that the signal is divided among the
available channels.  Thermal noise and uncorrelated electronics noise
satisfy these conditions.  Coherent noise is a special case ---
filters may be applied either online or offline to reduce its impact.

The signal-to-noise ratio requirement is set so that zero-suppression
can function without elaborate noise filtering.  A high
signal-to-noise ratio improves pattern-recognition performance,
calorimetric PID performance and $dE/dx$-based $e-\gamma$ separation.
It is also important for detecting sub-MIP signals, such as nuclear
de-excitation photons, and it is important when adding up the energy
of hits on the edges of showers or on the ends of stubs initiated by
supernova neutrino interactions.

The signal-to-noise ratio is expected to be higher in the collection
plane than in the two induction planes.  The need to deconvolve the
bipolar signals while filtering noise in the induction planes means
that the collection plane will be the most reliable in performing
$dE/dx$ measurements, though for tracks that travel in a plane
containing a collection wire and the electric field, the induction
planes will be critical for recovering PID efficiency.  Reducing the
spacing between wires will have an adverse impact on the detector
performance parameters that depend on the signal to noise, with the
effect seen more prominently in the induction-plane data.

The main benefit of a finer wire pitch is the ability to obtain higher
resolution measurements of the ionization density left by events in
the detector.  The spatial resolution in the plane perpendicular to
the electric field is also affected by transverse diffusion of the
drifting electrons.  It can be argued that the wire spacing does not
need to be much smaller than the typical width induced by diffusion,
though deconvolution and/or fitting techniques that combine signals
from nearby wires can recover some of the resolution lost to
diffusion.  Noise may limit the amount of resolution gained using
these techniques.

The separation of electrons from photons using the $dE/dx$ measured in
the initial part of an electromagnetic shower is described in \anxreco~\cite{cdr-annex-reco}.  The first 2.5~cm of a shower is the most important, since 
subsequent showering stages have not yet taken place, leaving one MIP
for an electron and two for a photon conversion to two electrons,
though sometimes the subsequent showering starts earlier.  As the
first hit cannot be used to measure $dE/dx$ (since it is not known
where in the volume of argon viewed by that wire the track started,
the second and subsequent hits must be used.  But the shower can be
aligned unfavorably along the wires of one view or another, resulting
in few usable hits.  If no hits are useful, then the $dE/dx$ method
cannot be used.  Reducing the spacing between wires in all three views
increases the precision of the measurement of the initial part of the
shower for $e-\gamma$ separation purposes.  The current study
described in~\cite{cdr-annex-reco} only uses the collection plane wires; 
thus with a more optimal strategy, some of the efficiency that is lost
with 5-mm wire spacing compared with 3-mm spacing can be recovered by
examining the other two views.

Separation of multiple close tracks is improved with more closely
spaced wires.  The position resolution of hits is expected to be much
better along the drift direction than in either of the axes
perpendicular to it as the sampling frequency times drift velocity is
much smaller than the wire spacing.  As long as the tracks that should
be separated from one another travel at an angle with respect to the
APA plane, then the fine time resolution will help with the pattern
recognition even if the wire pitch is large.

The reconstruction of short tracks, such as low-energy protons ejected
by the struck nucleus at the primary vertex of a neutrino scattering
event, is improved with higher spatial resolution.  A reduced wire pitch also
allows more precise measurements of the distance between the
primary vertex and photon conversion points, which is the other
component of $\pi^0\rightarrow\gamma\gamma$ separation from electrons.
Topological identification of two EM showers and their displacement
from the primary vertex is expected to provide a factor of ten to
twenty in NC background rejection while retaining at least 90\%
efficiency for $\nu_e$CC events.  A hand-scan study comparing liquid
argon TPC detector performance in topological separation of NC events
from $\nu_e$CC events in a detector with 5-mm wire pitch and a
detector with 10-mm wire pitch\cite{2008-hand-scan} showed no
degradation in performance with the coarser wire pitch.  Automated
topological selection has yet to be developed, though it is
anticipated that this finding will remain true, given that the
radiation length in liquid argon is typically $\sim$30 times the
typical wire spacing.

The gains in physics from a finer wire pitch must be balanced against
the increased cost of the electronics and online computing resources
needed to read out the additional wires.  In addition, the additional
cold electronics components would likely create a higher heat load in
the liquid and manufacturing the APAs would likely take longer due to
the higher density of wires.

The spacing between the planes is customarily chosen to be similar to
the spacing between the wires within the planes, though this, too, is
an optimizable parameter.  Narrowing this spacing improves the
sharpness of the signals (in time) in both the induction and
collection planes, though electronics shaping and diffusion will limit
the ultimate signal sharpness.  These functions can be deconvolved,
though deconvolution and noise filtering produce artifacts in the
signals.  Studies varying the spacing between the planes can be
performed to estimate the impact on two-hit time resolution.

Using automated tools, future studies will estimate the performance of the 
$dE/dx$-based $e-\gamma$ separation as a function of the wire pitch for 
events that do not contain showers identified topologically as $\pi^0$ decay 
candidates. Evaluation of the wire pitch will need to be done separately for 
this class of events since $dE/dx$-based separation and topological separation 
are correlated.

\subsubsection{Wire Angle}
\label{v4:fd-ref-wireangle}

Like the wire pitch, the choice of the angles of the induction-plane
wires relative to the collection plane wires affects the physics
performance of the detector.  Because the wires wrap from one side of
each APA to the other, a discrete ambiguity is added to the
continuous ambiguity of identifying where along the wire the charge was deposited.

Reconstruction of 3D objects based on 2D data (channel number
vs. time) requires associating hits in one view with those in at least
one other.  If two wires cross only in one place, the ambiguity is
removed once the hit is associated in the two views.  If the wires
cross more than once, three views are required in order to break the
ambiguity of even isolated hits.

This association can most easily be done using the arrival time of the
hits.  If the time of a hit is different from that of all other hits
in the event, then the association is easy.  In more complex cases,
where dense showers produce many hits at similar drift distances,
misassociation can happen.  In this case, the discrete ambiguity makes
it possible to displace a reconstructed charge deposition by multiple
meters from its true location.  In the case that the $U$ and $V$
angles are the same and the number of times a wire wraps around an APA
exceeds one, then even a single, isolated hit can be ambiguous.  A
small difference in the $U$ and $V$ angles breaks this ambiguity,
though misassociation still occurs in events with multiple nearby hits
close in time.  The use of clustering methods assists in obtaining the
correct ambiguity choices for hits in dense environments.

Reducing the wire angle reduces the number of crossings, but does not
eliminate the possibility of misassociating hits in events with
multiple hits arriving simultaneously on their respective wires.
Reducing the angle aligns the
shapes of features in the different views making it easier to correlate them.
The angle chosen for the DUNE far detector
reference design ensures that no induction wire crosses any collection
wire more than once.

On the other hand, reducing the wire angle worsens the resolution of
3D reconstruction of hits in the vertical direction. 
It also worsens the resolution on the measured separation between the
primary vertex and photon conversion points, though as pointed out
above, the radiation length is much longer than the wire spacing.  
The number of hits available for $dE/dx$ separation of electrons from
photons degrades for vertically-going showers if the wire angle is
reduced.  A parametric study of a figure-of-merit based on the
measurability of the two photon-conversion lengths as a function of
the induction-plane wire angles is provided
in\cite{wire-orientation}.

The impact on the physics sensitivity of a small number of hits that
are misreconstructed by many meters from their true locations is
estimated to be larger than that incurred by the degraded resolution
of each hit in the vertical dimension\cite{docdb-8981}, though studies
have yet to be performed to estimate the impact on event detection
efficiency, particle ID performance and energy resolution.  Hits with
locations that are misreconstructed by large distances are not
expected to contribute properly to energy sums and may cause pattern
recognition failures, such as missing tracks or splitting of tracks
into multiple pieces.  Advances in algorithms to break ambiguities in
complex environments can allow for steeper wire angles.

The fact that the channel count must be an integer multiple of 128 in
an APA also constrains the wire angles as functions of the APA frame
dimensions, though a procedure to find the proper arrangement of
channels that most closely approximates the optimized wire pitches and
angles and meets the channel count constraint may cause small
deviations in the parameters.

\subsubsection{Wire Length}

The length of the collection-plane wires is determined by the APA
dimensions (or vice-versa) and the length of the induction-plane
wires is determined by their angle and the APA dimensions.  The APA
dimensions are largely constrained by transport and handling needs as
well as stiffness and production cost issues as they get larger.
Capacitance and noise increase with wire length; this effect would
likely not be masked by electronics noise since the cold electronics
is expected to have very low noise.  Therefore, in order to meet the
signal-to-noise requirement with a finer wire spacing, the wires may
need to be made shorter.

On the other hand, longer wires lower the cost of the detector, as
fewer electronics channels and APA frames --- and winding time and
materials --- are needed to instrument a given volume of liquid argon.

It is anticipated that much of the work needed to study the impact of
the wire pitch will inform the wire length choice due to its impact on
the signal-to-noise ratio.

\subsubsection{Maximum Drift Length}

The maximum drift length is another optimizable parameter.  In
this case, the driver for longer lengths is the cost of the detector.
A longer drift length assigns more liquid argon to be read out by any
given channel, reducing the APA count and the channel count.  A longer
drift length also increases the fraction of liquid argon that is
fiducial.  Fiducial cuts will be made around the APA planes
to ensure containment and to minimize the impact of dead argon inside the APA
planes.  Reducing the APA count thus reduces the amount of non-fiducial
liquid argon.  

A longer drift length, and the correspondingly longer electron drift time, increase the
likelihood that the electron will attach to impurities as it drifts towards the
anode plane.  Once the drift length has been specified, therefore, the requirements on the 
signal-to-noise ratio, the electron lifetime and the dynamic range are coupled.

Increasing the drift length also degrades position resolution due to
diffusion, where the spread of a drifting packet of charge increases
in proportion to the square root of the drift time.  Charge deposited
near the APAs remains well measured, though charge deposited near the
CPAs will suffer from both attenuation and diffusion, lowering the
signal-to-noise ratio.  Small-signal detection efficiencies and PID
performance may decrease for events near the CPAs.  Sophisticated
reconstruction and analysis algorithms can be used to recover
resolution that is thus lost, but the resolution may be limited by
noise.  Simulation studies in advance of CD-2 will address the impact
of diffusion and noise on the particle reconstruction performance.

%% file: volume-detectors/fd-ref-sections/fd-ref-tpc.tex
%%%%%%%%%%%%%%%%%%%%%%%%%%%%%%%% 
\section{The Time Projection Chamber (TPC)} 
\label{sec:detectors-fd-ref-tpc}

\subsection{Overview}

The scope of the Time Projection Chamber (TPC) subsystem includes the
design, procurement, fabrication, testing and delivery of anode plane
assemblies (APAs), cathode plane assemblies (CPAs), the field cage and
the high voltage system.

The TPC is the key active detector element of each DUNE
far detector module. It is located inside the cryostat vessel and is
completely submerged in liquid argon at 88~K. 
%The TPC consists of
%alternating anode plane assemblies (APAs) and cathode plane assemblies
%(CPAs), with field-cage modules enclosing the four open sides between
%the anode and cathode planes.  
The TPC is constructed of modular APAs, CPAs and field-cage modules. The APAs and CPAs are assemblies of
wire planes and are tiled into alternating APA-CPA rows along the length of
the cryostat. The resulting rows are called \textit{anode planes} and \textit{cathode planes}, respectively.
(Note the different uses of the word \textit{plane}.)  Field-cage modules enclose the four open sides between
the anode and cathode planes.  
When proper bias voltages are applied
to the APAs and CPAs, a uniform electric field is created in the volume
between the anode and cathode planes. A charged particle traversing
this volume leaves a trail of ionization in the ultra-pure liquid
argon.  The electrons drift toward the anode wire planes, inducing
electric current signals in the front-end electronic circuits
connected to the sensing wires.  The current-signal waveforms from all
sensing wires are amplified and digitized by the front-end electronics
and transmitted through cold (immersed) cables and feedthroughs to the data
acquisition (DAQ) system outside of the cryostat. While electrons drift
toward the APAs, positive ions drift toward the CPAs at a velocity five orders of 
magnitude slower than that of the electrons and therefore contribute little to the signal on the wires.

\begin{cdrfigure}[Cross section of the TPC inside the cryostat]{tpc-xsect1}
{Cross sections of the LBNE \ktadj{5} TPC (left) and the DUNE \ktadj{10} TPC (right).  
The exchange of the APA and CPA positions significantly reduces the energy 
stored in the TPC by eliminating the two ground-facing cathode planes. 
This allows an increase in the detector's fiducial volume given the same cryostat volume. %with the same cryostat.  
The length of the DUNE TPC is  58~m along the direction of the neutrino beam (into the page).}
\includegraphics[width=\linewidth]{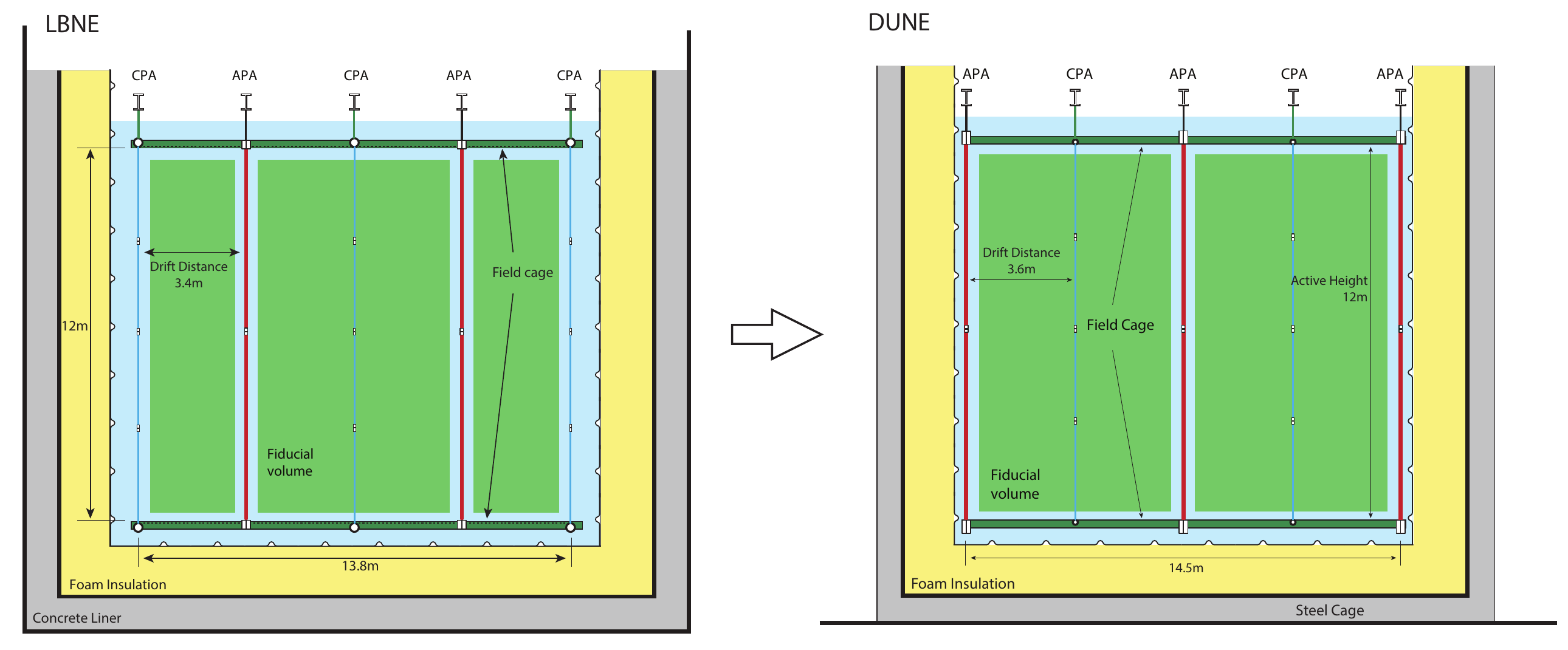}
\end{cdrfigure}
The TPC active volume (Figure~\ref{fig:tpc-xsect1}) is 12~m high,
14.5~m wide and 58~m long in the beam direction.  
Its three rows of APA planes interleaved with two rows of CPA planes
are oriented vertically, with the planes parallel to the beamline. The
electric field is applied perpendicular to the planes.  The maximum
electron-drift distance between a cathode and an adjacent anode is
3.6~m. This requires a $-$180~kV bias voltage on the cathode plane to
reach the 500~V/cm nominal drift field. The anode plane assemblies are
2.3~m wide and 6~m high. Two 6~m modules are stacked vertically to
instrument the 12~m active depth. In each row, 25 such stacks are
placed edge-to-edge along the beam direction, forming the 58~m active
length of the detector.  Each CPA has the same width, but half the
height ($\sim$3~m) as an APA, for ease of assembly and transportation.
Four CPAs will be stacked vertically to form the full 12-m active
height.  Each cryostat houses a total of 150~APAs and 200~CPAs.  Each
facing pair of cathode and anode rows is surrounded by a field
cage assembled from panels of FR-4 glass-reinforced epoxy laminate
sheets with parallel copper strips connected to resistive divider
networks.  The entire TPC is suspended from five mounting rails under the
cryostat ceiling (see Figure~\ref{fig:tpc-floor-view}).

\begin{cdrfigure}[A view of the partial assembled TPC]{tpc-floor-view}
{A view of the partially installed TPC inside the membrane cryostat.
  The APAs are shown in red, CPAs are in cyan, field-cage modules in
  yellow/green.  Some of the field-cage modules are in their folded
  position against the cathode (providing aisle access during installation).}
\includegraphics[width=\linewidth]{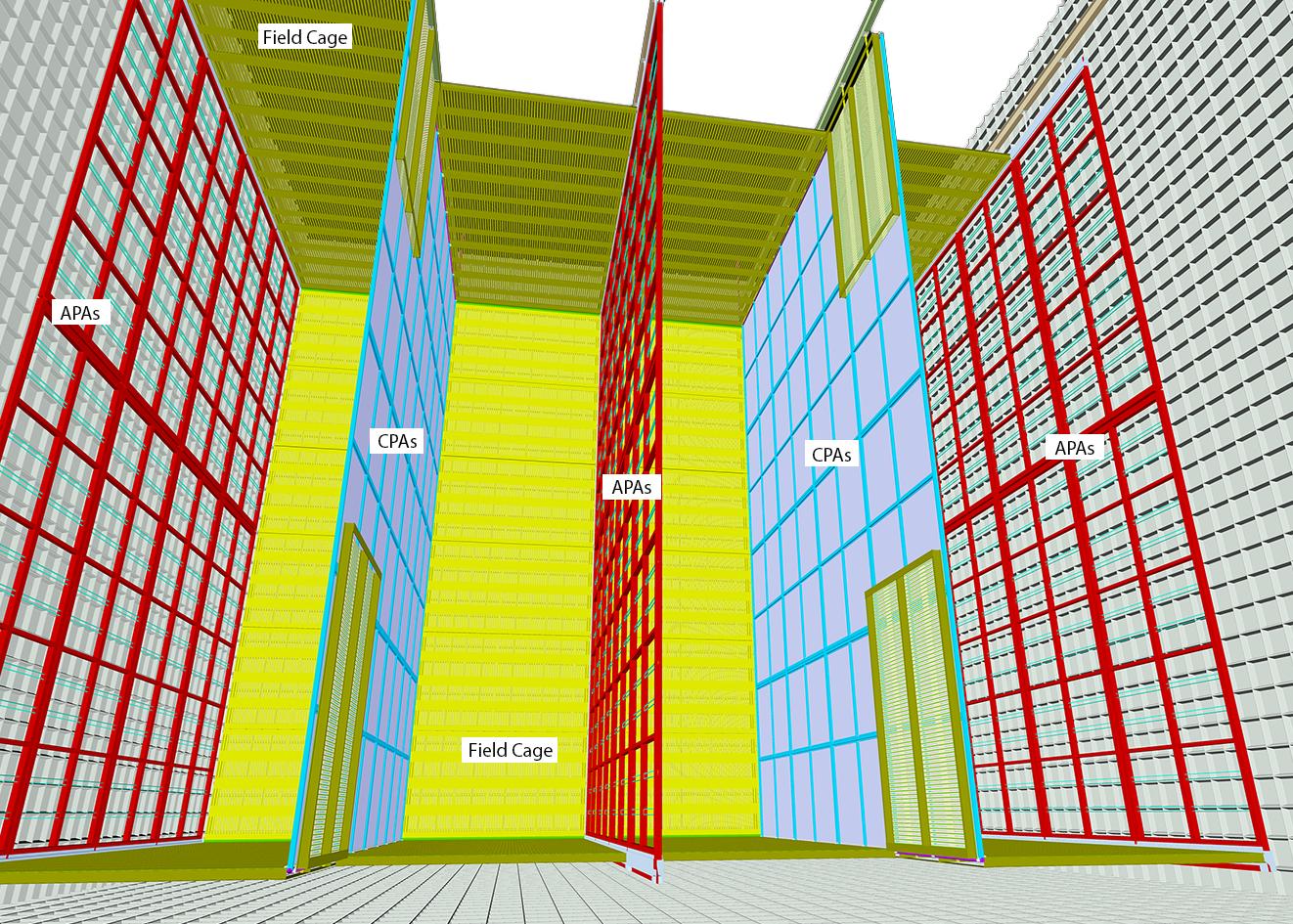}
\end{cdrfigure}

The units of construction of the active detector are the APAs, CPAs
and field-cage modules. These are modular elements of a size optimized
to simplify their manufacture, satisfy commercial highway and
underground transport requirements, and facilitate handling and efficient
installation in the cryostat.   The units will be
fully tested in LN$_2$ (or LAr) at the assembly site. They will be
tested again at the far detector site before installation and will be
monitored continuously during and after installation to detect any
failures.

%%%%%%%%%%%%%%%%%%%%%%%%%%%%%%%% 
\subsection{Anode Plane Assemblies (APA)}
\label{subsec:fd-ref-apa}

An APA is constructed from a framework of lightweight, rectangular
stainless steel tubing, with four layers of wires wrapped on each side
of the frame.  From the outside in, the first wire layer is a
shielding (grid) plane, next are two induction planes and the
collection plane.  The front-end electronics boards are mounted on one
end of the APA frame and protected by a metal enclosure.  The APAs are
2.3~m wide, 6.3~m high, and 12~cm thick. The height is chosen for
fabrication purposes and compatibility with underground transport
limitations. The 2.3-m width is set to fit in a standard High Cube
container for storage and transport with sufficient shock absorbers
and clearances.

%%%%%%%%%%%%%%%%  
\subsubsection{Wire Planes}
\label{subsec:fd-ref-wireplanes}

APAs are strung with wire of 150~$\mu$m diameter copper beryllium
(CuBe) alloy for high tensile strength, good electrical conductivity,
excellent solderability and a thermal-expansion coefficient compatible
with that of the stainless steel frame.  The wires will be epoxied to
fiberglass wire-bonding boards and then soldered to copper traces on
the boards for electrical connections.

Four planes of wires cover each side of an APA frame as shown in
Figure~\ref{fig:tpc-wire-frame-xsect}.
\begin{cdrfigure}[Illustration of the APA wire wrapping scheme]{tpc-wire-frame-xsect}{Illustration of the APA wire wrapping scheme (left), and three cross sectional views (right). At left, small portions of the wires from the three signal planes are
shown in color: magenta (U), green (V), blue (X). The fourth wire
plane (G) above these three, parallel with X, is present to improve
the pulse shape on the U plane signals. At right, the placement of the
wire planes relative to the stainless steel frame is shown for
different locations on the APA. See Table~\ref{tab:wire-parameters} for
wire plane parameters.}
  \includegraphics[width=0.8\linewidth]{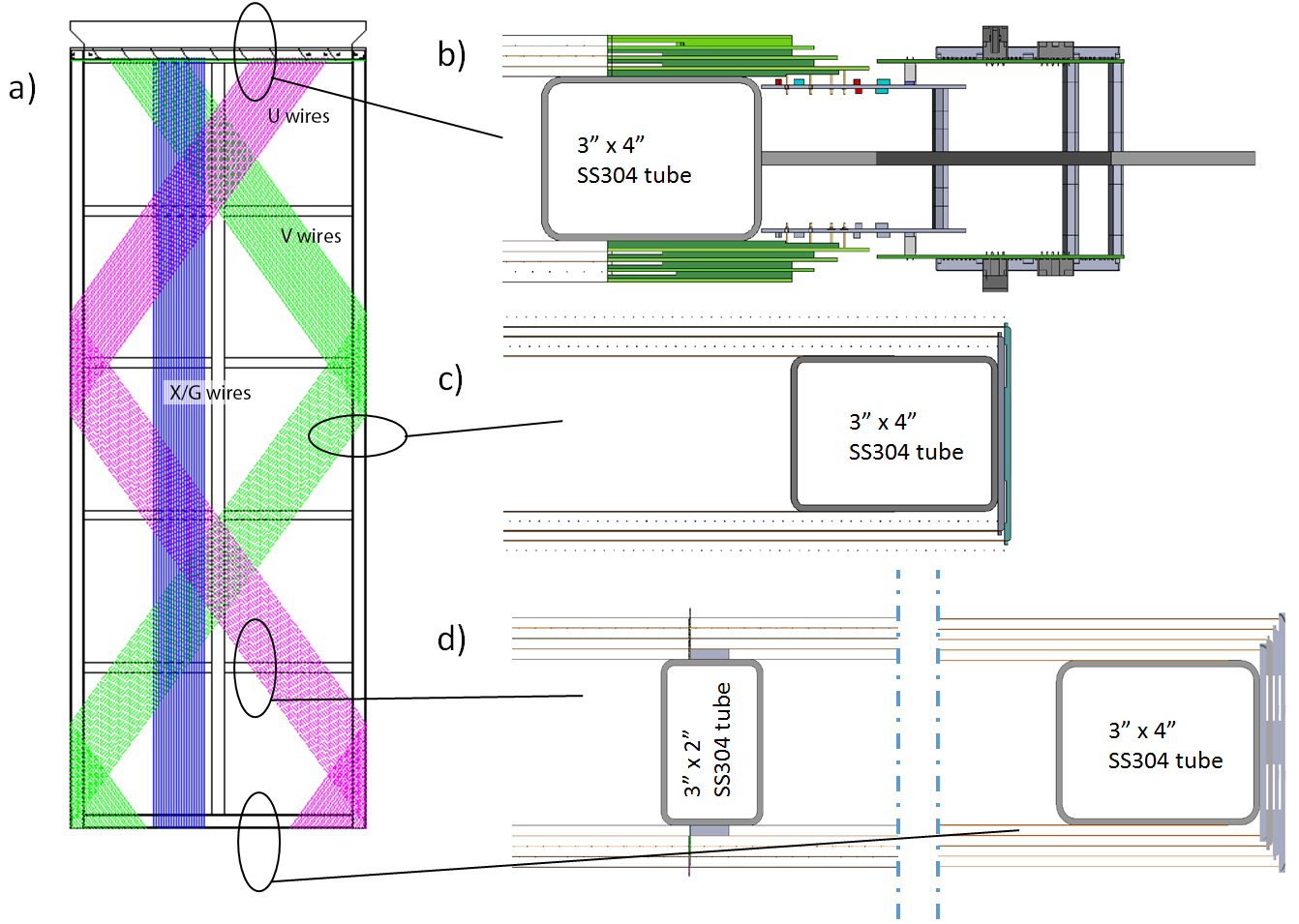}
\end{cdrfigure}
These four planes of wires are labeled, in order from the outside in:
G (for grid), U, V and X.  Table~\ref{tab:wire-parameters} summarizes
the key parameters of each of the wire planes.  
\begin{cdrtable}[Wire parameters]
  {llrrrr}{wire-parameters} {Parameters of the four planes of wires on an APA}
  
   Label & Function & Orientation &  Pitch & Number  & Bias Voltage 		\\ \rowtitlestyle
      			&						& (from vertical) 		& {(mm)}   	&   			& {(volt)} 	\\ \colhline
    G    		& Shield/grid plane 			&0$^\circ$  			& 4.79		& 960 		& $-$655   \\ \colhline
    U            	&  1$^{st}$ induction plane 	& +35.7$^\circ$  		& 4.67		&  800  		& $-$365 	\\ \colhline
    V            	&  2$^{nd}$ induction plane	& $-$35.7$^\circ$  	& 4.67	 	&  800  		& 0 			\\ \colhline
    X            	&  Collection plane			& 0$^\circ$ 			& 4.79 		&  960  		& +860 		\\

\end{cdrtable}
The distance between wire planes is 4.76~mm (3/16 inch, a standard
printed circuit board thickness).  Each wire plane is biased to a
particular voltage such that the ionization electrons from charged
particle tracks will drift past the first three wire planes and be
completely collected by the collection wire plane (X).  The V wires
are DC-coupled to the readout electronics to minimize the maximum
voltage on the other wire planes.  A grounded mesh plane with good
optical transparency, located 4.8~mm behind the collection plane,
prevents the electric field around this set of wires from being
distorted by the metal frame structure and the wires on the opposite
side of the frame. It also shields the sensing wires from potential EM
interference from the silicon photomultipliers (SiPMs) on the photon
detectors, mounted within the frame.  Each wire on the U, V and X
planes is connected to a front-end readout channel. The grid plane
wires are not read out, but serve the important purpose of shielding
the U wires from responding to distant moving charges. The total
number of readout channels in an APA is 2560, for a total of 384,000
in each cryostat.

The wires on the two induction planes (U and V) are wrapped in a
helical pattern around the long edges of the wire frame
(Figure~\ref{fig:tpc-wire-frame-xsect}a). This technique makes it
possible to place readout electronics only at one short edge of an APA
frame, and allows tiling of the APAs on the other three sides with
minimal dead space ($\sim$1.3\% of active area).  Although wires on
the induction planes are sensitive to tracks on both sides of an APA,
the vertical collection-plane wires are only sensitive to one side,
and therefore able to resolve this ambiguity.  The upper APAs in the
cryostat will have their readouts at the top edge of the frame (as
shown in Figure~\ref{fig:tpc-wire-frame-xsect}), while the lower APAs
will mount their electronics at the bottom edge.  These readout
electronics are located within the LAr volume but outside of the TPC
active volume.  On the readout end of an APA, 20 sets of front-end
readout boards with 128 channels each (40U+40V+48X) are distributed on
both sides of the APA, reading out the \SI{2560} sense wires.

With the APA length and width constrained by transportation and handling
limitations, the angles on the induction plane wires are chosen so
that they wrap less than one full revolution around the APA.  This
avoids an ambiguity problem where three wires from three readout
planes intersect more than once on an APA face (discussed in 
Section~\ref{v4:fd-ref-wireangle}).  Precise values of
wire angle and wire pitch (see Table~\ref{tab:wire-parameters}) were
chosen to give an integral number of wires across the boards at the
electronics end of the APA as well as an integral number of wire slots
in the boards along the sides of the APA.  A preliminary
study\cite{wire-orientation} has shown that this wire layout meets the
physics requirements.

The APAs facing the cryostat walls are sensitive on both sides, similar to
 those in the middle of the TPC.  However, the negative bias
voltage on their outer grid planes prevents any electrons drifting
from the cryostat walls toward the sensing wires.  The electronics for
the outer X wire plane can be eliminated to save cost.  Alternatively, 
these double-sided APAs can be utilized by adding another cathode plane
with a small negative bias between the cryostat wall and the anode plane to
form a very shallow veto region.

%%%%%%%%%%%%%%%%
\subsubsection{APA Frame}
\label{subsec:fd-ref-apaframes}

At the nominal wire tension of 5~N, the total of \SI{3520} wires
exerts a force of $\sim$7.0~kN/m on the short edges of the APA, and a
$\sim$1.5~kN/m force on the long edges. The wire frame must be able to
withstand the wire tension with minimal distortion, while minimizing
dead space due to the thickness of the frame.  The wire frame is
constructed from stainless steel tubes welded in a jig.  Structural
analysis has shown that the maximum distortion of the frame due to
wire tension is less than 0.5~mm. The total mass of a bare frame is
$\sim$260~kg.

%% Lengthwise buckling is not an issue, both because of the strength of the frame and because the wires are maintained at an approximately uniform distance from the frame by periodic comb-like structures.

All tube sections are vented to prevent the creation of trapped LAr
volumes. The three long tubes have slots cut in them so the photon
detectors can be inserted into the APAs after the wires are installed.
The two long outer members of the frame are open-ended, so the photon
detector cables can be threaded through them to reach the signal
feedthroughs on the cryostat roof.  These long tubes can potentially
be used to carry signal and power cables from the bottom APA cold
electronics boards to the signal feedthroughs.  This could
significantly reduce the cable length compared to, e.g., running the
cables from the middle bottom APAs along the floor and then up the
wall.

%%%%%%%%%%%%%%%%
\subsubsection{APA Wire Bonding and Support}
\label{subsec:fd-ref-wirewrap}

The wire bonding boards physically anchor the wires at the edges of an
APA and provide the interface between the wires and the cold
electronics at the readout end of the APA.  The four planes of wires
are attached to their respective wire bonding boards through a
combination of epoxy and solder. During winding of the X layer onto
the APA, the wires are placed across the top surface of the X wire
board. The wires are then glued down with a strip of epoxy at the
leading edge of the board.  After the epoxy has cured, the wires are
soldered onto the copper pads under each wire, and then the wires are
cut beyond the pads. The V, U and G planes are attached on top of the
X boards and similarly populated with wires, one layer at a time. An
array of pins is pushed through holes in the stack of wire bonding
boards, making electrical connections between the wires and the
capacitor-resistor (or CR) board, which is located between the wire
boards and the front-end electronics boards.  The CR boards distribute
the bias voltages to each wire through current-limiting 20~M$\Omega$
resistors, and bring the charge signal through high voltage AC
coupling capacitors to the cold electronics.

These readout boards, as described in
Section~\ref{sec:detectors-fd-ref-ce}, generate an estimated
$\sim$160~W of heat per APA which may produce a small quantity of
argon bubbles.  Stainless steel covers are placed over the readout
boards to contain the bubbles and direct them to the gas volume of the
cryostat. This is particularly important for the bottom APAs where the
bubbles must be contained and funneled through the vertical hollow
frame members to the top of the cryostat to prevent the bubbles
entering the TPC active volume.

Comb-like wire support structures (see Figure~2.9 in \anxlbnefd~\cite{cdr-annex-lbne-design}) are
located on each of the four cross beams (see Figure~\ref{fig:tpc-wire-frame-xsect})
 so that the longitudinal wires
are supported every 1.2~m and the angled wires about every 1.5~m while
introducing only millimeter-scale dead regions. The support structure
is composed of strips of thin G10 sheet, with notches machined at
correct intervals.  These wire supports play a key role in minimizing
wire deflection due to gravity and electrostatic force, enabling the
use of a moderate wire tension and reducing the risk of wire breakage.
They also maintain the correct wire pitch and wire plane separation
even if the APA frame has a small amount of twist and warp.  If a wire
breaks after installation, these intermediate wire support will limit
the movement of the broken wire such that it will not travel too far
into the drift volume and make contact with the field cage.  To
further reduce the risk and impact of a broken wire, a new wire
support scheme is being developed that can be applied to the outer
wire planes near the bottom of the TPC to prevent a broken wire from
contacting the field cage.

%%%%%%%%%%%%%%%%
\subsubsection{Wire-winding Machines}
\label{subsec:fd-ref-wirewinding}

A winding machine will be constructed to lay the \SI{3520} wires onto
each APA. It has sufficient versatility that the same mechanism can
wind both the angled and the longitudinal layers. Its working concepts
are illustrated in Figure~\ref{fig:tpc-winding-machine}.
\begin{cdrfigure}[Winding machine concepts]{tpc-winding-machine}
{Illustration of the wire winding machine concept.  The tensioner 
head is passed from one side of the APA to the other as it is moved 
around the APA to wind wire onto the APA frame.  The horizontal/vertical 
positioning systems on each side of the APA are made of commercial linear motion components. }
\includegraphics[width=0.9\linewidth]{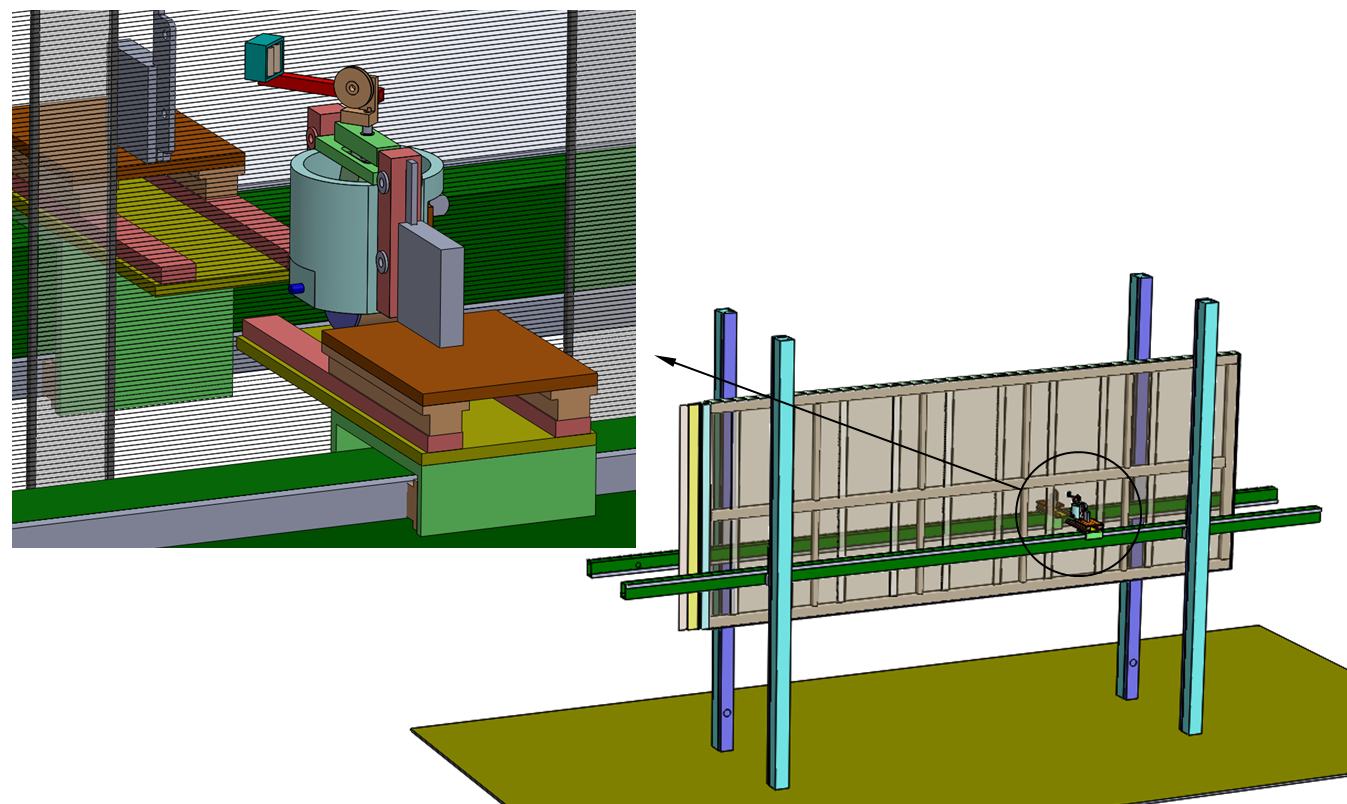}
\end{cdrfigure}
The wire tensioner is a self-contained unit that includes the wire
spool.  It is designed so that correct wire tension is maintained,
independent of the wire-feed rate and direction.  The APA, oriented
horizontally in the device (with one of its long edges down), is held
off the ground by a couple of posts.  There are X-Y positioners on
either side of the APA; the tensioner is moved across the face of the
APA by one of these positioners, unspooling tensioned wire as it
moves.  When the tensioner arrives at the edge of the APA it is passed
across to the positioner on the other side of the APA while placing
the wire into the appropriate slots of the edge boards.  In this way
the entire layer of wire can be placed on the frame.

Although a large part of an entire plane of wires can be wound in one
continuous process, a more fault-tolerant procedure will be adopted in
which the winding machine will be paused periodically to solder the
last wire winding. This intermediate soldering step will prevent the
unraveling of a large section due to an accidental broken wire.  An
automatic soldering robot will solder the wire ends after the wires
have been laid down on the APA. A wire-tension measuring device will
scan the newly placed wires and record the wire tension of each
wire. Any wires with abnormal tension will be replaced manually.

%%%%%%%%%%%%%%%%%%%%%%%%%%%%%%%%
\subsection{Cathode Plane Assemblies (CPAs)}
\label{subsec:fd-ref-cpa}

There are two cathode planes in each detector module.  Each cathode plane is 
tiled from a four-unit-high by 25-unit-wide array of CPAs. Figure~\ref{fig:tpc-cathode-model} shows the
building blocks of a cathode plane.  
\begin{cdrfigure}[Conceptual design of cathode plane components ] 
{tpc-cathode-model}{Conceptual design of the cathode plane 
components near a corner.  Two flavors of CPAs (outer 
and inner unit) are used to make up the entire cathode plane; all
CPAs are roughly 2.3~m wide by 3~m tall. 
The cathode plane is terminated at
right and left ends by the end pieces (cyan colored).  An HV receptacle 
(orange) connects with the HV feedthrough from the cryostat ceiling. }
\includegraphics[width=0.9\linewidth]{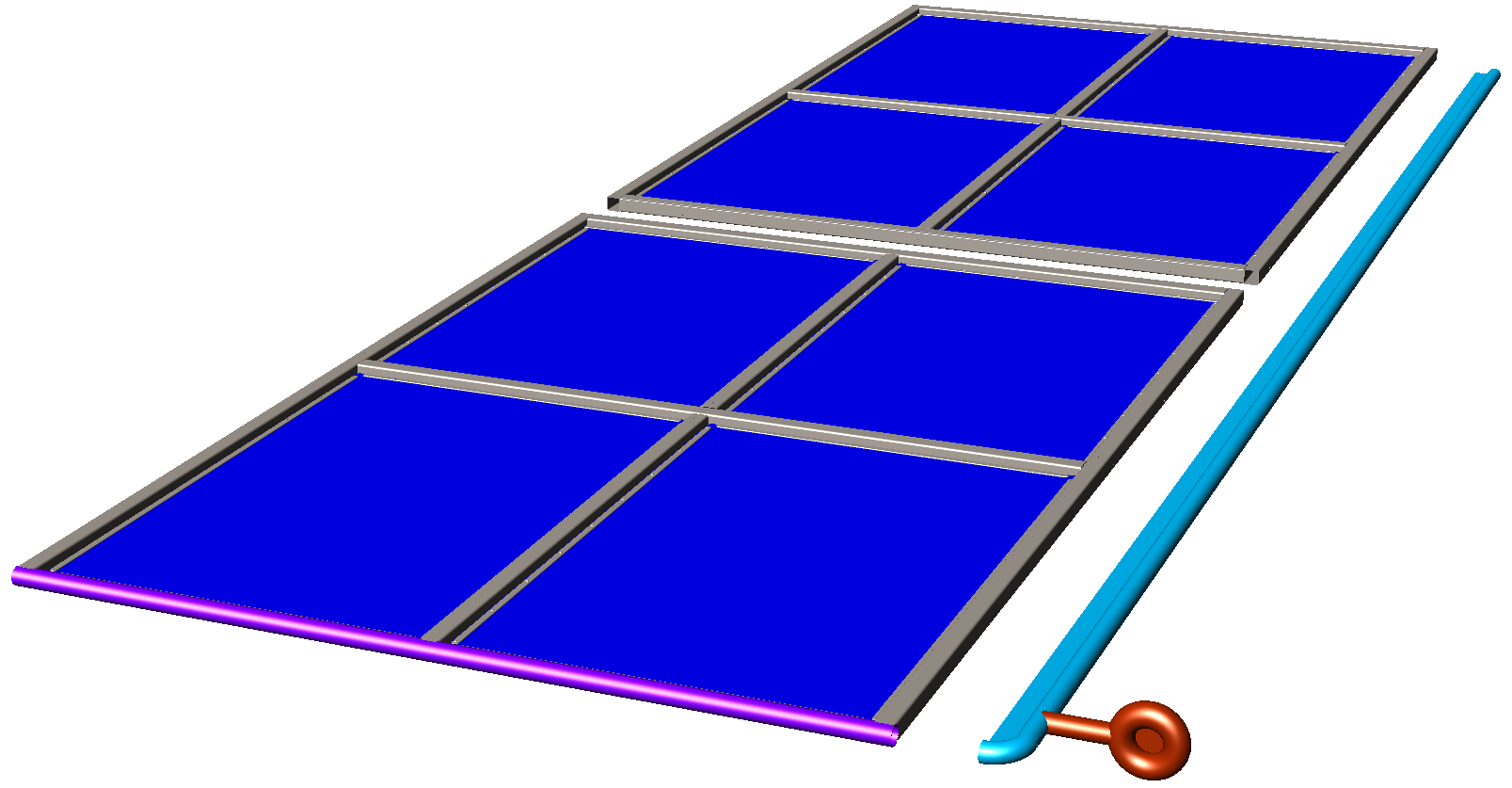}
\end{cdrfigure}

Each CPA is 2.3~m wide (identical to the APA width) and 3~m tall (half
of APA height) for ease of fabrication, assembly and handling.  Each
CPA is made of a stainless steel framework, with panels of solid
stainless steel sheets mounted between the frame openings.  Along each
vertical column of four CPAs, there are two slightly different
versions: the outer CPAs (top and bottom rows), and the inner CPAs
(2$^{nd}$ and 3$^{rd}$ rows).  The inner CPAs use all rectangular
tubes for the frame structure, while the outer CPAs use 5-cm OD round
tubes on the outside edge of the CPA facing the floor or ceiling of
the cryostat to minimize the surface electric field (shown in magenta in
  Figure~\ref{fig:tpc-cathode-model}).  Two sets of field-shaping
end pieces are installed at the two ends (e.g., right and left) of a
CPA wall to properly terminate the cathode wall with rounded edges.
All CPAs are suspended from the ceiling using G10 hangers under
fiberglass rails to insulate the CPAs from the cryostat.

A recent design decision exchanged the positions of the CPAs and APAs
in the detector relative to the earlier LBNE reference design, placing
the APAs adjacent to the cryostat wall instead of the CPAs.  This
change reduces the stored energy on each cathode plane by about 60\%.
Nevertheless, due to the enormous area of the stainless steel cathode
plane, there is still nearly 100~Joules of energy when biased at
180~kV, risking physical damage to the thin membrane structure as well
as to the CPA structure in the event of a high-voltage discharge.  In
addition, in such an event, a huge voltage swing could occur on the
cathode plane in tens of nanoseconds, injecting a charge pulse to the
sensing wires with a large peak current that could damage the
front-end electronics.

To mitigate this risk, analysis of the electrical properties of the
cathode has been carried out with the goal of developing a cathode
design that will substantially slow down the total energy release in
case of a discharge.  The best solution appears to be replacing the
metallic cathode structure by non-conductive materials with a robust
and highly resistive surface coating.  Many choices of 
resistive/anti-static coating and commercially produced
anti-static sheet materials are available.  Studies are underway to identify a
suitable coating and base material for this application.  Since the
electrical current feeding the field cage resistive dividers is
supplied through the cathode, a special current-distribution feature
must be designed to minimize voltage drop along this 58-m-long, highly
resistive structure.

%%%%%%%%%%%%%%%%%%%%%%%%%%%%%%%%
\subsection{Field Cage}
\label{subsec:fd-ref-fieldcage}

In the TPC, each pair of facing cathode and anode planes forms an
electron-drift region. A field cage must completely surround the four
open sides of this region to provide the necessary boundary conditions
to ensure a uniform electric field within, unaffected by the presence
of the cryostat walls.

Each \ktadj{10} detector module requires $\sim$2000~m$^2$ of field
cage coverage. In the current reference design, the field cages are
constructed using multiple copper-clad FR-4 sheets reinforced with
fiber glass I-beams to form modules of 2.3~m $\times$ 3.6~m in
size. Parallel copper strips are etched on the FR-4 sheets using
standard printed circuit board fabrication techniques. Strips are
biased at appropriate voltages provided by a resistive-divider 
network. These strips create a linear electric-potential gradient in
the LAr, ensuring a uniform drift field in the TPC active volume. 
Simulations have shown that the non-uniformity of the drift field quickly
drops to about 1\%, roughly a strip pitch away from the field-cage
surface.

Since the field cage completely encloses the TPC drift region on four
sides, while the solid cathodes block the remaining two, the field
cage sheets must be perforated to allow LAr recirculation in
the middle third of the TPC volume. The ``transparency'' of the
perforation will be determined by a detailed LAr computerized fluid
dynamic (CFD) study.

The resistor-divider network will be soldered directly onto the
field-cage panels. Multiple resistors will be connected in parallel
between any two taps of the divider, in order to provide fault
tolerance.  One end of the divider chain is connected directly to the
cathode, while the other end is connected to ground at the APA through
resistors of the appropriate value.  A pair of field-cage modules will
be pre-attached to the outer CPA modules through hinges, such that the
field-cage modules can be rotated into their final position during
installation, or folded back if aisle access is needed (see
Figure~\ref{fig:tpc-floor-view}).  In addition to the resistor-divider
network, surge suppressors such as varistors or gas discharge tubes
will be installed between the field-cage strips to avoid the
occurrence of an over-voltage condition between field-cage electrodes
and the cathode in the event of a high voltage discharge.

The major challenge of this field-cage design is minimizing the
electric field near the edges of the thin copper strips.  One solution
is to cover all copper edges with a thick layer of solder mask (an
acrylic-based polymer with a high dielectric strength) as part of the
standard PCB fabrication process.  This construction is currently
being implemented in the 35-t prototype TPC (see Section~7.5 of~\cite{cdr-annex-lbne-design}).  Figure~\ref{fig:tpc-field-cage} shows a section of this
partially constructed field cage.
\begin{cdrfigure}[35-t field cage]{tpc-field-cage}{A corner of the 35-t TPC 
field cage during construction}
\includegraphics[width=4in]{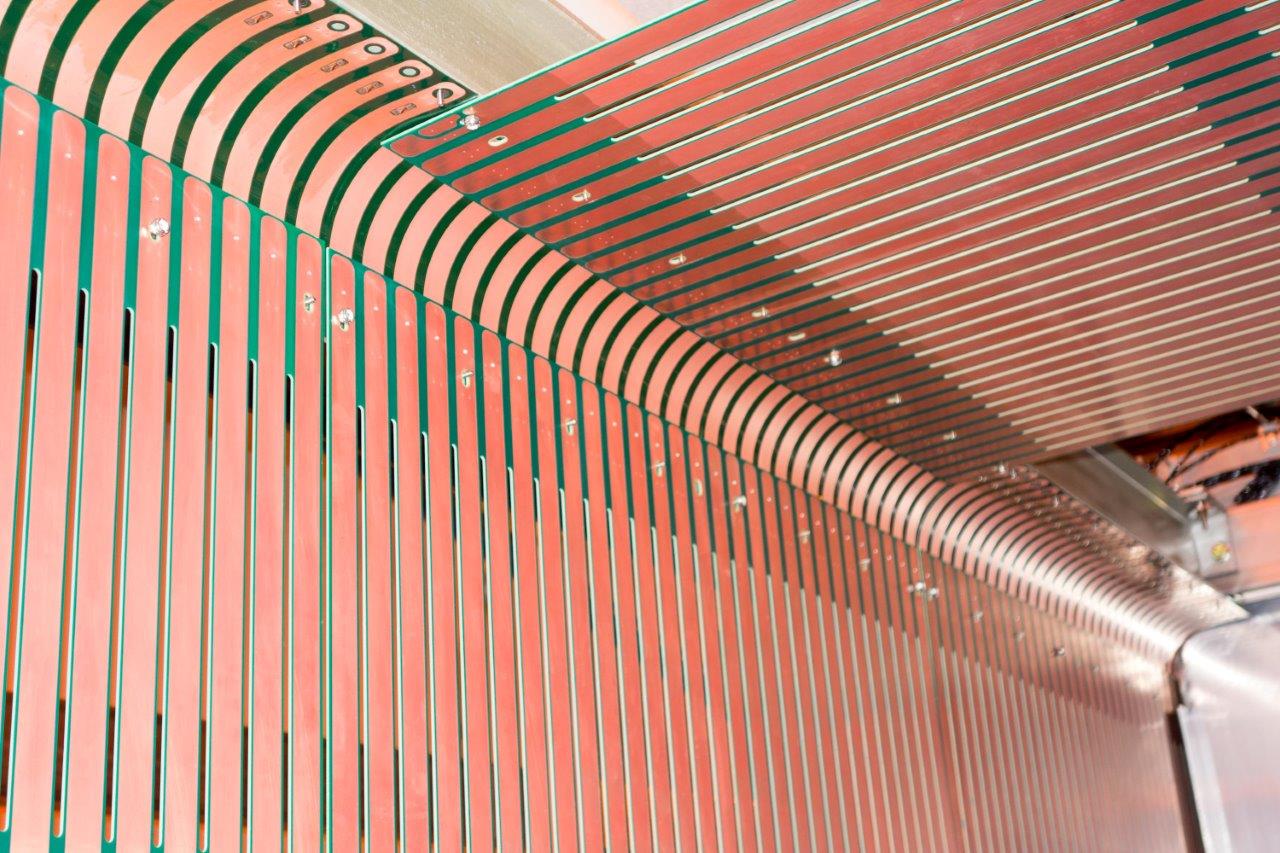}
\end{cdrfigure}
The 35-t prototype test results will be evaluated to determine if this
technique is suitable for the much larger far detector modules.
 
In the meantime, alternative concepts are being actively developed to
further minimize the electric field on the field cage.  One example is
application of a very high-resistance coating on the outside surface
of the field cage such that the surface potential distributes
uniformly across the gaps between conductors and therefore eliminates
the high-field region near the conductor edges.  The challenge here is
ventilation of this field cage surface without significantly
increasing the field at the edge of the perforations.  Another concept
is to use roll-formed metal profiles as the field-cage electrodes and
support them with insulating beams.  These profiles have large edge
radii; this makes their surface electric field relatively low, which
in turn makes it possible to place them even closer to the cryostat
walls to improve the efficiency of LAr use.  A sample profile is shown
in Figure~\ref{fig:tpc-field-cage-roll-form}.
\begin{cdrfigure}[FCA with roll-formed metal profile]{tpc-field-cage-roll-form}
{Left: electrostatic simulation of a field cage design that uses roll-formed 
metal profiles as the field-cage electrodes.  Right: a conceptual design of a 
field-cage module using this profile.}
\includegraphics[width=\linewidth]{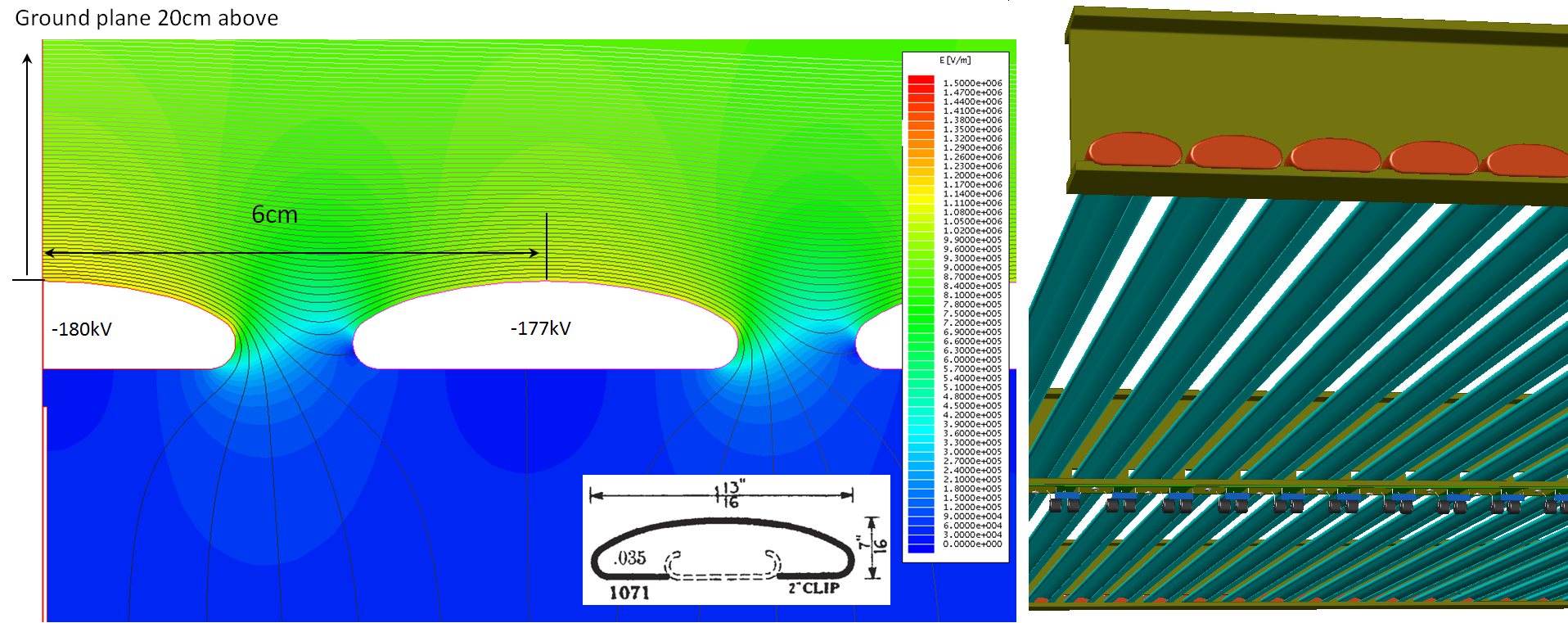}
\end{cdrfigure}
With only a 20-cm distance separating the field cage from a ground
plane, the electric field on the field cage is still under 12~kV/cm.
The ends of the profiles still have high electric field, however; a
possible solution is to cover the ends with UHMW polyethylene caps.
This design may also cost significantly less than the reference design
with PCBs.

%%%%%%%%%%%%%%%%%%%%%%%%%%%%%%%%
\subsection{High Voltage Components}  
\label{subsec:fd-ref-hv}
   
The two cathode planes are biased at $-$180~kV to provide the required
500~V/cm drift field. Each cathode plane will be powered by a
dedicated HV power supply through an RC filter and feedthrough.

The power supplies for the cathode planes must be able to provide
$-$200~kV at 1~mA current. The output voltage ripple must not
introduce more than 10\% of the equivalent thermal noise from the
front-end electronics.  The power supplies must be programmable to
trip (shutdown) their output at a certain current limit.  During power
on and off, including output loss (for any reason), the voltage ramp
rate at the feedthrough must be controllable to prevent damage to the
in-vessel electronics from excess charge injection.  High-voltage
feedthroughs must be able to withstand $-$250~kV at their center
conductors in a 1~atm argon gas environment when terminated in liquid
argon.

The current candidate for the high-voltage power supplies is the
Heinzinger PNC\textit{hp} series, which has the lowest output ripple
specification.  Additional filtering of the voltage ripples is done
through the intrinsic HV cable capacitance and series resistors
installed inside the filter box. Established techniques and practices
will be implemented to eliminate micro-discharges and minimize
unwanted energy transfer in case of an HV breakdown.
  
To ensure safe and reliable operation, the feedthroughs will be tested
at a much higher voltage than expected in routine operation
($\sim$250~kV) in LAr. The feedthroughs will be mounted on
the ceiling of the cryostat, their cold ends reaching through the gas
ullage space and submerging into the LAr. The center
conductor on the cold side of a feedthrough will be insulated and
shielded by a grounded shroud at least 50~cm below the surface of the
liquid to ensure bubble free operation at the
tip. Figure~\ref{fig:tpc-UCLA-feedthrough} shows an example of the
feedthrough and filter box made by the UCLA group for the 35-t prototype TPC,
as well as the conceptual design of a feedthrough suitable for the far
detector TPCs.
\begin{cdrfigure}[Concept of new feedthrough]{tpc-UCLA-feedthrough}
{Top: The high voltage feedthrough and filter developed by the UCLA 
group for the 35-t TPC. The total length of the feedthrough is about 2~m, and its main body has an OD of 3.8~cm.  It was tested up to 150~kV.  
Bottom: a conceptual design of a new feedthrough for the far detector.}
\includegraphics[width=\linewidth]{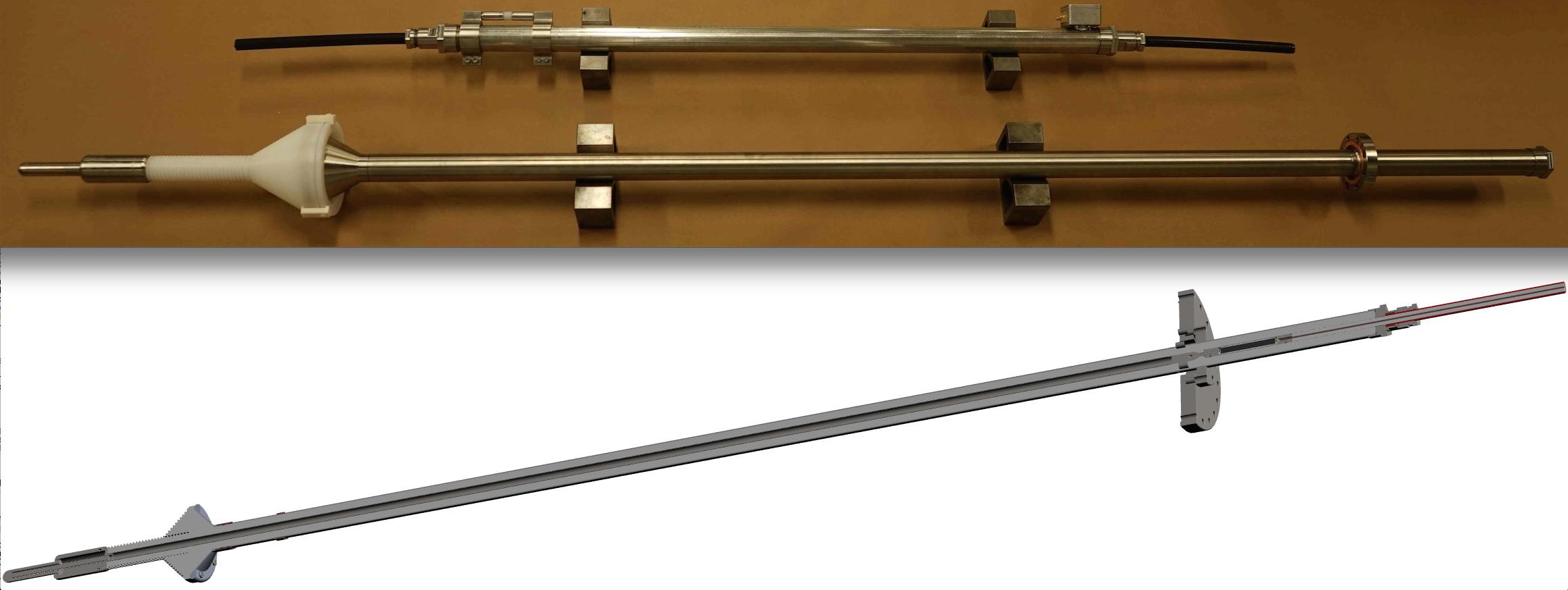}
\end{cdrfigure}

%%%%%%%%%%%%%%%%%%%%%%%%%%%%%%%%
\subsection{TPC Prototyping and Tests}
\label{subsec:fd-ref-tpc-proto}

Several prototype TPC modules were constructed during the LBNE project
design phase.  The initial prototypes were fractional-scale or partial
models of the APA and CPA. The CPA prototype was used to evaluate
field-shaping electrode attachment techniques. A 40\% scale APA
prototype was constructed earlier on to study the placement of the
wire-wrapping boards and wire-support structures. It was also used to
develop the prototype winding machines. The prototypes were subjected
to numerous thermal cycles down to liquid-nitrogen temperature to test
the integrity of the wire-to-board and board-to-frame bonds.

The second set of prototypes are scale models of the APA and CPA. They
are being used to validate the designs and to evaluate production
procedures. These functional prototypes will be installed in the 35-t
prototype cryostat and are expected to be operational in 2015.

A TPC prototype proposed for installation in a CERN test beamline
requires six full-size APAs with fully instrumented readout
electronics, six full-size CPAs and complete field cage coverage. This
prototype will be constructed using identical APAs, CPAs and
field-cage panels as designed for the far detector. Additional
features will be installed to ensure proper TPC operation given the
half-height cryostat configuration. The construction and assembly of
all TPC mechanical components will use the same materials and
techniques as designed for the far detector, with the exception of the
degree of automation for wiring the APAs, which will be reduced.

%% file: volume-detectors/fd-ref-sections/fd-ref-daq.tex
%%%%%%%%%%%%%%%%%%%%%%%%%%%%%%%% 
\section{Data Acquisition (DAQ) System and Monitoring} 
\label{sec:detectors-fd-ref-daq}

The scope of the Data Acquisition (DAQ) system and monitoring includes design, procurement, fabrication, testing,
delivery and installation of the following components (see Figure~\ref{fig:fddaqblock}):
\begin{itemize}
  \item LArTPC detector readout,
  \item Photon detector readout,
  \item Computer farm,
  \item artDAQ software toolkit,
  \item Run Control and Slow Control,
  \item Timing system, and
  \item Calibration system.
\end{itemize}
The DAQ system is required to
\begin{itemize}
\item collect data with a very high uptime (the goal is $>$99\%);
\item collect beam neutrino, atmospheric neutrino and proton
  decay candidates (generally all interactions with total energy deposition
  above about 100~MeV) with a high resolution (smart or no
  zero-suppression) with no dead-time; \fixme{It would be nice to reference a section; `smart' suppression
  is not mentioned in the S\%C chapter; zero supp is. How best to handle?}
\item collect interactions with total energy
  below 100~MeV with some
 low amount of zero-suppression loss;
\item trigger at the time of the beam pulses,
  irrespective of how little energy is deposited in the detector;
\item collect data with the most favorable zero-suppression possible over a
  period of $>$10~s (supernova trigger);
\item assemble data from sub-detectors into a unified
  event for offline analysis; and
\item provide access to the shift operator and others to control and
  monitor the data collection and detector status, view online
  histograms and monitor (and provide for offline use) historical
  status of measured detector parameters. 
\end{itemize}

This section outlines a conceptual design intended to meet the
required performance for the DAQ for the DUNE far detector.  To reduce
considerably the times when none of the far detector modules are collecting
data (particularly important for supernova detection) and to allow
the designs of the DAQ in the different \ktadj{10} modules to be entirely
different if desired, the DUNE DAQ employs a decoupled design.
%\fixme{(but less coupled or more uncoupled to/from what? -AH, original phrasing: a less coupled design than has ever been used in data acquisition before is employed. } 
% GB - I am trying to say the parts within the DAQ are less coupled
% with each other
The synchronization, triggering, data collection and run-state in the
different \ktadj{10} modules are completely independent in real time and
are only coupled by processes running asynchronously (in the same fashion
as a batch queue) in the hour or so after data collection (see Section
4.11 of~\cite{cdr-annex-lbne-design}).  This allows one %portion of the
detector module to be restarted without interrupting data collection in the
others. %rest.

The layout of the DAQ is shown schematically in
Figure~\ref{fig:fddaqblock}.  
\begin{cdrfigure}[DAQ subsystem block diagram]{fddaqblock}{Block diagram layout of the main
    components of the DAQ subsystem.}
\begin{tikzpicture}[
  every matrix/.style={ampersand replacement=\&,column sep=0.4cm,row sep=0.6cm},
  to/.style={->,>=stealth',shorten >=1pt,semithick,font=\sffamily\footnotesize},
  data/.style={thick},
%  box/.style={draw,thick,rounded corners,fill=yellow!20,inner sep=.3cm},
  box/.style={draw,inner sep=.1cm},
  boxa/.style={box,align=center}]

% Lay the nodes out using a matrix
  \matrix{   
% 1st row
  \node[boxa] (ce) {{\rm Cold}\\{\rm electronics}};
 \&
 \& \node[boxa] (rce) {{\rm RCE}\\{\rm LArTPC data}\\{\rm processors}};
 \&
 \& \node[boxa] (trigfe) {{\rm Trigger}\\{\rm front-end}\\{\rm computers}};
 \& \node[boxa] (trigbe) {{\rm Software}\\{\rm trigger}\\{\rm farm}}; \\

% 2nd row
 \& \node[box,inner sep=0.4cm] (fthru) {}; 
 \& \node[boxa] (time) {{\rm Time}\\{\rm sync}};
 \&
 \& \node[boxa] (sc) {{\rm Slow}\\{\rm control}};
 \& \\

% 3rd row
  \node[boxa] (pd) {{\rm Photon}\\{\rm detectors}};
 \&
 \& \node[boxa] (ssp) {{\rm SSP}\\{\rm Photodetector}\\{\rm digitizers}};
 \&
 \& \node[boxa] (datafe) {{\rm Data}\\{\rm front-end}\\{\rm computers}};
 \& \node[boxa] (databe) {{\rm Full-data}\\{\rm collection}\\{\rm farm}}; \\
 };

\coordinate (fthruWL) at ($ (fthru.south west)!0.3!(fthru.north west) $);   % West Low edge of fthru 
\coordinate (fthruWH) at ($ (fthru.south west)!0.7!(fthru.north west) $);   % West High edge of fthru 
\coordinate (fthruEL) at ($ (fthru.south east)!0.3!(fthru.north east) $);   % East Low edge of fthru 
\coordinate (fthruEH) at ($ (fthru.south east)!0.7!(fthru.north east) $);   % East High edge of fthru 

\coordinate (trigfeWL) at ($ (trigfe.south west)!0.3!(trigfe.north west) $);   % West Low edge of trigfe 
\coordinate (trigfeWH) at ($ (trigfe.south west)!0.7!(trigfe.north west) $);   % West High edge of trigfe
\coordinate (datafeWL) at ($ (datafe.south west)!0.3!(datafe.north west) $);   % West Low edge of datafe
\coordinate (datafeWH) at ($ (datafe.south west)!0.6!(datafe.north west) $);   % West High edge of datafe

\coordinate (rceEL) at ($ (rce.south east)!0.3!(rce.north east) $);  % East low edge of RCE 
\coordinate (rceEH) at ($ (rce.south east)!0.7!(rce.north east) $);
\coordinate (sspEL) at ($ (ssp.south east)!0.3!(ssp.north east) $);  % East low edge of SSP
\coordinate (sspEH) at ($ (ssp.south east)!0.8!(ssp.north east) $);

\coordinate (rceSR) at ($ (rce.south)+(0.7cm,0) $);  % South right edge of RCE 
\coordinate (sspNR) at ($ (ssp.north)+(0.7cm,0) $);  % North right edge of RCE 
\coordinate (scWL) at ($ (sc.south west)!0.3!(sc.north west) $);  % West low edge of SC 
\coordinate (scWH) at ($ (sc.south west)!0.7!(sc.north west) $);  % West high edge of SC 

\draw[data] (ce) -| ($(fthruWH)-(0.1cm,0)$) -- (fthruWH);
\draw[data] (pd) -| ($(fthruWL)-(0.1cm,0)$) -- (fthruWL);
\draw[data] (fthruEH) -- ($(fthruEH)+(0.1cm,0)$) |- (rce);
\draw[data] (fthruEL) -- ($(fthruEL)+(0.1cm,0)$) |- (ssp);

\draw[data] (rceEH) -- ($ (rceEH)+(0.40cm,0) $) |- (trigfeWH); 
\draw[data] (rceEL) -- ($ (rceEL)+(0.40cm,0) $) |- (datafeWH); 
\draw[data] (sspEH) -- ($ (sspEH)+(0.55cm,0) $) |- (trigfeWL); 
\draw[data] (sspEL) -- ($ (sspEL)+(0.55cm,0) $) |- (datafeWL); 

\draw[data] (trigfe) -- (trigbe);
\draw[data] (datafe) -- (databe);

\draw[data,dashed] (time) -- (rce);
\draw[data,dashed] (time) -- (ssp);

\draw[data,dashed] (scWH) -| (rceSR);
\draw[data,dashed] (scWL) -| (sspNR);

\draw[data,dotted] (fthru.north) -- ($(fthru.north)+(0,2.7cm)$) node [left] {\rm In cryostat} node [right] {\rm Room temp.};
\draw[data,dotted] (fthru.south) -- ($(fthru.south)-(0,2.7cm)$);
\end{tikzpicture}
\end{cdrfigure}
To collect full detail of the
most important events (deadtimeless detection of all those with energy
deposition of 100~MeV and above), while avoiding tricky communication
to flag neighboring channels to those that have large hits, a model is used
in which the digitized data are collected twice from the initial data
storage that is associatated with each peice of hardware. 
% GB - I can't think of anything better, unless we expand this section
% a lot, which whould give it too much emphasis.  Ideally we DO want
% to flag neighboring channels to those with big hits, but it is
% difficult because it would entail a big complicated network of
% comunications between all the boards which is very tricky.
%\fixme{Previous sentence not clear: you want to flag channels that have large hits and NOT flagneighboring channels?  What's the `tricky communication'?}

The first collection
supplies a centralized software-based trigger farm continuously with
zero-suppressed information (a threshold on each channel detects hits
above the noise level; this causes the ADC samples to be kept in a
time window around the hit).  
The second collection reads the full data at the times and
in the regions %of interest that have been 
selected by the
trigger. This is similar to the multi-level triggering in a collider experiment,
but with only one level of triggering.
%\fixme{in other words, here we have one trigger + two collections; in
%  collider expt, two of each? Please clarify sentence.} 
The software trigger farm also
records the continuous stream of zero-suppressed hits in an archival
ring buffer on disk to allow later analysis of supernova candidates.

{\bf LArTPC detector readout} The LArTPC signals are digitized in the
cold in a
continuous flash-ADC stream at 2~MHz, (not zero-suppressed) and
serialized on 12,000 high-speed links per \ktadj{10} module that exit the
cryostat.
% GB - agreed, lets just leave this stuff about costs out
% (costed separately in the cold-electronics WBS). \fixme{Do we want to refer to costs here?
%And which thing falls under CE, the links? Maybe say `(the links are
%covered in Section~\ref{sec:detectors-fd-ref-ce})'} 
The data
are received by LArTPC data processors called RCEs (Reconfigurable
Computing Elements) %which are 
housed in
industry-standard aTCA crates on COB (cluster-on-board) motherboards that are designed at SLAC
for a wide range of applications.  The RCEs are part of a network of
field programmable gate arrays (FPGAs) that buffer the full raw data,
zero-suppress it for passing to the trigger and accept requests for
data-fetching from the trigger.  The FPGAs in the RCEs are from the
Xilinx Zynq family and contain a full Linux processor system on the
chip.  They facilitate the high-speed data transfer from firmware into
DRAM memory that is accessible from Linux.  A fast data-transfer
network using the Ethernet protocol is used on the COBs and in the
aTCA crates to allow for development of more sophisticated zero-suppression algorithms
for improved supernova acquisition.

{\bf Photon detector readout} is performed by the SiPM signal
processor (SSP) described in
Section~\ref{sec:detectors-fd-ref-pd-refsystem}.  The additional
buffering required for the separate trigger and data collection paths
is implemented in the SSP front-end computers.

{\bf Computer farm} Both the LArTPC and photon detector data are
received in commodity Linux computers, with no deadtime, from where
the data are routed to the trigger and full-data collection farms of
computers.  Although the front-end computers are logically distinct as
shown in Figure~\ref{fig:fddaqblock}, one physical computer is
sufficient for all the processes for each rack of APA readout
electronics. 

{\bf artDAQ software toolkit} (from Fermilab) supplies the real-time
data collection functionality (buffering, matching event parts,
synchronization, inter-process communication, etc.) in a modern
design that facilitates the efficient use of multicore commodity
computers running on the Linux computer farms.  The multicore
functionality is crucial in the high data-rate environment on DUNE.  
The detector-specific
code is supplied by DUNE groups (this is centered in the UK), along
with detector-specific triggering (also UK).  The architecture
provided by artDAQ can be tailored for each experiment and is entirely
suitable for the ``collect-twice'' architecture envisioned for DUNE.

{\bf Run Control and Slow Control software framework} manages the
control, status display and status archival of the experiment.  It is
based on the design of the DAQ of the ICECube experiment\cite{Abbasi:2008aa}. This
exploits a combination of readily available, well-supported packages
[message passing (zeroMQ), a web-framework (Django), databases (postgres)] 
to give the shift operator a unified view of the status
of the running experiment, and views of the monitoring data including
customized views and historical views.  The database of slow-control
measurements is exported to the host lab to give access for offline
programs.

{\bf Timing system} To acomplish the software-based deadtimeless data
acquisition in DUNE, it is necessary to synchronize the clock across
all readout boards.  This is accomplished in two stages. The main
cavern-wide distribution uses the design from the NO$\nu$A experiment,
which distributes a 64-MHz clock, synchronization pulses and
cable-delay correction on RJ45 cables.  The overall clock is
synchronized using a GPS receiver and transferred from the surface
over optical fiber.  The synchronization is transferred from the COBs
to the electronics in the cryostat over the same cabling as provides
the data links.

{\bf Calibration system} The calibration is done in three stages: (1)
Pedestal and charge-injection pulser events are used to calibrate and
remove drifting of the individual electronic channel responses. (2)
External UV lasers are directed into the cryostat through glass-tube
ports, and swivelling mirrors select the trajectories of the beams
through the cryostat to provide calibration of the field
non-uniformities and attenuation. (3) Cosmic-ray muons are used for
the determination of the energy scale and for calibration
cross-checks.

\begin{cdrfigure}[DAQ time diagram]{fddaqtime}{Main DAQ steps displayed relative to the event time.}
\begin{tikzpicture}[scale=0.81,
  every matrix/.style={ampersand replacement=\&,column sep=0.4cm,row sep=0.6cm},
  to/.style={->,>=stealth',shorten >=1pt,semithick,font=\sffamily\footnotesize},
  timeline/.style={thick},
  transline/.style={dashed,thick},
  pointline/.style={thick},
  proline/.style={thick},
  driftline/.style={shorten <=1pt,decorate,decoration={snake,pre length=4pt}},
  data/.style={thick},
%  box/.style={draw,thick,rounded corners,fill=yellow!20,inner sep=.3cm},
  box/.style={draw,inner sep=.1cm},
  boxa/.style={box,align=center}]

\draw [timeline] (-8.2,0) -- (8.5,0);
\foreach \i in {-8,-6,-4,-2,2,4,6,8}
  \draw [timeline] (\i,0.05) -- (\i,-0.3) node [below] {$10^{\i}$};
\foreach \i in {-7,-5,-3,-1,1,3,5,7}
  \draw [timeline] (\i,0.05) -- (\i,-0.2);
\draw [timeline] (0,0.1) -- (0,-0.3) node [below] {$1\mathrm{s}$};

\draw [pointline,->] (-8.6,6) node [right,align=left] {{\rm Event}\\{\rm time}} -- (-8.6,0.2);
\draw [timeline] (-8.6,0) -- (-8.3,0);
\draw [timeline] (-8.28,0.1) -- (-8.32,-0.1);
\draw [timeline] (-8.18,0.1) -- (-8.22,-0.1);

\draw [pointline,->] (1.77815,0.5) node [above,font=\small] {\rm 1min} -- (1.77815,0.1);
%\draw [pointline,->] (2.77815,-0.9) node [anchor=north] {$10\mathrm{m}$} -- (2.77815,-0.1);
\draw [pointline,->] (3.5563,0.5) node [above,font=\small] {\rm 1h} -- (3.5563,0.1);
%\draw [pointline,->] (4.5563,-0.9) node [anchor=north] {$10\mathrm{h}$} -- (4.5563,-0.1);
\draw [pointline,->] (4.9365,0.5) node [above,font=\small] {\rm 1d} -- (4.9365,0.1);
\draw [pointline,->] (7.4988,0.5) node [above,font=\small] {\rm 1y} -- (7.4988,0.1);

\draw [pointline,->] (-7.8061,0.5) node [above,font=\footnotesize] {\rm 64MHz} -- (-7.8061,0.1);
\draw [proline,|-|] (-7.8,1.7) -- node [above,font=\small] {\rm Form PD pulses} (-4,1.7);
\draw [proline,|-|] (-4,1.3) -- node [below,font=\small] {\rm Form PD packets} (-2.5,1.3);
\draw [proline,|-|] (-3,1.7) -- node [above,font=\small] {\rm Ring buffer store} (2,1.7);

\draw [driftline] (-6.3,5.4) -- node [above,font=\small] {\rm LArTPC drift time} (-2.3,5.4);

\draw [pointline,->] (-6.3010,0.5) node [above,font=\footnotesize] {\rm 2MHz} -- (-6.3010,0.1);
\draw [proline,|-,dotted] (-6.3010,4.4) -- (-5,4.4);
\draw [proline,-|] (-5,4.4) -- node [above,pos=0.2,font=\small] {\rm ZS \& packetize} (-2.5,4.4);
\draw [proline,-|] (-5,3.7) -- node [below,pos=0.3,font=\small] {\rm Non-ZS ring buffer} (-0.6,3.7);
\draw [proline,|-,dotted] (-6.3010,3.7) -- (-5,3.7);

\draw [transline,<-] (-1.5,7.95) .. controls (-1.5,5.4) .. node [below,sloped,pos=0.35,font=\scriptsize] {} (-2.5,4.4);  
\draw [transline,<-] (-1.5,7.95) .. controls (-1.5,2.3) .. node [below=-3pt,sloped,pos=0.18,font=\scriptsize] {\rm send to trigger} (-2.5,1.3);  
\draw [transline,->] (-2.5,7.4) node [anchor=east,align=right,font=\footnotesize] {{\rm Beam spill}\\{\rm time updated}}.. controls (-1.0,7.4) .. (-1.0,8.4);  
\draw [proline,|-|] (-1.5,8.4) -- node [above,font=\small] {\rm Trigger} (-0.8,8.4);
\draw [proline,|-|] (-1.5,8.1) -- node [below,font=\small] {\rm Offline archive of trigger hits} (7.4,8.1);

\draw [dotted,ultra thick] (-0.8,8.1) node [right,font=\small] {} -- (-0.8,0.);

\draw [proline,|-|] (-0.3,6.) -- node [below,font=\small] {\rm Module file} (3.1,6.);
\draw [proline,|-] (3.1,7.) -- node [below,pos=0.5,font=\small] {\rm Merged file} (8.5,7.);
\draw [transline,->] (3.1,6.) -- (3.1,6.85);

\draw [transline,<-] (-0.3,5.85) .. controls (-0.3,2.7) .. node [below=-3pt,sloped,pos=0.25,font=\scriptsize] {\rm send data} (-0.8,1.7);  
\draw [transline,<-] (-0.3,5.85) .. controls (-0.3,4.7) .. node [above,sloped,pos=0.5,font=\small] {} (-0.8,3.7);  

\end{tikzpicture}
\end{cdrfigure}
The time sequence, trigger deadlines and buffering of the readout are all
shown in Figure~\ref{fig:fddaqtime}.  This figure shows time in the
horizontal direction on a logarithmic scale to indicate how long after
the particles appear in the detector each process can start and must
finish.  There is one level of triggering, with a trigger deadline
(latest time a trigger decision can arrive) of
0.16\,s after the event has occurred, indicated by the vertical dotted
line.  This is long enough that in 99\% of the cases, a message will
arrive from the Fermilab site to update the predicted time of a
neutrino spill with the actual time in order that the detector can be
triggered independently of any signals in the detector.  This
operation has been used successfully on MINOS for many years.  

As shown in Figure~\ref{fig:fddaqtime}, prior to the trigger time and
independently in each detector module, the detector readout assembles
packets of data corresponding to fixed time intervals and sends them
to the trigger event builders.  Both the LArTPC readout and the SSP
readout send zero-suppressed data, suitable for triggering, to a set
of event-builder processes that run in parallel, each accepting all
the data from the entire \ktadj{10} module for a specific time interval and
performing trigger algorithms on them.  All time intervals are
processed so that the trigger has no dead time.  The data from the LArTPC and SSP
are also stored in ring buffers
  to await collection after an
affirmative trigger decision --- in the case of the LArTPC, this data
is not zero-suppressed.  The data are built into events in a ``10kt-module
file'' that is written to disk.  About an hour later, an offline process merges
the data from the separate \ktadj{10} module files and archives them at the host lab.

To maximize data collection for a supernova, the continuous
zero-suppressed trigger data is kept in a large buffer area on disk.
%Ways to collect data that are cropped more gently than the zero-suppressed 
%data stream during the extended period of a supernova are
%under study: 
During the extended period of this event, it is desirable to collect data that are cropped 
more gently than the zero-suppressed data stream. %; ways to do this are under study: 
The non-zero-suppressed ring buffers on the current
design of RCEs are
sufficient for 0.4~s of buffering; during a supernova, the trigger farm can 
send
instructions in the trigger messages to manage storage of data in
these buffers.  Two possible extensions to this method
are (a) extend the memory to buffer longer, either on the RCEs or
elsewhere in the aTCA crate, or (b) use the powerful intercommunications between RCEs to ensure that  information read from neighboring channels around the time of a
potential supernova event candidate is kept.

At the 4850L, the data rate in the trigger is dominated by the unwanted 
radioactivity from $^{39}$Ar and $^{85}$Kr.  The cosmic rays that
occur about once per minute, on the other hand, are the major source of events that should
be collected on the data stream; 
 the physics beam events, atmospheric
neutrinos and other candidate events 
are included in this data budget.  
Table~\ref{tab:daqrefrates}) gives estimates of the
rate of occurence of these events and the expected steady-state data rate
(after the derandomization provided by the buffering in the front-end
computers).  
\begin{cdrtable}[Estimated data rates]{lcc}{daqrefrates}{Estimated data
    rates in the DAQ system}  %The third argument (reads {cc}) can use c, l, r or p{some length} % but please do not include lines like “|c|l|l|”. It CAN look like {cll} or {llp{3cm}}, for instance.
Process & Rate (kHz/APA) & Data rate (MBytes/s) \\ \toprowrule
Generic 2.3\,ms interval & 0.43 & 6,000\\ \colhline
Cosmic-ray muons (4850L) & $6\times 10^{-7}$ & $1\times 10^{-5}$ \\ \colhline
Radioactivity & $\sim 64$ & 1.9 \\ \colhline
Electronics noise (not common mode) & $\sim 1$ & 0.03 \\
\end{cdrtable}
The requirements are met in this conceptual design by
the very high throughput provided by modern off-the-shelf components 
and the parallelism provided by artDAQ, which
makes the design extendable to avoid bottlenecks.  The triggering
allows the zero-suppression to be tuned to optimize for the final
level of noise, while retaining the maximum level of information for
the important physics events.

This DAQ design is being prototyped for the 35-t tests in 2015 (see Section~\ref{sec:proto-35t-ph2}), with
two COBs (containing 16 RCEs) and eight SSPs.  The artDAQ software
toolkit is being used to implement the readout, event-building and
triggering, although the ``collect-twice''
model will not be implemented for the 35-t test.

%Here is a sample table:
%

%% file: volume-detectors/fd-ref-sections/fd-ref-ce.tex
%%%%%%%%%%%%%%%%%%%%%%%%%%%%%%%% 
\section{The Cold Electronics (CE)} 
\label{sec:detectors-fd-ref-ce}

The TPC read-out electronics are mounted on the APA front-end
(Figure~\ref{fig:elec_CMBonAPA}) in LAr.  
\begin{cdrfigure}[The front-end electronics as mounted on an APA]{elec_CMBonAPA}
 {The front-end electronics, shown in the red circle, as mounted on an APA. (Note that this figure
 was not updated to show the current photon detection system scheme.)}
\includegraphics[width=0.70\linewidth]{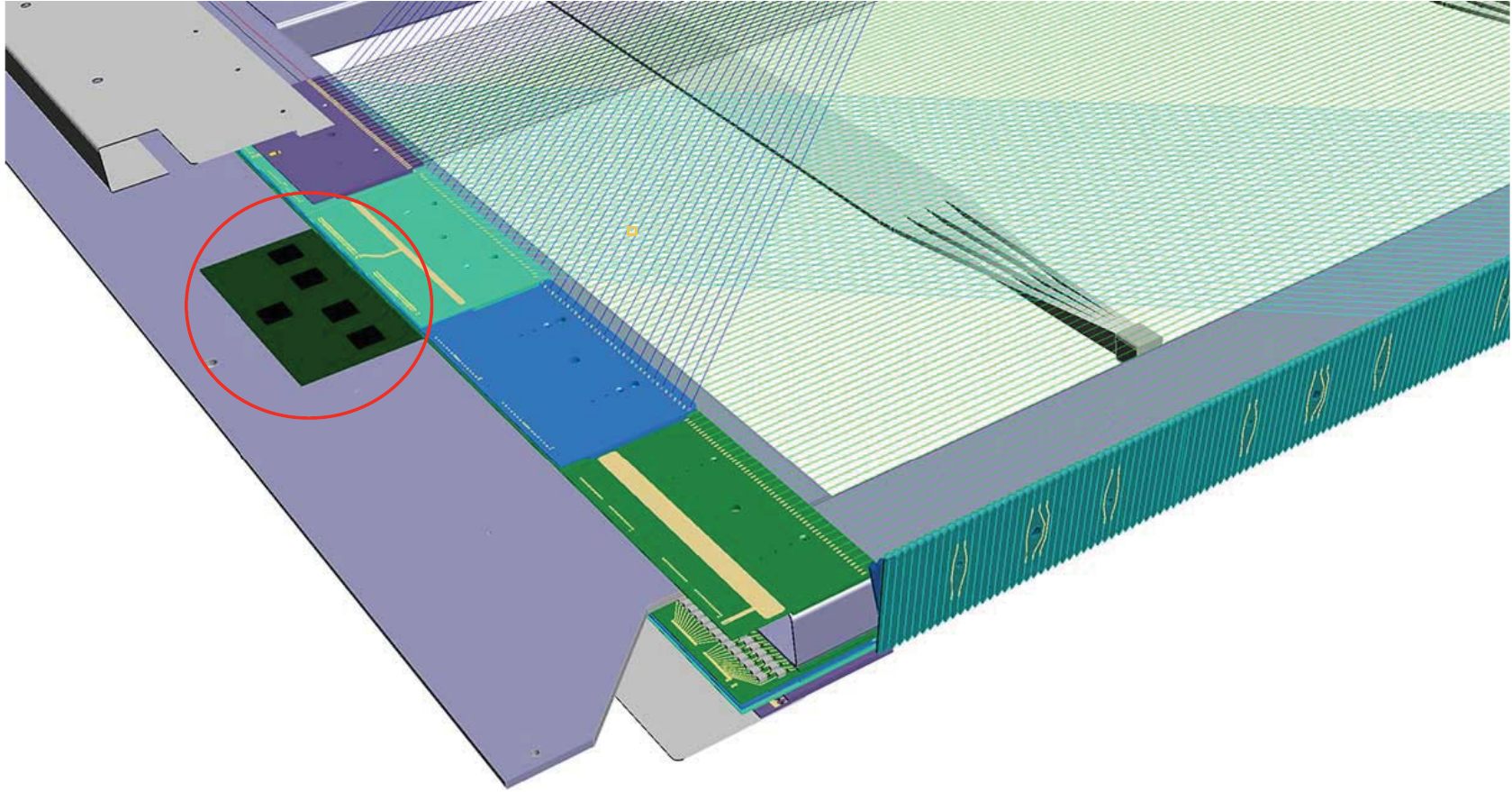}
\end{cdrfigure}
These electronics are
therefore referred to as the ``Cold Electronics'' (CE) subsystem.  The
scope of the CE subsystem includes design, procurement,
fabrication, testing, delivery and installation of
\begin{itemize}
\item front-end electronics cards installed on the APAs,
\item signal feedthroughs mounted on the cryostat,
\item power supplies and cabling.
\end{itemize}
The following are the most significant requirements for the CE:
\begin{itemize}	
\item minimize channel capacitance and noise,
\item minimize dissipated power per channel in the LAr,
\item read out the TPCs and transmit their data to the DAQ,
\item record the channel waveforms continuously without dead time,
\item provide sufficient precision and range to satisfy the Key Physics Parameters,
\item operate for the life of the facility without significant loss of function, and
\item use only materials that are compatible with high-purity liquid argon and that have minimal natural radioactivity.
\end{itemize}

The CE are implemented as ASIC chips using CMOS technology, which
performs well at LAr temperatures\cite{ThornEtAl:CELAr}, and provides
amplification, shaping, digitization, buffering and multiplexing (MUX)
of the signals.  The CE architecture is manifested in the Front End
Mother Board (FEMB), a 128-channel board which uses eight 16-channel
Front End (FE) ASICs and eight 16-channel ADC ASICs
(Figure~\ref{fig:elect_schem}).  
\begin{cdrfigure}[The Cold Electronics architecture]{elect_schem}
{
  The CE architecture. The basic unit is the 128-channel FEMB. FEASIC: Front End ASIC.
  PA: Pre-amplifier.  $\int$: Shaper.  MUX: Multiplexer.  Drv: Driver.
}
\includegraphics[width=0.80\linewidth]{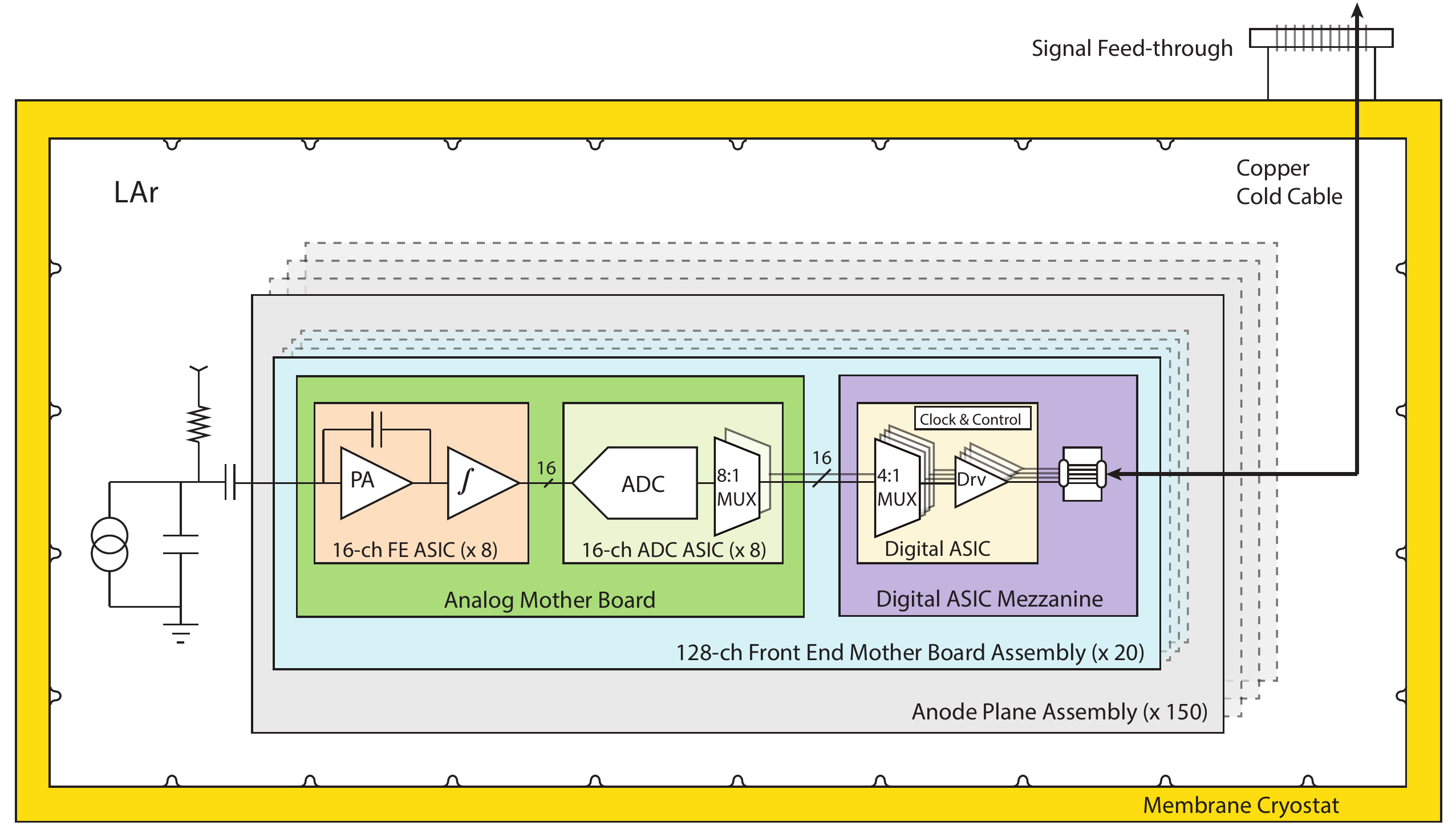}
\end{cdrfigure}
The FE ASIC provides amplification
and pulse shaping, while the ADC ASIC comprises a 12-bit digitizer and
an 8:1 MUX stage with two pairs of serial readout lines in parallel.
A Cold Digital Data (COLDATA) ASIC chip
(Figure~\ref{fig:elec_COLDATAfig}) mounted on each FEMB provides an
additional MUX of 4:1 and is capable of driving the data at 1~Gb/s
through 30~m of copper cable to the feedthrough and on to the DAQ.
Tables~\ref{tab:DeviceCounts} and \ref{tab:KeyParameters} list the CE
device counts and key parameters.
\begin{cdrfigure}[Functional block diagram of the COLDATA ASIC]{elec_COLDATAfig}
{Functional block diagram of the COLDATA ASIC. PLL: Phase Locked Loop.  VCXO: Voltage-Controlled Crystal Oscillator}
\includegraphics[width=0.70\linewidth]{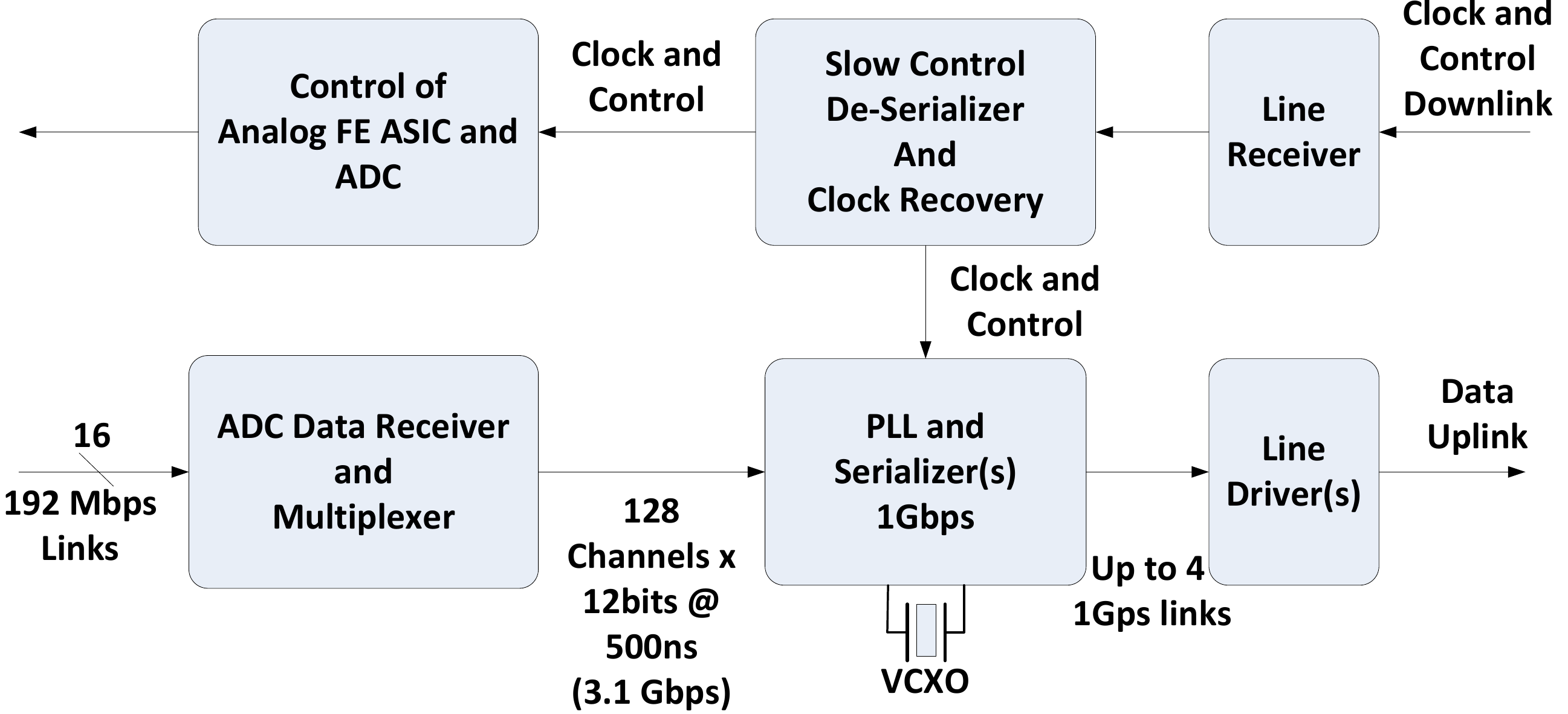}
\end{cdrfigure}
%\vskip -10pt % This prevents the last table from appearing in the PD section by pulling the figure up, which is easily accomodated.
\begin{cdrtable}[Cold Electronics device counts]{lrrr}{DeviceCounts}{CE device counts}
 Parameter           & per FEMB& per APA& per 10~kt Detector\\ \toprowrule
 Channels            & 128    & 2560   & 384,000           \\ \colhline
 FE \& ADC ASIC Chips&   8    &  160   &  24,000           \\ \colhline
 COLDATA ASIC Chips  &   1    &   20   &   3,000           \\ \colhline
 FEMB                &   1    &   20   &   3,000           \\ \colhline
 Signals-Out         &   4    &   80   &  12,000           \\ \colhline
 APA                 & ---    &    1   &     150           \\
\end{cdrtable}
\begin{cdrtable}[Cold Electronics key parameters]{ll}{KeyParameters}{CE key parameters}
 Parameter                &  Value                               \\ \toprowrule
 Signal-to-Noise (in LAr) &  9:1 for 1~$\mu$s peaking time       \\ \colhline
 MUX Level                &  32                                  \\ \colhline
 Sampling Frequency       &  2~MHz                               \\ \colhline
 ADC Resolution           &  12~bits                             \\ \colhline
 FE Peaking Time          &  0.5, 1, 2, 3~$\mu$s (selectable)    \\ \colhline
 FE Gain                  &  4.7, 7.8, 14, 25~mV/fC (selectable) \\ \colhline
 Calibration Precision    &  1\%                                 \\ \colhline
 Power dissipation        &  11~mW/channel                       \\
\end{cdrtable}

An important aspect of CMOS technology is that the lifetime at
cryogenic temperatures is well understood and can be well controlled,
{\em provided that close control is maintained over the implementation
  details}.  This precludes the use of commercial devices, which are
not intended for use in LAr, and which are produced by proprietary
processes over which DUNE has no control.
% This requirement of close control cannot be satisfied by any commercial device,
% where the manufacturing process is subject to change.
% Strict requirements in industry to deliver a device with equivalent performance and lifetime,
% despite a change in the proprietary manufacturing process or technology,
% only applies to a range of temperatures which does {\em not} include LAr temperatures.
% Performance and accelerated-lifetime testing of commercial devices will not protect against 
% a substantially different performance or shorter lifetime in the cold following a manufacturing process change
% over which we have no control, nor necessarily any knowledge.
% It is because of this serious concern that we are undertaking the development and fabrication of our own devices, 
% so that tight control of the process can be maintained.
It is worth noting that the FEMB, together with the FE and ADC ASIC
chips, has already been prototyped and tested using a commercial FPGA
to perform the role of the COLDATA ASIC, which is currently under
development.  The 1-Gb/s data rate can be achieved with copper links
and without zero-suppression or data compression.
This greatly reduces the complexity of the COLDATA ASIC, with a
corresponding decrease in overall risk, including risk of
failure-to-implement (within a fixed schedule and budget).  The
COLDATA work is especially challenging, with final production not
scheduled to begin until late 2019.  Alternative approaches are
currently under study.

Good reliability of cold electronics has been achieved in several
previous experiments. More than 8,000 cold electronics boards are used
to read out $\sim$180,000 channels in the ATLAS Liquid Argon
Calorimeters with a failure rate of $\sim$0.02\% in 11 years of
operation so far. The failure rate is less than 0.2\% for $\sim$13,000
channels of cold preamplifiers installed in the Liquid Krypton
Calorimeter of the NA48 experiment for 17 years of operation. Lifetime
studies of CMOS cold electronics for DUNE have been carried out to
understand the aging mechanism due to channel hot carrier
effects\cite{Li:CELAr}.  The design of the analog front end and ADC
ASICs in 180-nm CMOS technology follows these design rules to secure
lifetimes two orders of magnitude higher than the 30 years of
experiment lifetime. 

%% file: volume-detectors/fd-ref-sections/fd-ref-pd.tex
%%%%%%%%%%%%%%%%%%%%%%%%%%%%%%%%
\section{The Photon Detection System}
\label{sec:detectors-fd-ref-pd}

The scope of the photon detector (PD) system for the DUNE far detector
reference design includes design, procurement, fabrication,
testing, delivery and installation of the following components:
\begin{itemize}
\item light collection system including wavelength shifter and light guides,
\item silicon photo-multipliers (SiPMs),
\item readout electronics,
\item calibration system, and
\item related infrastructure (frames, mounting boards, etc.).
\end{itemize}

LAr is an excellent scintillating medium and the photon detection
system will exploit this property in the far detector.  With an
average energy of 19.5~eV needed to produce a photon (at zero field),
a typical particle depositing 1~MeV in LAr will generate
40,000~photons with wavelength of 128~nm. At higher fields this will
be reduced, but at 500~V/cm the yield is still $\sim$20,000~photons
per MeV. Roughly 1/4 of the photons are promptly emitted with a
lifetime of about 6~ns while the rest have a lifetime of
1100--1600~ns. Prompt and delayed photons are detected in
  precisely the same way by the photon detection system. LAr is
highly transparent to the 128-nm VUV photons with a Rayleigh
scattering length of (66~$\pm$~3)~cm~\cite{Rayleigh} and absorption
length of $>$200~cm; this attenuation length requires a LN2
  content of less than 20~ppm. The relatively large light yield makes
the scintillation process an excellent candidate for determination of
$t_0$ for non-beam related events. Detection of the scintillation
light may also be helpful in background rejection and triggering on
non-beam events.

The photon detection system reference design described in this section
meets the required performance for light collection for the DUNE far
detector. This includes detection of light from proton decay
candidates (as well as beam neutrino events) with high efficiency to
enable 3D spatial localization of candidate events. The TPC will
provide supernova neutrino detection. 
The photon system will provide the $t_0$ timing of
events relative to TPC timing with a resolution better than 1~$\mu$s
(providing position resolution along drift direction of a couple of mm). 

Alternative photon detector designs are under investigation to improve
both light detection for the low-energy supernova events and their
momentum resolution through determination of the event t$_0$.

Figure~\ref{fig:PD_overview} shows the layout for the photon detector
system, which will be described in the following sections.
\begin{cdrfigure}[Photon detection system overview]{PD_overview}{Overview of the PD
    system showing a cartoon schematic (a) of a single PD module
    in the LAr and the channel ganging scheme used to reduce the
    number of readout channels. Panel (b) shows how each PD module
    will be inserted into an APA frame. There will be 10 PDs instered
    into an APA frame.}
\includegraphics[width=1.0\linewidth]{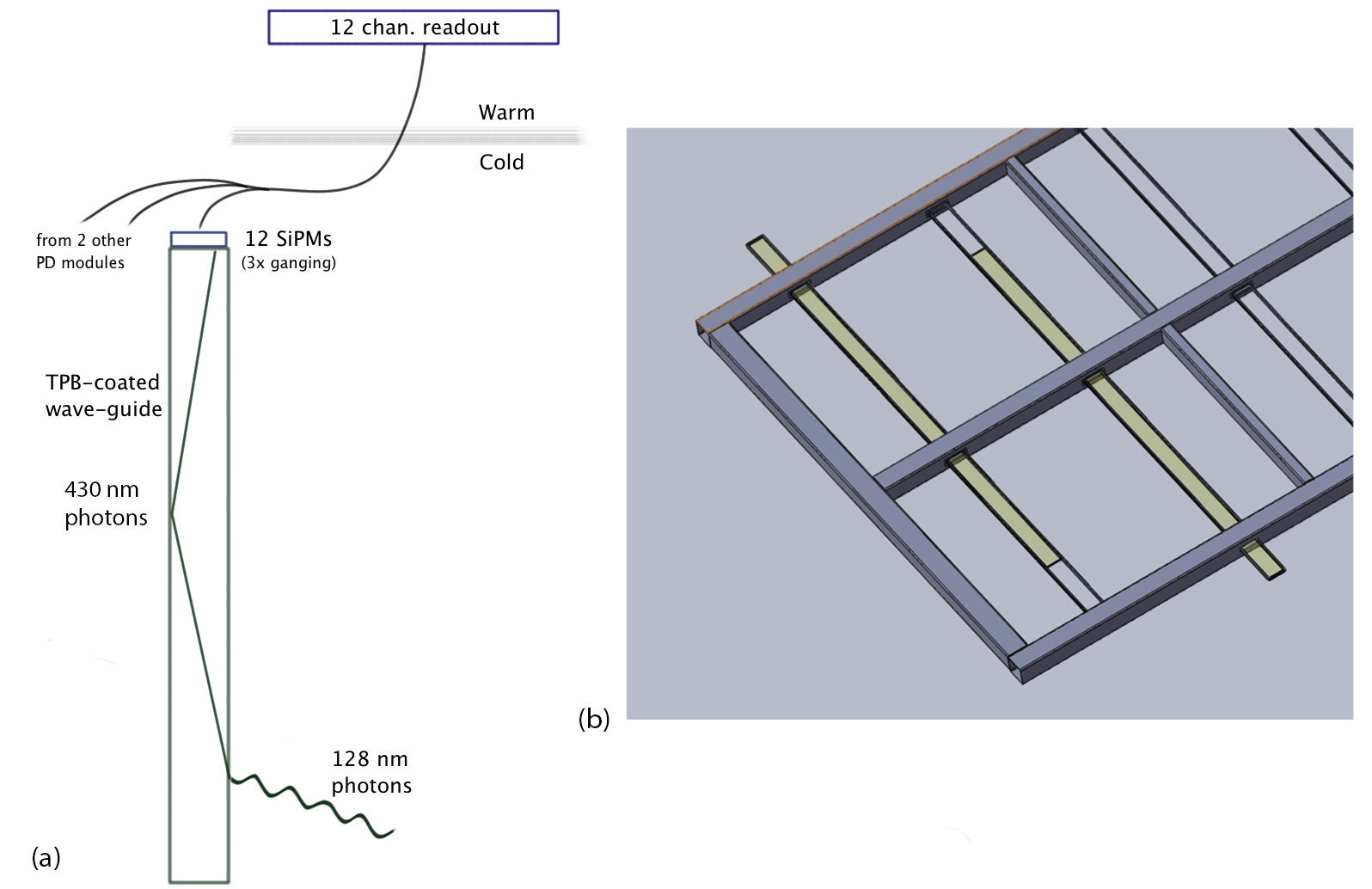}
\end{cdrfigure}

\subsection{Reference Design}
\label{sec:detectors-fd-ref-pd-refsystem} 

The PD system is mounted as modules on the APA frames.  A PD module is
the combination of one light-guide (also called a ``bar'' due to its
shape) and 12 SiPMs, as shown in Figure~\ref{fig:PD_overview}~(a).  To
enable this, the the reference design for mounting the PDs onto the
APA frames calls for ten PD modules per APA, approximately 2.2-m long,
83-mm wide and 6-mm thick, equally spaced along the full length of the
APA frame, as shown in Figure~\ref{fig:PD_overview}~(b). 

The 128-nm scintillation photons from LAr interact with the wavelength
shifter on the surface of the bar, and the wavelength-shifted light,
with a peak intensity around 430~nm, is re-emitted inside the bar and
transported through the light-guide to 12 silicon photo-multipliers
(SiPMs) mounted at one end of the bar.

The wavelength shifter converts the scintillation photons striking the
bar surface and directs them into the bar bulk with an efficiency of
$\sim$50\%.  A fraction of the wavelength-shifted optical photons are
internally reflected to the bar's end where they are detected by SiPMs
with quantum efficiency well matched to the wavelength-shifted
photons. The light guides are coated with TPB
(1,1,4,4-tetraphenyl-1,3-butadiene). A testing program is currently
underway to determine the absolute performance of the light guides in
LAr.

The SiPMs used in the reference design are SensL C-Series 6~mm$^2$
(MicroFB-60035-SMT) devices. These SiPMs have detection efficiency of
41\%; the detection efficiency combines QE and effective area
  coverage accounting for dead space between pixels. While the
C-Series SensL SiPMs are not rated for operation below
$-$40$^{\circ}$~C their performance has been excellent for this
application. At LAr temperature (89~K) the dark rate is of order 10~Hz
(0.5 p.e. threshold) while after-pulsing has not been an
issue. Extensive testing is underway to ensure that the SiPMs can
reliably survive the stresses associated with thermal cycling in LAr
and long-term operation at LAr temperature.

The SiPMs are read out using shielded twisted-pair cable, one per SiPM,
but the expected final design will have three SiPMs ganged together and
each readout cable will contain four individual channel cables to keep
the cost and cable packing density down. During the R\&D phase of the
project each SiPM was read out individually in order to maximize the information
gathered.  

The front-end electronics reside outside of the cryostat in
instrumentation racks. A custom module for receiving SiPM signals has
been designed and built. The module also performs signal processing in
the front-end as preprocessing for trigger and DAQ.  The module is
called the SiPM Signal Processor (SSP) and consists of 12 readout
channels packaged in a self-contained 1U module.  Each channel
contains a fully-differential voltage amplifier and a 14-bit, 150-MSPS
analog-to-digital converter (ADC) that digitizes the waveforms
received from the SiPMs. There is no shaping of the signal, since the
SiPM response is slow enough relative to the speed of the digitization
to obtain several digitized samples of the leading edge of the pulse
for the determination of signal timing. Digitized data is processed by
a Xilinx Artix-7 Field-Programmable Gate Array (FPGA).  The use of the
FPGA processing allows for a significant amount of customization of
the SSP operation. 

Once the DUNE collaboration arrives at a refined set of physics
requirements for the photon detection system a set of criteria for a
calibration system will be determined. In the absence of such criteria
two calibration systems are being explored, and will be tested in
the 35-t phase-II test. The first system, developed by ANL, utilizes five
fiber-fed diffusers mounted on the TPC CPA which uniformly illuminate
the photon detectors. An alternative design is employed on the IU
prototypes and uses LED-driven fibers mounted alongside the
waveguides. 

\subsection{Alternative Designs} 

Three alternative photon detector designs are currently being considered for the PD
system. More extensive descriptions of the alternative designs can be found in 
Section 5.3 of~\cite{cdr-annex-lbne-design}.

The first alternative is based on a TPB-coated acrylic panel with an
embedded S-shaped wavelength-shifting fiber. The Louisiana State
University (LSU) group has developed prototypes based on this design
in an attempt to allow an increase in detector size and hence increase
geometric acceptance of the PD system and reduce overall system
cost.

In this design, a single acrylic panel PD module has the same
dimensions as the reference design and consists of a TPB-coated
acrylic panel with an embedded multi-lobed ``S-shaped'' wavelength
shifting (Y11) fiber. The fiber is read out by two SiPMs (one on the
top edge, and the other on the bottom edge of the plate), which are
coupled to each end of the fiber and serve to transport the light over
long distances with minimal attenuation. The double-ended fiber
readout has the added benefit of providing some position dependence to
the light generation along the panel by comparing relative signal
sizes and arrival times in the two SiPMs. The WLS fiber converts the
430-nm light from the TPB to light with a peak intensity of
480--500~nm, which is well-matched to the peak photon-detection
efficiency of typical SiPMs.

A prototype of a second alternative, under investigation by the Colorado
State University (CSU) group, is intented to address an issue with the reference
design, in which the application of TPB to acrylic, or other base
materials, has been found to cause a significant decrease in
attenuation length (down to about 30~cm) of the light guide.
This prototype has a thin TPB-coated acrylic radiator located in front
of a close-packed array of blue-green (Y11) WLS fibers.  The prototype
is two-sided and has two identical fiber arrays and radiators mounted
back-to-back with a tyvek reflector between. This design allows for a
reduction in the number of SiPMs required per PD module. Three SiPMs
per side are needed per PD module (again the same dimensions in length
and width as the reference design) for a total of six SiPMs per PD
module.

The Indiana University (IU) group has advanced the CSU design and
arrived at the third alternative design by replacing the Y11 fiber
with a reference-design-dimensioned cast bar doped (by Eljen
Technology) with the same wavelength shifter as the Y11 fiber. The
TPB-coated acrylic radiator has also been replaced with a thin-fused
silica plate coated with TPB. This prototype has demonstrated an
attenuation length greater than 2.5~m and early indications point to
it meeting the SN neutrino energy requirement. It has the highest
light collection efficiency of any prototype tested so far.

\subsection{Technology Selection}

The alternative designs have all demonstrated the ability to detect
LAr scintillation light, and development work is continuing. A testing
program is underway comparing the various alternatives against the
reference design.  The Tall Bo large dewar at Fermilab and the
cryogenic detector development facility at CSU are being used to
compare the performance of full-scale and near-full-scale prototypes
in LAr utilizing alpha sources and cosmic muons. Data from the 35-t
prototype will also provide input into the technology decision. %In Fall 2015 a decision will be made regarding which design to adopt and optimize for the first 10-kt far detector module.

%% file: volume-detectors/fd-ref-sections/fd-ref-install.tex
%%%%%%%%%%%%%%%%%%%%%%%%%%%%%%%%
\section{Installation and Commissioning}
\label{sec:detectors-fd-ref-install}

The scope of the Installation and Commissioning  (I\&C) task includes the
design, procurement, fabrication, labor, testing and delivery of
equipment and infrastructure to support installation and commissioning
of the detector at the far site. The following are included in the
scope:
\begin{itemize}
\item detector installation planning;
\item installation equipment design and procurement;
\item construction of a full scale mockup, consisting of four APAs, two CPAs and associated
  field cage, to test installation operations and
  equipment;
\item procurement of support rails for the TPC;
\item procurement and installation of relay racks to house the
  electronics provided by other subsystems for detector operation;
\item material receipt, storage and transport to underground at the far site;
\item installation of the TPC, photon detection and DAQ systems at the
  far site (support for detector checkout will be provided by the
  subsystems); and
\item coordination of the commissioning for the detector and personnel
  to support detector operations.
\end{itemize}

I\&C will have many interfaces with LBNF, and LBNF
has certain responsibilities of its own, including the following.
%however, it will not be responsible for the following installation activities at the far site:
\begin{itemize}
\item Excavation and outfitting of the the caverns is the responsibility of
  the Conventional Facilities (CF) subproject of LBNF.
\item Construction and installation of the cryogenics system and
  cryostats is the responsibility of the cryogenics system subproject
  of LBNF.
\end{itemize}
%The design presented here meets the required performance for the
%Installation and Commissioning of the DUNE far detector.

\subsection{Equipment and Services}

The I\&C system provides 
permanently installed equipment that is used by multiple detector
systems and/or is integral to the installation process. This 
includes the relay racks, cable management and support rails for the TPC.
I\&C is also responsible for several detector-specific aspects of the cavern outfitting, including a clean area enclosure near the cryostat hatch to 
isolate  the open hatch and the TPC components
from the cavern environment. 

The TPC elements (APAs and CPAs, described in Section~\ref{sec:detectors-fd-ref-tpc}), are
supported by a set of five support rails permanently mounted at the
top of the cryostat. The rails are supported
by rods spaced at 5-m intervals from anchor points at the cryostat
roof. The rods are installed with an angle bias that allows the
rails to return to a level condition after the cryostat and TPC are
cooled. 
%\fixme{Is the following relevant? It's more for the TPC section}  <--- Jack said ok to remove
%The APAs are arranged in three rows
%with two of the rows near the cryostat walls. Relative to the LBNE configuration, 
%this represents an
%increase of 50\% in the number of APAs and a corresponding decrease of the number of
%CPAs (see \anxlbnefd). 

The TPC and photon detectors require various electrical services
for operation, including bias voltages, power and control signal. These services
follow a grounding plan (still under development) 
discussed in Section~\ref{sec:detectors-fd-ref-install-ground}.  The
electrical services pass through a total of 78 feedthrough ports located on top of the
cryostat, above every other APA junction.  A relay rack is located
adjacent to each port. The rack space is shared between the TPC,
photon detection system readout and power supplies. For the upper APA of each pair,
the cables are pre-installed and routed from the cold side of the
feedthrough down and along the support rail to the location where they
connect to their corresponding APA. For the lower APAs with the
electronics near the floor of the cryostat, the cables are routed
from the cold side of the feedthrough down the sides of the
cryostat. The lower APA cables are routed in cable trays supported
from the cryostat walls.

\subsection{TPC Installation Process}

Once detector components arrive at the far site they are put in a
facility used for both storage and testing/checkout, called an integration facility.  
Material is moved from the
integration facility to the cavern (after undergoing checkout)  via the Yates shaft. 
As installation space in
the cavern is very limited, the moves will take place at the rate of installation. 
 Most items can be
transported inside the Yates shaft cage, however, the APAs are too
long to fit in the cage and therefore are slung underneath it in a special %APA transportation 
container that holds four APAs 
in an internal rack. % for movement in the Yates shaft.The APA transportation container includes 
This cleaner rack %inside the container 
can be extracted from the outer container into the clean area used for installation.
%thus avoiding moving the outer
%container into .
The TPC components will have been cleaned and
protected to a level suitable for installation into the cryostat as
part of the TPC production process, and will %be delivered to 
have arrived at the far
site in clean containers.

The clean area enclosure, in the range of class 100,000 (ISO 8
equivalent), %will be constructed near the entrance to the cryostat. The enclosure will have 
provides an area for personnel to gown with the appropriate
clean-room clothing and safety shoes. A large closable door is
located at the drift junction where TPC storage containers can be
parked to allow unloading of the TPC components 
directly %from the container 
into the clean area. 

The TPC installation process requires the temporary installation of several 
items in the cryostat before it is filled with argon.
\begin{itemize}
\item A lighting system with emergency backup lighting will be installed and then removed in sections
as the TPC installation proceeds. (This lighting will be filtered
to the appropriate spectrum to protect the photon detection system
installed in the APAs.) 
\item A filtered air ventilation system
with air-monitoring sensors and alarms will be installed to ensure adequate air
quality for work inside the cryostat. The system will also include a
high-sensitivity smoke-detection system that is interlocked to the
power for all devices inside the cryostat. 
\item A raised floor will be
installed at the bottom of the cryostat to protect the cryostat
membrane and provide a flat surface above the corrugations of the
cryostat. A modular design will allow it to be removed
in sections as the TPC installation progresses.
\end{itemize}

A combination of commercial and specially designed tooling will be
required for TPC installation. All of the detector components and
equipment inside the cryostat will be inserted through hatches
located at one end of the cryostat.  Temporary fixed scaffolds with integral
stair towers will  provide personnel access
into the cryostat. A rolling scaffold, on the cryostat raised floor, also with an integral stair tower,
will provide access to the top
of the cryostat where the TPC connections are made.  Special fixtures
and commercial gantry hoists are required to move APAs from a
horizontal orientation in storage racks to a vertical orientation at
the cryostat hatches. Special platforms located at the
cryostat hatches will support each lower APA section while its upper APA
section is connected. The platform will have multiple levels to allow
personnel to access the connection points at the top of an APA and at the
junction between an upper and lower APA. The installation equipment
and installation procedures will be tested with a full-height mock TPC
section at a suitable location at Fermilab.

%The TPC installation is a highly repetitive process. % with most steps repeated many times. 
%
Installation of the TPC is preceded by installation of the DAQ, including relay racks and TPC cables, 
 in order to allow immediate testing of APAs upon their placement in the cryostat.

The TPC installation starts with installation of the cathode planes, one 
row at a time, starting with the top row of a plane, and progressing, one CPA at a time, 
from the far end of the
cryostat to the hatch end.  As each cryostat-length row is completed, 
it is lifted, and the next row
is attached below it in the same manner; this is repeated until all four rows of the cathode plane are in place.
At this point, the end-wall field cage is installed at the non-hatch end of
the cathode plane. 

APA installation begins next. %The first step of installing an APA pair is to move an 
An APA, electronics side down, is first moved %towards the bottom 
into the cryostat and held temporarily in the area of the
hatch. A second APA, its pair, is positioned above 
the first, electronics side up. The two APAs are joined at the
center, lifted and attached to the support rail. The connected pair is 
moved along the support rail to its designated position and, except for the first pair,
connected to the previously installed adjacent stacked pair.  
At this point the power and signal
cables are connected to the APAs for testing. After proper functioning
is confirmed, the field cage
sections between the APAs and their facing CPAs are installed and the raised floor
sections in that area are removed.  This APA and field cage installation process is repeated progressing towards 
the hatch end of the cryostat until the entire anode plane is in place; the field cage is then
installed. The process is then repeated
for the other anode planes. Once TPC installation is complete, the installation equipment and
the scaffolding is removed from the cryostat.

Once the TPC is installed and all temporary equipment is removed from
the cryostat, the hatches are closed and all channels of the
detector are tested for expected electronics noise. After
successful testing, the cryostat hatches are sealed and the purge
proceeds, followed by a cool-down of the cryostat and detector.
At this point extensive detector testing will be conducted prior to 
filling with LAr. Filling each \ktadj{10} cryostat 
requires approximately six months, after which a several-month-long
detector commissioning phase begins.

\subsection{Grounding}
\label{sec:detectors-fd-ref-install-ground}

The detector will have approximately 300,000 channels of electronics with
an intrinsic noise level less than 1,000 electrons. The channels will
be connected to signal collection wires that are up to 7~m long, thus
grounding, shielding and power distribution are critical to the success
of the experiment.
The installation and commissioning group will develop a detector
grounding plan that coordinates between the CF
power distribution, cryostat design and the detector systems.   The grounding will be configured such that each
detector is on an isolated and separate detector ground that is
referenced to building ground through a safety saturable inductor.
Dielectric breaks will be used on all conductive piping/services that
penetrate the cryostat.  A copper ground plate under the steel top
plate of each cryostat will be provided as part of the cryostat and
used to serve as a central star ground point.

%% file: volume-detectors/chapter-fd-alt.tex
\chapter{Far Detector Alternative Design: Dual-Phase LArTPC}
\label{ch:detectors-fd-alt}

%%%%%%%%%%%%%%%%%%%%%%%%%%%%%%%%
\input{volume-detectors/fd-alt-sections/fd-alt-overview}
\input{volume-detectors/fd-alt-sections/fd-alt-charge-read}

\input{volume-detectors/fd-alt-sections/fd-alt-fieldcg-hv}
\input{volume-detectors/fd-alt-sections/fd-alt-elec}
\input{volume-detectors/fd-alt-sections/fd-alt-dcs-dss}
\input{volume-detectors/fd-alt-sections/fd-alt-light-read}
\input{volume-detectors/fd-alt-sections/fd-alt-install}

%% file: volume-detectors/fd-alt-sections/fd-alt-overview.tex
%%%%%%%%%%%%%%%%%%%%%%%%%%%%%%%%
\section{Overview}
\label{sec:detectors-fd-alt-ov}

This chapter describes an alternative far detector design for
DUNE. The first detector module to be installed will use the reference
design described in Chapter~\ref{ch:detectors-fd-ref}, however this alternative
 design is under consideration for one or more
subsequent modules. This design implements a dual-phase liquid argon
time projection chamber (LArTPC) augmented with a light-readout
system. ``Dual-phase'' refers to the extraction of ionization
electrons at the interface between liquid and gas argon and their
amplification and collection in the gas phase.

This dual-phase design is the result of 13 years of R\&D consisting of two
consecutive design study programs funded since 2008 by the European
Union: LAGUNA and LAGUNA-LBNO. The LAGUNA-LBNO design study was
concluded in August 2014.  In collaboration with industrial partners,
LAGUNA-LBNO designed an innovative, optimized and cost-effective
configuration for a long-baseline experiment.

The studies focused on the underground implementation of a very large
LAr detector (GLACIER) and produced many technical advances with respect to
the tank, field cage and cathode, charge multiplication, collection and readout, as well as 
advances in the areas of assembly sequencing and logistics for the detector and 
full costing. The full design and the related
technical developments are described in \anxlbnob~\cite{cdr-annex-lbno-2}, which was submitted to the
EU at the conclusion of the design study. This chapter describes the detector
configuration and components, and describes how the design meets the DUNE
far detector physics requirements.

The scope of a dual-phase far detector module for DUNE includes the design,
procurement, fabrication, testing, delivery, installation and
commissioning of the detector components:
\begin{itemize}
\item Charge-Readout Planes (CRP), including extraction grid, Large Electron Multiplier (LEM) and anode and readout planes;
\item Cathode, field cage and high voltage system;  
\item Electronics and data acquisition; 
\item Chimneys, isolated volumes used for electronics feedthroughs;  
\item Slow Controls; and
\item Light-readout system.
\end{itemize}

The detector components and the liquid argon (LAr) will be housed in cryostats
provided by LBNF, described in \vollbnf~\cite{cdr-vol-3}.  Similar to the reference design, 
this alternative design satisfies the performance 
requirements on the DUNE far detector, described in Chapter~\ref{ch:detectors-fd-ref} for the reference design.
Parameters specific to the dual-phase design are listed in
Table~\ref{tab:FD_req}.

\begin{cdrtable}[Dual-phase (alternative) far detector performance parameters]{llll}{FD_req}{Performance parameters specific to the dual-phase far detector design}  
Parameter & Requirement & Achieved Elsewhere & Expected Performance \\ \toprowrule
Gas phase gain & 20 & 200 & 20-100  \\ \colhline
Electron Lifetime & 3~ms &  $>3$~ms 35-t prototype  & $>5$~ms \\ \colhline 
Minimal S/N after 12 m drift & 9:1 &  $>100$:1 & 12:1-60:1  \\ 
\end{cdrtable}

\section{Highlights of the Design}

This innovative dual-phase design is similar in many ways to the single-phase design,
but implements some unique features and offers several advantages over it, in particular
\begin{itemize}
\item  higher gain, leading to a larger signal-to-noise ratio (S/N);
\item  larger fiducial volume, enabling very long drift paths;
\item  lower detection threshold;
\item  finer readout pitch (3~mm), implemented in two identical collection views, $x$ and $y$;
\item  fewer readout channels (153,600 vs 384,000 for a reference design \ktadj{10} module); and
\item  the absence of dead material in the LAr volume.
\end{itemize}

Following the GLACIER concept\cite{LAGUNA-LBNO-deliv} (see
Figure~\ref{fig:LBNO_50}), the DUNE dual-phase LArTPC detector design 
has a fully homogeneous liquid argon volume, in which electrons
drift upwards vertically towards an extraction grid just below the liquid-vapor interface. From there they
are extracted from the liquid into the gas phase, amplified, and
collected on a finely segmented
anode\cite{Badertscher:2013wm,Badertscher:2012dq,Badertscher:2010zg}. 
\begin{cdrfigure}[The \ktadj{50} LBNO detector, GLACIER]{LBNO_50}{The \ktadj{50} LBNO detector, GLACIER}
\includegraphics[width=0.9\linewidth]{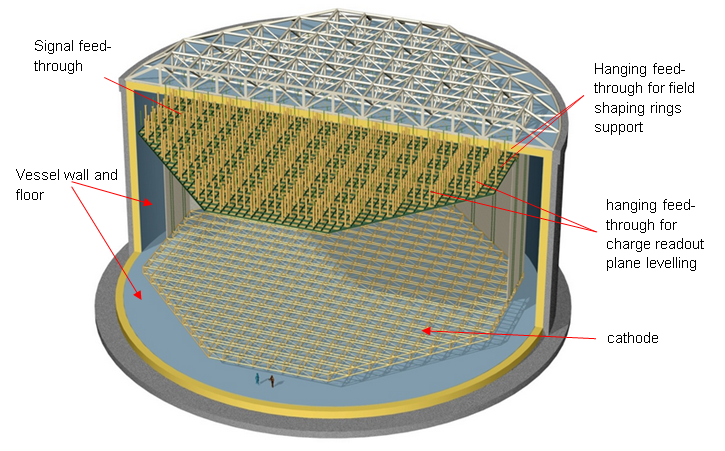}
\end{cdrfigure}

The electron amplification in the gas phase enables a robust and tunable signal-to-noise ratio. 
The detector configuration is similar to a single-phase LArTPC. The features of the dual-phase design, e.g., high gain, 
allow achieving very long drift paths and large detector dimensions while minimizing the number of readout channels.

\subsection{Charge Collection, Amplification and Readout}

An extraction efficiency of 100\% of the electrons from the liquid to
the gas phase is achieved with an electric field of the order of
2~kV/cm across the liquid-gas interface, applied between an 
extraction grid submersed in the liquid and charge amplification 
devices situated in the ultra-pure argon gas. 

These amplification devices, called Large Electron Multipliers (LEMs), are horizontally 
oriented 1-mm-thick printed 
circuit boards with electrodes on the top and bottom surfaces. They are drilled
through with many holes that collectively form a micro-pattern structure;  
when a 3-kV potential difference is applied across the electrodes
the ionization electrons are amplified by avalanches (Townsend multiplication) occurring in the 
pure argon gas in this micro-pattern structure\cite{Bondar:2008yw} due to the high electric field (30 kV/cm).

The use of avalanches to amplify the charges in the gas phase increases
the S/N ratio by at least one order of magnitude with a typical gain of 20--100, significantly
improving the event reconstruction quality. It also lowers the
threshold for small energy depositions and provides a better
resolution per volumetric pixel (voxel) compared to a single-phase
LArTPC. 

The charge is collected in a finely segmented 2D ($x$ and $y$) readout anode
plane at the top of the gas volume and fed to the front-end electronics.   

The  collection, amplification and readout components are combined in an array of 
independent (layered) modules called Charge Readout Planes (CRPs). A CRP is 
composed of several 0.5$\times$0.5-m$^2$ units, each of which is composed 
of a LEM/anode sandwich. 
These units are embedded in a mechanically reinforced frame of FR-4 and stainless steel. The CRP structure also integrates
 the submersed extraction grid, which is an array of $x$ and $y$ oriented stainless steel wires, 0.1~mm in diameter, with 3.125-mm
pitch. Thicknesses and possible biasing voltages for the different layers are indicated in Figure~\ref{fig:CRP_struct}.

\begin{cdrfigure}[Charge Readout Plane (CRP) structure]{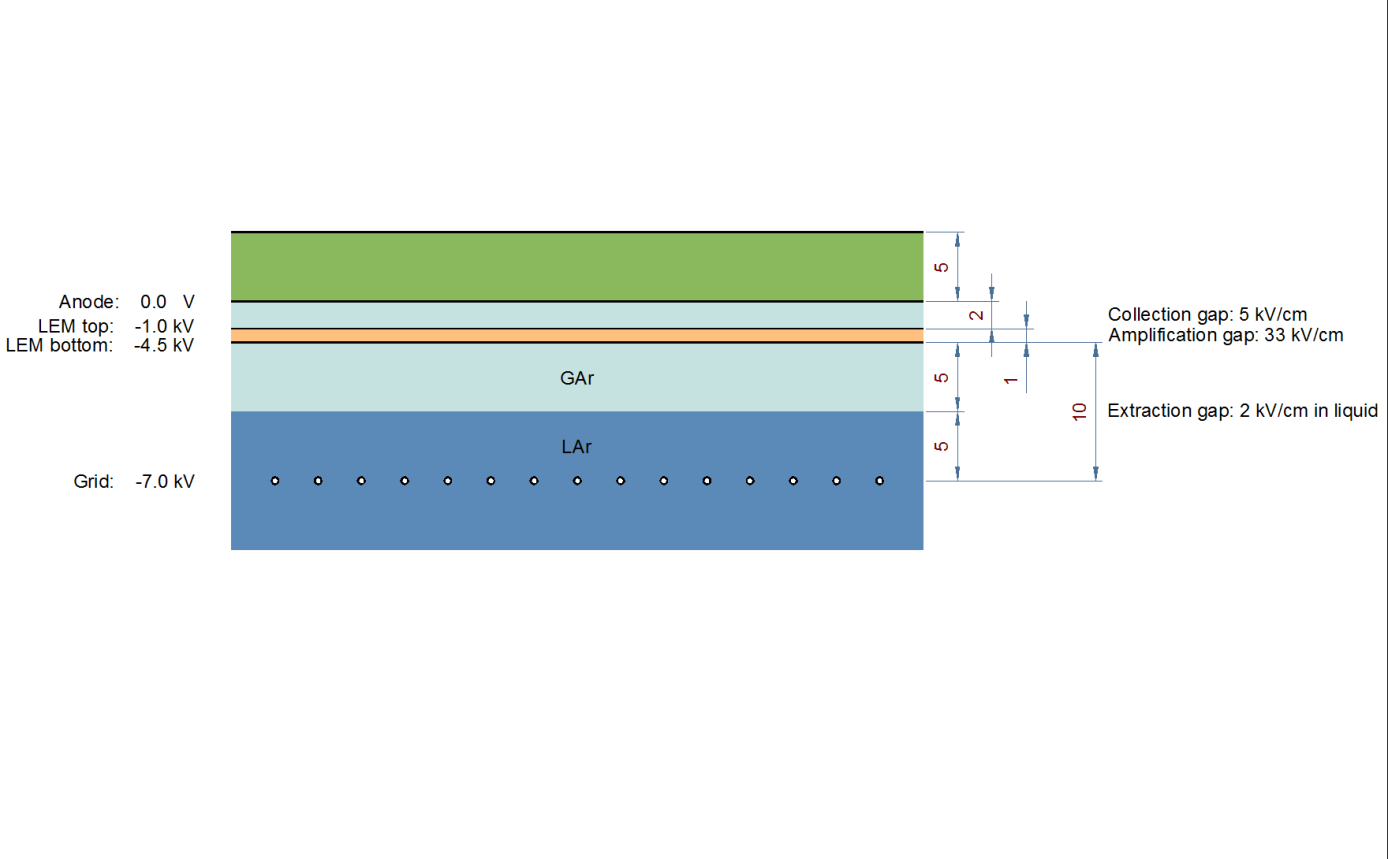}
{Thicknesses and HV values for electron extraction from liquid to gaseous Ar, their 
multiplication by LEMs and their collection on the $x$ and $y$ readout anode plane. The 
HV values are indicated for a drift field of 0.5~kV/cm in LAr.}
\includegraphics[width=\linewidth]{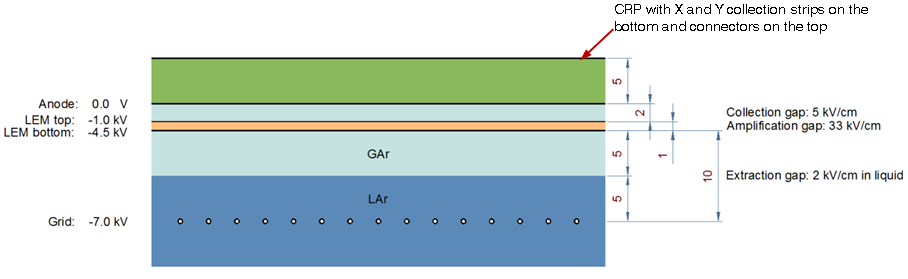}
\end{cdrfigure}

Each CRP  is independently suspended with stainless-steel ropes linked to the tank top deck. This suspension system allows adjustment of the CRP distance and parallelism with respect to the LAr surface, and keeps the extraction grid immersed.

Figure~\ref{fig:CRP_unit1} shows the top of two side-by-side units and Figure~\ref{fig:CRP_unit2} shows the
$x$ and $y$ readout views.

\begin{cdrfigure}[DUNE CRP unit, $x$ and $y$ view (3 $\times$ 3~m$^2$)]{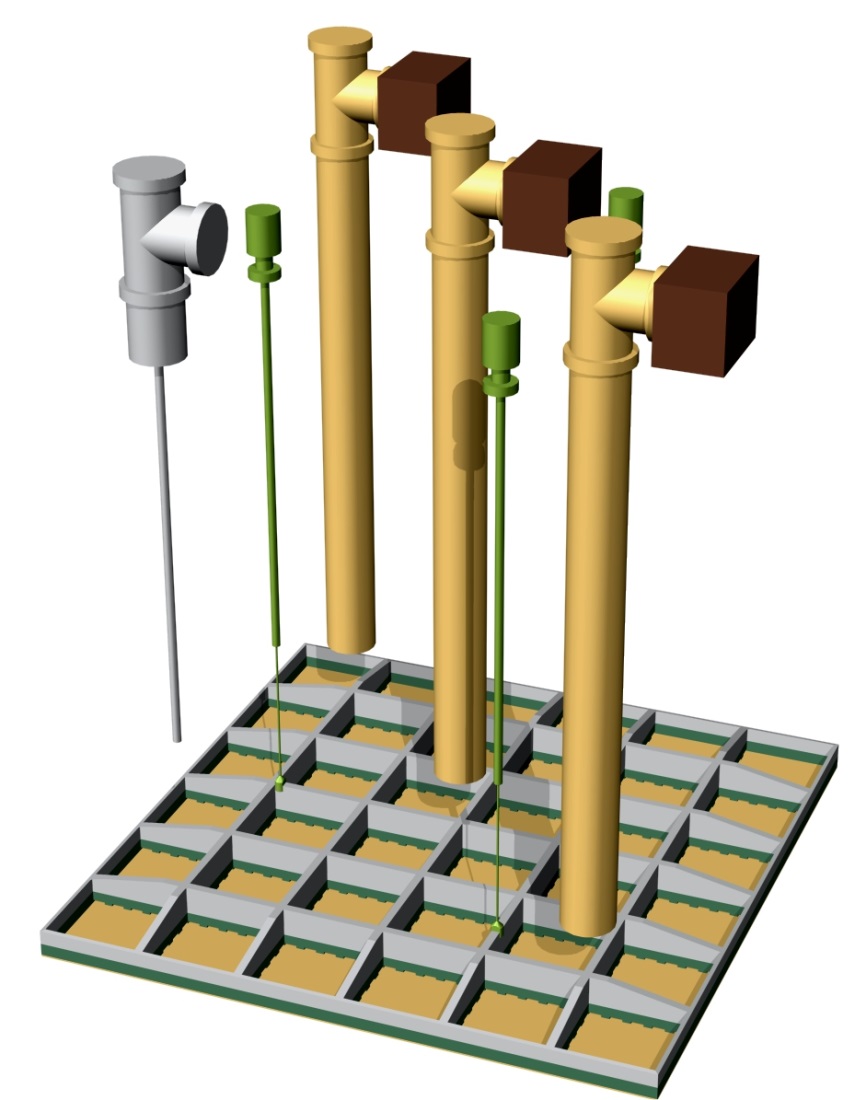} {Two DUNE CRP 3~m$\times$3~m units side by side. On the left one of the 79 equal 
CRP units, on the right the 1$^{st}$ CRP unit with a chamfered LEM/Anode Sandwich  for the insertion of the high voltage feedthrough.}
\includegraphics[width=.7\linewidth]{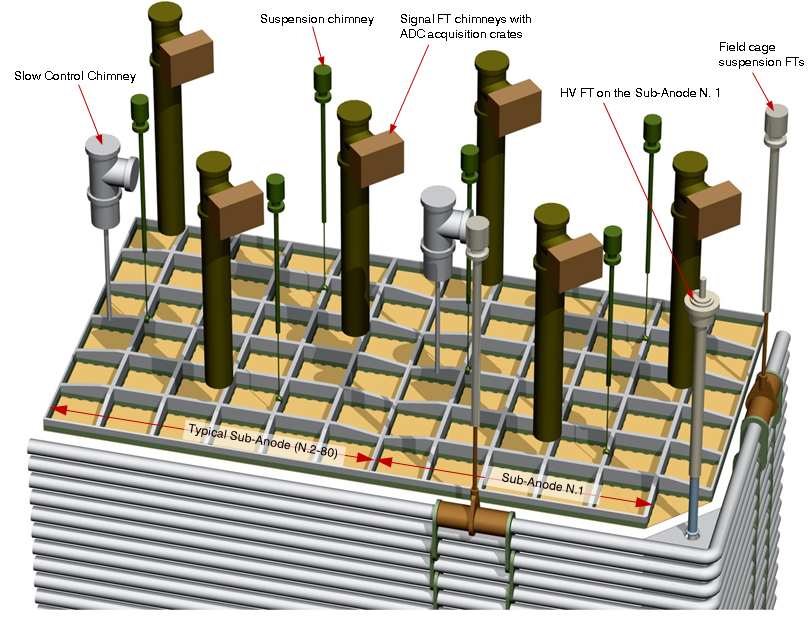}
\end{cdrfigure}
\begin{cdrfigure}[Signal collection in the $x$ and $y$ views by the three signal FT chimneys.]
{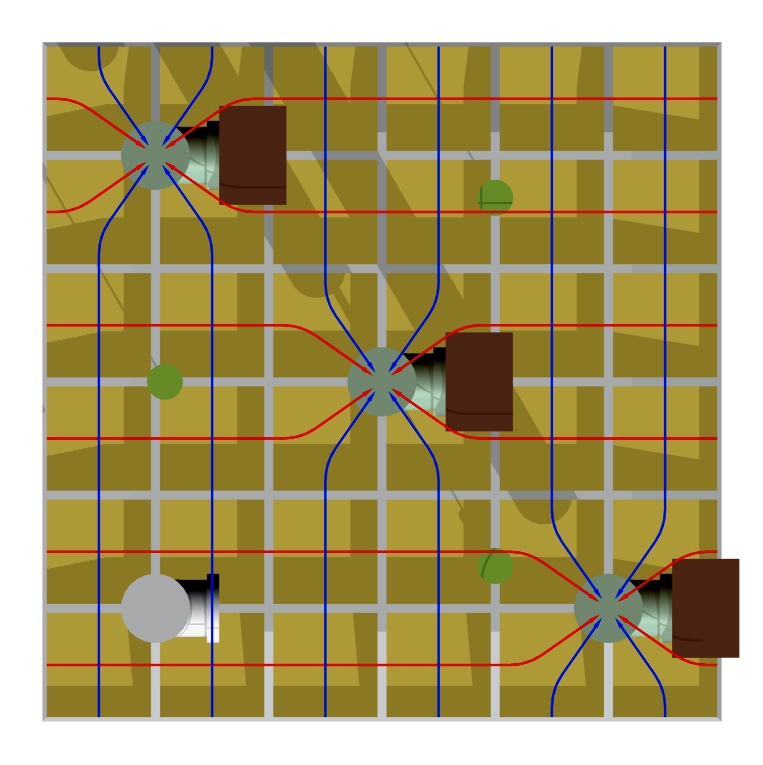}{Signal collection in the $x$ and $y$ views of the  3$\times$3~m$^2$ DUNE CRP unit by the 3 SFT chimneys.}
\includegraphics[width=.5\linewidth]{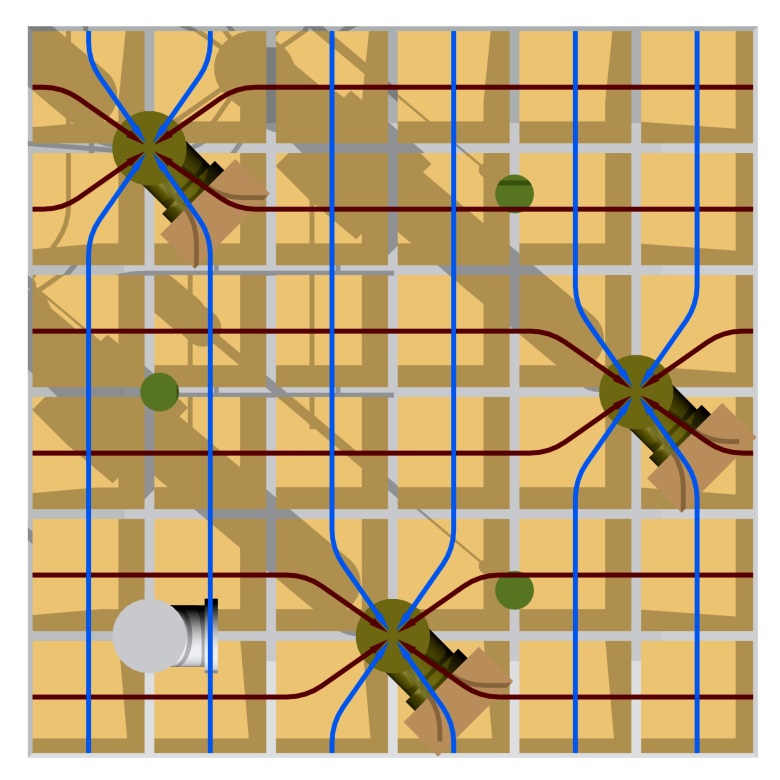}
\end{cdrfigure}

A CRP provides an adjustable charge gain (with a minimal required gain of 20) and two
independent, orthogonal readout views, each with a pitch of 3.125~mm.  The LEM/anode sandwiches 
in the same CRP unit are interconnected with short flat cables so that each readout
channel corresponds to a total strip length of 3~m.

Combined with the time information coming from the LAr scintillation readout by
the PMT arrays ($t_0$), a CRP provides 3D track imaging with $dE/dx$ information. 
The CRPs and their components are described in Section~\ref{sec:detectors-fd-alt-chg-readout}.

The typical amplification achieved by this design, between 20--100, improves the S/N ratio and thus 
compensates for the charge losses that occur along the very long drift paths due to the presence of 
electronegative impurities. Therefore, despite the longer drift length, this design requires no higher 
purity of the LAr than does the reference design, around 0.1~ppb (or 100~ppt) of oxygen equivalent,
and yields a 3-ms electron lifetime. The required level of purity can be reached by starting from 
commercially available ppm-level bulk argon and filling a non-evacuated vessel\cite{WA105_TDR}.

The S/N ratio can exceed 100 for a minimum
ionizing particle (MIP) after a drift path of 12~m (given an
electron lifetime of 3~ms, a drift field of 0.5~kV/cm and a LEM gain
of 180). With the same drift field, the same electron-lifetime conditions and a
LEM gain of 25, the S/N is larger than 50:1 for tracks up to 6~m from
the anode; it reaches 14:1 for MIP tracks that are 12~m from the
anode.

\subsection{Electronics and ``Chimneys''}
 
The electrical signals from the collected charges
are passed to the outside of the tank via a set of dedicated signal
feedthrough ``chimneys'' (insulated volumes filled with nitrogen
that pass through the top layer of insulation). 
The cryogenic front-end (FE) electronics cards, housed at the bottom of the
chimneys, are based on analog preamplifiers implemented in CMOS ASIC circuits for high integration and large-scale
affordable production. Within the chimneys, the cards are actively cooled to a temperature of about 110 K and
isolated with respect to the LAr vessel by a cold feedthrough.  This
feedthrough is connected to the CRP via short flat cables of (0.5~m length) in order to minimize the
input capacitance to the preamplifiers. Each chimney collects 640 readout channels.

The chimney design allows access to and replacement of the FE from the
outside without contaminating the LAr volume. The digital electronics
and DAQ system are completely outside the cryostat and are housed in
microTCA racks mounted on each signal feedthrough chimney. 

Other feedthroughs are planned for the cathode HV connection, the
CRPs' suspension and level adjustment, the high voltage and signal
readout of the PMTs, and the monitoring instrumentation (level meters,
temperature probes, strain gauges, etc.).

\subsection{Cathode, Field Cage and HV System}
\label{v4:fd-alt-ov:cathode}

The drift field (E ${\simeq}$ 0.5~kV/cm) inside the fully
active LAr volume is produced by applying high voltage to the cathode
plane at the bottom of the cryostat and is kept uniform by the field cage, a stack
of 60 equally spaced field-shaping electrodes, %in the form of tubes (diameter 140~mm,  vertical pitch 200~mm), 
polarized at linearly decreasing voltage from the cathode 
voltage to almost ground potential, reached at the level of the charged readout plane.
The electrodes are rectangles made of stainless-steel tubes  (diameter 140~mm,  vertical pitch 200~mm)
with rounded corners, running horizontally (and stacked vertically) around the
active volume (see Figure~\ref{fig:DP_det2}). 

\begin{cdrfigure}[Dual-phase detector 3D view (partially open)]{DP_det2}
{The DUNE dual-phase detector (partially open) with cathode, PMTs, field cage and anode plane with chimneys.}
\includegraphics[width=\linewidth]{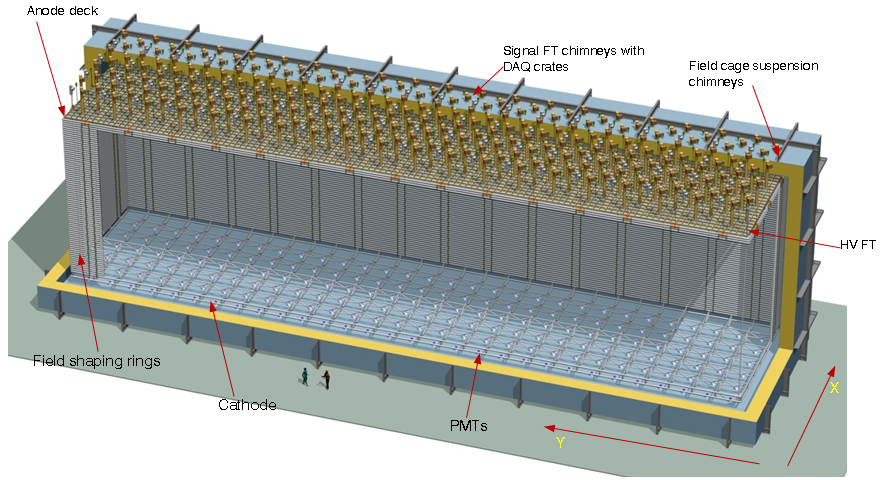}
\end{cdrfigure}

The field cage is held in place by mechanical structures hung from the
top deck of the vessel that also provide insulation.  The cathode
structure, constructed of a reinforced frame to 
guarantee its planarity, is suspended from the field cage and hangs near the 
bottom of the cryostat. It is a segmented structure of tubes of different sizes 
arranged in a grid to minimize weight, limit sagging and avoid high electric field
regions in its proximity.  The segmented structure allows scintillation light to pass
through and be detected by uniform arrays of photomultipliers (PMTs) mounted
1~m below it at the bottom of the tank.

\section{Detector Configuration}

The detector for the \ktadj{12.1} active mass module is built as a single
active volume 60~m long, 12~m wide and 12~m high, with the anode at the
top, the cathode near the bottom and an array of 180 photon detectors (PMTs, 1 per 4~m$^2$)
located at the bottom of the vessel underneath the cathode. 
The active volume (see Figure~\ref{fig:DP_det1}) is surrounded by the
field cage. These components are described in Section~\ref{v4:fd-alt-ov:cathode}.

%The cathode plane (on the bottom) made by a reinforced frame filled by a tubular grid (not visible in
%Figure~\ref{fig:DP_det2}) allows optical transparency for the scintillation light towards .

The proposed design optimally exploits the
cryostat volume of 14(w)$\times$14.1(h)$\times$62(l)~m$^3$ with an
anode active area of 12$\times$60~m$^2$ and a drift length of 12~m,
corresponding to an active mass of 12.096~kt of LAr (10.643~kt
fiducial).  \fixme{Do we want this many significant digits? In next pgraph, the number
of sig figs doesn't even match.}

 The design is based on the \ktadj{20} LAGUNA-LBNO design study
with a CRP unit size adapted to the dimensions on the active area. The
cryostat height could be increased to achieve 15-m drift, resulting in
an active mass of 15.12~kt (13.444~kt fiducial).  This \ktadj{15.1}
configuration, apart from the longer drift distance and field cage,
would have the same characteristics of the \ktadj{12.1} configuration,
given that the covered active area is exactly the same. With these
transverse dimensions, every additional meter of drift length provides
a \ktadj{1} increase in the active mass at a moderate additional cost.

\begin{cdrfigure}[Dual-phase detector 3D view]{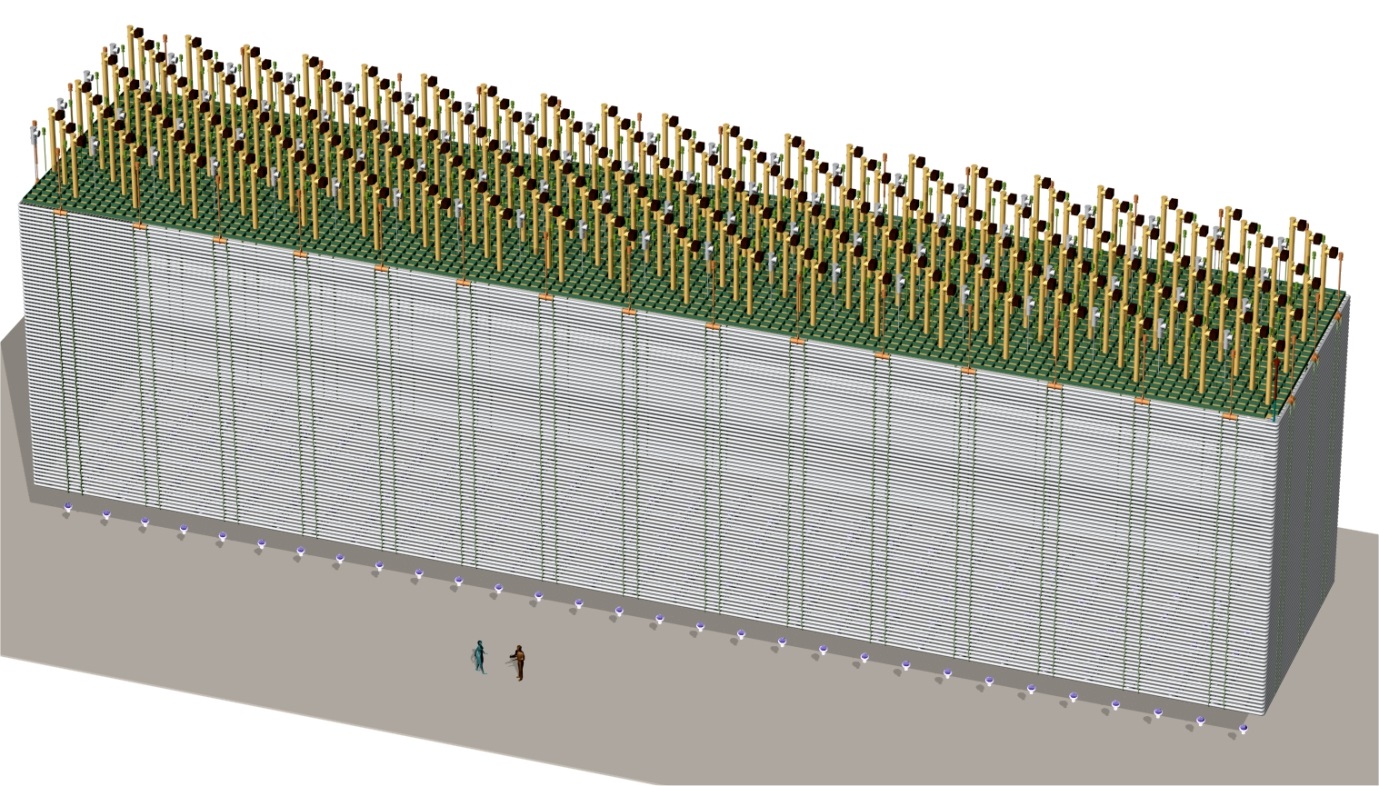}{The DUNE dual-phase 
detector with cathode, PMTs, field cage and anode plane with chimneys.}
\includegraphics[width=\linewidth]{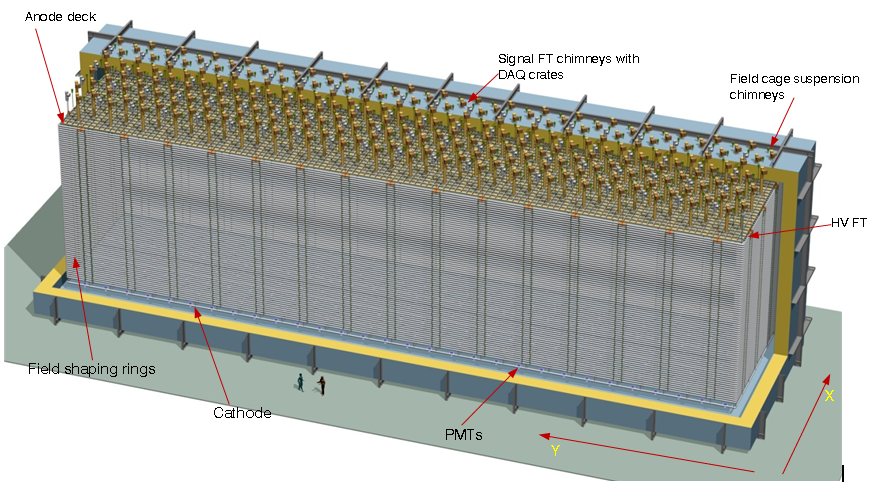}
\end{cdrfigure}

The ionization electrons in the liquid phase drift  in a uniform electric field towards the anode plane at the top of the active
volume. This is made by by an array of 80 independent CRP modules, 3$\times$3~m$^2$ each.
The extraction of the electrons from the liquid to vapor phase is performed thanks to the submersed horizontal extraction grid, 
integrated in each CRP structure. A CRP unit includes 36 (0.5~m$\times$0.5~m) LEM/anode sandwiches, providing tunable
amplification and charge collection on two independent views organized in strips of 3-m length and 3.125-mm pitch. There are
1920 readout channels for each CRP. Signals in each CRP unit are collected via three signal feedthrough chimneys hosting the
the front-end cards with the cryogenic ASIC amplifiers (640 channels/chimney) which are accessible and replaceable without
contaminating the pure liquid argon volume. Each chimney is coupled to a microTCA crate ensuring the signals' digitization and 
data acquisition. These crates are connected  via optical fiber links to the DAQ back-end. The total number of readout channel 
per \ktadj{10} module is 153,600.
 
Each CRP unit is independently suspended by three stainless steel ropes. The vertical level of each CRP unit can then be automatically
adjusted with respect to the LAr level via three suspension feedthroughs, electrically operated from outside. A Slow Control feedthrough, 
one per CRP unit, is used for the signals readout for level meters and the temperature probes and to apply the HV bias on the two
sides of the LEMs and on the extraction grid. The number of components and parameters for the \ktadj{12} (\ktadj{15})
dual-phase LArTPC are summarized in Tables~\ref{tab:DP_params}
and~\ref{tab:DP_numbers}.

\begin{cdrtable}[Sizes and dimensions for the \ktadj{12} (\ktadj{15}) dual-phase LArTPC]{lll}
{DP_params}{Sizes and Dimensions for the \ktadj{12} (\ktadj{15}) dual-phase  LArTPC}  Item & Value(s) &  \\ \toprowrule
Active volume width and length & W = 12~m &  L = 60~m \\ \colhline
Active volume height &  H = 12~m (H = 15~m)  &  \\ \colhline
Active volume/LAr mass & 8,640 (10,800)~m$^3$ &  12,096 (15,120) metric ton \\ \colhline
Field ring vertical spacing & 200~mm  \\ \colhline
Field ring tube diameter & 140~mm \\ \colhline
Anode plane size & W = 12~m & L = 60~m \\ \colhline
CRP unit size & W = 3~m & L = 3~m  \\ \colhline
HV for vertical drift & 600--900~kV \\ \colhline
Resistor value & 100~M$\Omega$ \\ 
\end{cdrtable}
\begin{cdrtable}[Quantities of items for the \ktadj{12} (\ktadj{15}) dual-phase LArTPC]{ll}{DP_numbers}{Quantities of Items for the \ktadj{12} (\ktadj{15}) dual-phase  LArTPC}  Item & Number    \\ \toprowrule
Field rings & 60  (75)  \\ \colhline
CRP units & 4 $\times$ 20 = 80 \\ \colhline
LEM/Anode sadwiches per CRP unit & 36 \\ \colhline
LEM/Anode sandwiches (total) & 2,880 \\ \colhline
SFT chimneys / CRP unit & 3 \\ \colhline
SFT chimneys (total) & 240 \\ \colhline
Readout channels / SFT chimney & 640  \\ \colhline
Readout channels (total) & 153,600 \\ \colhline
Suspension FT / CRP unit & 3  \\ \colhline
Suspension FTs (total) & 240  \\ \colhline
Slow Control FT / sub-anode & 1  \\ \colhline
Slow Control FTs (total) & 80 \\ \colhline
HV feedthrough & 1  \\ \colhline
Voltage degrader resistive chains & 4 \\ \colhline
Resistors (total) & 240 (300)  \\ \colhline
PMTs (total) & 180 (1/4~m$^2$) \\ 
\end{cdrtable}

A number of factors make the dual-phase TPC concept as described in this chapter 
well suited to large detector sizes like the DUNE far detector.

In this design, the charge attenuation on the long drift paths is compensated by the
charge amplification in the CRPs.  This configuration also simplifies
construction by optimally exploiting the long vertical dimensions of
the cryostat, providing a large homogeneous fiducial volume 
free of embedded passive materials (effectively increasing the detector size),
reducing the number of readout channels,  and ultimately lowering costs.  
The CRPs collect the charge in a projective way,  with practically no dead region and read the signals out 
in two collection views, eliminating the need for  induction views, which 
simplifies the reconstruction of complicated topologies. The tunable high S/N provides operative margins
with respect to the noise and electron lifetime conditions and lowers the threshold on the minimal
detectable energy depositions .

The dual-phase readout scheme has been successfully demonstrated on
several prototypes through R\&D over a span of more than 10 years.  The
design of very large (20--50~kt) underground detectors based on this
concept has been developed in great detail in the context of the
LAGUNA and LAGUNA-LBNO design studies.  The CERN WA105 demonstrator, 
described in Section~\ref{sec:proto-cern-double}, is intended to prototype 
a full-scale implementation of this technique, as
well as demonstrate other technologies developed for the construction of large
underground TPC detectors.  

 A complete configuration, based on the double-phase design, been optimized for the \ktadj{10}
detector module of the DUNE far detector.

%% file: volume-detectors/fd-alt-sections/fd-alt-charge-read.tex
%%%%%%%%%%%%%%%%%%%%%%%%%%%%%%%% 
\section{The Charge Readout System} 
\label{sec:detectors-fd-alt-chg-readout}

In the dual-phase LArTPC concept, the ionization electrons are multiplied in avalanches 
occurring inside detectors, the Large Electron Multipliers (LEMs), located in the argon gas 
phase above the liquid argon level. The drift field of the TPC brings the electrons up to the liquid argon surface where they can  be   
extracted into the gas using a 2-kV/cm electric field defined across the liquid-gas interface.
This extraction field is applied between a submersed extraction
grid (stainless steel wires tensioned in both $x$ and $y$
directions) and the bottom side of the LEMs.
The LEMs are printed circuit boards oriented horizontally, with
conductive layers (electrodes) on the top and bottom surfaces, and many holes drilled
through. The holes form a micro-pattern structure within which the amplification occurs.  
By applying voltages across the two
electrodes of the LEM, a 30-kV/cm electric field region is defined in the holes\cite{Bondar:2008yw}.
Electrons transiting these high electric field regions in the holes trigger Townsend multiplication in the
pure argon gas.

The amplified charge is then collected and recorded on a 2D anode
consisting of two sets of 3.125-mm-pitch gold-plated copper strips that provide the $x$
and $y$ coordinates (and thus two views) of the event.

Typical electric fields between each stage of the readout are
illustrated in Figure~\ref{fig:setup}. Table~\ref{tab:crp_dist} shows
the inter-stage distance and the tolerances required to obtain
uniformity of gain to within $\sim$5\%.
\begin{cdrfigure}[Dual-phase readout]{setup}{Illustration of the electric fields in the amplification region of a dual-phase LArTPC. The simulated field lines in dark blue indicate the paths followed by the drifting charges (without diffusion).}
 \includegraphics[width=.8\textwidth]{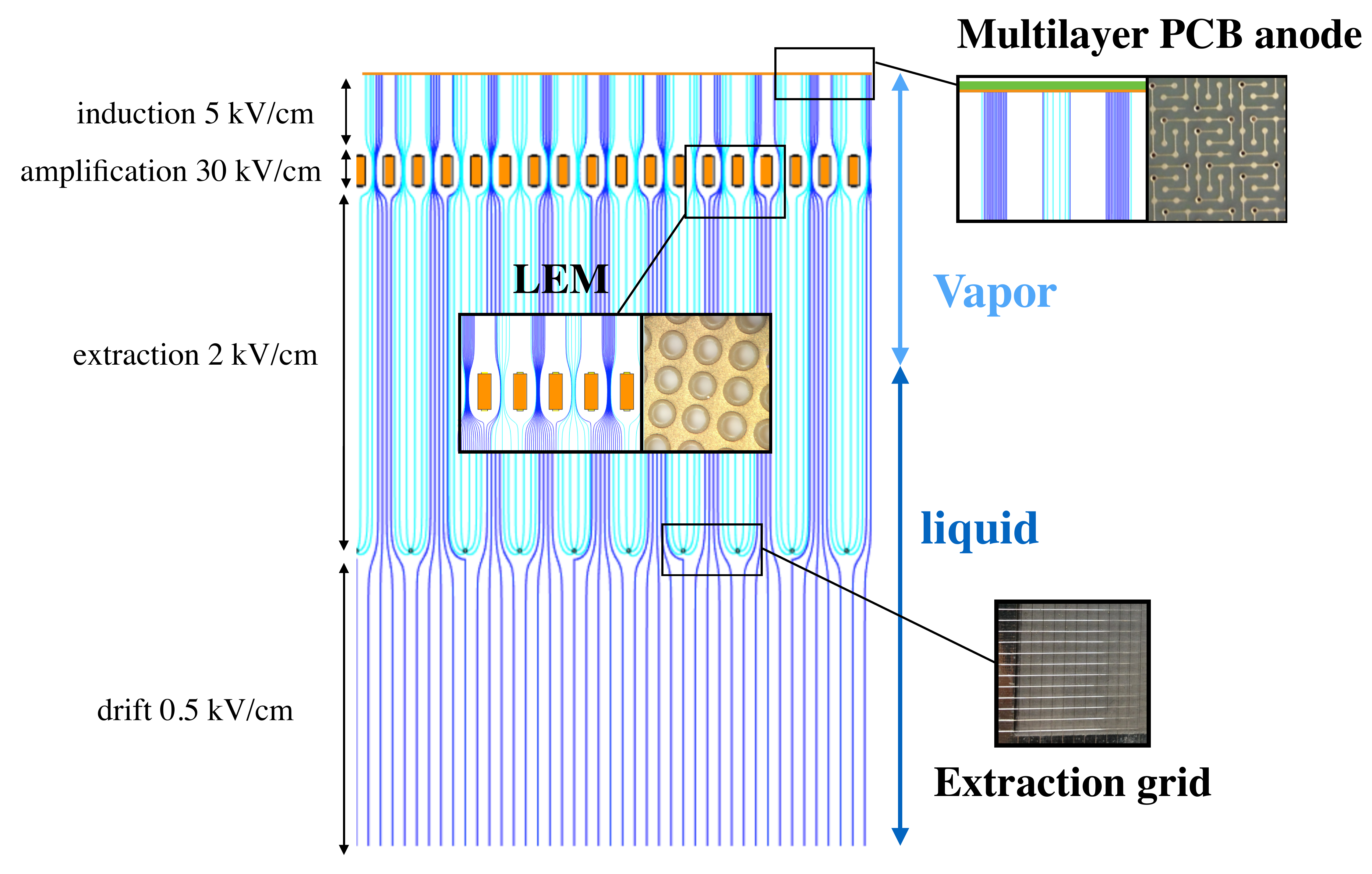}  
\end{cdrfigure}
\begin{cdrtable}[Interstage distances and electric field settings of the dual-phase readout components]{lp{2cm}p{2cm}l}{crp_dist}{Interstage distances and electric field settings of the dual-phase readout components.} 
 Component & Distance [mm] & Tolerance [mm] & Electric field [kV/cm]  \\ \toprowrule
 Anode-LEM top electrode  & 2 & 0.1 & 5\\ \colhline
 LEM top-bottom electrode   & 1 & 0.01 & 30-35\\ \colhline
 LEM bottom electrode-grid        & 10 & 1 & 2 (in LAr) and 3 (in GAr)\\
 \end{cdrtable}

The extraction grid, LEM and anode are assembled into three-layered ``sandwiches'' with precisely defined inter-stage distances and inter-alignment,  which are then connected together horizontally into
modular units of area \num{9}~m$^2$. These units are called Charge Readout Planes (CRPs).

\subsection{The Charge Readout Plane (CRP)}

Each CRP is an independent detector element that performs charge
extraction, multiplication and collection, and has its own high
voltage system and independent signal feedthroughs. The entire area of
the LEM and anode in a CRP is active.

The LEM and corresponding anode are pre-mounted in units of 50$\times$50~cm$^2$, called
LEM/Anode Sandwich (LAS) modules, before being assembled with an extraction
grid into a CRP. Each
anode in a LAS is segmented in 50-cm long $x$ and $y$ strips . Adjacent LAS anodes
are bridged together to form readout strips of the required length by
connecting short flat cables to KEL connectors soldered onto the top
sides of the anodes. The signals from the last anode in each 
strip chain are brought to feedthroughs
mounted on the other side of the front-end electronics embedded inside
dedicated signal-feedthrough chimneys using 50-cm-long flat cables.

The LBNO 20-kt detector design (described in~\cite{cdr-annex-lbno-2}) featured
modularized CRPs of dimensions of 4$\times$4~m$^2$, with 2-m long
anode strips. For the DUNE cryostat geometry, a size of 3$\times$3~m$^2$
with a strip length of 3~m is found to be optimal. The description in
this section is based on the LBNO 4$\times$4~m$^2$ CRP.

Each CRP is independently hung from the vessel deck through its three
suspension feedthroughs. It has its own high voltage system and 
independent signal and slow-control feedthroughs.
Figure~\ref{fig:4_4CRP_FRONT} illustrates the 4$\times$4 m$^2$ CRP;
its characteristics are summarized in Table~\ref{tab:crp_para}.

\begin{cdrfigure}[Side and top views of the $4\times4$~m$^2$ LBNO CRP]{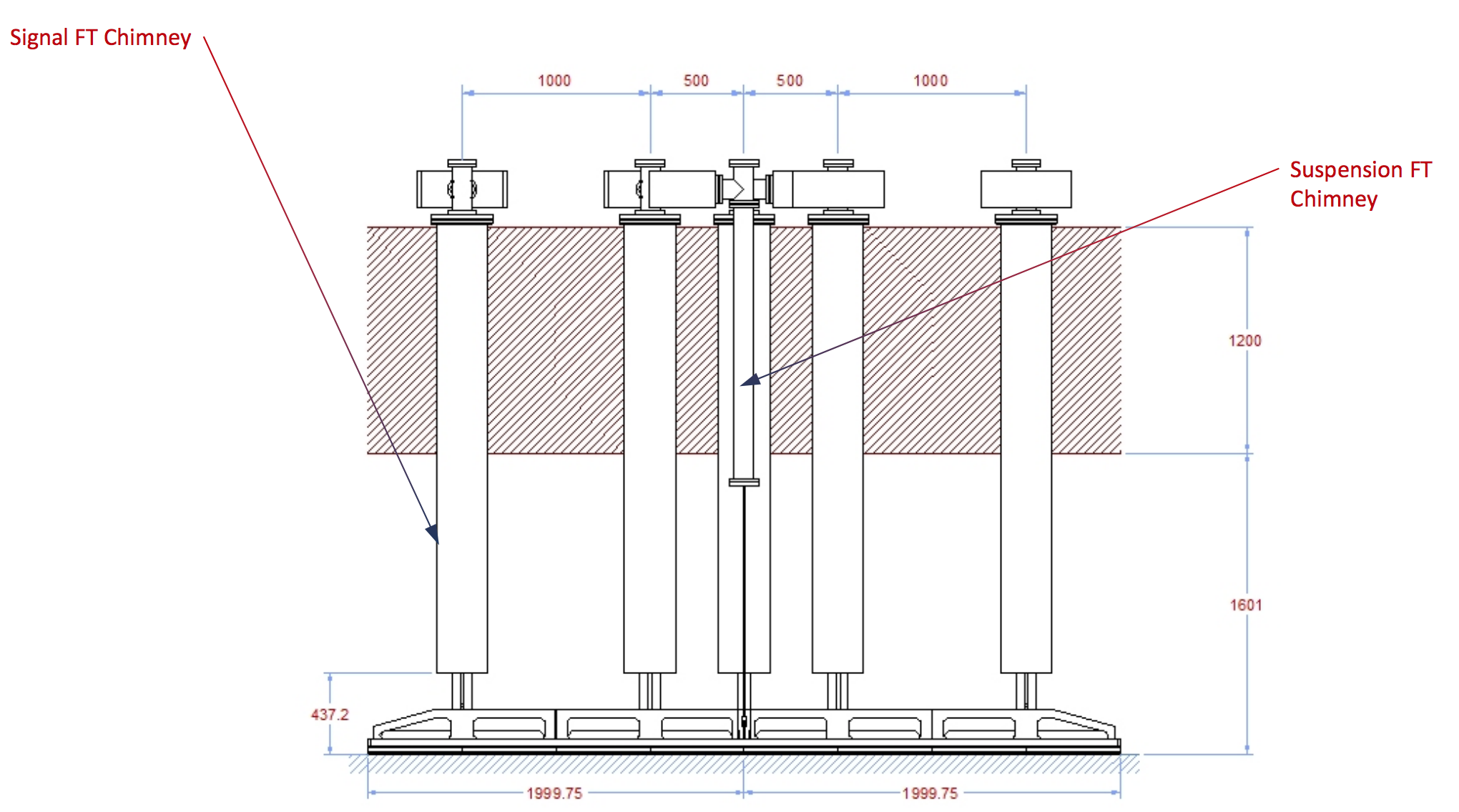}{Side and top views of the $4\times4$~m$^2$ CRP designed for LBNO (units in mm).}
 \includegraphics[width=\textwidth]{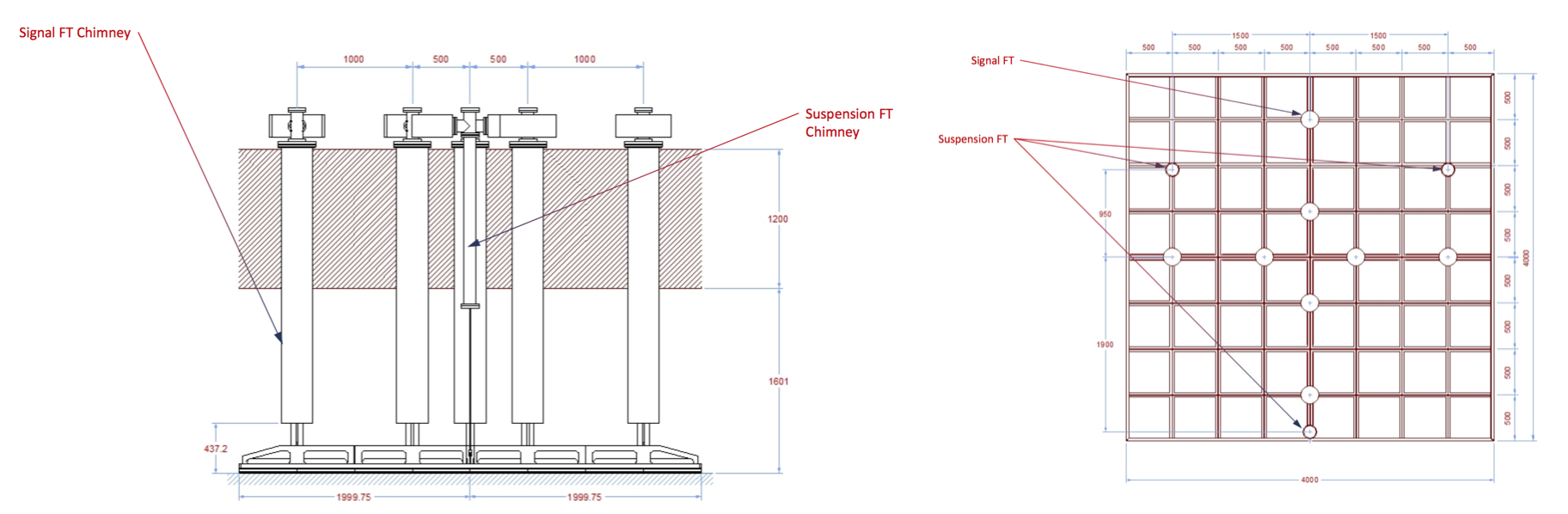}  
\end{cdrfigure}
\begin{cdrtable}[Numbers of components of the 4$\times$4 m$^2$ CRP]{lr}{crp_para}{Numbers of components of the 4$\times$4 m$^2$ CRP designed for LBNO} 
Component & Number \\ \toprowrule
$50\times50$ cm$^2$ anode panels & 64\\ \colhline
$50\times50$ cm$^2$ LEM  panels&  64\\ \colhline
Signal  feedthroughs & 8\\ \colhline
Suspension  feedthroughs & 3\\ \colhline
Readout strip length (m)& 2\\ \colhline
Number of channels & 5120\\
\end{cdrtable}

The entire area of the LEM and anode is active, as noted earlier, and
each adjacent 50$\times$50~cm$^2$ LAS module has a gap of only 0.5~mm.
Therefore, the 4$\times$4~m$^2$ area of the CRP is fully 
active; the 0.5-mm edge gaps occurring every 50~cm
do not interfere with the
charge collection in the  
anode, given its readout pitch of 3.125~mm.

The extraction grid consists of 100~$\mu$m diameter stainless steel
wires tensioned in both $x$ and $y$ directions over the entire 4-m
length/width of the CRP  with 3.125~mm
pitch. They are soldered into groups of 32 on independent
wire-tensioning pads oriented perpendicularly to the side of the CRP
frame.  Each wire-tensioning pad consists of a printed circuit board
(PCB) for HV-connection that is fixed very precisely to a mechanical 
wire holder. The PCB has 32 soldering pads with 200-$\mu$m grooves for
precise positioning of the wires. During the wire-soldering process
each wire is tensioned by 150-g lead weights and positioned in a
groove.  (With this method better than 50~$\mu$m precision on the wire
pitch, measured under the microscope, was achieved for the LBNO-WA105
prototypes.) The PCB is then mounted on the wire holder and the
tension of the group of 32 wires can be precisely adjusted by pushing
the holder against the CRP's FR4 frame with two screws.

The wires, $\sim$3~m long in both $x$ and $y$ directions,
have their sags minimized to $\sim$0.1~mm thanks to $x$ and $y$ oriented
supporting comb-teeth blades (see Figure~\ref{fig:Wires_comb}) inserted
between anode planes of 1~m$\times$1~m size. The array of blades
penetrates the liquid surface and has the additional
benefit of maintaining the liquid level still. 

\begin{cdrfigure}[Comb for hanging extraction grid wires]{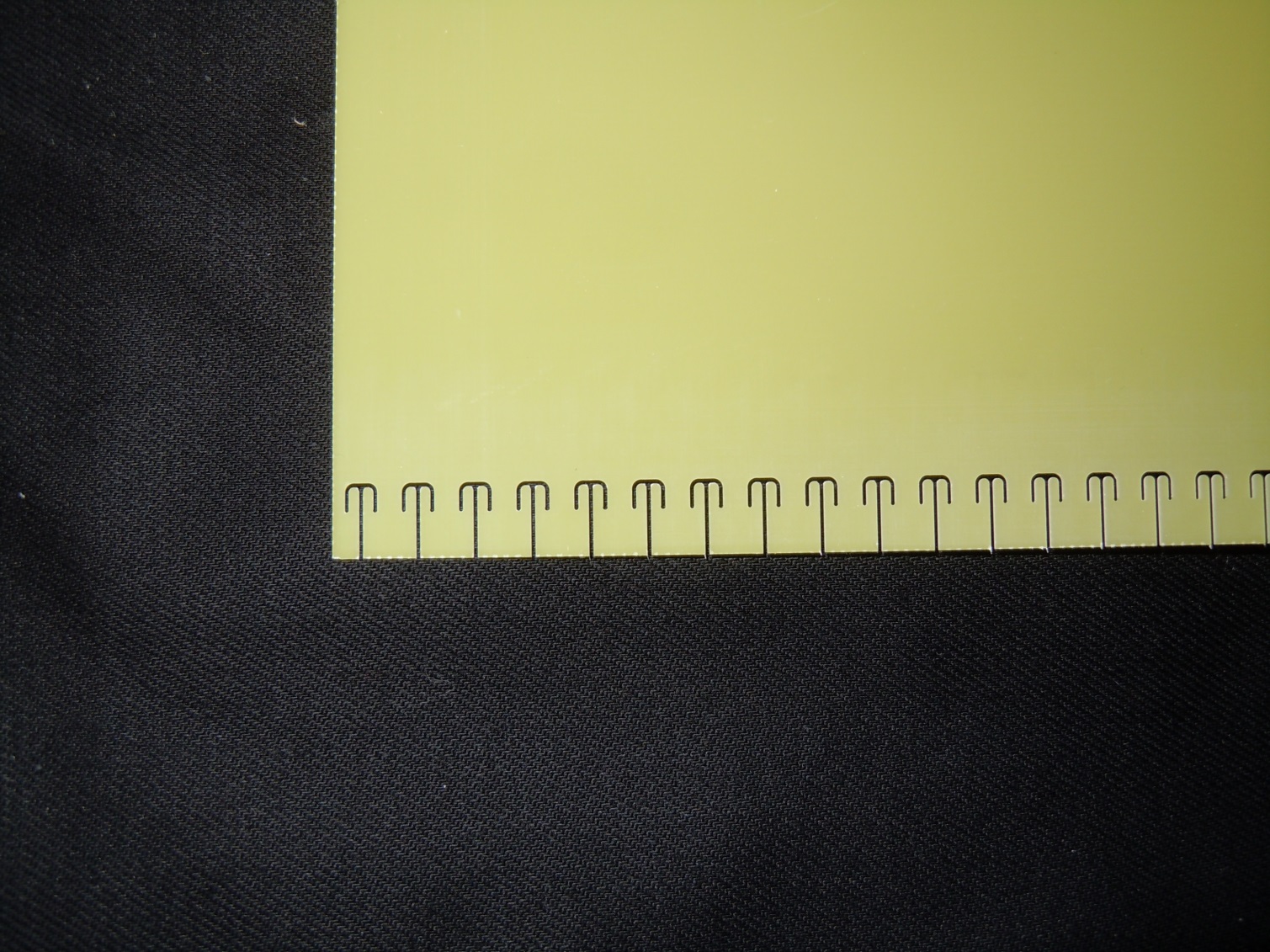}{Comb for hanging extraction grid wires}
\includegraphics[width=.6\linewidth]{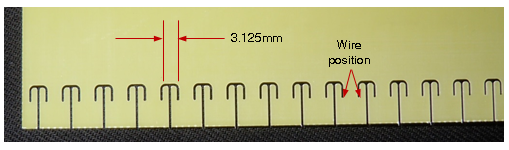}
\end{cdrfigure}

The 4$\times$4~m$^2$ CRP has 5120 readout channels in total. The
signals from the CRP are read out through eight signal feedthroughs
(SFTs) chimneys
at the bottom of which the front-end electronics cards are mounted. Amplified signals
are transmitted to the DAQ system located on top, outside of the vessel.
Each chimney groups 640 channels. 

The 3$\times$3~m$^2$ DUNE CRP, is a down-sized version of the LBNO
CRP; it has three signal feedthrough chimneys and 1920 readout
channels.

Three suspension
feedthroughs are arranged as an equilateral triangle whose barycenter
coincides with that of the CRP; they suspend the CRP at the required
position and precisely adjust the CRP level with respect to the liquid
argon surface. Figure~\ref{fig:4_4CRP_3D} shows a 3D view of the CRP,
where the signal chimneys  (discussed in 
Section~\ref{sec:detectors-fd-alt-elec}) and the stiffening frame are
visible.
\begin{cdrfigure}[3D view of the $4\times4$~m$^2$ CRP.]{4_4CRP_3D}{3D view of the $4\times4$~m$^2$ LBNO CRP.}
\includegraphics[width=0.8\textwidth]{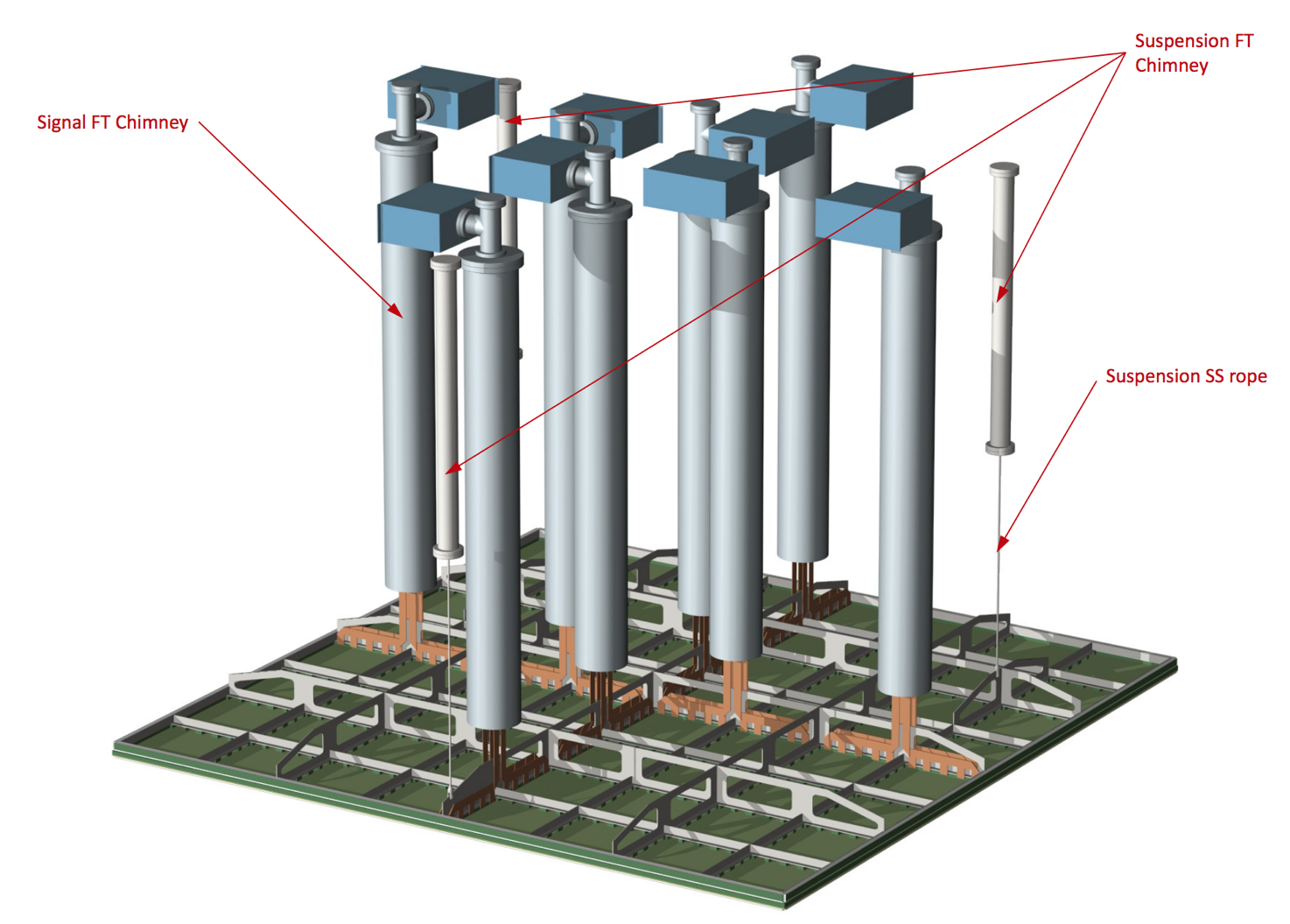}  
\end{cdrfigure}

\subsection{The LEM/Anode Sandwich (LAS)}

LAS modules, the CRP building blocks composed of 50$\times$50-cm$^2$ LEM-anode 
sandwiches,  have been extensively studied as part of the ongoing CERN
WA105 prototyping efforts (see~\ref{sec:proto-cern-double}). The LEMs
and the anodes are produced by a PCB manufacturing company called
ELTOS (\url{www.eltos.it}). Their designs are the outcome of
intensive R\&D effort over the last few years, aimed at maximizing the
S/N ratio for the large-area readouts envisioned for use in giant
dual-phase LArTPCs.  Figure~\ref{fig:LEM_anode} shows the LEMs and
anodes.  This section summarizes key features of the LAS.

\begin{cdrfigure}[Pictures of the LEM and anode along with microscope views]
{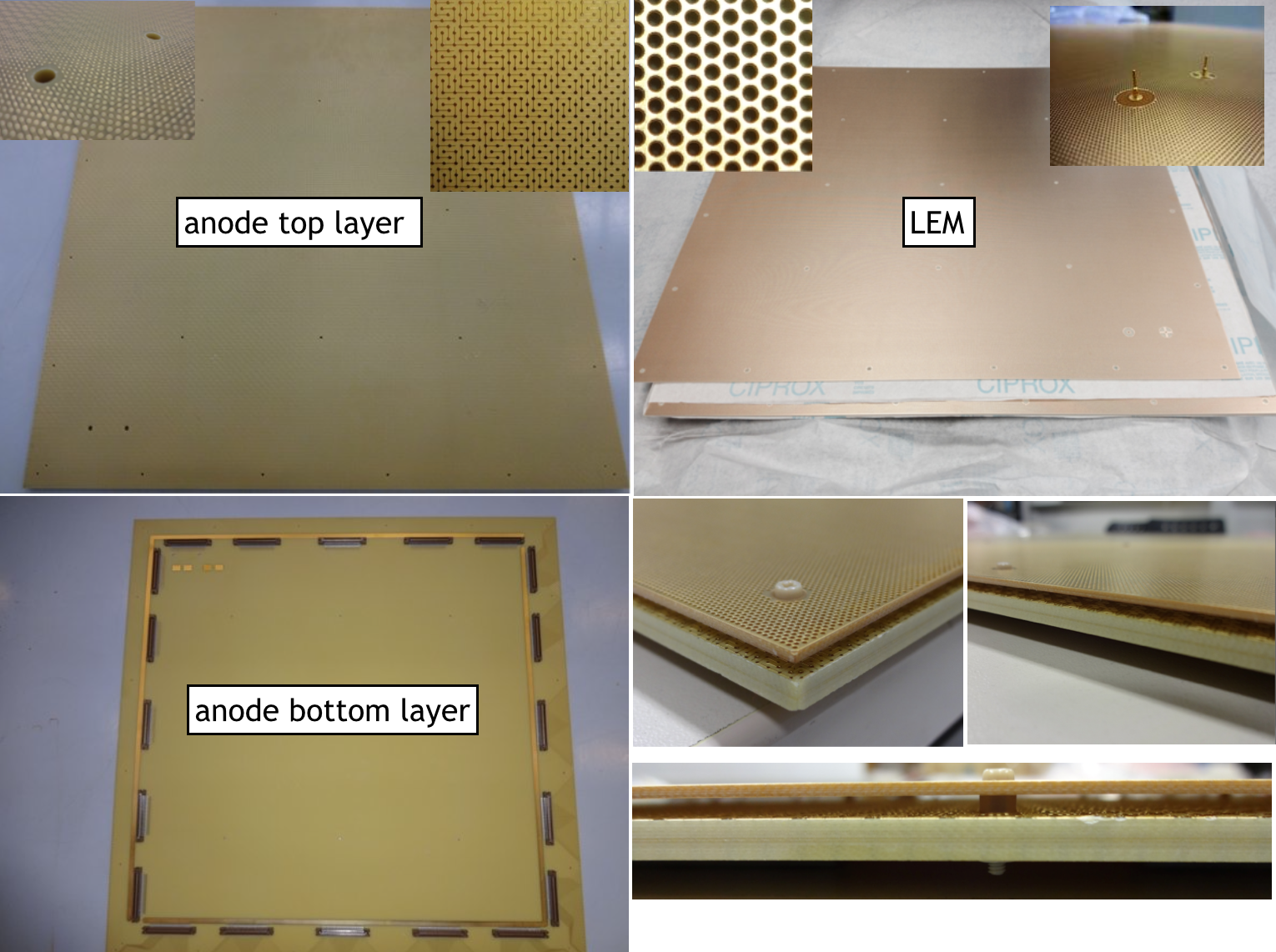}{Top: pictures of the LEM and anode along with microscope
  views. Bottom: close up of the LEM HV connectors and back view of the anode 
with the KEL signal connectors to bridge to the adjacent LAS or to connect 
flat cables going to the signal feedthrough}
 \includegraphics[width=.8\textwidth]{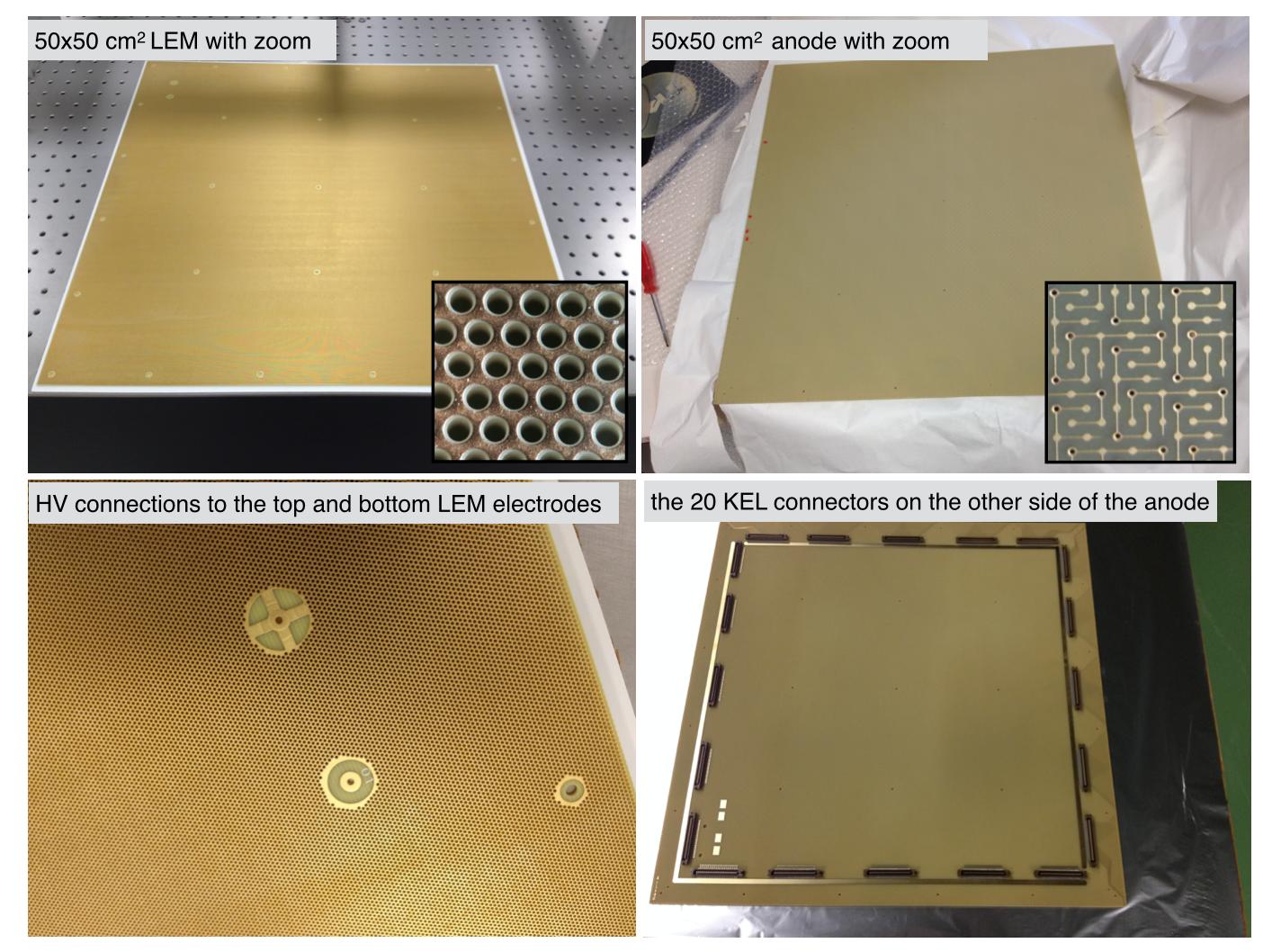}  
 \end{cdrfigure}

 \paragraph{The 50$\times$50~cm$^2$ anode:}

Each  50$\times$50~cm$^2$ anode is manufactured from a single multilayer printed circuit
board (PCB). The readout strips for both $x$ and $y$ views consist of a pattern of
gold-plated copper tracks 
with a 3.125-mm readout pitch. The two views have superimposed track patterns
that are  electrically insulated from one another. Electrical insulation in the points where
the $x$ and $y$ tracks would superimpose is achieved by having tracks crossing
over and under each other using a system of vias between the top and bottom layers of the PCB.
 
The design of the track patterns forming the strips is such that both $x$ and $y$ views
collect the same amount of charge, independent of the angle of
charged-particle tracks with respect to the readout strip
orientation. The tracks pattern should then ensure a uniform and isotropic coverage of the strip surface while 
minimizing the strip capacitance. These criteria have driven a thorough design optimization.  
Various PCB layouts were tested in order the achieve the best performance, as described in~\cite{Cantini:2013yba}. 
The final layout and schematic of the anode are shown in Figure~\ref{fig:anode_sch}. 

\begin{cdrfigure}[The 2D anode]{anode_sch}
{The 2D anode (left) and its schematic showing the  interconnections  ensuring continuity within each view while preserving insulation
with respect to the perpendicular view  (right). One view is filled  in red and the other in white.}
\includegraphics[scale=0.2]{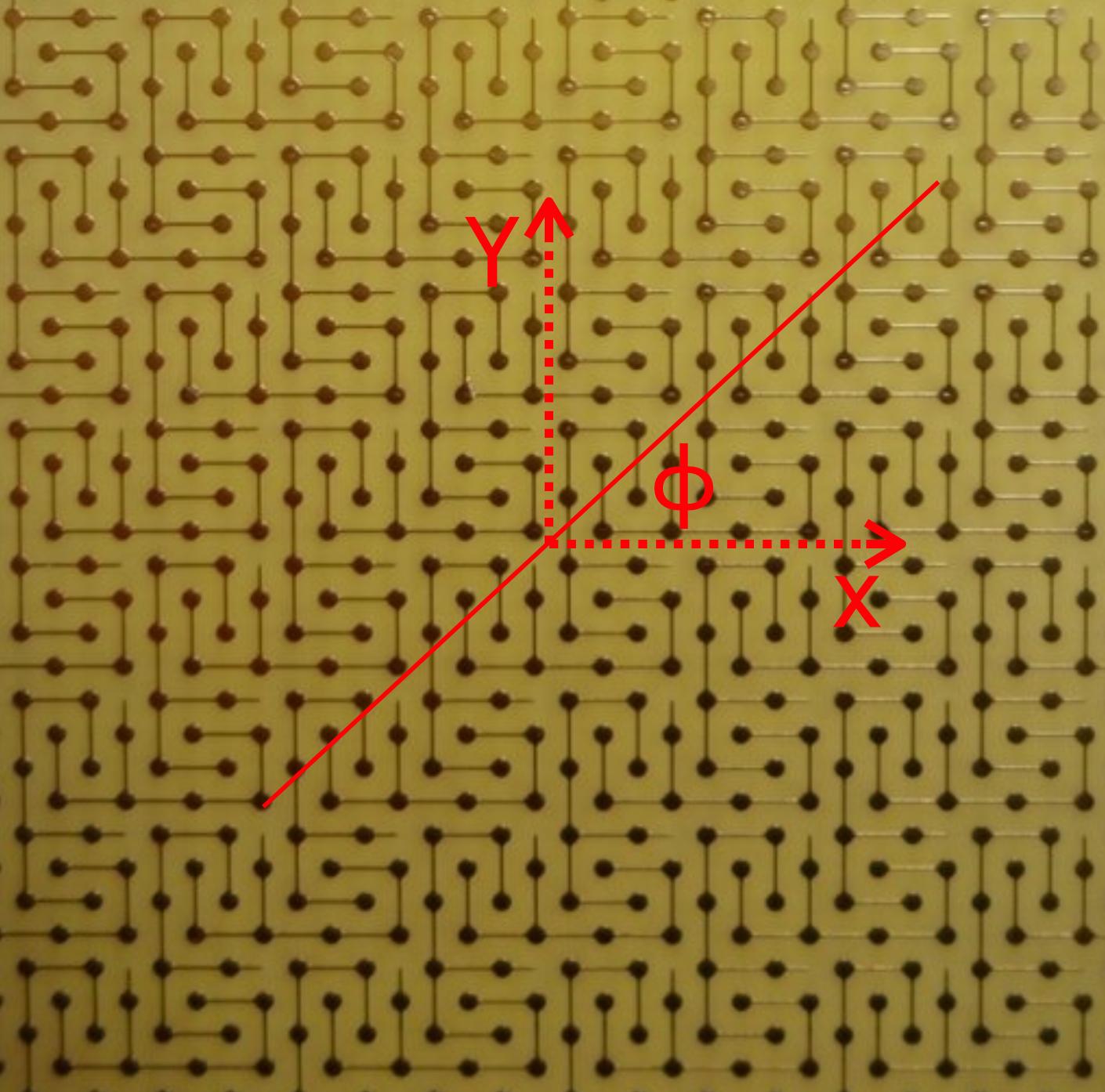} \hspace{0.2cm} \includegraphics[scale=0.2]{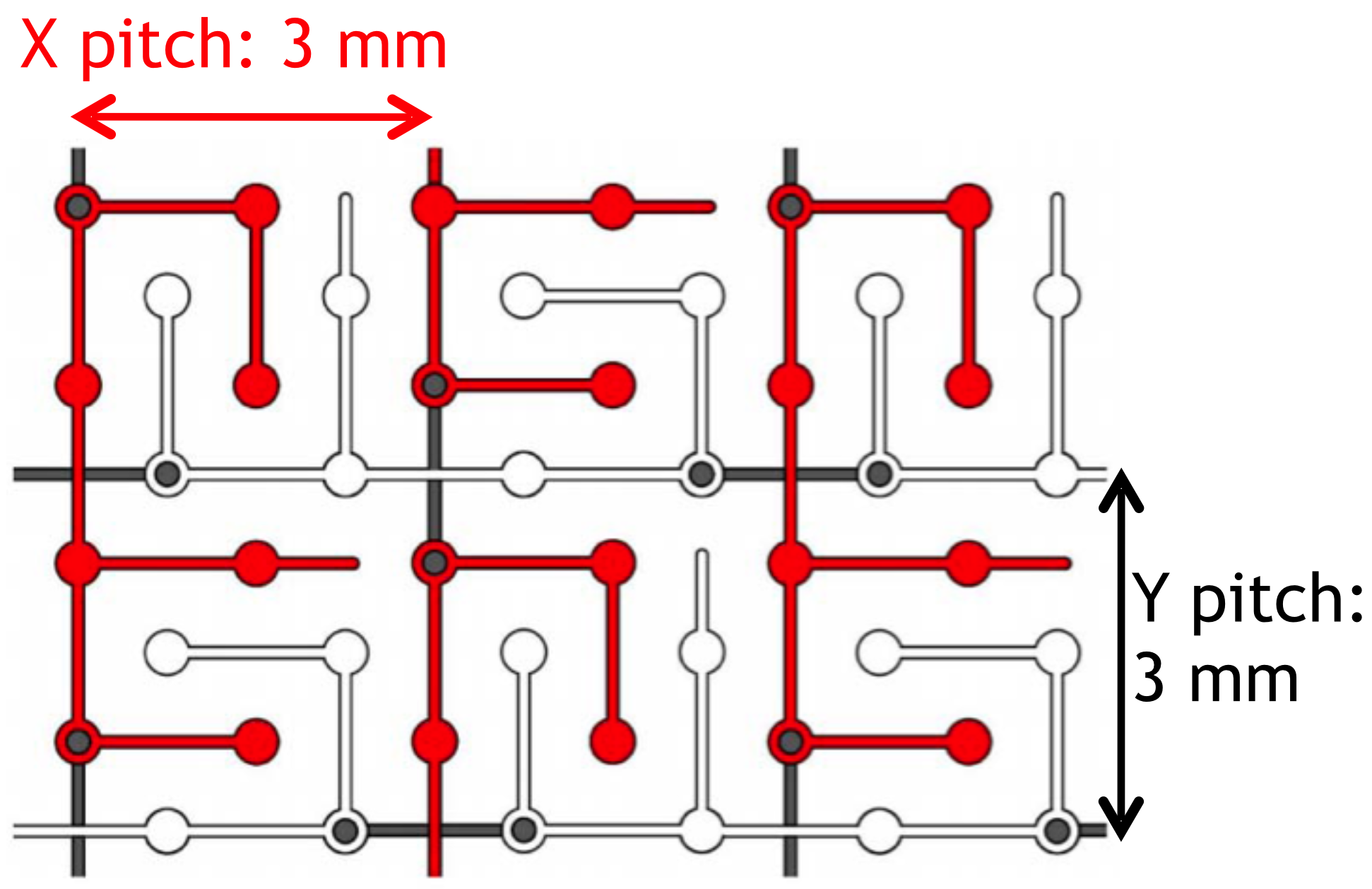}
\end{cdrfigure}

As result of this optimization, the electrical capacitance of the
readout strips has been limited to only 150~pF/m, which translates
into an electronic noise of about $\sim$1000 electrons for a 2-m readout
length.  Figure~\ref{fig:anode_res} (right) shows that the
charge-sharing asymmetry between the two views is kept within 1\%. 
The two views can thus be treated in a completely equivalent way
from the point of view of the reconstruction. The response in terms of
the charge collection per unit pathlength $\Delta Q/\Delta s$ is
independent of the charged-particle tracks' azimuthal angle $\phi$
(see Figure~\ref{fig:anode_res} left and middle).
\begin{cdrfigure}[Charge deposition as function of track angle ]{anode_res} {Charge deposition per unit of pathlength measured on LEM view 0 
($\Delta Q_0/\Delta s_0$) as a function  of the track angle $\phi$ (left) and  projection of the  $\Delta Q_0/\Delta s_0$ distribution in three $\phi$ intervals (middle).  The right plot  shows the distribution of the difference between the total charge  collected on both views normalized to their sum}
\includegraphics[width=.9\textwidth,scale=1]{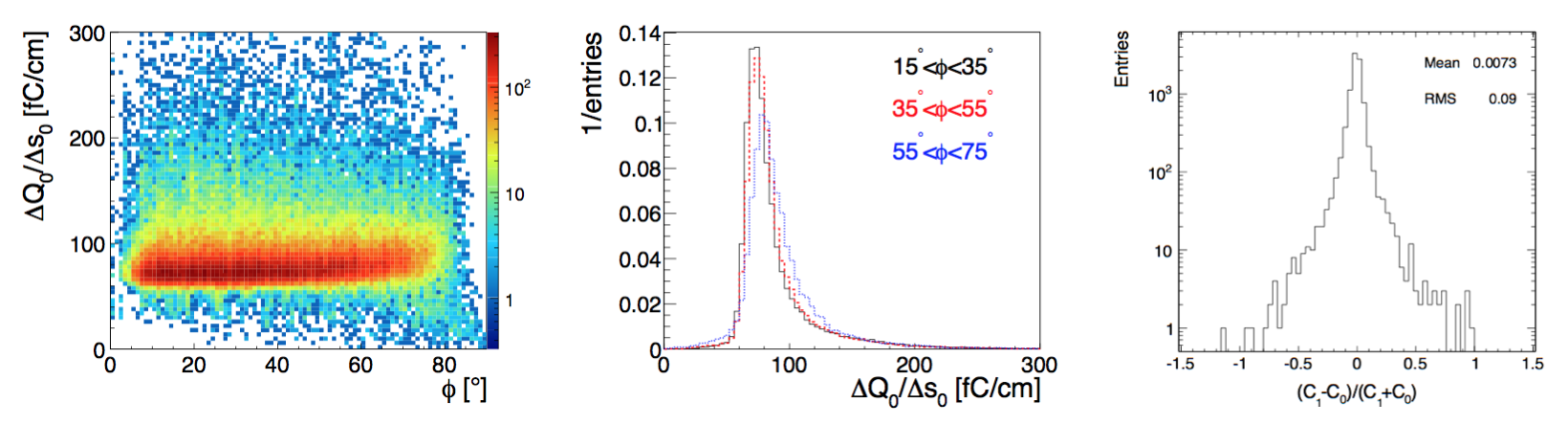}
\end{cdrfigure}

\paragraph{The 50$\times$50~cm$^2$ LEM:}

Each LEM is built from a 1-mm-thick copper-clad epoxy PCB with
500~$\mu$m diameter holes drilled through, surrounded by a
40-$\mu$m dielectric rim. The holes are arranged in a honeycomb
pattern with a pitch of 800~$\mu$m, resulting in about 200 holes per
cm$^2$ and $\cal O$(500,000) holes over the entire 50$\times$50~cm$^2$
area. The holes provide confinement for the UV photons produced during
the avalanche process and thus act as a mechanical quencher to prevent
photon feedback. This property makes the LEM suitable for operation in
ultra-pure argon vapor without the addition of a quenching gas. 

The amplification of the drifting charges in pure argon vapor at 87~K with
LEMs has been extensively demonstrated on a chamber with
10$\times$10~cm$^2$ area readout (see e.g.,
\cite{Badertscher:2008rf,Badertscher:2010fi}) as well as on
a larger device consisting of a 40$\times$80~cm$^2$
readout\cite{Badertscher:2013wm}.  Both setups were successfully and
stably operated at constant gains of at least 15, corresponding to
S/N $\approx$ 60 for MIPs. Recent studies\cite{Cantini:2014xza}
systematically characterize the impact of the rim size, insulator
thickness, hole diameter and hole layout on 10$\times$10~cm$^2$ area
LEMs. The response in terms of maximal reachable gain and influence on
the collected charge uniformity, as well as the long-term stability of
the gain, has been thoroughly compared for these different
layouts. Some results are shown in Figure~\ref{fig:LEM}.  Gains of
almost 200 were reached and the LEMs could be operated at stable gains
of at least $\sim$15 after a charging up period of about a day.
\begin{cdrfigure}[LEM performance vs geometry]{LEM}{Performance of the LEMs with different geometry parameters. Left: effective gain vs. LEM electric field; right: the stabilizations of the effective gain over time.}
\includegraphics[scale=0.35]{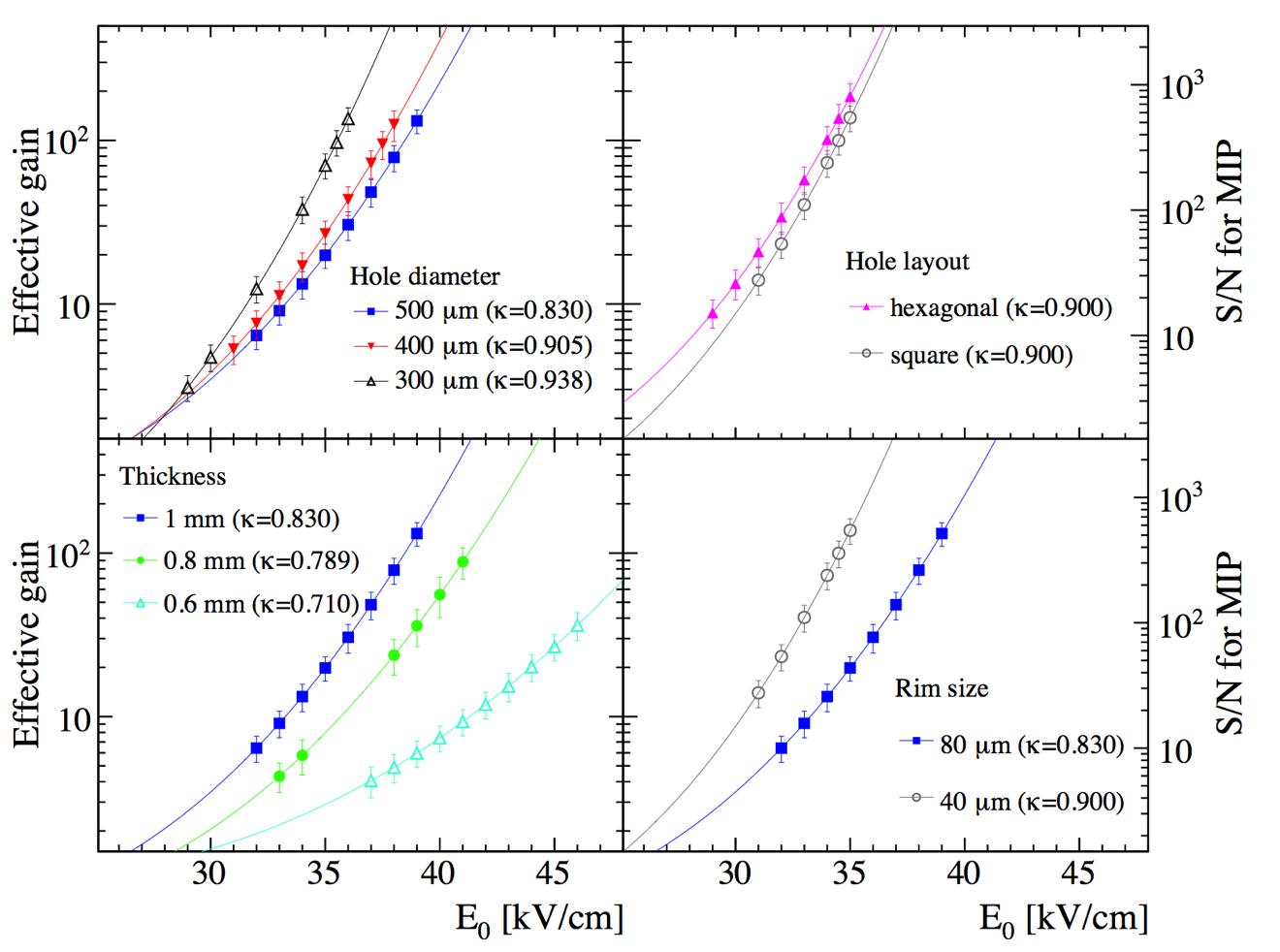}
\includegraphics[scale=0.32]{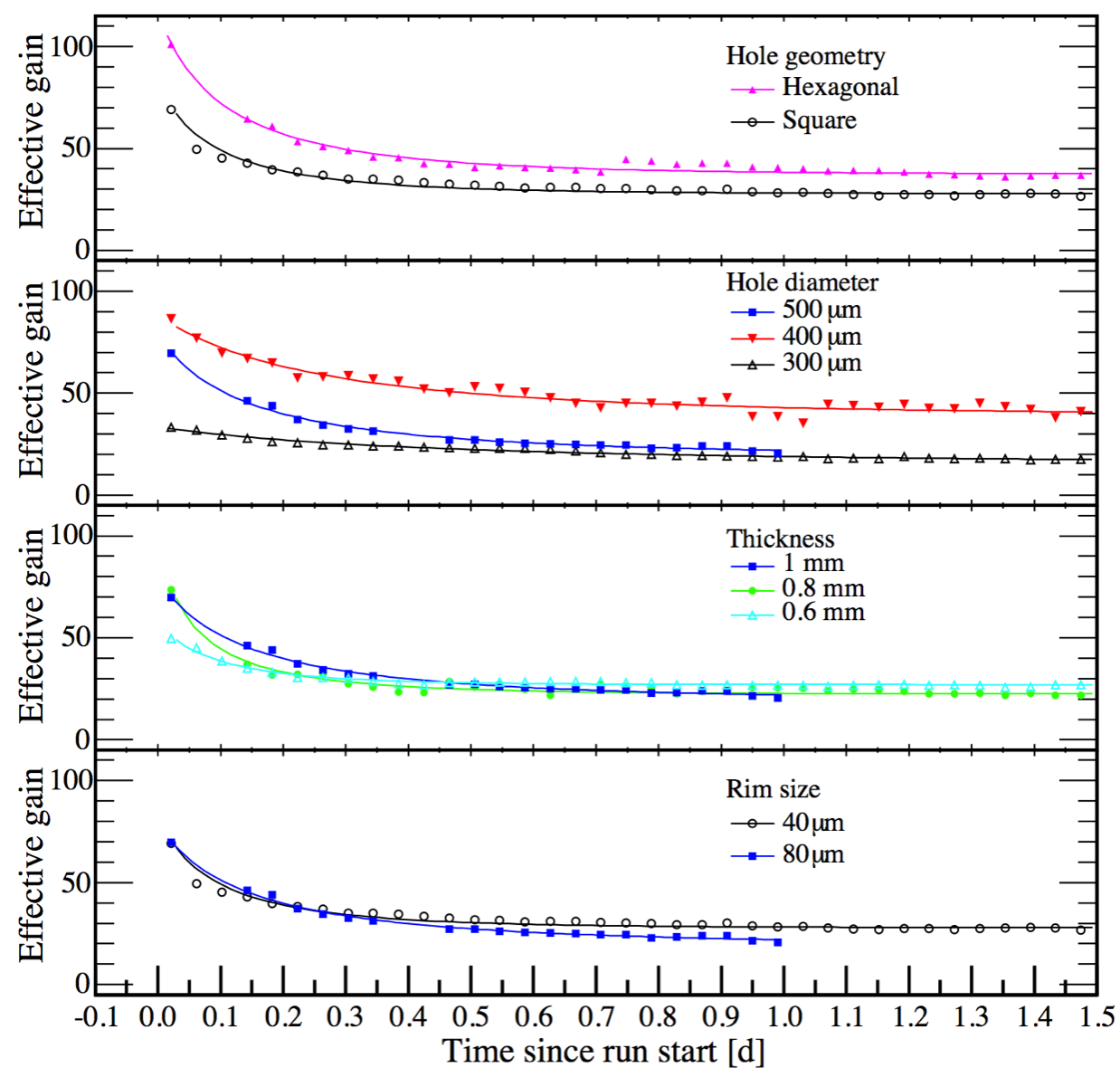}
\end{cdrfigure}

\paragraph{LAS Assembly:}

Figure~\ref{fig:LEM_metro} shows the LEM/anode sandwich (LAS).  A LAS is 
fixed together with 29 M2 PEEK screws, each containing a precisely
machined 2-mm-thick pillar to guarantee a constant inter-stage
distance between LEM and anode on the entire $50\times50$~cm$^2$
area.  The dead zones caused by the supporting pillars and the two HV
pins on the LEMs are minimized and make up less than 0.5\% of the
total area. The inter-stage distance between the LEM and anode in the
LAS has been measured at many points. The results are shown in
Figure~\ref{fig:LEM_metro} and are described
in~\cite{EDMS_metro_lem_anode}. They indicate that the planarity is
within the required tolerance of 2~mm $\pm$ 100~$\mu$m .
\begin{cdrfigure}[LEM/anode sandwich metrology]{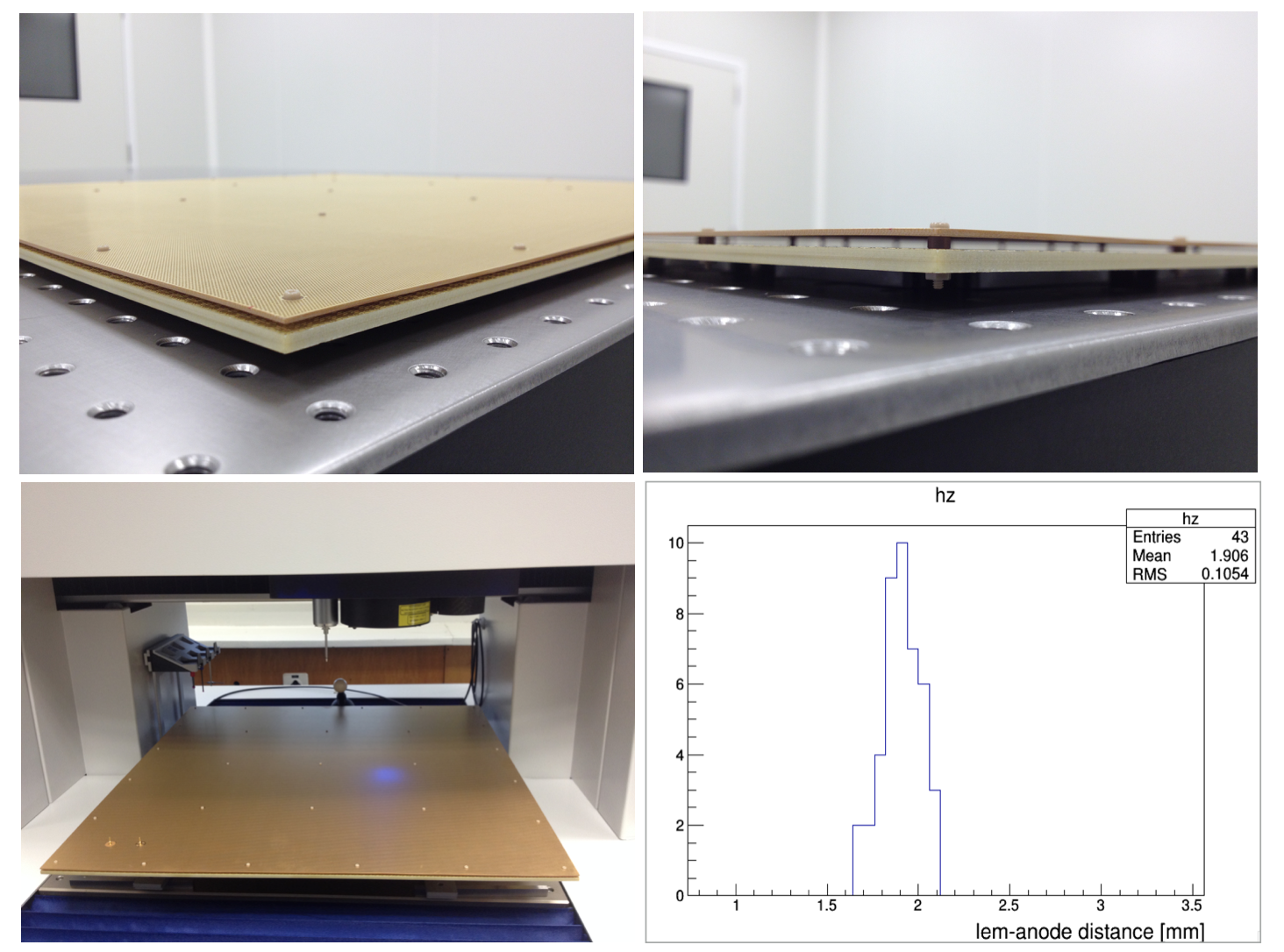}{Close up pictures of the LEM/anode sandwich. The two
       bottom figures show a the measurement at the CERN metrology lab
       and a histogram illustrating the measured gap between the LEM
       and anode in various points. As can be seen the distribution is
       centered on the nominal distance of 2~mm and has an RMS of
       about 100$\mu$m.}
     \includegraphics[width=.7\textwidth]{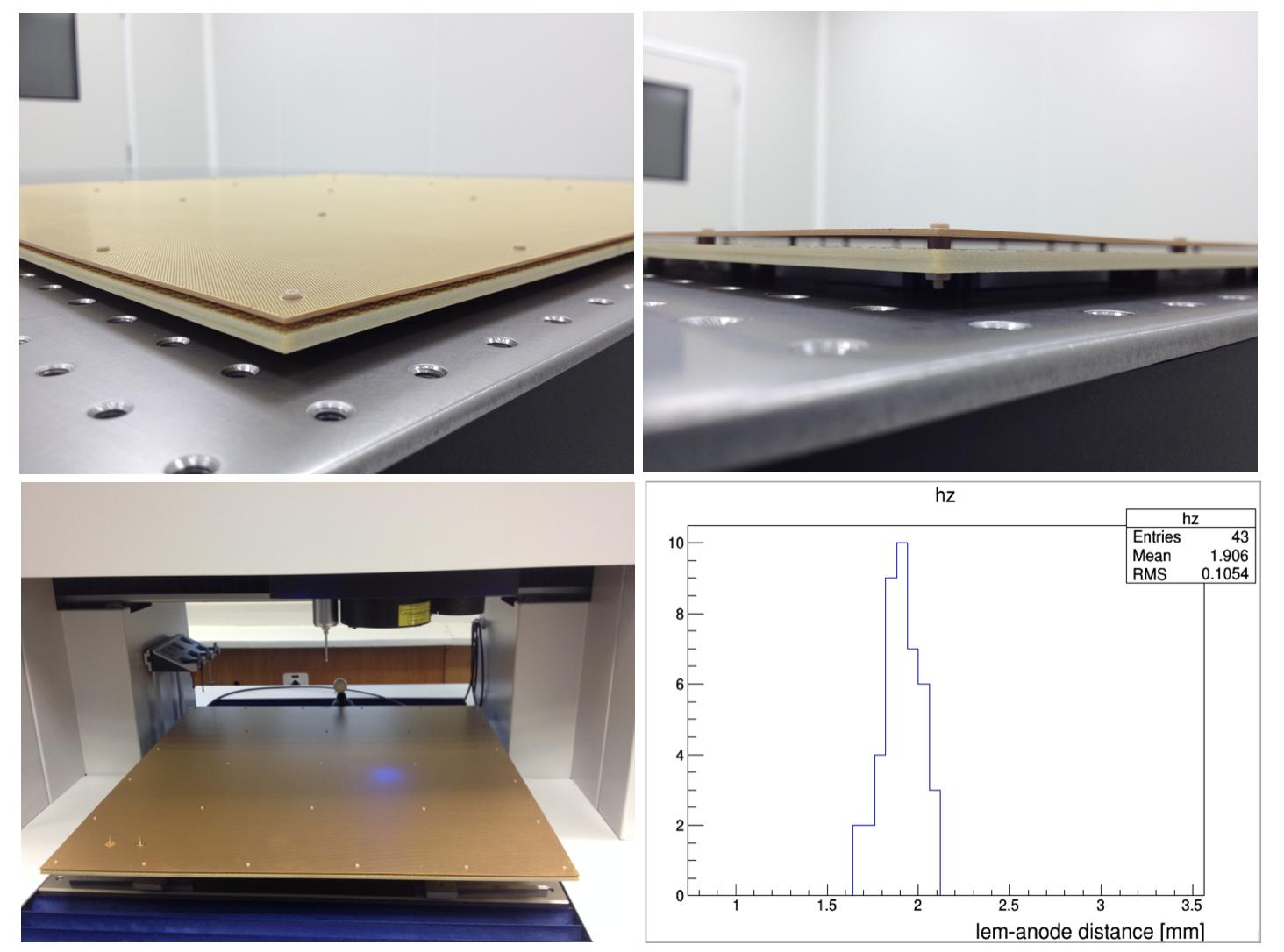}
\end{cdrfigure}

The entire mounting sequence of the sandwich as well as that of the
different elements of the CRP are being addressed in the WA105
prototype detectors. An example of a sandwich assembly on a
3 $\times$ 1~m$^2$ CRP is shown in Figure~\ref{fig:CRP_assembly}.
\begin{cdrfigure}[Pictures of the assembly of a $3\times1$m$^2$ CRP]{CRP_assembly}{Pictures of the assembly of a $3\times1$m$^2$ CRP}
     \includegraphics[width=\textwidth]{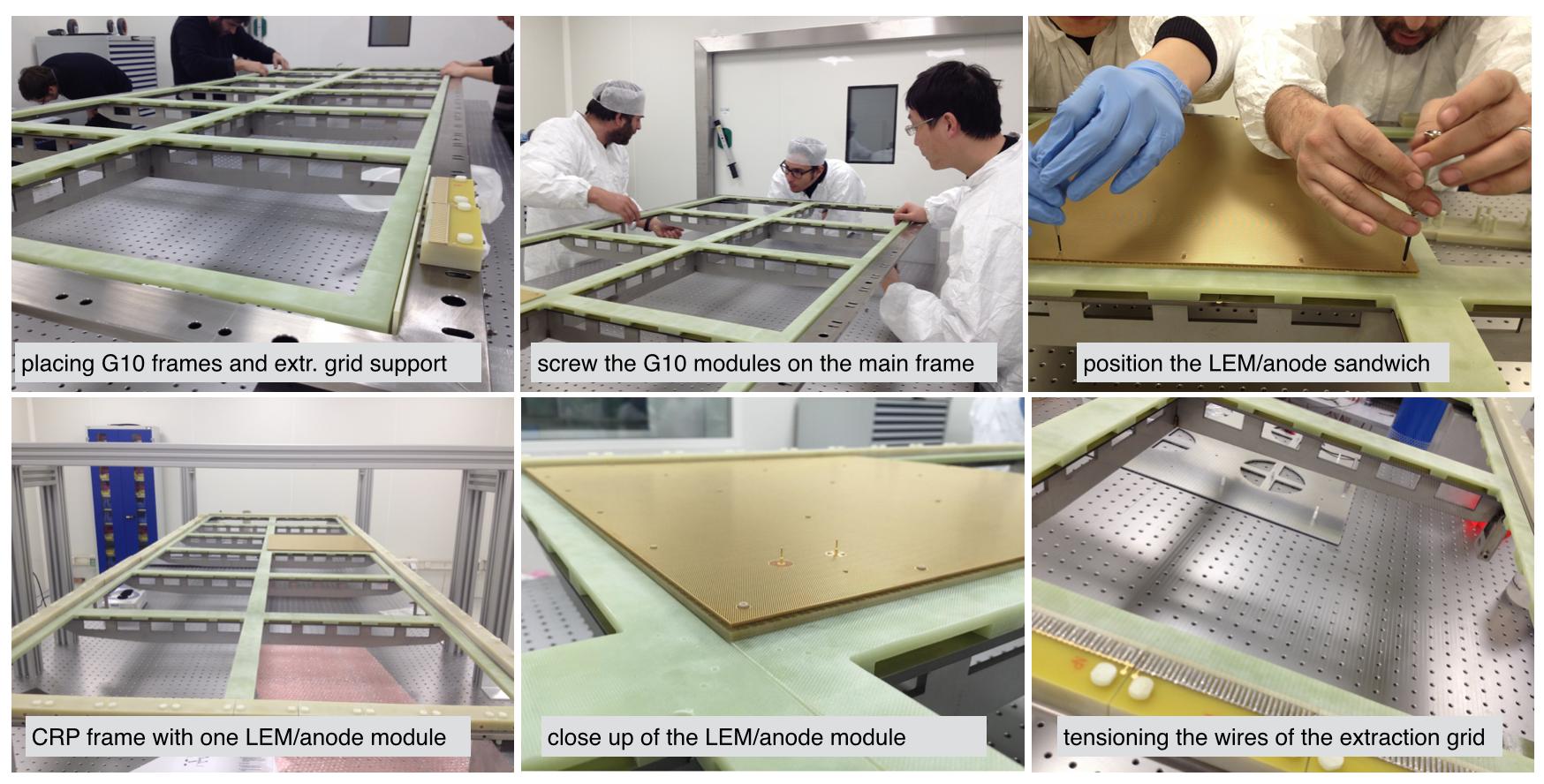}  
\end{cdrfigure}

%% file: volume-detectors/fd-alt-sections/fd-alt-fieldcg-hv.tex
%%%%%%%%%%%%%%%%%%%%%%%%%%%%%%%% 
\section{The Field Cage, High Voltage System and Cathode} 
\label{sec:detectors-fd-alt-hv}

This section describes the design of the high voltage system, field
cage and cathode for the TPC.  It is inspired by the LBNO 20-kt and
50-kt detector designs, described in~\cite{cdr-annex-lbno-2}, which can
be simplified and down-sized for the DUNE detector (12~kt or 15~kt),
due to the shorter drift path and the rectangular aspect ratio of the
detector. The much shorter transverse dimension of DUNE's with respect
to the LBNO design (12~m vs 40~m span)  permits a lighter cathode structure (less sag
requires less compensation) and a simpler hanging system for the drift
cage and the cathode.

In the LBNO design the field cage is composed of equally spaced
octagonal rings stacked around the active volume 
that create a uniform drift field; they have an intensity 
 that is adjustable in the range 500 to 1000~V/cm. This
leads to a cathode voltage of up to 2~MV when operating at the maximal field 
intensity of 1~kV/cm over a drift distance of 20~m.

Two different approaches have been developed for the drift-field high voltage
generation system. 
The first one uses an external HV power supply and uses
HV feedthroughs to penetrate into the detector volume. The WA105
demonstrator will use this approach for its drift of 6~m and a cathode
voltage up to 600~kV.  The second approach places a HV generator, the Greinacher HV
multiplier, directly inside the LAr volume. It is an innovative technique, with some advantages
relative to the first approach. This technique particularly suits
giant-scale detectors that require a very high voltage of
$\sim$1--2~MV.

The DUNE detector (12-m drift) requires a voltage of 600~kV in order
to operate at a field intensity of 0.5~kV/cm. This voltage is a factor
of 3.3 higher than that of the reference design
(Chapter~\ref{ch:detectors-fd-ref}) and it will be tested during the
WA105 detector operation at 1~kV/cm over a 6-m drift.

The field cage designed for the LBNO 20-kt detector (20-m drift path)
is composed of 99 octagonal field-shaping coils manufactured from 316L
stainless-steel tubing and long radius elbows to EN 10217-7 shop.  The
straight pipes and the elbows are assembled to form the coils
by using a combination of welded and clamped joints.

The coils are supported by 32 off-hanging columns of G-10CR glass
fibre/epoxy-laminated sheet insulating material, built in the form of
chains, and suspended from the tank deck structure.

Each coil is designed as a series of fully welded infill tubes
intended to fit between pairs of hanging support columns to form one
section of the field cage.  Short sections of the field-shaping coils
are integrated as pins into links to assemble each chain.  Longer
sections and corner sections of the field-shaping coil are then fixed
between the hanging columns to complete each coil
(Figure~\ref{fig:LBNO_FC}). The combined assembly of 99 sets of field
shaping coils within the 32 off-hanging columns provides a complete
field cage.
\begin{cdrfigure}[Assembly concept of the field cage]{LBNO_FC}
{\small Left: assembly of the elements of a hanging chain. Right: 
construction of a field cage section from the hanging chains and the field shaping coil elements.}
\includegraphics[width=.5\linewidth]{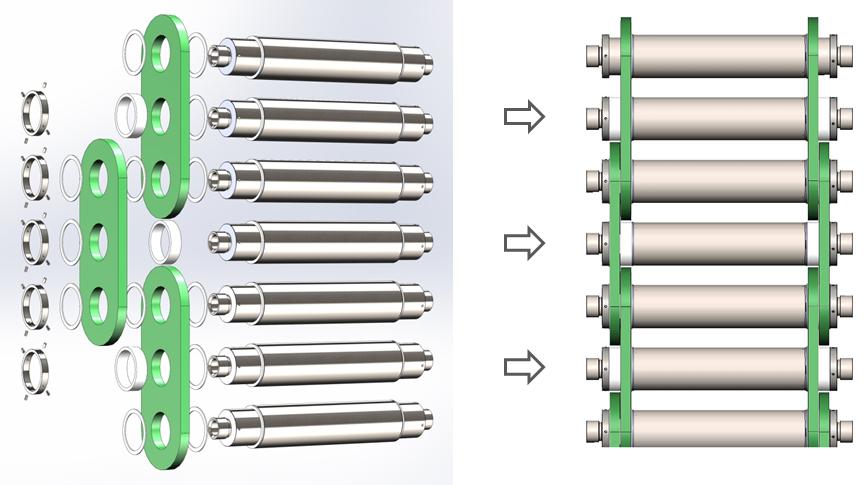} \hfill
\includegraphics[width=.4\linewidth]{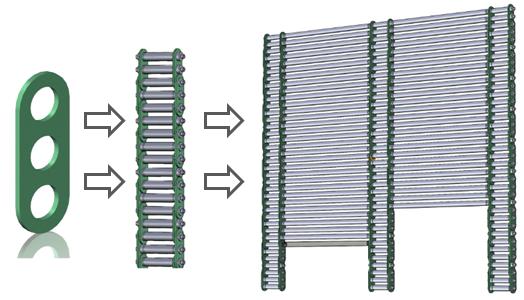}
\end{cdrfigure}

The infill tube specifications assume an outer tube diameter of 139.7~mm,
which is common to the cathode structure.  This allows a thinner wall,
1.6~mm, to be used in non-structural parts of the coils.  Although a
non-standard size, the total length of this tube can be manufactured
as a special mill run.  This will make it possible to save 21~tons of
material relative to the standard tubes (wall thickness of
2.0~mm). The wall thickness of the link pins is 2.6~mm; this will
provide sufficient stiffness to resist the bending torques across
the link pins.  All specialist preparation and welding of the link
pins will be carried out in shop facilities under controlled
fabrication.  This includes the rough machining of the end fittings,
preparation of the tube ends and the jig-welding of the complete
assemblies. Further machining, after welding, will be carried out to
ensure correct alignment and tolerance levels in conjunction with the
hanging columns.  Vent holes will be incorporated into the tubes as
required to facilitate construction and to allow purging with GAr/LAr
on commissioning.

Manufacturing, transportation and underground construction
considerations were a fundamental part of the field-shaping coil
design process, in collaboration with the LAGUNA-LBNO industrial
partners. The requirement of construction in a clean-room environment
within the completed membrane tank presented considerable challenges
in terms of logistics and the development of the overall concept for
fabrication.  It was concluded that a modular construction approach
would be required in order to (1) maximize off-site shop fabrication
and minimize on-site assembly, and (2) ensure the cleanliness of
construction and minimize the installation time.
Each field-shaping coil is broken down into sets of three main construction modules
(for more details see~\cite{cdr-annex-lbno-2}). 

Although separate components, the field shaping coils and the cathode
structure share identical features and dimensions.  Thus, the
maintenance of common interfaces is an important advantage of the
overall field cage design.
 
The DUNE 12-kt detector field cage would require 
60 stacked rectangular rings
used to cover the 12 vertical meters of drift volume.

The LBNO cathode design for the 50-kt detector follows an extensive
review of options and analysis. The design incorporates features to
minimize the static deflection of the cathode and to maximize the
electrostatic performance. (To avoid regions with high electric
fields, the electric field is limited to 50~kV/cm.)  Similar to the
field-shaping coil, and for the same reasons, the cathode is designed
as a modular structure ensuring a minimal on-site assembly
time. The cathode is designed as a fully welded tubular rectangular
 3D-grid structure in 2~m$\times$2~m modules, 
 1~m deep in the vertical
direction supported only from the periphery
(Figure~\ref{fig:LBNO_cathode}).
\begin{cdrfigure}[Cathode design]{LBNO_cathode}
{\small Left: cathode plane design for the LBNO detector. Right: breakdown of 
the cathode structure in construction modules.}
\includegraphics[width=.4\linewidth]{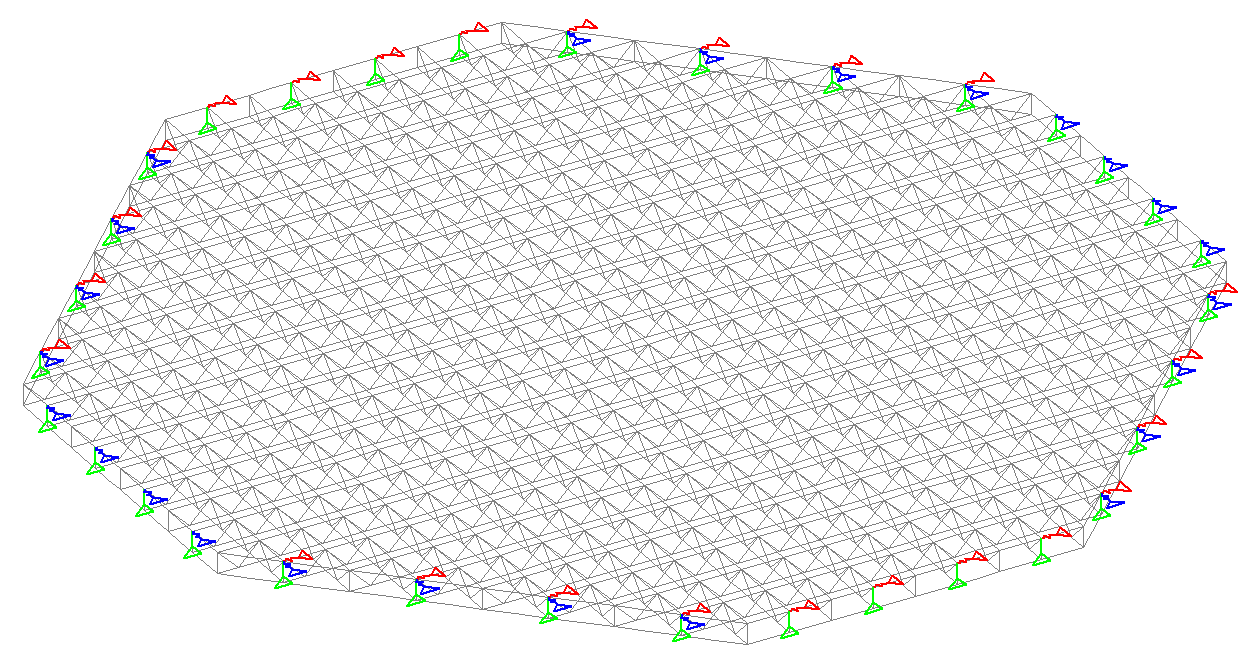} \hfil
\includegraphics[width=.5\linewidth]{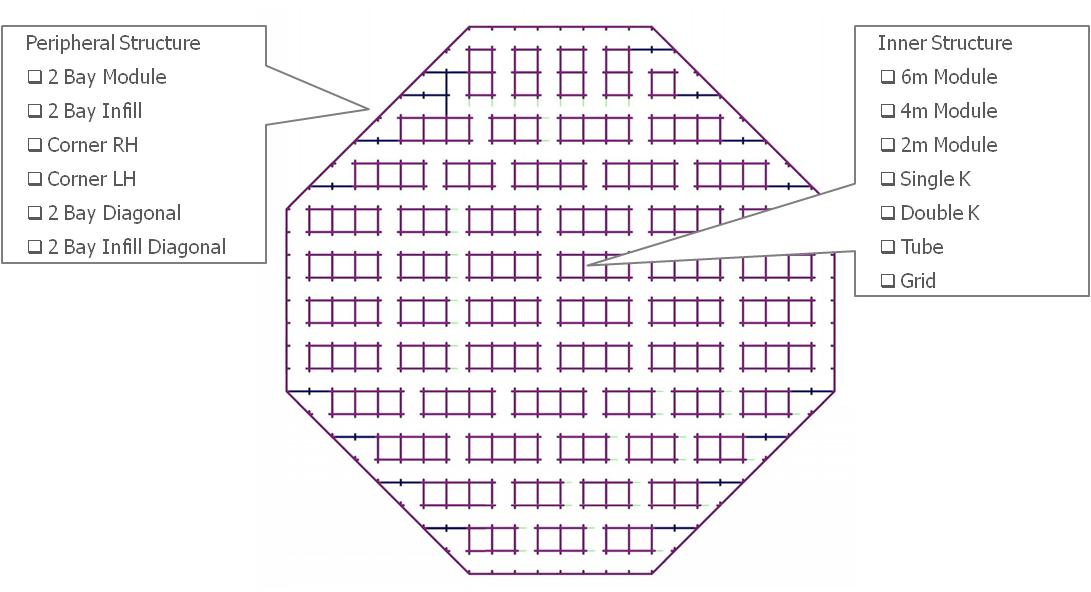}
\end{cdrfigure}

The top and bottom grid structures for the cathode are manufactured from 139.7-mm OD
tubes with wall thickness 2.6~mm to EN 10217-7 in 316 stainless steel.
The bracing structure is manufactured from 60.3-mm OD tubes with wall
thickness 2.6~mm, also to EN 10217-7 in 316 stainless steel.  A grid
structure comprising 10-mm OD tubes with wall thickness 1~mm, arranged
in a single plane at 100-mm centers, is fitted to the top of the
cathode.  The maximum module size for this structure is 6~m,
comprising three full 2~m$\times$2~m$\times$1~m deep modules of the
cathode structure.  The  preparation and welding of
the modules will be carried out in controlled facilities at the
fabrication shop.  Vent holes are incorporated into the grid structures to
facilitate construction and to allow purging with GAr/LAr on
commissioning. High levels of quality control will be possible with
the modular construction design, and following inspection, each module
will be cleaned to ISO 8 cleanliness standard and double-wrapped prior
to dispatch and transportation to the site for installation and final
assembly.

The cathode outer top tubular structure is identical to the bottom-most
field-shaping coil; they use tubes of the same outer diameter
(139.7~mm).  The spans are the same (48~m for the 50-kt LBNO detector)
and the vertical distance separating these components is the same as
for the remaining field-shaping coils (200-mm centers). The cathode
will be attached at the bottom of each hanging column by a split link
in G-10CR. The cathode attachment points will also incorporate locally
thickened sections of tube (as in the hanging chains) included as part
of the peripheral structure nodes.
 
The complete assembly procedure, logistics and tooling for the field
cage and cathode is described in~\cite{cdr-annex-lbno-2}.  It is expected that the
general design, adapted to the rectangular geometry, and the basic
elements for the cathode construction would be similar for the 12-kt
DUNE detector, but down-sized to the less challenging requirements.

%% file: volume-detectors/fd-alt-sections/fd-alt-elec.tex
%%%%%%%%%%%%%%%%%%%%%%%%%%%%%%%%
\section{The Electronics, Chimneys and DAQ}
\label{sec:detectors-fd-alt-elec}

\subsection{Overview}
\label{sec:fd-alt-elec-ovvw}

The LBNO 20--50-kt detector designs, developed in the LAGUNA-LBNO
design study have channel counts in the range of 500,000 to 1,000,000.
This large number has spawned extensive R\&D over the last few years
into large-scale charge-readout solutions optimized for double-phase detectors.  
The solutions that have been developed provide high integration levels, and significant cost
reduction and performance improvement. They can be fully adopted for
the DUNE far detector double-phase alternate design, which  foresees
153600 channels for each 10-kt far detector module.

The R\&D activities (ongoing since 2006) have focused effort on two
main areas:
\begin {itemize} 
\item{development of cold front-end ASIC electronics, and}
\item{development of low-cost, largely scalable data 
acquisition systems (DAQ) based on modern telecommunications technologies.}
\end{itemize}

One of the goals of the WA105 6$\times$6$\times$6~m$^3$ demonstrator,
a LArTPC with \num{7680} charge-readout channels described in Section~\ref{sec:proto-cern-double}, is to test the
large-scale readout system developed in the LAGUNA-LBNO design
study. \anxdualtdr{}~\cite{WA105_TDR} and~\cite{WA105_SREP} provide detailed descriptions
of the charge-readout electronics, including the cold front-end ASICs
and the DAQ.

The LAGUNA-LBNO design was driven both by cost and by the particular use of the
electronics for the dual-phase readout, which implies larger signals
from the detector relative to single-phase, effectively releasing requirements on noise. 
Section~\ref{sec:detectors-fd-alt-chg-readout} describes
the charge readout system.  Recall that the Large Electron Multipliers (LEMs) 
amplify the ionization charges by at least a factor 20, and that by 
adjusting its voltage, the LEM gain is tunable from there
up to 200. When the charges reach the segmented anode, they are
equally shared among two perpendicular collection views.
The front-end amplifier connected to the anode, given the capacitance
of the anode strips, would have a S/N ratio of 14 at unitary LEM gain. 
Considering a minimal LEM gain of 20, the amplifier provides an
overall S/N ratio of 140. The S/N ratio is boosted by the LEM gain, thus 
implying less stringent requirements on the preamplifier noise.

The electronics consists of front-end amplifiers that are implemented as cryogenic CMOS ASICs
connected to the anode with 50-cm cables; they are completely
accessible from the outside while remaining very low-noise. The
amplifiers must be housed in separate volumes that
are completely distinct from the tank volume
in order to replace them as needed without
contaminating the pure argon inside the cryostat. These volumes are called chimneys.

Flat, 2-m-long cables inside each chimney connect front-end cards
to digitization electronics, housed in microTCA ($\mu$TCA)  crates \cite{mTCA-standard} at the chimney exits, outside the cryostat.
This design provides a way to maintain the front-end at cryogenic temperatures but keep
it accessible, and maintain the digital electronics externally, at room temperature.
 This provides risk mitigation and significant flexibility.

The digitization units in the $\mu$TCA
crates are synchronized with the White Rabbit (WR) time-distribution
standard\cite{WR-standard}, which was originally designed to achieve
sub-ns accuracy.
This built-in accuracy, while not a critical aspect of the system
design, is much better than what is needed to align the 400-ns samples
of the charge readout.  WR was adopted for its practical
integration aspects and for cost reduction; its very high timing accuracy is a
bonus.  WR is also used in this design as a dedicated network
network for the trigger distribution. 

The light-readout digitization
electronics (see Section~\ref{sec:detectors-fd-alt-light}) is also
implemented in $\mu$TCA and provides triggers from the
photomultipliers (PMTs) that are distributed to the DAQ via the WR
network. 

Commercial high-bandwidth and high-computing-power back-end
cards are used for event-building and are coupled to a farm for online
processing, which is implemented for event filtering, data
reconstruction, calibrations and data-quality assessment.

\subsection{Front-end Cryogenic Amplifiers and Chimneys}

In the framework of the R\&D related to LAGUNA-LBNO since 2006,
several generations of prototypes of cryogenic ASIC 0.35~microns CMOS
multi-channel preamplifier chips have been developed. 
The capability to operate at cryogenic temperatures means that cables can be shorter 
(in WA105 these cables are just 50~cm long), which reduces the associated 
capacitance for the connection to the detector. It also makes it possible to reach
 an optimal amplifier S/N ratio at a
temperature around 110~K, which can be easily achieved in the GAr at
the top of the cryostat.

Another significant feature of the design is the ad-hoc designed chimneys, which 
enable the front-end electronics to remain a very short
distance from the detectors in the CRP and accessible for repairs without
opening the cryostat.
Chimneys are
separate, insulated, cylindrical volumes that penetrate the cryostat top, 
their lower half immersed in GAr, their upper half outside the tank at room temperature.
The front-end electronics cards are installed inside the chimneys near the bottom. 
The chimneys are filled with inert gas and have a cooling
system to keep the electronics at the optimal temperature.  A cold feedthrough at
the bottom of each chimney isolates the cards from the inner volume of the
vessel and allows connection from the anode to the electronics;
a warm feedthrough (FT) at the top allows connection to the digitization electronics 
on the outside.

The first ASIC versions were designed principally for the readout of
charge from collection and induction wire planes, and could also handle
bipolar signals. Since 2012 some versions of the dual-phase
have been developed specifically to match the dynamic range of signals
coming from the two collection views of the anode PCB after LEM
amplification. As described in Section~\ref{sec:detectors-fd-alt-chg-readout},  
each collection view is instrumented
with strips of 3.125-mm pitch and 3-m length (150~pF/m capacitance).
For this pitch, simulations of electromagnetic showers predict that a single channel
will collect a maximum 40~MIP. The design of the dual-phase cryogenic ASIC
is based on a LEM minimal gain of 20, which corresponds to
1200~fC for this maximal signal.

Two versions of the dual-phase ASIC chips have been produced
for WA105, both with 16 readout channels. The first version has a
constant gain in the region 0--40~MIP. The second is
characterized by a double-slope gain. This second solution optimizes the
resolution while preserving a large dynamic range. It is characterized
by a high-gain region extending up to signals of 10~MIP, after which the
gain is reduced by a factor of three in order to enable a better overall match for a dynamic
range of 40~MIP.  It provides the best resolution in the MIP
region ($dE/dx$ measurements) without limiting the dynamic range for
showers, which can still reach up to 40~MIP (see
Figure~\ref{fig:FE_ASIC1}). This double-slope regime has been
optimized on the basis of simulations of hadronic and electromagnetic
showers. Both ASIC versions, compatible with the LEM signal dynamics,
are implemented in the CMOS 0.35~$\mu$m technology; they have 16 channels,
18-mW/channel thermal dissipation or less, about 1300 electrons ENC at
250-pF input detector capacitance, and operate with this best S/N ratio at about 110~K.
\begin{cdrfigure}[Dual-phase cryogenic ASIC amplifiers]{FE_ASIC1} 
{Front-end 16 channels cryogenic ASIC amplifier with the double-slope gain implementation}
\includegraphics[width=.3\linewidth]{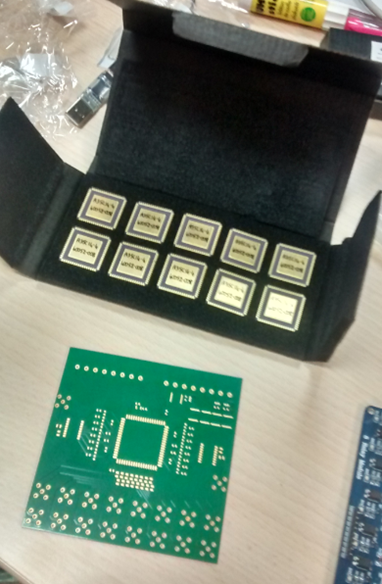}
\end{cdrfigure}

The implementation of the double-slope gain regime is obtained by
replacing the feedback integration capacitor of the OPAMP with a MOS
capacitance, which changes its value above a certain threshold
voltage. This effect is also present during the discharge phase and it
can be corrected with the inclusion in the feedback loop of an
additional branch with a diode and a resistor designed to keep the RC
value roughly constant during discharge. This branch can be
selected/deselected with an internal switch for all the channels in
the ASIC (see Figure~\ref{fig:FE_doubleslope}).
\begin{cdrfigure}[Double-slope ASIC response]{FE_doubleslope}
{Response of the double-slope ASIC amplifier to progressively larger 
pulses with and without the diode/resistor feedback branch}
\includegraphics[width=\linewidth]{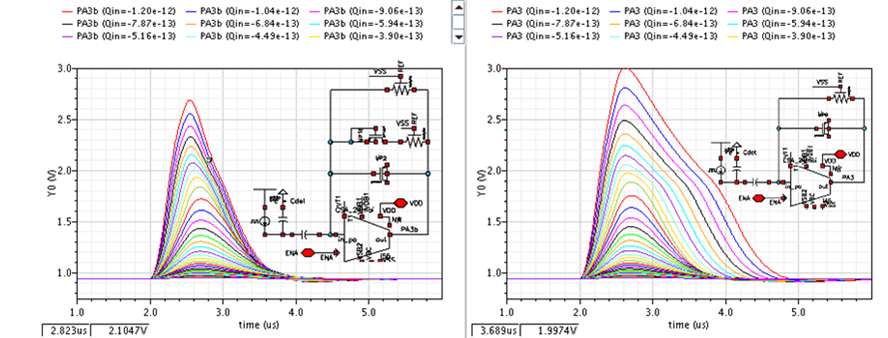}
\end{cdrfigure}

In the design under implementation in WA105 and proposed for DUNE,
there are 640 channels per chimney. The 40 ASIC amplifiers needed for
the readout of each group of 640 channels will be arranged on 10 pairs
of front-end cards plugged into the FT at the bottom of each chimney.
Each front-end card holds two ASIC chips and a few discrete
components. Particular care has been taken in testing several options
(gas discharge tubes, metal oxyde varistors, double diodes) for the
surge-arrestor components, which have to protect the ASICs from
occasional sparks occurring in the CRP.  This study was aimed at
maximizing the protection efficiency, testing the components'
durability for a very high number of sparks and minimizing the input
capacitance seen by the pre-amplifiers. Double-diodes have been
selected as the best solution given their performance and
capacitance. The total dissipation of the front-end electronics will
be about 11.5~W per chimney. This heat source is minor with respect
to the heat conduction from the flat cables going to the digitization
electronics and from the walls of the chimney. The front-end cards are
kept at low temperature by a cooling system installed at the bottom of
chimney 
that compensates for this overall heat flow. The front-end
electronics is coupled to the DAQ system, described in Section~\ref{sec:fd-alt-elec-daq},
that is based on 12-bit ADCs, matching the needed dynamic
range quite well. Figure~\ref{fig:chimneys_scheme} shows the 3D model of the
signal FT chimneys hosting the cryogenic ASIC amplifiers.
\begin{cdrfigure}[3D model of the signal feedthrough chimneys]
{chimneys_scheme}{3D model of the signal feedthrough chimneys}
\includegraphics[width=.5\linewidth]{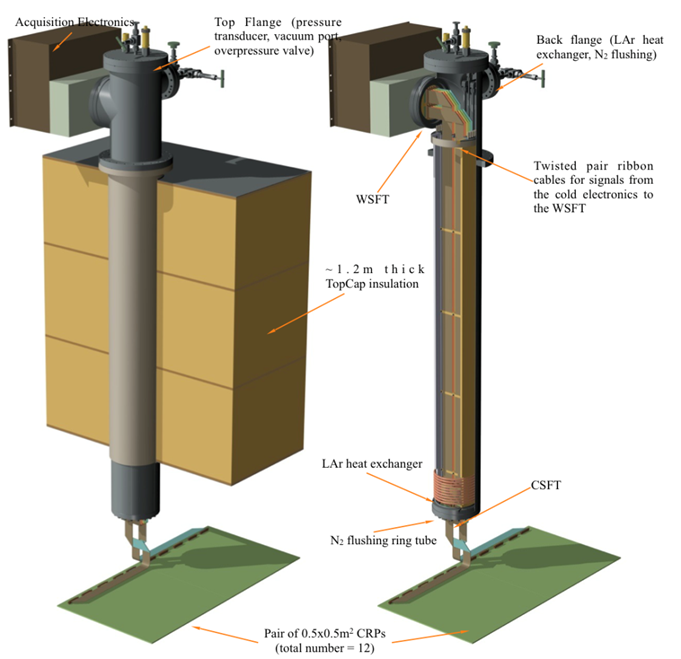}
\end{cdrfigure}

A signal FT chimney prototype for 320 channels built for the
3$\times$1$\times$1~m$^3$ WA105 prototype is shown in
Figure~\ref{fig:chimneys_proto}.
\begin{cdrfigure}[Prototype of the signal feedthrough chimneys]
{chimneys_proto}{Prototype of the signal feedthrough chimney built 
for the WA015 3$\times$3$\times$1~m$^3$ prototype}
\includegraphics[width=\linewidth]{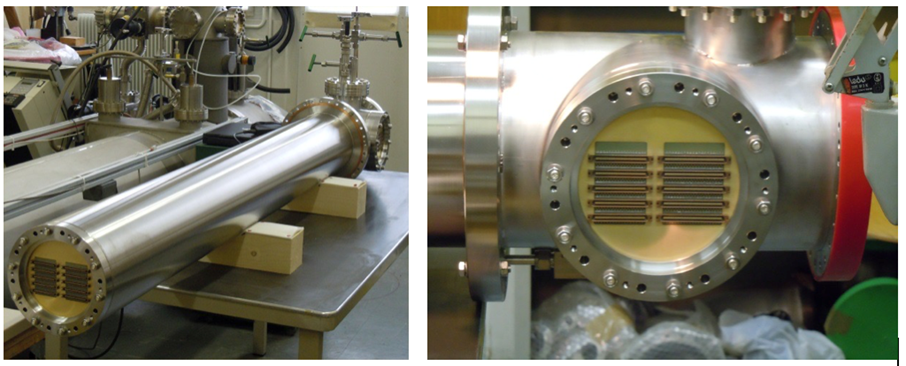}
\end{cdrfigure}

Pairs of cryogenic electronics front-end cards are mounted at the end of
sliding G10 blades which can be extracted from the top of the
chimney. The blades, which also carry the flat cables for the
connections, slide on guides mounted inside the chimney. By
moving the blades, the front-end cards can be plugged/unplugged to/from 
connectors on the top side of the cold FT at
the bottom of the chimney. This FT completely isolates the
chimney from the LAr vessel. Connectors are mounted on the bottom of the FT 
for the 50~cm flat cables coming from the CRPs.

\subsection{Digital Electronics and DAQ Architecture}
\label{sec:fd-alt-elec-daq}

The DAQ system  proposed for the dual-phase DUNE
detector design is based on two industrial standards:
\begin{itemize}
\item MicroTCA ($\mu$TCA) standard for the distributed data network\cite{mTCA-standard}
\item White Rabbit (WR) standard for the distributed clock network\cite{WR-standard}
\end{itemize}

The analog electrical signals from the front-end electronics ASICs transit through the signal chimneys up to the
digitization boards in the $\mu$TCA crates (DAQ L1).
The backplane of each $\mu$TCA crate (a \textit{shelf}) is connected through a 10GbE up-link
to the next level (DAQ L2). The L2 directly connects the $\mu$TCA crates
to FPGA-based high-performance back-end processing boards. 
The design calls for a lossless transmission scheme all the way down to the
back-end processing board, which applies all filtering algorithms and
does the event building. The Huffman lossless algorithm is easy to
implement and typically provides a factor  of 10 compression on LAr events.

Recorded data are sent to a local storage level where Object Storage
Servers (OSS) and MetaData Servers (MDS) are connected with the 
event-building workstations via a 10/40-GbE network (Ethernet or
InfiniBand). In parallel, signals from a high-stability common clock and time
synchronization signals are distributed (using the WR standard) 
to the L1 digitization cards, through a dedicated, deterministic
network. The WR network is also used to transmit the trigger time-stamp signals,
which can be generated either by the PMTs' readout electronics or by
additional sources, e.g., for WA105 operation in the charged particle beam, 
the beam trigger counters. The clock is derived
from a Master Clock generator connected to the WR
Grand-Master switch. WA105 will implement this DAQ scheme, and 
additional details may be found in~\cite{WA105_TDR}.

\subsection{MicroTCA Standard and Applications}

The MicroTCA ($\mu$TCA) standard offers a very compact and easily scalable
architecture to handle a large number of channels at low cost. The
$\mu$TCA or related standards --- such as ATCA or xTCA --- are now
well known in the HEP community and have been integrated into various
designs at CERN (e.g., LHC upgrades), DESY, etc.  $\mu$TCA fulfills
requirements of the telecommunication industry and offers the
ability to interconnect distributed applications while providing a
standard, compact and robust form factor with simplified power supply
management, cooling, and distribution of internal clocks. The backplane of
a $\mu$TCA crate (the $\mu$TCA shelf) accommodates high-speed
serial links; they are arranged in a variety of topologies to support a 
variety of protocols, including Ethernet 1GbE or 10GbE, PCI Express, SRIO,
etc. Use of Ethernet-based solutions is proposed for both data
and clock distribution, through the $\mu$TCA backplanes. This choice
obviously optimizes the connections between the various components of the system.
 Constraints imposed on the data transfer bandwidth point towards use of 
the 10GbE protocol. For the
clock distribution, dedicated lanes on the backplane must be defined by the user. The $\mu$TCA
standard also offers so-called \textit{clock} lanes, which distribute the timing signals to all boards
hosted in the crate and which may be used for any type of signal.

The signal digitization boards plugged into a $\mu$TCA shelf are
called Advanced Mezzanine Cards (AMCs)\cite{picmg-2006}. Each AMC
is connected to one or two $\mu$TCA Carrier Hub (MCH) boards through
the backplane serial links. The MCH provides a central switch function
allowing the AMCs to communicate with each other or with external
systems through an up-link access. The MCH manages both the 10GbE
uplink and the WR bi-directional clock
distribution. Figure~\ref{fig:mTCA-features} provides a sketch of the
backplane layout and its implementation in a particular shelf type selected in the design of the system.

\begin{cdrfigure}[MicroTCA crate organization]{mTCA-features}
{\small Left: global microTCA crate organization. AMCs 
(providing basic ADC functions) are connected to the crate 
controller or MCH which up-links the external systems. A dedicated 
AMC for the clock receives dedicated signals (master clock, trigger 
signals) from the timing distribution system and transcript them onto 
the backplane. Right: backplane layout of the Schroff 11850-015 reference.}
\includegraphics[width=.5\linewidth]{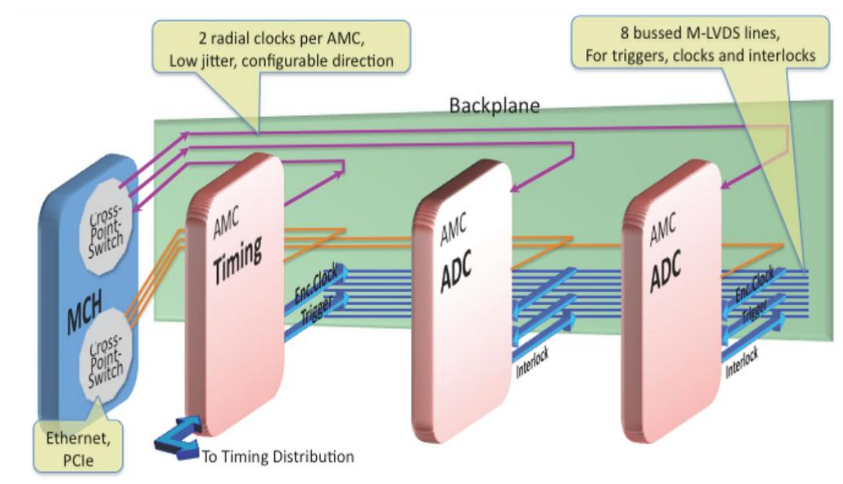}\hfill
\includegraphics[width=.4\linewidth]{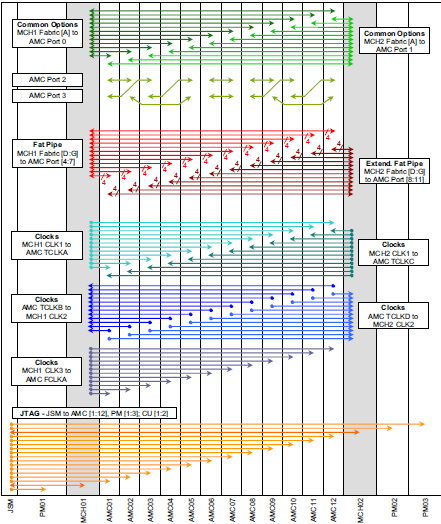}
\end{cdrfigure}

The production version DAQ designed for the WA105 DAQ is based on the
$\mu$TCA.1 standard, with connections to the user input signals from 
the front side only. These connections are made with VHDCI
cables in order to minimize the number of cables. One $\mu$TCA shelf
is connected to each signal chimney, reading out 640 channels corresponding to 10 AMCs.

%To increase the channels density one may profit of a different $\mu$TCA standard: $\mu$TCA.4. This standard offers the possibility to connect the AMCs hosted in a crate from both the front and rear sides (Figure~\ref{fig:schroff-mTCA-4}). The front card, still called AMC, is connected to the backplane of the shelf, while the read card, so-called $\mu$RTM (Rear Transition Module) is only connected to the AMC. This standard allows to double the number of connections per slot, at the cost of a slightly asymmetric design for the 2 boards. 

Many types of  $\mu$TCA shef are available on the market, e.g.
11850-015 8U from Schroff for $\mu$TCA.1
standard, NATIVE-R9 from NAT for $\mu$TCA.4 standard. The cost of
these items is quickly decreasing due to the fast pace of
developments by the internet providers. These items all have redundant power
supplies, redundant MCHs and offer different segmentations to connect
the AMCs.

%\begin{cdrfigure}[General organization of AMC]{schroff-mTCA-4}{\small Left: general organization of AMC and $\mu$RTM boards in $\mu$TCA.4 standard. %Right: picture comparing $\mu$TCA.4 boards and a VME board.}
%\includegraphics[width=.5\linewidth]{fig-mTCA-4-1.png} \hfill
%\includegraphics[width=.4\linewidth]{fig-mTCA-4-2.png}
%\end{cdrfigure}

The advantage of this architecture is that it limits the DAQ electronics
developments to the AMC only since that is the component that provides
the functionalities for the digitization, data formatting and
compression, event time-stamping and data transfer through the
backplane. For WA105, the AMC is a double-size module (also compatible
with $\mu$TCA.4 standard) with a single input connector and a 10GbE
link to the backplane. The input stage performs the 64-channel
digitization through eight 8-channel, 14-bit ADC chips read out at a
2.5-MHz frequency. The ADC readout sequence is controlled by two FPGAs
that make the data available on a double port memory. Readout of the
data is performed continuously and they are stored in a local
buffer. The recorded samples, each corresponding to a drift window, are
selected in coincidence with the received trigger. When a trigger
occurs, the samples written in the memory can be treated with
compression algorithms (such as Huffman or RLE) or zero-suppression
(if required) and transmitted over the network until the end of the
drift window, which closes the event. These operations are managed by a
third FPGA, which sends the data on the shelf backplane in order that they can be transmitted to L2.  
This readout scheme and hardware implementation have been validated for WA105 on a
Stratix 4 prototype board, shown in Figure~\ref{fig:AMC-bloc-diag}.

\begin{cdrfigure}[Charge readout AMC prototype]{AMC-bloc-diag}
{\small Prototype of AMC, using $\mu$TCA.1 standard, and hosting 
64 ADC channels on a mezzanine board. This prototype is used as a validation 
of the full and final ADC chain in WA105.}
\includegraphics[width=.5\linewidth]{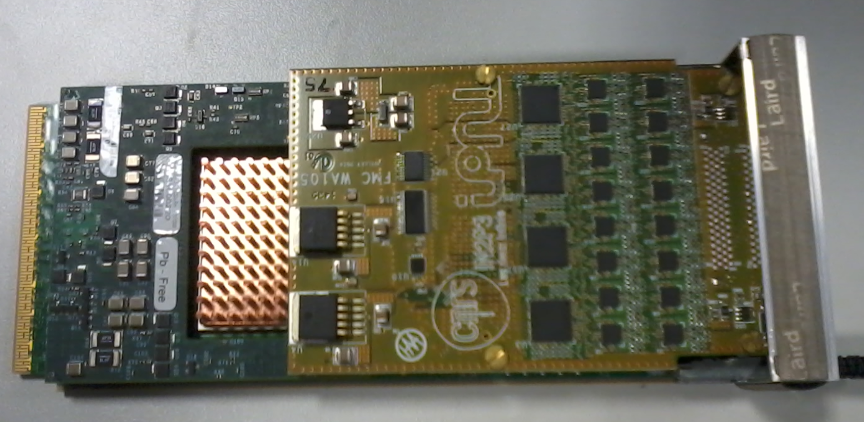}
\end{cdrfigure}

\subsection{Back-end and Event Builder}

A network hierarchical structure is implemented for the back-end and event builder in which all crates are
interconnected to a dedicated back-end FPGA processing board (such as
S5-PCIe-HQ, Figure~\ref{fig:Bittware-board}).

\begin{cdrfigure}[FPGA processing board]{Bittware-board}
{\small FPGA processing board based on Stratix V from Altera. The board 
features a dual QSFP+ cages for 40GigE or 10GigE links, 16 GBytes DDR3 SDRAM, 
72 MBytes QDRII/II+, two SATA connectors and is programmable via OpenCL.}
\includegraphics[width=.4\linewidth]{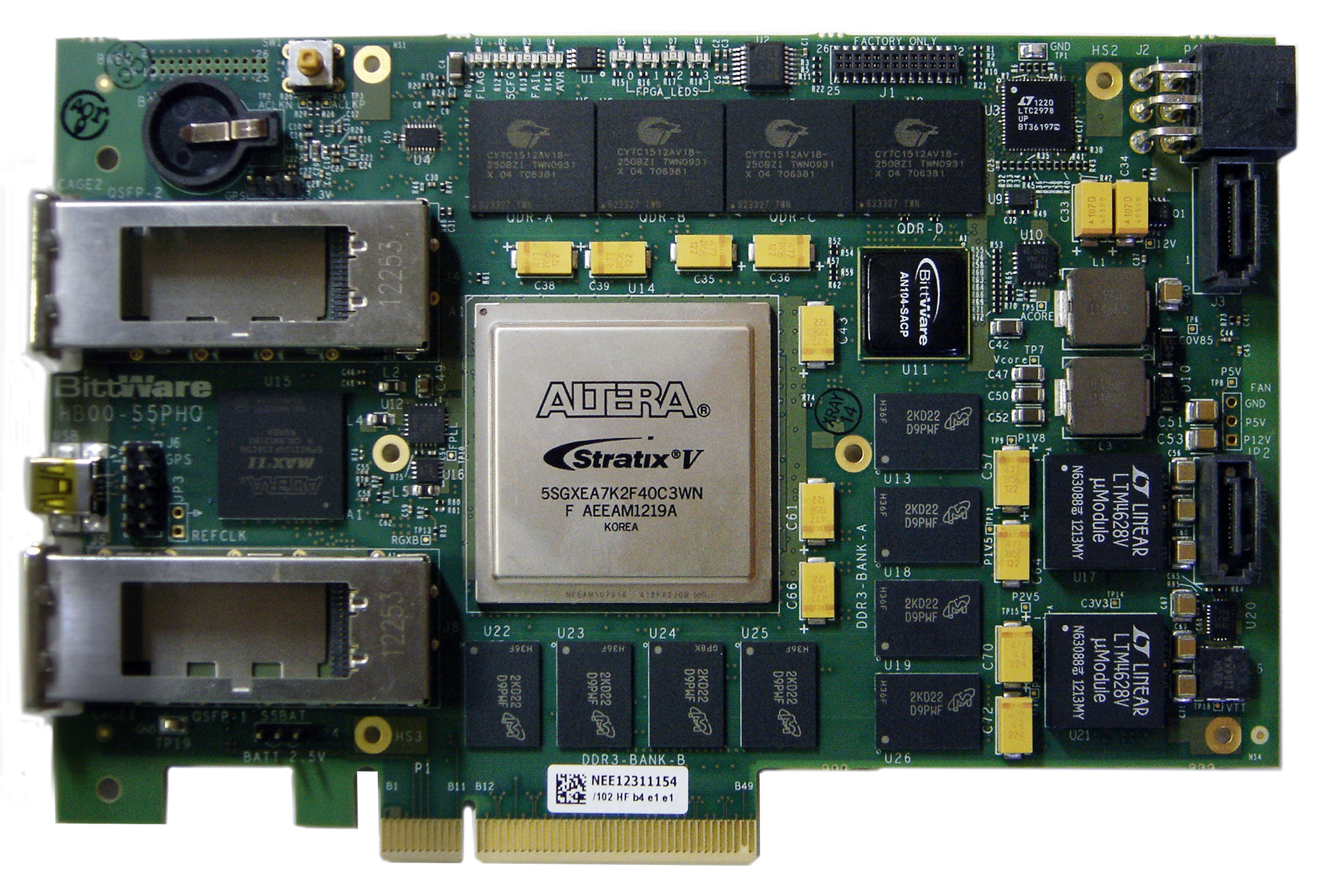}
\end{cdrfigure}

This kind of board is used for massive data processing in many fields
(medical imaging, stock market, etc.) that require parallel
processing with reduced power consumption (it dissipates only 10\% of the power
dissipated by an equivalent CPU for a comparable number of
operations). This particular board has two QSFP+ cages to bring the
data directly to the FPGA for the lowest possible latency. Up to 8$\times$10GbE
links without data loss are available per board.  The board performs
further data processing, filtering and transmission to the highest
level for storage. This type of board is widely used and the present
generation, based on the Altera Stratix V, will evolve to the Aria X
and the Stratix X. Stratix X will be probably available at the time
of construction of the DUNE DAQ system. Programming of the back-end
processing board is achievable through the OpenCL software suite where
a kernel code, on top of a host code, allows programming  the FPGA directly in a
high-level language without a classical VHDL synthesis
chain. OpenCL applications are transparent to the hardware used for
procesing (FPGAs, CPUs, GPUs). This highly flexible feature is fully
adapted to the requirements of large DAQ systems, where conditions
of filtering, event building, etc., may evolve with time.
 
\subsection{Timing Distribution System and White Rabbit (WR) Standard}

The clock distribution will use a parallel, independent
network to distribute the signal from a GPS-disciplined Master Clock down to each $\mu$TCA
shelf, through specific switches. Technically, the WR standard is based
on a combination of Synchronous Ethernet (Sync-E) and Precision Time
Protocol (PTP, IEEE1588), where the Ethernet clock is generated by a
GPS-disciplined clock. At the level of each shelf, this high-accuracy
clock is made available to each AMC through dedicated lines off
the backplane. As discussed before, the $\mu$TCA standard accommodates
special lanes for clock transmission. The trigger signals
(time stamps) are encoded and sent through this dedicated WR network
which has enough bandwidth for this transmission without interfering
with the PTP synchronization signals. The requirements on the
synchronization for the charge readout are quite loose since the
typical readout frequency is of the order of a few MHz. The
requirements for the PMT readout on the contrary are more
stringent. The goal is to provide a nanosecond synchronization at the
level of all L1 elements. This goal is typically achievable with the
WR standard\cite{WR-standard}.

The WR provides an extension to the Ethernet network with Gb/s data
transfer speed and allows for accurate synchronization among the different
network nodes. 
It provides a common clock for the physical layer in the entire
network, allowing sub-nanosecond synchronization accuracy and 20-ps
jitter time. The WR network is designed to host up to thousands of
nodes and to support distance ranges around 10~km using fiber cables. It
ensures that all the Ethernet frames sent are delivered
after no more than a fixed delay  (controlled latency). The order of the frames should be
preserved.  A typical application scheme is displayed
in Figure~\ref{fig:WR_elements}.
\begin{cdrfigure}[White Rabbit network organization]{WR_elements}
{\small Left: general organization of a typical WR network. Right: standalone WR switch.}
\includegraphics[width=.5\linewidth]{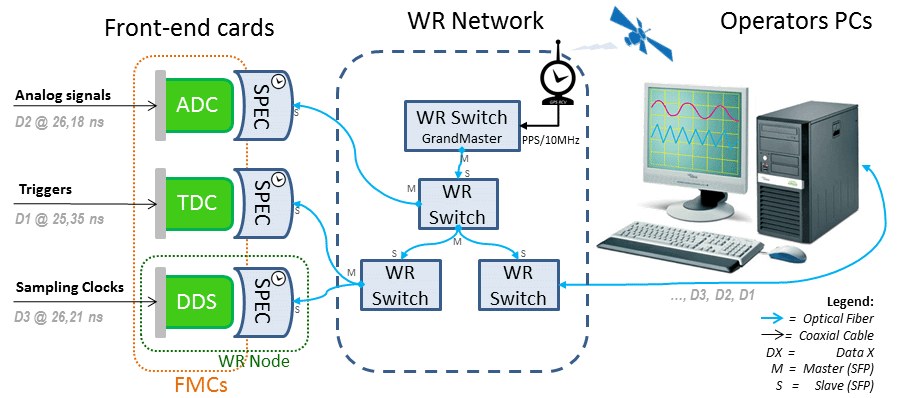}
\includegraphics[width=.3\linewidth]{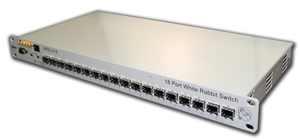}
\end{cdrfigure}

The WR application in the $\mu$TCA standard is currently engineered 
to easily interconnect $\mu$TCA cards.  The WR  switch
can therefore be connected directly to its different nodes in the same
rack to facilitate 
maintenance and limit the space occupied. 
The implementation scheme is based on the integration of a WR mezzanine board on the MCH
of each shelf. 
In the future  a full integration of the WR within the $\mu$TCA DAQ system would be very powerful. 
The current WR implementation scheme is based on the integration of a WR mezzanine board on the MCH of each shelf. 
The development of this WR MCH is in progress for the WA105
demonstrator (on a MHC produced by the company NAT); the companies
producing the MCH and the WR slave to be integrated show clear interest in this development.

%% file: volume-detectors/fd-alt-sections/fd-alt-dcs-dss.tex
%%%%%%%%%%%%%%%%%%%%%%%%%%%%%%%% 
\section{The Slow-Control System} 
\label{sec:detectors-fd-alt-dcs}

The slow-control system for the far detector is designed  to
monitor the detector operation conditions, in particular, the following physical
quantities inside the tank:
\begin{itemize}
 \item temperatures (with platinum resistors),
 \item pressures (with commercial piezoelectric sensors),
 \item LAr levels (with custom-made capacitive sensors and electronics), and
 \item deformations of materials (with resistive strain gauges).
\end{itemize} 

In addition, the slow-control system provides the hardware
infrastructure needed to monitor traces of O$_2$, N$_2$ and H$_2$O
impurities in the tank, to monitor and control the high- and
low-voltage power supplies, heaters, lighting system and cryogenic
video system. It will also interface to the cryogenic system and to
the motorized system that adjusts the position of each Charge Readout
Plane (CRP).

The design of the slow-control system is part of a
continued, progressive prototyping effort aimed at developing a
control system dedicated to multi-kiloton LAr dual-phase detectors. It
has been designed in the framework of the LAGUNA-LBNO design study and
 the WA105 experiment. WA105 represents a first, fully engineered,
implementation of this design, which can be extrapolated to larger
detector scales. The design also benefited from the successful example
and the expertise developed in the context of the ArDM
experiment\cite{Badertscher:2013ygt} which is currently operating a
LArTPC for dark matter searches in an underground laboratory (LSC,
Spain).

The slow-control system introduces the use of National Instruments
Compact RIO (Reconfigurable Input Output) modules for acquisition of
all the physical quantities of interest.  Figure~\ref{fig:NI_proto}
shows a rack prepared for the WA105 3$\times$1$\times$1~m$^3$
prototype that is ready to be tested at CERN.
\begin{cdrfigure}[Slow Control prototype rack]{NI_proto}{The rack is a  prototype of the entire Control System; it  embeds modules for resistive 
temperature sensors, pressure  sensors, strain gauges, liquid argon level  meters, control for  heaters. On the upper part a redundant 24 V power supply 
provides fault tolerant power to the National Instrument controller and modules.  Calibration of modules and sensors is ongoing.}
\includegraphics[scale=0.4, angle=0]{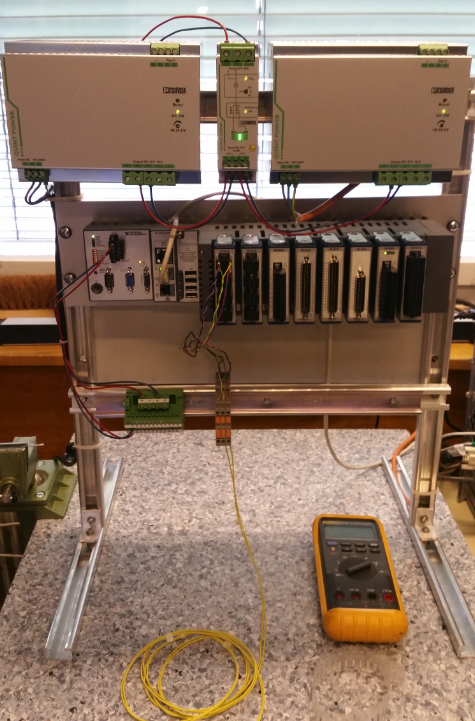}
\end{cdrfigure}

The entire slow-control system of the WA105 demonstrator will be
managed through a single LabView interface\cite{WA105_SREP} which
will provide access to all the sensors, control the actuators and
provide the platform for the video monitoring system both inside and
outside the tank.  Supervisory software will be implemented in the
CERN UNICOS (UNified Industrial Control System)
framework~\cite{unicos} to provide the operator interface for the
monitoring of all the physical quantities and the handling of alarms.

As discussed in Section~\ref{sec:detectors-fd-ref-ov}, the
charge-readout system is implemented via CRP modules of
3$\times$3~m$^2$. Each CRP is an independent detector, hence its
instrumentation can also be treated as independent. A complete list of
the sensors planned for use in both the 3$\times$3 m$^2$ DUNE CRP
module and and the 3$\times$1$\times$1~m$^3$ WA105 prototype is
provided in Figure~\ref{fig:sc_sensors}.
\begin{cdrfigure}[Dual-phase slow control sensors]{sc_sensors}
{List of the slow control sensors for the 3$\times$1$\times$1~m$^3$ WA105 prototype and far detectors CRP}
 \includegraphics[scale=0.7, angle=0]{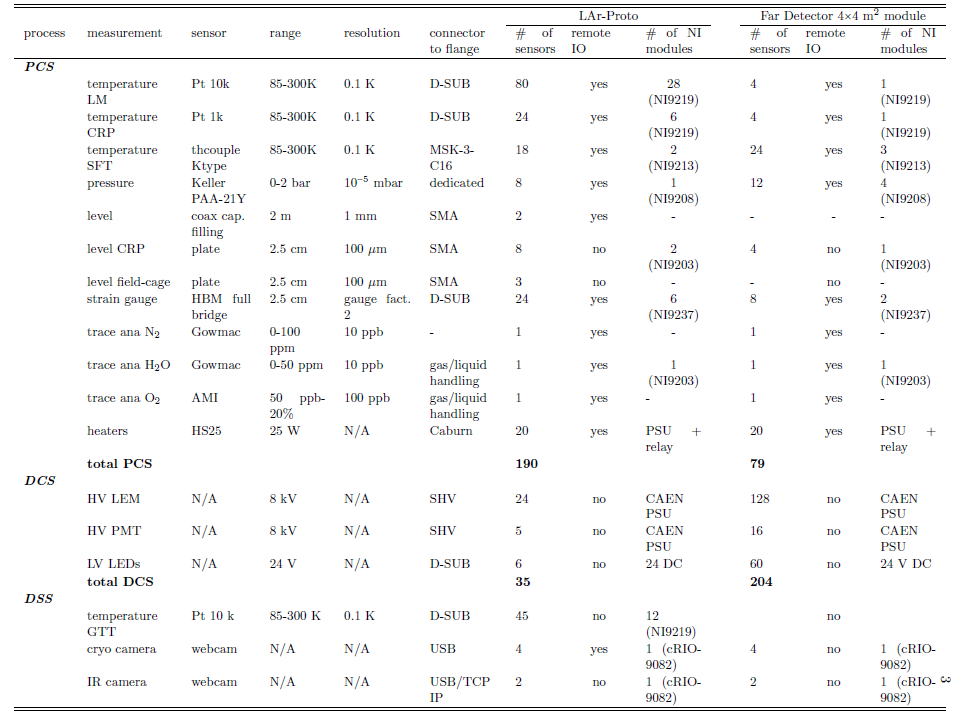} 
 \end{cdrfigure}

The number of sensors for the far detector CRP is extrapolated from
the number for the prototype and is not yet final but it should be
considered as an upper limit. The sensor instrumentation of the
3$\times$1$\times$1~m$^3$ WA105 prototype has also led to the design
of a custom Slow-Control Feedthrough (SCFT), based on the use of
weldable connectors for high vacuum (see
Figure~\ref{fig:SC_flange}). A specific SCFT for the DUNE
3$\times$3~m$^2$ CRP would be based on this design.
\begin{cdrfigure}[Slow control feedthroughs]{SC_flange}{The 3 SCFTs providing 
weldable connectors for all the instrumentation inside the 3$\times$1$\times$1~m$^3$  WA105 tank. 
The number of sensors per module in the DUNE far detector will be drastically 
reduced with respect to this WA105 prototype.}
  \includegraphics[scale=0.52, angle=0]{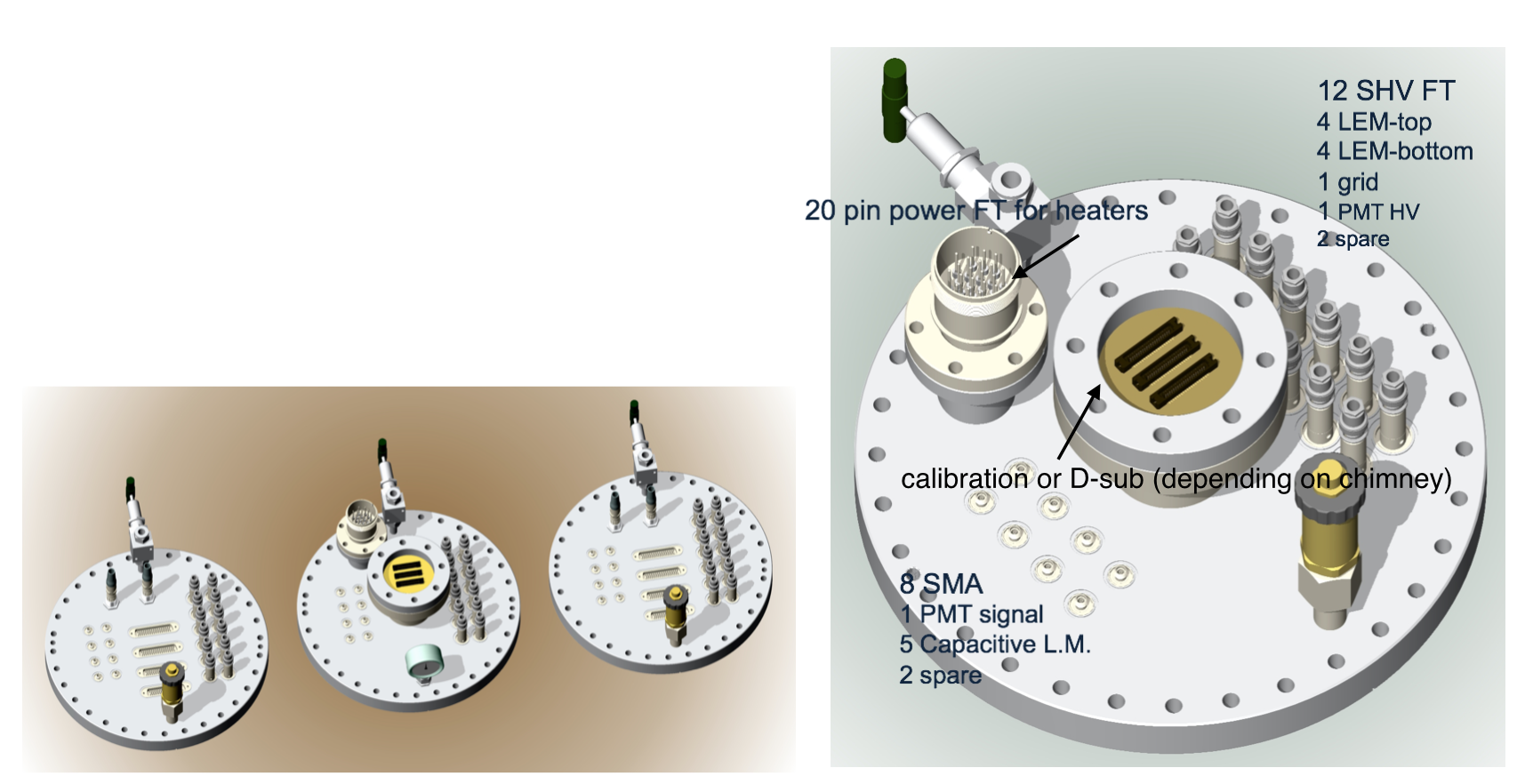}
 \end{cdrfigure}

%% file: volume-detectors/fd-alt-sections/fd-alt-light-read.tex
%%%%%%%%%%%%%%%%%%%%%%%%%%%%%%%% 
\section{The Light-Readout System} 
\label{sec:detectors-fd-alt-light}

The light-readout system developed in the LAGUNA-LBNO design study
detects the scintillation light using 8-inch cryogenic
PMTs (Hamamatsu R5912-02mod) with TPB coating. The PMTs
are placed about 1\,m below the cathode.  The application of the TPB
coating can be done at the level of the glass itself or on a
transparent plate mounted over the photocathode surface. The WA105
demonstrator will use 36 PMTs R5912-02mod (1/m$^2$ of
cathode surface). The mechanics for the PMTs' attachment has been
carefully studied; it must counteract the PMT buoyancy while avoiding
stress to the PMT glass (due to differentials in the thermal
contraction between the support and the PMT itself).

The dual-phase LAr detectors designed for LBNO are equipped with a
large number of PMTs. Depending on the size of the DUNE detector module and on the
density of PMT on the detection surface, this number may be as high as
1000. The large number of photosensors called for a large integration
scale solution for the front-end electronics.

Solutions of this kind have been studied, in the framework of the
R\&D program PMm2 \cite{PMM2-1, PMM2-2}, for the instrumentation of
giant water Cherenkov detectors. The signal
digitization is performed by grouping the PMTs in arrays
of 16. Each PMT array is read out by an ASIC chip in AMS
SiGe 0.35\,$\mu$m technology. The ASIC, which is called PARISROC
(PMT ARrray Integrated in Si-Ge Read Out
Chip)\cite{Parisroc}, provides a complete readout system for
triggerless data acquisition. The solution developed by this program
represents an important handle for cost reduction.

%\begin{cdrfigure}[Layout of the PARISROC ASIC]{parisroc}{Layout of the PARISROC ASIC used for the production of the second iteration of the chip (from Nucl.Instrum.Meth. A623 (2010) 492-494).}
% \includegraphics[width=.5\textwidth]{PARISROC2}  
%\end{cdrfigure}

The front-end electronics for the light readout of the WA105
demonstrator will be based on the solution developed by the PMm2
R\&D. The PARISROC ASIC is currently
adapted to the time structure of the scintillation light produced in
the interactions of secondary particles in neutrino interactions in
LAr. The detection of the direct scintillation light (S1) that
provides the absolute event time is the main task of the electronics.
The electronics will also process the signals from the scintillation
light (S2) produced by the electrons as they are extracted and
amplified in the gaseous phase.

The PARISROC chip reads the signals of 16 PMTs
independently of one another. Each analog channel consists of a
low-noise preamplifier with variable and adjustable gain (8 bits) to
compensate for the relative PMT gain differences when
powered by a single high voltage. The preamplifier is followed by a
slow channel for the charge measurement in parallel with a fast
channel for the trigger output. The slow channel includes a variable
(50--200\,ns) slow shaper followed by an analog memory with depth of 2
to provide a linear charge measurement up to 50\,pC; this charge is
then converted by a 10-bit Wilkinson ADC. The fast channel is composed
of a fast shaper (15\,ns) followed by a low offset discriminator to
auto-trigger down to 10\,fC. This auto-trigger feature makes the PMT
array completely independent of the other PMT arrays. The
threshold is loaded by an internal 10-bit DAC common to the 16
channels.

%\begin{cdrfigure}[Block diagram of the PARISROC ASIC.]{block-diag}{Block diagram of the PARISROC ASIC.}
 %\includegraphics[width=.7\textwidth]{parisroc_blockdiag.pdf}  
%\end{cdrfigure}

The variable gain of each preamplifier provides the flexibility to
adapt the system to the characteristics of each PMT after
it has been correctly calibrated. In the timestamping process, there
are two TDC ramps working in phased opposition in order to reduce the
dead time (i.e., when the ramp goes to zero) by selecting the other ramp
that will be in its active state.

When an event occurs, the value of the correct ramp is digitized and
inserted into the data stream that includes: a 24 bit counter that
goes much more slowly than the ramps in order to have a coarse time
measurement, a flag that indicates which ramp has been digitized, an
ID of the channel triggered, and the timestamp and charge information.

%\begin{cdrfigure}[MicroTCA rack and Bittware S4AM board]{LRO-rack}{MicroTCA rack and Bittware S4AM board}
 %\includegraphics[width=.6\textwidth]{rack}  
%\end{cdrfigure}

The light readout is fully integrated in the WA105 DAQ scheme. A
microTCA crate housing the light-readout digitization cards is
naturally integrated into this architecture by taking into account the
common time distribution and data transmission systems.  During the
data-taking outside the beam spills, a trigger that can be generated
by the light-readout microTCA crate plus its timestamp can be
transmitted over the White Rabbit network similarly to the beam
triggers.

On the light-readout front-end board there is also an ADC (AD9249 from
Analog Devices) that digitizes the PMT charge information on every
channel independently. The charge measurement can be correlated with
the timing information coming from the PARISROC for better precision
on both quantities. A prototype of the light-readout AMC is also being
developed by using the Bittware development card S4AM (the same used
for the LAr TPC ionization readout developments) and a mezzanine card
including the ADC and the trigger circuit (see
Figure~\ref{fig:DAQ_LRO}).
\begin{cdrfigure}[Block diagram of the light readout AMC demonstrator]{DAQ_LRO}{Block diagram of the light readout AMC demonstrator}
 \includegraphics[width=.6\textwidth]{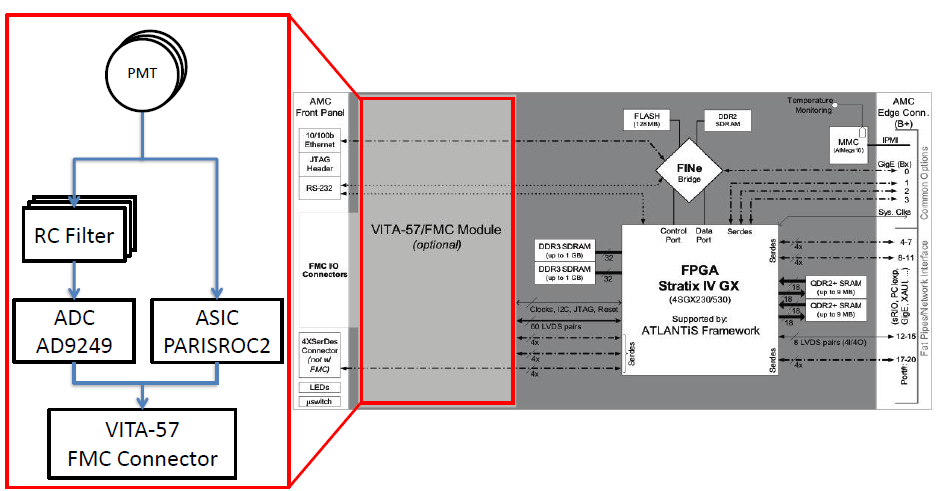}  
\end{cdrfigure}

The PARISROC chip provides the timestamp at nanosecond precision and
generates the trigger. In parallel, an ADC AD9249 continuously
digitizes all channels independently at 65\,MSPS with 14 bits. The FPGA
Stratix IV from the Bittware S4AM card controls and programs the ADC
and PARISROC. It also receives the data coming from both PARISROC and
ADC and controls which information is passed through the network to
store for analysis.  The charge digitization of the light signals is
performed at 14\,bits in samples of 15\,ns. Considering a 5-$\mu$s
light-signal duration, the digitization will produce more than 300
samples. With this amount of data it is possible to measure not only
the total charge generated but also the shape of the signal and the
long-component decay of the light.

%The firmware of the FPGA is in its early stage of development and there is still an open debate about which information will be useful. It will definitively include the time stamp of every event on each channel as well as the digitization of the charge received in each of them. Some questions concern the duration of the PMT sampling and the feasibility of including some real-time processing. This could include an inverse filter of the RC circuit to compensate its contribution.

With this light readout architecture and the White Rabbit
time-distribution system, it is possible to achieve 1-ns accuracy at
the level of trigger generation for the $t0$ and timestamping of the
events. This trigger is sent to the general DAQ and the event
builder. Since the charge-readout system of the TPC runs at 2.5\,MHz,
the minimal time unit in the reconstruction of the drift is 400\,ns.

%% file: volume-detectors/fd-alt-sections/fd-alt-install.tex
%%%%%%%%%%%%%%%%%%%%%%%%%%%%%%%%
\section{Installation and Commissioning}
\label{sec:detectors-fd-alt-install}

\subsection{Preparatory Work}

Before the installation of the detector can begin, a number of
preparations must be conducted and boundary conditions must be met in
order to enable a safe and efficient process that minimizes the
manpower required for the underground detector construction. A careful
study of the installation sequence and tools and an optimization of
the size and characteristics of the detector elements has been
performed in order to facilitate the installation procedures. Some of the
most important preparation and installation steps are listed in this
section. More detailed explanation and figures are available in~\cite{cdr-annex-lbno-2}.

\subsection{Detector Installation Sequence}

This installation procedure has been extensively studied in the
framework of the LAGUNA-LBNO design study. A construction sequence for
the dual-phase DUNE LArTPC detector module installation was defined based on
the use of a slightly modified scaffolding arrangement with respect to
the LBNO design, discussed in~\cite{cdr-annex-lbno-2} (Design: courtesy Technodyne
Ltd., Eastleigh, UK). The proposed construction sequence assumes
completion of the membrane and insulation system installation together
with completion of all tank internal pipework and cable trays.  At
this stage, sections of the scaffolding will be removed and replaced
by the Alimak-Hek or similar climbing access platforms to provide
increased functionality to the installation.  On completion of the
scaffolding revisions and the climbing access platform installation,
the entire tank and scaffolding systems will be cleaned in preparation
for the detector module installation. The proposed far detector module construction
sequence consists of:
\begin{enumerate}
\item{Complete installation of insulation and membrane, install cable trays from top to bottom for photomultipliers (PMTs) electrical cables.}
\item{Adjust scaffolding platforms, add Alimak-Hek platform and floor protection.}
\item{Air purge top level installation.}
\item{Install hanging columns for detector.}
\item{Install lowest field shaping coil to stabilize columns, and then install first top 15 levels of field shaping coils.}
\item{Thoroughly clean top level assembly.}
\item{Install Charge Readout (CRP) from top scaffolding platform (detailed sequences described in~\cite{cdr-annex-lbno-2}.}
\item{Thoroughly clean top level assembly.}
\item{Screen off top level to protect Anode (protective screen 1).}
\item{Air purge top level (allow bleed air into middle and lower levels).}
\item{Continue installing field shaping coils.}
\item{Complete installation of field shaping coils.}
\item{Thoroughly clean field shaping coils, remove protective screen (screen 1) to top level and progressively remove all scaffolding and Alimak-Hek platforms.}
\item{Screen off field shaping coils (including CRO): protective screen 2.}
\item{Air purge top and middle levels.}
\item{Construct Cathode from Modules. Cathode to be raised 300\,mm off tank bottom during construction.}
\item{Thoroughly clean cathode and space used for fabrication.}
\item{Remove protective screen 2.}
\item{Fit cathode to field cage using suitable jacks.}
\item{Screen off entire detector: protective screen 3.}
\item{Remove floor protection.}
\item{Add cable trays, junction boxes and cables for PMTs.}
\item{Install PMTs to tank Bottom (pre-assembled L-flanges). Check out and test PMTs.}
\item{Clean air purge bottom level.}
\item{Install temporary enclosure around TCO inside and outside with air lock within the enclosure.}
\item{ Remove protective screen (screen 3) using air lock system to prevent contamination of detector}.
\item{Close temporary construction openings.}
\item{Thoroughly clean TCO areas.}
\item{Remove temporary enclosures.}
\item{Remove all tools, equipment, etc., through tank roof.}
\item{Exit via room manways.}
\item{Close all tank roof openings.}
\end{enumerate}

\subsection{Detector construction program and installation schedule}

Both the 3$\times$3$\times$1\,m$^3$ prototype detector and the
6$\times$6$\times$6\,m$^3$ WA105 demonstrator are planned to be built in
advance of the larger 4$\times$10-kt modules of the experiment at
SURF. It is expected that valuable information will be gathered
from the construction of both the prototype and demonstrator detectors
that will ultimately benefit both the planning and construction
forecasting for the larger far detector. 

For comparison the 20-kt LAr detector for Pyhasalmi was designed for a
drift surface (roughly the CRP and cathode area) of 824\,m$^2$ in an
octagonal shape with a drift length of 20\,m. The dual-phase DUNE
LArTPC experiment at SURF has an equivalent area of 12$\times$60 =
720\,m$^2$ (for one module of 12\,kt) with a drift length of 12--15\,m
(fiducial mass 1\,kt / 1\,m drift, total 12--15\,kt). The construction
program calculated by Rockplan Ltd, Alan Auld Ltd and Rhyal Eng. Ltd
can be seen as a conservative approach for the SURF site, as most
of the time corresponds more closely to
the instrumented surface required  rather than to 
the drift length, but neither specific SURF site-related
effects nor effects of US legal procedure are  taken into account
in this construction program.

The DUNE detector module installation program has been divided into three distinct (and separate)
stages:
\begin{enumerate}
\item{Design,}
\item{Manufacture/Fabrication, and} 
\item{Assembly.}
\end{enumerate}  

The detector design must be done together with the tank deck design,
as the complete detector is suspended from the deck.  Fabrication and
manufacturing can be started while the tank construction is still
on-going.  The total time for manufacture/fabrication and construction
is calculated to be around six years for a DUNE far detector module, of which
\begin{itemize}
\item{14 months	is for	for manufacture/fabrication off-site, }
\item{20 months	is for	for construction/installation + testing, and }
\item{32 months	is for	total works (with partial overlap). }
\end{itemize}

%% file: volume-detectors/chapter-synergy.tex
\chapter{Synergies Between Far Detector Designs}
\label{ch:detectors-synergy}

\section{Overview}

As discussed in Section~\ref{sec:detectors-strategy-FD}, two
technologies for LArTPCs are being pursued. There are a number of
synergies between these development efforts.

Both the reference and alternative designs for the DUNE far detector
are liquid argon TPCs. The designs assume nearly identical cryostats
(with some differences in the cryostat roof) installed in identical chambers and
supported by identical cryogenic systems. The designs differ mostly in
their approaches to collection and readout of the ionization
signals. In the reference design, the ionization charge is measured by
successive wire planes, two induction and one collection, all immersed
in the LAr. In the alternative design, the charge is extracted from
the liquid to the vapor and then amplified and finally collected on a
2D anode, providing two independent views.

%A list of synergies between far detector subsystems designs includes
Several of the far detector subsystems offer great potential for synergy between the reference and alternative designs, including the
\begin{itemize}
\item Interface to the cryogenics system,
\item High voltage, 
\item Photon detection,
\item Calibration,
\item Underground installation strategies,
\item Local computing infrastructure and DAQ, and
\item Detector modeling and simulation.
\end{itemize}

\section{Interface to the Cryogenics System}

In both designs, ionization electrons have drift lengths on the order
of several meters. In order to reach the required millisecond scale for
electron lifetime, the electronegative impurities in the LAr must be
maintained below the ppb level. The contamination will
come primarily from impurities adsorbed onto the tank and detector element surfaces.
Given that the detector modules will be housed in cryostats of the same design,
using the same industrial LNG cryostat technology and the same cryogenics systems,
the process of understanding these sources and minimizing them
is to a great extent the same for either TPC design. This leads to %potential 
expected synergies in the areas of
\begin{itemize}
\item Electronegative contamination mitigation, 
\item Modeling contamination sources,
\item Contamination migration modeling,
\item Material properties,
\item Filtration	,
\item Design of the cryogenics system,
\item Purity monitoring,	
\item Roof interfaces (hatch, feedthrough, mounts),
\item Grounding and shielding, and
\item Installation spaces and cryogenics system needs.	
\end{itemize}

%\subsection{Drift Electron Lifetime}  <--- can't have one subsection in a section; whole sec is about e drift
%\label{sec:detectors-synergy-lifetime}

For detecting interactions of beam neutrinos, the requirement for
electron lifetime of $>$\,3\,ms derives from the minimum signal to noise
ratio (S/N$\,>\,$9) required for MIP signals on induction plane wires from
interactions near the cathode to be above the zero suppression
thresholds.  Initial studies of energy resolution for supernova physics
also requires an electron lifetime above 3\,ms.  The 3\,ms lifetime
detector requirement is the same for both the single-phase reference
design and the dual-phase alternate design.  As the argon purity goal
is similar, work on contamination mitigation can be done jointly.

There is much experience in the community to justify confidence that
high levels of argon purity can be achieved.  Careful design of gas
ullage and the recirculation system is vital to avoid trapped pockets
of gas and to minimize the mixing of the gas and the liquid at the
interface.  ICARUS achieved a lifetime above 15\,ms after modification of
the cryosystem to extend the lifecycle of the recirculation
pump.\cite{Antonello:2014eha} The materials test stand (MTS) at Fermilab has
demonstrated that contaminants in the liquid argon originate from
materials in the gas space in the ullage, where the warmer
temperatures allow for outgassing of exposed surfaces and that
materials immersed in the liquid argon are not a source of
contamination.\cite{andrewsNIM} The MTS has measured the
contamination rate for many materials and is available for
continued testing of additional materials.  The Liquid Argon Purity
Demonstrator (LAPD) acheived lifetimes above 14\,ms without evacuating
the cryostat and with a functioning TPC inside the
cryostat.\cite{Bromberg:2015uia}

The phase-1 run of the 35-t achieved a peak 3\,ms electron lifetime;
however, the purity was still improving when the run ended.  The
engineers and scientists from LBNF and both DUNE detector options will
work together to optimize the cryogenic design for high purity. Two
examples are understanding the sources of contamination in the ullage
and how this contamination migrates to the liquid and developing a
fill procedure that preserves the purity of the incoming liquid.
There are several membrane cryostats that will be designed and built
over the next 10 years by a common engineering team: the
1$\times$1$\times$3\,m$^3$ dual-phase prototype, Short Baseline
Neutrino Detector (SBND), WA105, the Single-Phase CERN Prototype
(SPCP) engineering prototype as well as the DUNE far detector. Each of
these will learn from its predessors and inform its successors. Based
on existing measurements and extrapolations to the \ktadj{10} design a 3\,ms
lifetime should be readily achievable.

\section{High Voltage}

Both LArTPC designs require a large HV to
produce an electric field of the order of 500\,V/cm in the drift
volume.  They both thus require a HV generator, HV
feedthoughs and a field cage to correctly shape the electric
field. While these elements differ in the two designs, they present a common
set of problems to solve, including
\begin{itemize}
\item Design rules for HV,
\item HV generation,
\item HV Feedthroughs,
\item Field cage structure, and
\item Arc mitigation (Stored energy and discharge).
\end{itemize}

\section{Photon Detection}

The approaches to photon detection in the two designs is
different.  The reference design uses TPB-coated light guides
instrumented with SiPMs, whereas the alternate design uses large PMTs
(also coated with TPB). Nevertheless, several aspects of and
techniques used in the development of these systems have strong synergies, including
\begin{itemize}
\item Requirements refinement and validation,
\item Development and evaluation of photosensors,
\item Impact of background light,
\item  Surface reflectivity, and
\item Photon detector calibration.
\end{itemize}

\section{Detector Calibration}

The challenging requirement on systematic uncertainties calls for a
robust program of calibration, which may include the use of calibration sources
deployed in the detector, complementary external measurements
and data-driven calibration procedures. It is expected that this effort
will have significant synergies between the two designs, including
\begin{itemize}
\item Active volume,
\item Energy scale,
\item Energy resolution,
\item PID likelihoods, and
\item Absolute light yield.
\end{itemize}

\section{Underground Installation Strategies}

A fundamental aspect of the detector cost optimization is related to
the development of a strategy for underground logistics, safety and detector
installation. Dimensioning of the components to be transported and
assembled underground is a common issue, affording concomitant synergies. 
Strategies for and requirements on implementing the clean
rooms, additional tooling and needs for temporary installations, such
as scaffolding, also present opportunities for potential synergies.

\section{Local Computing Infrastructure and DAQ}

Once the electrical signals from the detector have been processed
(e.g., by front-end preamplifiers), the treatment of the digitized raw data and
their compression can be strongly unified and will therefore provide
synergies. The online computing farm will have a very similar layout
for both reference and alternative designs. The software triggering and
filtering algorithms will be based on similar local computing
architectures, offering strong synergies. Finally, the local data
storage and transmission to offsite tier centers will be common.

\section{Detector Modeling and Simulation}

Accurate and detailed detector modeling is required.  Simulations are
needed for both ionization electrons and scintillation photons. The
%basis of the synergies in this area is the 
common detection medium of
LAr provides a basis for synergies in this area that include
\begin{itemize}
\item Charge generation and transport,
\item Charge diffusion and attenuation studies,
\item Noise and its impact on the detector performance, and
\item Optical model and light propagation.
\end{itemize}

%\section{Further Steps}
\section{Summary}
A large set of possible synergies exists between the reference and
alternative TPC designs. These synergies will be exploited and
developed within the DUNE collaboration and the LBNF team as the
program of prototypes, demonstrators and other development activities
continues and as the detector modules and the accompanying facilities
are constructed. The CERN neutrino platform, in particular, will
provide an excellent opportunity for joint detector development.

%% file: volume-detectors/chapter-nd-ref.tex
\chapter{Near Detector Reference Design}
\label{ch:detectors-nd-ref}

%%%%%%%%%%%%%%%%%%%%%%%%%%%%%%%%
\input{volume-detectors/nd-ref-sections/nd-ref-overview}
\input{volume-detectors/nd-ref-sections/nd-ref-fgt}

\input{volume-detectors/nd-ref-sections/nd-ref-fgt-req}
\input{volume-detectors/nd-ref-sections/nd-ref-alts}
\input{volume-detectors/nd-ref-sections/nd-ref-blm}

\input{volume-detectors/nd-ref-sections/nd-ref-daq}

%% file: volume-detectors/nd-ref-sections/nd-ref-overview.tex
%%%%%%%%%%%%%%%%%%%%%%%%%%%%%%%%
\section{Overview} %Near Detector Systems}
\label{sec:detectors-nd-ref-ov}

This chapter describes the reference design of the DUNE Near Detector
Systems (NDS). The scope includes the design, procurement,
fabrication, testing, delivery and installation of the NDS components:
\begin{itemize}
\item Fine-Grained Tracker (FGT) near neutrino detector (NND),
\item Beamline Measurement System (BLM), and
\item NDS Data Acquisition system (DAQ).
\end{itemize}
Detailed descriptions of the NDS subsystems are provided in \anxndref~\cite{cdr-annex-nd}.

The concept and design of the reference DUNE-ND evolved from
experience with MINOS, the first generation long-baseline
neutrino experiment at Fermilab; NOvA, the second generation
experiment; the high resolution NOMAD detector at CERN; and the T2K
detector at JPARC. MINOS and NOvA employ functionally identical
detectors that fulfill the missions of these experiments, given the
statistics and resolution of the respective far detectors.  DUNE, the
third generation experiment at Fermilab, has more ambitious goals:
discovery of CP-violation, discovery of mass hierarchy, and a search
for physics beyond PMNS with unparalleled precision. DUNE will have a
more intense neutrino source and a higher-resolution massive far detector.  To
meet the ultimate systematic precision needed to fulfill these goals,
the ND must thoroughly characterize the beam composed of muon
neutrinos/antineutrinos and electron neutrinos/antineutrinos. It must
precisely measure the cross sections and particle yields of various
neutrino processes.  The particle yields include multiplicity and
momentum distributions of pions, kaons and protons produced in the
hadronic jet.
%An ``identical ND", like those in MINOS or NOvA with the experience gained therein, cannot accomplish these requirements. 

%As pointed out in Chapter 6 of CDR \volphys %Chapter~\ref{ch:physics-nd} of Volume-2
%, because not a single spectrum at FD is identical to that at the ND 
%the concept of `identical' detectors is an over-simplification.
%Furthermore 
The need to precisely quantify the neutrino source and
cross sections, including the hadronic composition, motivates
a high-resolution ND. The NOMAD experience suggests that a high-resolution 
detector, capable of measuring $e^{\pm}$, $\mu^{\pm}$,
$\pi^{-+0}$, proton and $K^{0}$, would partially meet the
challenges of DUNE --- the detector must be augmented to measure, and
thereby precisely model, the neutrino-nuclear effects. The reference
DUNE-ND is a next-generation near detector concept
compared to the T2K experiment, %. Such a detector 
and will enrich the physics potential of the DUNE/LBNF program.  Complementary
LAr detector(s) upstream of the high-resolution ND will enhance the
capability of the ND complex.

The reference DUNE-ND, a Fine-Grained Tracker (FGT),
consists of a straw-tube tracking detector (STT) and electromagnetic
calorimeter (ECAL) inside of a 0.4-T dipole magnet. In addition, Muon
Identifiers (MuIDs) are located in the magnet steel, as well as
upstream and downstream of the STT. The FGT is designed to make
precision measurements of the neutrino flux, cross section, signal
and background rate at the percent level.  \fixme{is it `signal and background' rates?}

The Beamline Measurement System (BLM) will be located in the region of
the Absorber Complex at the downstream end of the decay region to
measure the muon flux from hadron decay. The absorber itself is part
of the LBNF Beamline.  The BLM will determine the neutrino
flux and spectra and monitor the beam profile on a spill-by-spill
basis, and will operate for the life of the experiment.

The Near Detector System Data Acquisition system (NDS-DAQ) collects
raw data from each NDS DAQ, issues triggers,
adds precision timing data from a global positioning system (GPS) and
builds events.  The NDS-DAQ is made up of three parts: NDS Master DAQ
(NDS-MDAQ), Beamline Measurements DAQ (BLM-DAQ) and Near
Neutrino Detector DAQ (NND-DAQ).

%% file: volume-detectors/nd-ref-sections/nd-ref-fgt.tex
%%%%%%%%%%%%%%%%%%%%%%%%%%%%%%%% 
\section{The Fine-Grained Tracker} 
\label{cdrsec:detectors-nd-ref}

The scope of the DUNE Fine-Grained Tracker (FGT) near neutrino
detector includes the design, procurement, fabrication, testing,
delivery and installation of all FGT subsystems:
\begin{itemize}
\item Central straw-tube tracker (STT),
\item Electromagnetic calorimeter (ECAL),
\item Dipole magnet (0.4-T) surrounding the STT and ECAL,
\item Muon identifiers (MuID): in the magnet steel and upstream/downstream of the STT, and 
\item Instrumentation for monitoring and control.
\end{itemize}

A schematic drawing of the FGT design is shown in
Figure~\ref{fig:STT_schematic}.
\begin{cdrfigure}[A schematic drawing of the fine-grained
tracker (FGT) design]{STT_schematic}{A schematic drawing of the fine-grained tracker design.}
\includegraphics[width=0.8\textwidth]{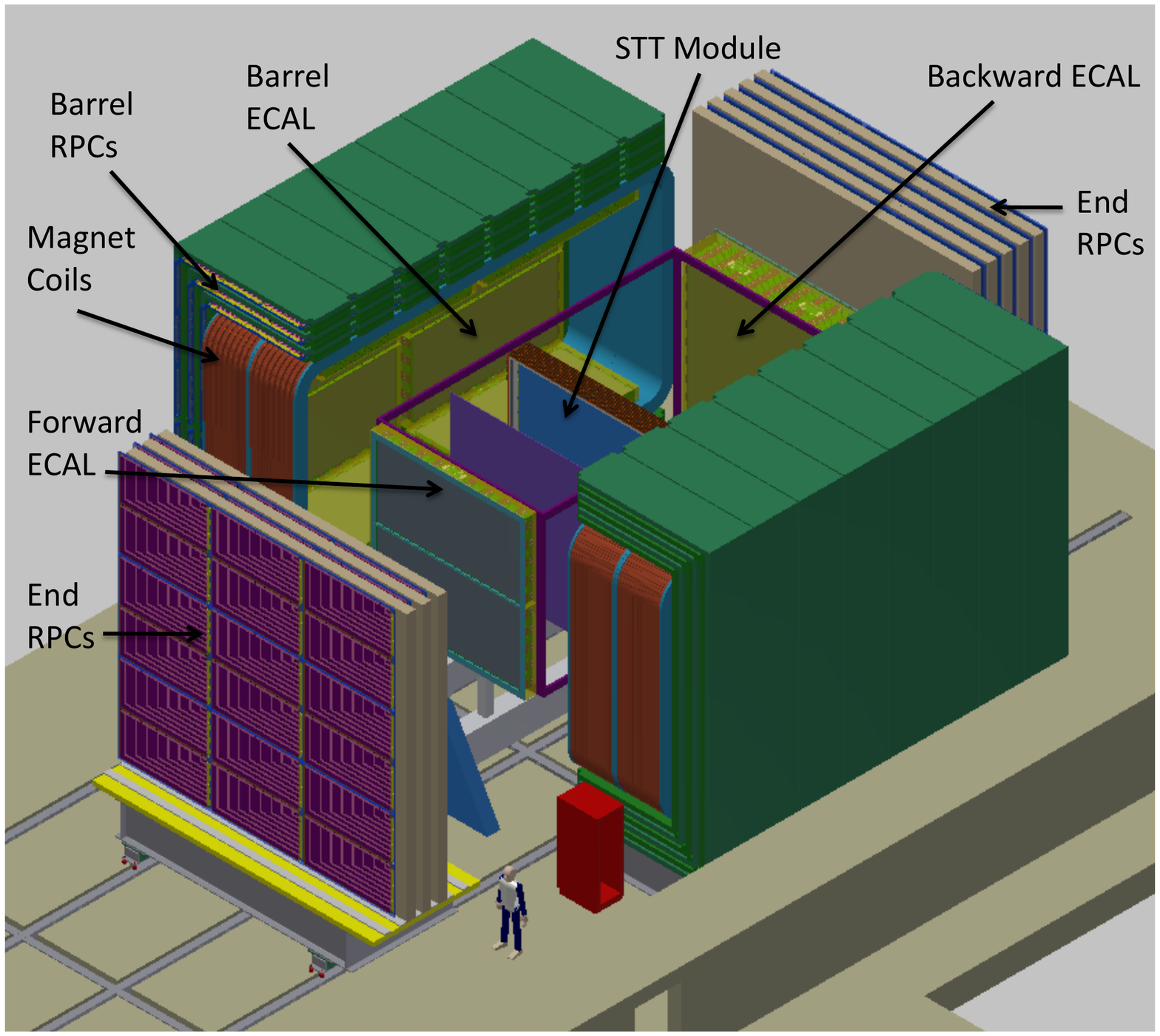}
\end{cdrfigure}
%The design presented here meets the required performance for making
%precision measurements of neutrino fluxes, cross sections, signal and
%background rates for the DUNE far detector.  ---- ALREADY SAID IN OVERVIEW
The DUNE ND is designed to 
meet the required performance and fulfill the physics program described in Chapter 6 of CDR \volphys~\cite{cdr-vol-2}, 
%Section~\ref{ch:physics-nd}, 
and in the DAE/DST Detailed Project
Report\cite{DPR}. The most significant
requirements\cite{ND-REQ1,ND-REQ2} for the FGT include:
\begin{itemize}
\item Muon energy scale uncertainty better than 0.2\% and hadronic
  energy scale uncertainty better than 0.5\% for the low-$\nu$ flux
  measurements;
\item Magnetized detector capable of separating $\mu^+$
  from $\mu^-$ as well as $h^+$ from $h^-$, where $h$ is a charged
  hadron;
\item Capability to to separate $e^+$ from $e^-$ for absolute and
  relative flux measurements;
%\item low density $\rho \sim 0.1$ g/cm$^3$, similar to that of liquid hydrogen
%\item dipole magnetic field of $B=0.4$ T with high detector granularity 
\item Excellent momentum ($<$5\%) and angular ($<2$\,mrad) resolutions
  for $\mu^{\pm}$, $e^{\pm}$, $\pi^{\pm}$ and proton, and
  $\pi^0$/$\gamma$ via decay/conversion and $K^0_S$/$\Lambda$
  produced in $\nu$-induced interactions;
\item 4$\pi$ ECAL coverage to ensure accurate determination of the
  momentum vector of the hadronic shower;
\item 4$\pi$ MuID coverage to identify muons with a wide range of
  energies and angles;
\item Electron/positron identification through the use of transition
  radiation (TR) in the entire tracking detector for low-energy and/or
  large angle tracks (e.g. $\gamma$ conversions);
\item $\pi/K/p$ identification by $dE/dx$ in the entire tracking
  detector;
\item Identification of $\pi^0$ and $\gamma$ that can mimic
  $\nu_e$ signals at the far detector;
\item Use of a variety of nuclear targets, (C$_3$H$_6$)$_n$, Ar, Ca,
  C, Fe, etc., to quantify the impact of nuclear effects in
  $\nu$($\bar \nu$)-nucleus cross sections; and
\item Provision of more than ten times the unoscillated statistics on Ar targets expected
  in a 40-kt far detector.
%\item  identification and measurement of processes such as neutral current $\pi^0$ production that can mimic oscillation signals at the FD
%\item comparability with measurements made in the FD
%\item measurement of nuclear effects, including short-range correlations, two-body currents, pion absorption, initial-state interactions, 
%and final-state interactions
\end{itemize}
The requirements listed above imply the use of a low density, $\rho
\sim$0.1\,g/cm$^3$, high-resolution magnetic spectrometer.  A summary
of the performance requirements is given in Table~\ref{tab:comparison}
Regardless of the process under study, the goal is to have the
systematic error less than the corresponding statistical error.
\begin{cdrtable}[A summary of the performance for 
the FGT configuration]{ll}{comparison}{A summary of the performance for 
the FGT configuration}
Performance Metric&FGT\\ \toprowrule
Dipole magnetic field & 0.4 T \\ \colhline 
Average target/tracker density & $\rho \sim 0.1$ g/cm$^3$ \\ \colhline 
Target/tracker Volume & 3.5m x 3.5m x 6.4m \\ \colhline
Target/tracker Mass&8\,t \\ \colhline
Vertex Resolution&0.1 mm \\ \colhline
Angular Resolution&2 mrad \\ \colhline
$E_e$ Resolution&5\% \\ \colhline
$E_\mu$ Resolution&5\% \\ \colhline
$\nu_\mu/\bar \nu_\mu$ ID&Yes \\ \colhline
$\nu_e/\bar \nu_e$ ID&Yes \\ \colhline
NC$\pi^0$/CCe Rejection&0.1\% \\ \colhline
NC$\gamma$/CCe Rejection&0.2\% \\ \colhline
CC$\mu$/CCe Rejection&0.01\% \\
\end{cdrtable}

%The FGT will measure the neutrino event rates and cross sections 
%on argon, water, and other nuclear 
%targets for both $\nu_e$ and $\nu_\mu$ charged current (CC) and
%neutral current (NC) scattering events. The FGT design 
%consists of a straw-tube tracker (STT), consisting of straw tubes, water targets, argon targets, 
%and radiator targets, and an electromagnetic calorimeter (ECAL), both inside a
%dipole magnet. In addition, muon detectors (MuID) consisting of resistive plate
%chambers (RPCs) will be embedded in the steel
%of the magnet. 

%%%%%%%%%%%%%%%%% 
\subsection{Straw-Tube Tracking Detector}
\label{cdrsec:detectors-nd-ref-fgt-stt}

\subsubsection{Straw Tubes} 

The Straw-Tube Tracking Detector (STT) at the center of the FGT 
is composed of straw tubes with an outer diameter of 1\,cm, as well as 
radiators and targets that reside next to the straw tubes as shown in Figure~\ref{fig:STT_Detail}.
\begin{cdrfigure}[A schematic drawing of the straw tube tracker (STT)]
{STT_Detail}{A schematic drawing of the Straw-Tube Tracking Detector (STT) design.}
\includegraphics[width=0.4\textwidth]{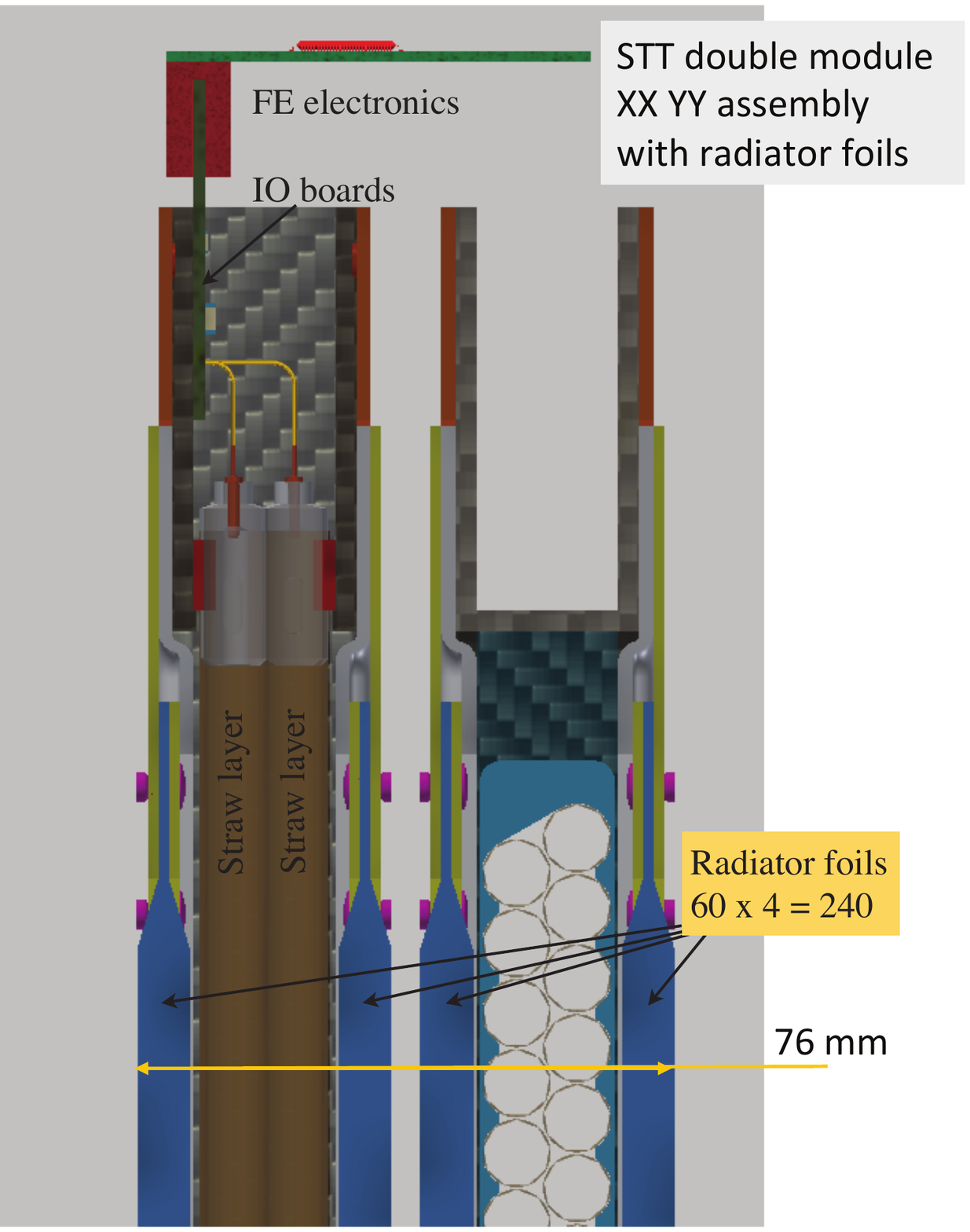}
\end{cdrfigure}

The straw walls are made by winding together a film of carbon loaded
Kapton XC (inner) and a film of aluminum coated Kapton HN (outer), for
a total thickness of about 70\,$\mu$m. The anode wire is gold-plated
tungsten with 20 $\mu m$ diameter.  Two consecutive layers of straws
are staggered by a half-diameter, glued, and inserted into a
carbon-fiber composite mechanical frame to form a single module.  One
vertical (YY) and one horizontal (XX) module are then assembled
together to form a double-module (XXYY) equipped with frontend
electronic readout boards.  Figure~\ref{fig:STT_Detail} shows a
schematic drawing of the STT module with four straw-tube planes and
radiators. Each XXYY module equipped with radiators is equivalent to
$1.45\times10^{-2} X_0$ ($0.2\times10^{-2} X_0$ without
radiators), with a radiation length $X_0 \sim5.5$\,m. The momentum
measurement requires that tracks are detected in at least six straw
layers. The staggered double layer design, high number of straw planes
and double-end readout will contribute to resolving short track
disambiguation.

The straw tubes will be filled with a gas mixture of either 70\% Ar
plus 30\% CO$_2$ (for modules with nuclear targets) or 70\% Xe plus 30\%
CO$_2$ (for modules with radiators).  The dimensions of each double module in
the reference design will be approximately
350\,cm$\times$350\,cm$\times$8.0\,cm, including a nuclear target or four radiator
planes and four straw planes. For ease of construction and
transportation, each double module is made up of two modules, with
two straw layers and dimensions of appproximately 350\,cm$\times$175\,cm$\times$4.0\,cm.  Each
module will have a carbon composite frame around the perimeter for
support and will have an attached target or radiator.

The modularity of the STT provides for successive measurements using
thin nuclear targets (thickness $< 0.1 X_0$), while the excellent
angular and spatial resolution allows a clean separation of events
originating in different target materials.

The STT will have a total of 107,520 straws --- corresponding to 336
straws per plane, 1344 straws per module --- and 80 modules. Both ends
of the straw tubes will be read out, leading to a total number of
215,040 electronics channels. The total mass of the STT, including
targets and radiators, is approximately 8\,t. Table~\ref{tab:STT_details} 
summarizes the main STT parameters.
\begin{cdrtable}[Straw Tube Tracker (STT) specifications]{ll}{STT_details}{Straw Tube Tracker  specifications}
Item&Specification \\ \toprowrule
Straw Tube Geometry & 1\,cm Diameter $\times$ 3.5\,m Long \\ \colhline
Number of Straw Tubes & 107,520 \\ \colhline
Number of Straw Tubes per Plane & 336 \\ \colhline
Number of Straw Tube Planes per Module & 4 \\ \colhline
Number of Straw Tube Sub-Modules per Module & 4 \\ \colhline
Number of Straw Tube Modules & 80 \\ \colhline
Number of Straw Tube Sub-Modules & 320 \\ \colhline
Length of Straw Tube Wire & 376.3 km \\ \colhline
Number of Electronics Channels & 215,040 \\ \colhline
Number of Modules with Radiators & 75 \\ \colhline
Radiator Thickness per Module & 3.6\,cm \\ \colhline
Radiator Mass per Module & 69.1 kg \\ \colhline
Number of Modules with Nuclear Targets & 10 \\ \colhline
C Mass per Target Plane & 192 kg \\ \colhline  
Number of Modules with C Target Planes & 2 \\ \colhline 
Ca Mass per Target Plane & 132 kg \\ \colhline  
Number of Modules with Ca Target Planes & 1 \\ \colhline 
Ar Target Geometry & 5.08\,cm Diameter $\times$ 3.5\,m long \\ \colhline
Number of Ar Targets per Plane & 68 \\ \colhline
Ar Mass per Target Plane & 112 kg \\ \colhline  
Number of Modules with Ar Target Planes & 1 \\ \colhline
Fe Mass per Target Plane & 96.5 kg \\ \colhline  
Number of Modules with Fe Target Planes & 1 \\  
%Water Mass per Target Plane & 95 kg \\
\end{cdrtable}
In addition to tracking charged particles and measuring Transition Radiation (TR), 
the STT provides $dE/dx$ measurement to identify particles. 
Figure~\ref{fig:PartID_dedx} provides a sample of pions, kaons and protons identified via $dE/dx$ in the STT.
\begin{cdrfigure}[Simulated distributions of $dE/dx$ for different particles in FGT]
{PartID_dedx}{Simulated distributions of $dE/dx$ for different particles in FGT.}
\includegraphics[width=0.7\textwidth]{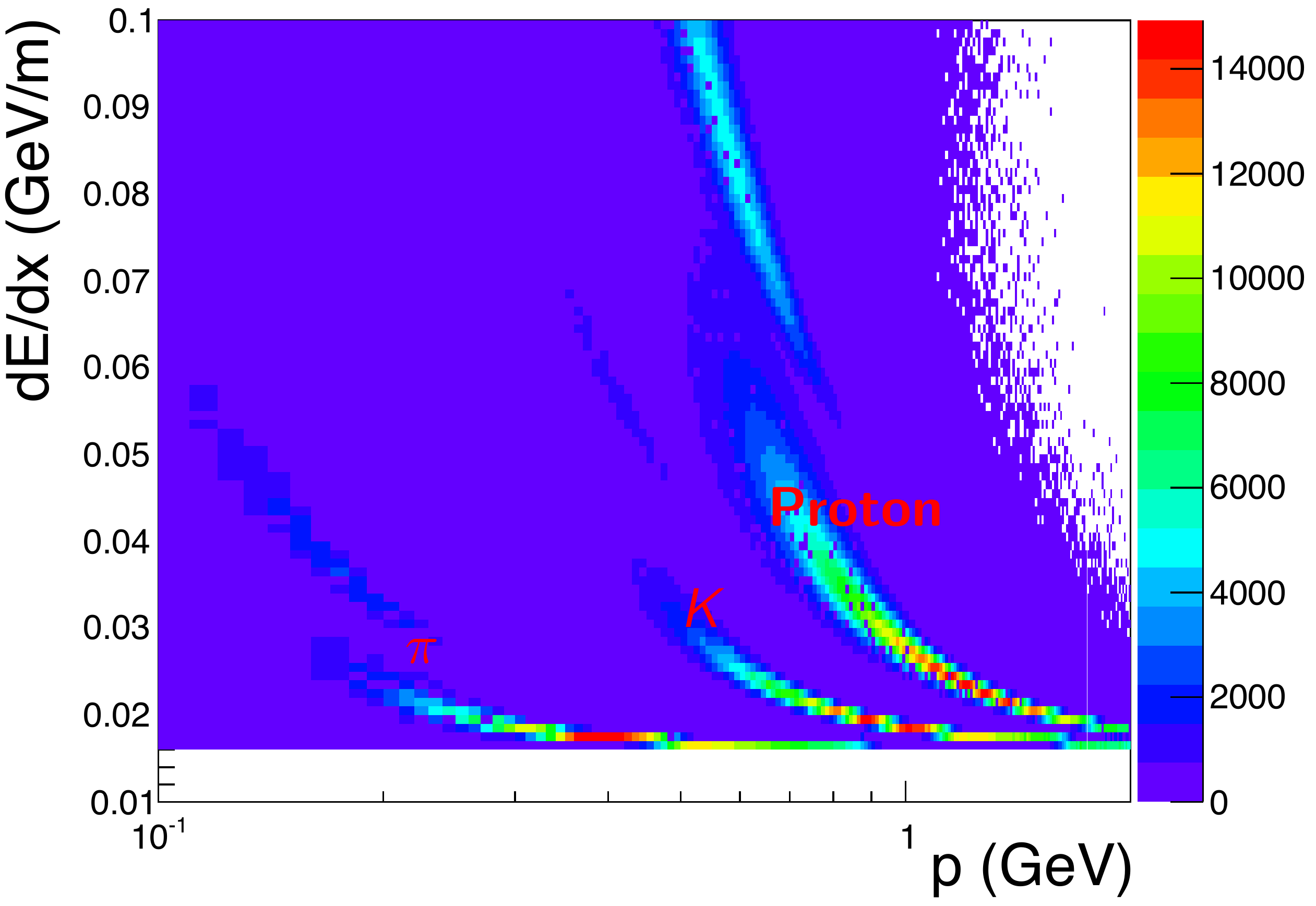}
\end{cdrfigure}

\subsubsection{Radiator Targets}

Radiators will be placed in the downstream STT modules and will serve
as targets for both neutrino interactions and Transition Radiation
(TR) production. Each STT module contains four radiators, where each
radiator consists of 60 layers of 25-$\mu$m polypropylene
(C$_3$H$_6$)$_n$ foils, which are embossed to keep 125-$\mu$m air gaps
between consecutive foils.  The mass of each radiator is $\sim$17\,kg
and the thickness is $\sim$9\,mm. The use of thin plastic foils
regularly spaced allows the emission of transition radiation for
electron/positron identification, which is detected by the Xe gas in
the straws. The plastic radiators account for about 83\% of the mass of each STT module and 
also provide the main (anti)neutrino target. Overall, a radiator mass of about 5 tons 
is required to achieve the physics sensitivity discussed in 
Chapter 6 of CDR \volphys{}~\cite{cdr-vol-2}  %Section~\ref{ch:physics-nd} 
and the DAE/DST Detailed Project Report\cite{DPR}. 

\subsubsection{Nuclear Targets} 

A set of different nuclear targets will be installed in front of the
most upstream STT modules, which will not be equipped with radiators.
The most important nuclear target is the argon target that matches
the DUNE far detector.  
This target will consist of planes of cylindrical tubes filled with 
argon gas pressurized to 140\,atm ($\rho = 0.233$), with
sufficient Ar mass to provide $\sim$10 times the unoscillated
statistics expected in a \ktadj{40} far detector. The proof-of-concept design 
consisted of 0.5-inch diameter, 3.5-m-long stainless steel tubes, 
with wall thickness 0.065-inch. In order to minimize the mass of the 
tube walls, the use of a single plane of C-composite 
tubes with 2-inch diameter and a wall thickness of 0.04-inch is under investigation.  

%This target will consist of planes of 0.5-inch
%diameter, 3.5-m-long stainless steel tubes, with wall thickness 0.065-inch,
%filled with argon gas pressurized to 140~atm ($\rho = 0.233$), with
%sufficient Ar mass to provide $\sim$10 times the unoscillated
%statistics expected in a 40~t far detector.  We are currently investigating the
%possibility to use carbon-composite tubes instead of stainless steel to contain 
%the pressurized Ar gas.

Relevant to argon, a crucial target is calcium which has the same
atomic weight ($A=40$) as argon but is isoscalar.  Since most nuclear
effects depend on the atomic weight $A$, inclusive properties of
(anti)neutrino interactions are expected to be the same for these two
targets.  This fact will allow the use of both targets to model
signal and backgrounds in the DUNE far detector (argon target), as
well as to compare DUNE results for nuclear effects on argon with the
extensive data on calcium from charged lepton scattering.

An equally important nuclear target is carbon (graphite), which is
essential in order to get (anti)neutrino interactions on free proton,
through a statistical subtraction procedure from the main
polypropylene target (C$_3$H$_6$)$_n$.  The availability of such a
free-proton target will allow accurate flux determinations and cross
section measurements, and, for the first time, a direct
model-independent measurement of nuclear effects --- including both
the primary and final-state interactions --- on the argon target
relevant for the far detector oscillation analysis. The required
carbon target mass is about 0.5\,t (in addition to the carbon in the
STT frames). The corresponding expected number of events on H target
are $5.0 (1.5) \times 10^6 \pm 13(6.6) \times 10^3$ $\nu(\bar \nu)$
CC, where the uncertainty is dominated by the subtraction procedure.

A stainless steel target in the form of a single thin slab will
provide service measurements of (anti)neutrino cross sections for the
INO experiment in India.

Finally, cylindrical tubes similar to those used for the pressurized Ar gas can
be filled with standard and heavy water (H$_2$O and D$_2$O). The
statistical subtraction of H$_2$O from D$_2$O will result in a
quasi-free neutron.

Table~\ref{tab:STT_details} gives a reference configuration of the
radiators and nuclear targets, listed according to their location from
downstream to upstream.  The final configuartion of the nuclear
targets will require detailed Geant4 simulations of FGT and
corresonding physics sensitivity studies.

%%%%%%%%%%%%%%%%% 
\subsection{Electromagnetic Calorimeter}
\label{cdrsec:detectors-nd-ref-fgt-ecal}

An electromagnetic calorimeter (ECAL) will surround the tracking
volume on all sides and consist of three separate pieces: Forward
ECAL, Barrel ECAL and Backward ECAL.  The expected energy resolution
is $\sim 6$\%/$\sqrt{E}$ for the forward ECAL.  The ECAL must provide
high segmentation in both the transverse and longitudinal directions
to reconstruct photons from $\pi^0$ decay and electron/positrons from
their Bremsstrahlung emissions.  The ECAL conceptual design consists
of layers of either 1.75-mm-thick (for the forward ECAL) or
3.5-mm-thick (for the barrel and backward ECAL) lead sheets and
2.5-cm-wide by 10-mm-thick plastic scintillator bars, as shown in
Figure~\ref{fig:ECAL_detail}.
\begin{cdrfigure}[Schematic drawing of the ECAL]{ECAL_detail}
{Schematic drawing of the forward ECAL equipped with the front-end and back-end readout boards (left), 
and a cross section of one barrel ECAL module (right), showing the details of the assembly of alternating planes
of plastic scintillator and Pb sheets.}
\includegraphics[width=0.9\textwidth,angle=0]{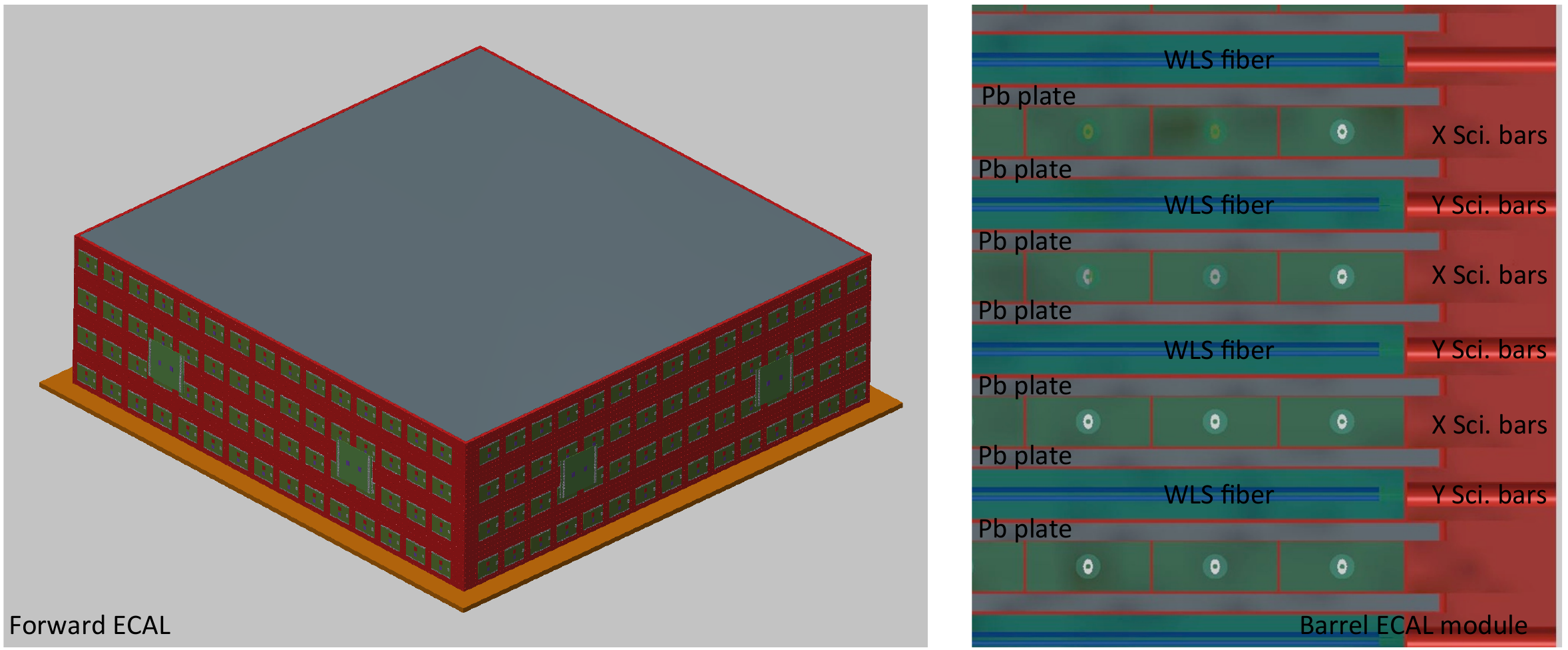}
\end{cdrfigure}
The scintillator layers for the Forward and Backward ECAL alternate as
XYXYXY..., while the scintillator layers for the Barrel ECAL are all
horizontal along the axis of the magnet.  The Forward ECAL will
consist of 60 layers of scintillator bars, where each bar has
dimensions 3.2\,m$\times$2.5\,cm$\times$1\,cm. The Backward ECAL will
consist of 16 layers of scintillator bars, where each bar has the same
dimensions, 3.2\,m$\times$2.5\,cm$\times$1\,cm. The Barrel ECAL will also
consist of 16 layers of scintillator bars, where each bar has the same
dimensions, 3.2\,m$\times$2.5\,cm$\times$1\,cm. The parameters of the
ECAL design will be further optimized with the help of a full Geant
simulation of FGT. \fixme{Geant4?}

The lead sheets and scintillator bars will be assembled and glued
together into complete modules of dimension
3.2\,m$\times$3.2\,cm$\times$81\,cm for the Forward ECAL and
3.2\,m$\times$3.2\,cm$\times$27.5\,cm for the Backward ECAL. For the
Barrel ECAL, the module dimensions will also be
3.2\,m$\times$3.2\,cm$\times$27.5\,cm. Two Barrel modules are placed
end-to-end along the sides of the inner surface of the magnet (eight
Barrel modules total) to provide full coverage of the barrel region.
The total numbers of scintillator bars in the Forward (7680), Backward
(2048) and Barrel ECAL (16384) is 26112 bars.

The scintillator bars will be extruded with holes in the middle of
each bar. The holes will then be fitted with 1-mm-diameter Kuraray
wavelength-shifting (WLS) fibers.  The fibers will be read out by SiPM
(silicon photomultiplier) photosensors at each end.  Detailed R\&D
studies will be performed to optimize the diameter of the scintillator
hole, the fiber diameter and the proper coupling between the
scintillator and the fiber for optimum light transmission.

The total mass of scintillator is 20.9\,t, the total mass of Pb is
70.8\,t, and the total number of readout channels is 52,224.
Table~\ref{tab:ECAL_specs} summarizes the specifications of the ECAL.
\begin{cdrtable}[ECAL specifications]{ll}{ECAL_specs}{ECAL specifications}
Item&Specification \\ \toprowrule
Scintillator Bar Geometry & 3.2m $\times$ 2.5cm $\times$ 1cm \\ \colhline
Number of Forward ECAL Scintillator Bars & 7680 \\ \colhline
Forward ECAL Pb thickness & 1.75mm \\ \colhline
Number of Forward ECAL Layers & 60 \\ \colhline
Number of Forward ECAL Radiation Lengths & 20\\ \colhline
Dimensions of Forward ECAL Module & 3.2m $\times$ 3.2m $\times$ 81cm \\ \colhline
Number of Barrel ECAL Scintillator Bars & 16,384 \\ \colhline
Barrel ECAL Pb thickness & 3.5mm \\ \colhline
Number of Barrel ECAL Layers & 16 \\ \colhline
Number of Barrel ECAL Radiation Lengths & 10 \\ \colhline
Number of Barrel ECAL Modules & 8 \\ \colhline
Dimensions of Barrel ECAL Modules & 3.2m $\times$ 3.2m $\times$ 27.5cm \\ \colhline
Number of Backward ECAL Scintillator Bars & 2048 \\ \colhline
Backward ECAL Pb thickness & 3.5mm \\ \colhline
Number of Backward ECAL Layers & 16 \\ \colhline
Number of Backward ECAL Radiation Lengths & 10 \\ \colhline
Dimensions of Backward ECAL Module & 3.2m $\times$ 3.2m $\times$ 27.5cm \\ \colhline
Total Length of 1-mm Diameter WLS Fiber & 83.6km \\ \colhline
Total Number of Scintillator Bars & 26,112 \\ \colhline
Total Number of Electronics Channels & 52,224\\ \colhline
Total Mass of Scintillator & 20,890 kg \\ \colhline
Total Mass of Pb & 70,800kg \\\end{cdrtable}

%%%%%%%%%%%%%%%%% 
\subsection{Dipole Magnet}
\label{cdrsec:detectors-nd-ref-fgt-magnet}

The STT and ECAL modules reside inside a 0.4-T dipole magnet for
the measurement of particle momentum and charge.  The magnet will have
inner dimensions (inside the coils) 4.5\,m wide $\times$ 4.5\,m high
$\times$ 8.0\,m long. The magnet has four vertical copper coils, stacked
horizontally, producing a horizontal magnetic field. The return yoke
will be divided into two halves along the longitudinal center line to
allow the magnet to be opened to service the detector inside, as
shown in Figure~\ref{fig:STT_schematic}.  Each half yoke will be built
from eight ``C'' (C-shaped) sections, and the thickness of the magnet
steel will be 60\,cm, consisting of 6$\times$10-cm-thick plates. The
magnet power requirement with copper coils is $\sim$2.4\,MW, corresponding
to 6\,kA at 400\,V. The water flow required for cooling is 20\,liter/s.
Figure~\ref{fig:Magnet_Bfield} shows the $B$ field results obtained
from detailed simulations performed at the Bhabha Atomic Research
Center (BARC) in India.
\begin{cdrfigure}[Magnetic field maps]{Magnet_Bfield}{Results from the detailed simulations of the 
magnetic field in the dipole magnet performed at the Bhabha Atomic Research Center (BARC) in India. 
The plots show the deviations (in percentage) from the nominal field in the (X,Y) plane (left plot)
and along the beam axis (right plot).}  
\includegraphics[width=\textwidth]{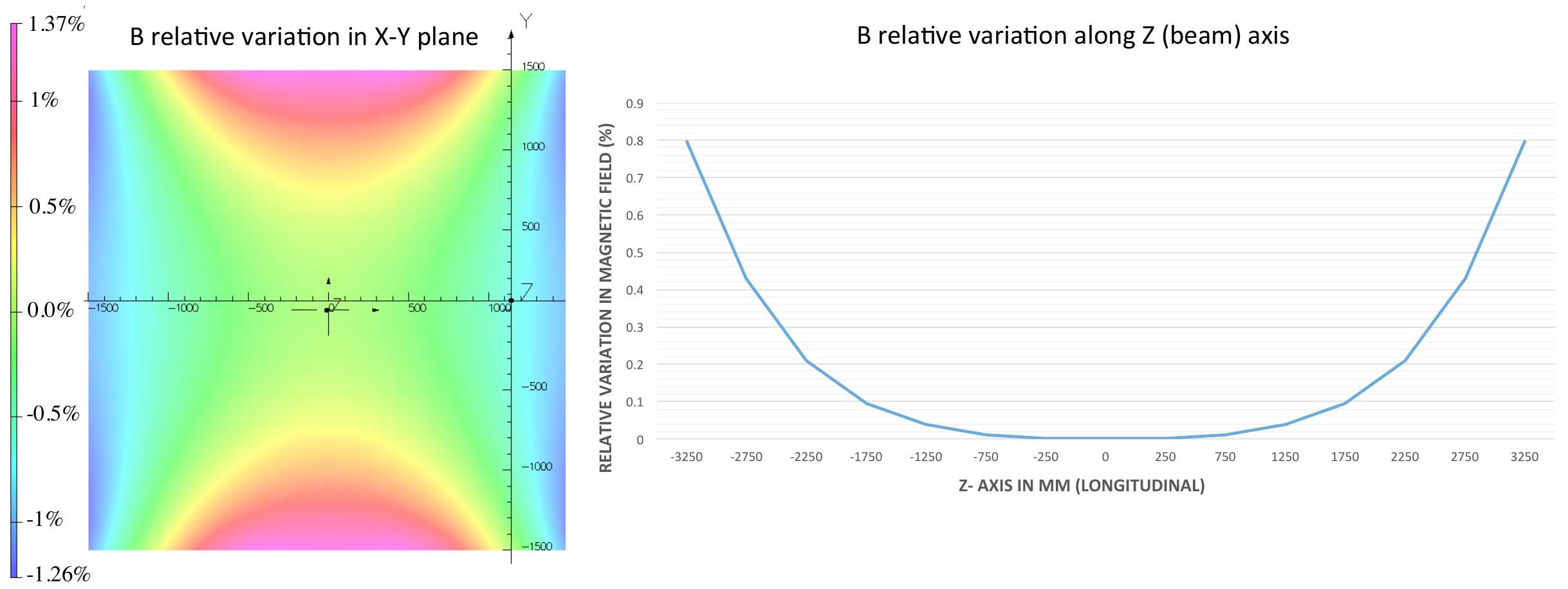} % [width=6in,angle=90] <-- was too small
\end{cdrfigure}
The magnet specifications are summarized in Table~\ref{tab:Magnet_specs}.
\begin{cdrtable}[Dipole Magnet specifications]{ll}{Magnet_specs}{Dipole Magnet specifications}
Item&Specification \\ \toprowrule
Inner Dimensions & 4.5\,m$\times$4.5\,m$\times$8.0\,m \\ \colhline
Magnetic Field & 0.4 T \\ \colhline
Number of ``C'' Sections & 16 \\ \colhline
Thickness of Steel in the ``C'' Sections & 60\,cm \\ \colhline
Mass per ``C'' Section & 60\,t \\ \colhline
Number of Coils & 4 \\ \colhline
Mass per Coil & 40\,t \\ \colhline
Magnet Current & 6 kA \\ \colhline
Magnet Voltage & 400 V \\ \colhline
Magnet Power Requirements & 2.4 MW \\ \colhline
Water Flow for Cooling & 20 l/s \\\end{cdrtable}

The momentum resolution is dominated by multiple scattering in the
STT. The momentum resolution is therefore given by $\delta p/p =
0.053/\sqrt(LX_0)B$. For B = 0.4T, L = 3m, and $X_0 = 4$m; the
expected momentum resolution is $\sim$3.8\%.

%%%%%%%%%%%%%%%%% 
\subsection{Muon Identifier}
\label{cdrsec:detectors-nd-ref-fgt-muonid}

The sides and ends of the dipole magnet will be instrumented with a
muon-identifier detector (MuID) that will distinguish muons from
hadrons by the ability of muons to penetrate the iron without
showering or interacting.  The task of the MuID is to reconstruct the
muon track segments and match them with the corresponding charged
track reconstructed in the STT.  The MuID will consist of 432
resistive plate chamber (RPC) modules interspersed between two
10-cm-thick steel plates of the dipole magnet and between 20-cm-thick
steel plates at the upstream and downstream ends of the magnet as is
detailed in Table~\ref{tab:MID_specs}. 
\begin{cdrtable}[MuID specifications]{ll}{MID_specs}{MuID specifications}
Item&Specification  \\ \toprowrule
Number of Barrel RPC Trays of Dimension 2.2\,m $\times$ 4\,m & 8 \\ \colhline
Number of Barrel RPC Trays of Dimension 2.5\,m $\times$ 4\,m & 16 \\ \colhline
Number of Barrel RPC Trays of Dimension 2.8\,m $\times$ 4\,m & 16 \\ \colhline
Number of Barrel RPC Trays of Dimension 3.1\,m $\times$ 4\,m & 8 \\ \colhline
Number of END RPC Trays of Dimension 2\,m $\times$ 6\,m & 24 \\ \colhline
Total Number of RPC Trays & 72 \\ \colhline
Total Number of RPC Modules & 432 \\ \colhline
Mass of Downstream Steel Planes & 283,500 kg \\ \colhline
Mass of Upstream Steel Planes & 170,100 kg \\ \colhline
RPC Thickness & 1.5\,cm \\ \colhline
Number of 7.65mm Pitch X Strips per Module & 256 \\ \colhline
Number of 7.5mm Pitch Y Strips per Module & 128 \\ \colhline
Total Number of RPC Strips and Electronics & 165,888 \\
\end{cdrtable}
The choice of RPC is motivated by the combineation of low cost,
ability to reach sub-mm space resolution, and existing expertise.
The MuID is only meant to provide identification of the muon; the muon
momentum will be measured by the STT inside the magnetic field.

The nominal dimensions of all RPC modules will be 1\,m $\times$ 2\,m
with active areas of 96\,cm $\times$ 196\,cm. Each module has 256 X
strips at 7.65-mm pitch and 128 Y strips at 7.5-mm pitch. This
fine-grained pitch allows to reach spatial resolution of $\sim$0.7\,mm
and to disentangle multiple hits resulting from events originated in
the magnet iron.  The modules will be grouped into trays, each
containing six modules, and the trays will be sufficiently wide to
allow overlapping modules.  The downstream MuID will contain five
steel planes of overall dimensions 6$\times$6$\times$0.2\,m$^3$
(283.5\,t) and five RPC planes, while the upstream MuID will contain
three steel planes (170.1\,t) of dimensions 6$\times$6$\times$0.2\,m$^3$
and three RPC planes. The barrel MuID will contain 24 planes (three
layers $\times$ eight sides) of RPCs. The RPCs will have a total
thickness of 15\,mm and a gap width of 2\,mm. One possible gas mixture
could be Ar (75\%), tetraflouroethane (20\%), isobutane (4\%) and
sulphurhexaflouride (1\%).  A full scale prototype of the RPC modules
was built at the Variable Energy Cyclotron Centre (VECC) in India.
Figure~\ref{fig:FGT_RPC} shows a picture taken during the assembly of
the prototype and the corresponding efficiency measurement with a
cosmic ray telescope.
%\begin{cdrfigure}[Schematic drawing of an RPC]{FGT_RPC}{Schematic drawing of an RPC.}
\begin{cdrfigure}[Fabrication and test of RPC prototype]{FGT_RPC}{Fabrication of a large 
(2.4\,m$\times$1.2\,m) RPC prototype at the Variable Energy Cyclotron Centre (VECC) in India 
(left) and corresponding efficiency tests (right).}
\includegraphics[width=\textwidth]{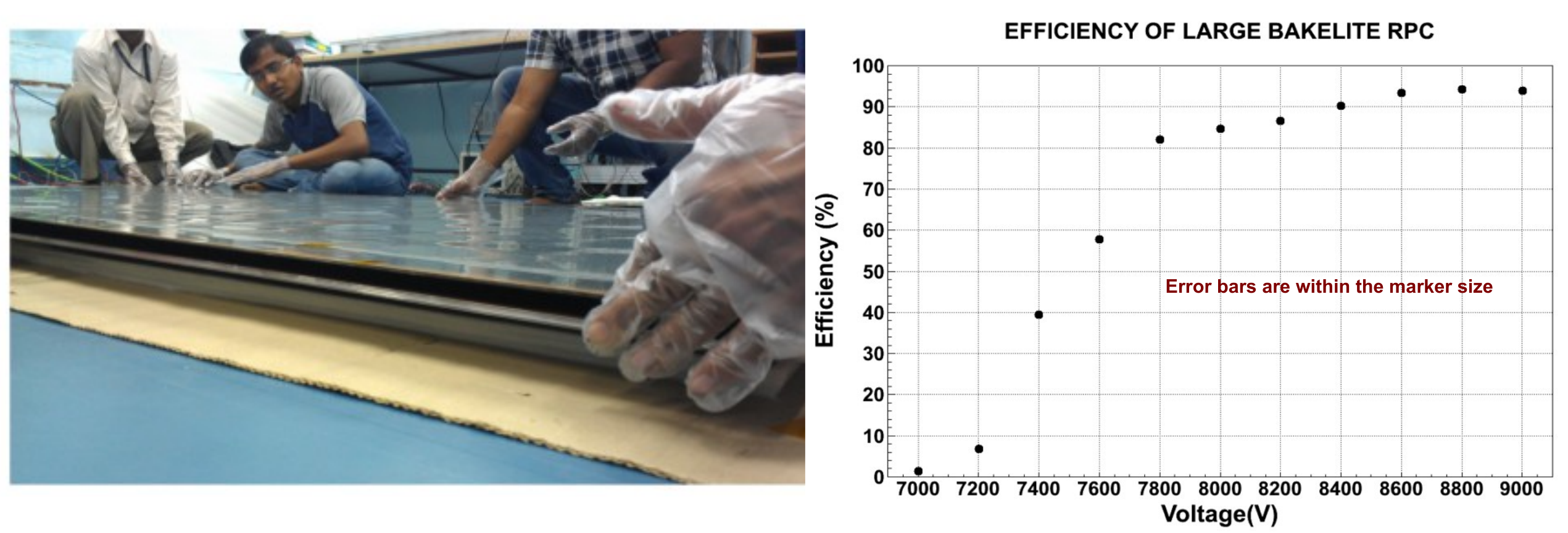} % [width=5in,angle=90] <-- was too small
\end{cdrfigure}

%%%%%%%%%%%%%%%%%e
\subsection{Instrumentation}
\label{cdrsec:detectors-nd-ref-fgt-instrum}

The instrumentation includes both fast readout electronics for the
subdetectors and slow control of the subdetectors, involving
monitoring the humidity, temperature, gas pressure, etc.  There is
considerable synergy in the information gathered in the STT, ECAL and
MuID.  Both the STT and ECAL are required to measure the total charge
and the time associated with a given hit. The MuID RPCs are required
to provide the position and time associated with a traversing
track. Similarly, the slow control of the subdetectors share many
features.

The electronics for the three subsystems, STT, MuID, and ECAL, are all
``fast'' systems, i.e., all of the signals are in the few-to-10
nanosecond range.  The requirements for each system are very similar:
a fast output and both an ADC and a TDC on each channel.
Additionally, for the STT straw tubes it is desirable to wave-form
digitize the analog signal in order to enhance the ability to separate
the ionization signal from the transition radiation signal.  The total
channel count in FGT is 433,152 channels, as shown in
Table~\ref{tab:elect_ch}.
\begin{cdrtable}[Number of electronics channels for each of the
three detector systems]{ll}{elect_ch}{The number of electronics channels for each of the
three detector systems}
Detector&\# of Channels\\ \toprowrule
STT & 215,040 \\  \colhline
ECAL & 52,224 \\  \colhline
MuID & 165,888 \\
\end{cdrtable}

Recently, an interesting new ASIC development for an upgrade to the
ATLAS muon system at the LHC has come out of BNL), the VMM2 chip.
%\fixme{add G. DeGeronimo at BNL ref} 
It handles 64 channels and produces both fast ADC and TDC outputs.  It
has been fabricated and tested and should be ready by 2017, long
before it will be needed for DUNE.

To maintain a low level of humidity and to maintain a desired
temperature, both STT and ECAL subdetectors will have dry nitrogen
circulating within their outer layers.  Magnet coils are cooled by
water, while the magnet yokes are instrumented with RPCs that must
remain dry. Continuous control of humidity in all these detectors is
needed.  Temperature must also be continuously monitored in all of the
subdetectors in order for the electronics to not overheat.
%To ensure that the magnet does not overheat, the water flow (pressure gradient) will be continuously monitored.  
Also, all power sources instrumenting the FGT and its readout need to be monitored for 
appropriate voltage and current.

Gas leaks need to be monitored in the STT and MuID.  The STT will
employ Xe gas, which helps with the measurement of transition
radiation.  Xe gas is expensive and, hence, will be recirculated;
leak-monitoring is particularly important here.  The requirement on
leaks is less stringent for the RPCs, which have less expensive gas.

The water flow (pressure gradient) will be continuously monitored in
order to ensure that the magnet does not overheat.

%% file: volume-detectors/nd-ref-sections/nd-ref-fgt-req.tex
%%%%%%%%%%%%%%%%%%%%%%%%%%%%%%%% 
\section{Matching the ND Requirements for DUNE/LBNF} 
\label{cdrsec:detectors-nd-ref-fgt-req}

The scope of the this section addresses how the the reference NND
described in Section~\ref{cdrsec:detectors-nd-ref} meets the
requirements of the oscillation studies described in
Section 3.8.2 of~\cite{cdr-vol-2} %~\ref{sec:physics-lbnosc-beamnd-req} 
and the short-baseline precision measurements described in Sections 6.1.2
and 6.1.3 of the same document. %~\ref{sec-nd-sbp} and ~\ref{sec-nd-np}.  
First,
we present the oscillation related systematics; the systematics
affecting the precision measurements and new-physics search program
follows.

%\subsection{Matching ND Requirements for the Oscillation Analyses} 
\subsection{Oscillation Analyses} 
\label{cdrsec:detectors-nd-ref-fgt-req-oscl}

The Table 3.8   % Anne updated 6/20
%~\ref{tab:nuesysts} 
in Section 3.6.2 of~\cite{cdr-vol-2}
%~\ref{sec:physics-lbnosc-beamnd-req} 
presents a conservative projection of systematic errors affecting
$\nu_e$ appearance. The FGT measurement alleviates the systematic
errors as enumerated below:
\begin{itemize}
    \item {\bf Beam $\nu_e$:} The FGT will offer an event-by-event
      measurement of the beam $\nu_e$ via the identification of the
      emergent $e^-$ in STT while rejecting the $\pi^0 \rightarrow
      \gamma \rightarrow e$ background via the determination of the
      missing-$P_T$ vector to a high degree.  In a 5-year neutrino-run
      (focus-positive) the FGT will accumulate $500,000$ $\nu_e$
      with $\simeq 55\%$ efficiency and $\geq 95\%$ purity.
      The resulting $\nu_e / \nu_\mu$ ratio will be determined to
      $\leq 1\%$ precision. Furthermore, by constraining the sources
      of $\nu_e$ ($\mu^+$, $K^+$ and $K^0_L$), the FGT will predict
      the the ratio of $\nu_e$ and $\nu_\mu$ spectra at the far
      detector (FD) with respect to the near detector (ND) as a
      function of the neutrino energy, FD/ND ($E_\nu$).

    \item {\bf Beam $\bar\nu_e$:} Although the FD does not distinguish
      $\nu_e$ from $\bar\nu_e$, the FGT will accurately measure the
      beam $\bar\nu_e$ by identifying the emergent $e^+$ in the STT
      with efficiency and purity similar to those for $\nu_e$.  (We
      point out that the dominant kaonic source of $\bar\nu_e$ is
      $K^0_L$; the neutrino spectra from $K^0_L$ are different from
      those of $K^+$ at the FD.)  In a 5-year neutrino-run, the FGT
      will accumulate 40,000 $\bar\nu_e$ and $5\times 10^6$
      $\bar\nu_\mu$ samples, providing a precise FD/ND prediction for
      the antineutrino to $\simeq 1\%$.  These constraints will be
      even more valuable during the antineutrino run (focus-negative)
      where the wrong-sign backgrounds are larger.

    \item {\bf Cross Sections:} First, the FGT will 
      measure the absolute flux, via $\nu$-e scattering,
      to $\simeq 2.5\%$ precision. Second, using radiator targets,
      the FGT will measure exclusive channels, such as
      quasi-elastic, resonance, coherent-mesons and the inclusive DIS
      channel with unparalleled precision. Since Argon is the
      far detector nuclear target, a set of various nuclear targets will
      allow translation of the cross-section measurements to $\nu$-Ar
      scattering.

    \item {\bf Nuclear Effects:} The FGT will employ a suite of
      nuclear targets including Ar-gas in pressurized tubes, a thin
      solid calcium target (which has the same A=40 at Ar), a
      C-target, etc.  Specifically, the number of $\nu$-Ar interaction
      will be 10 times larger than that expected in a \ktadj{40} FD,
      without oscillations. Additionally, comparisons of calculations
      of elastic and inelastic interactions in Ar versus Ca, including
      FSI effects, indicate negligible differences between the two
      targets. Thus, the combination of the Ar and Ca targets will
      provide a strong constraint on the nuclear effects from both
      initial and final state interactions.  Finally, FGT's ability to
      isolate $\nu$ ($\bar\nu$) off free-hydrogen, via subtraction of
      hydrocarbon and carbon targets, will provide a 
      {\em model-independent} measurement of nuclear effects.

    \item {\bf Hadronization:} A notable strength of FGT is to
      identify the yield of $\pi^0$ $separately$ in neutral-current
      (NC) and charged-current (CC) interactions. (The estimated
      $\pi^0$ detection efficiency is $\simeq 50\%$.) In addition, FGT
      will determine the yields of $\pi^-$ and $\pi^+$, the dominant
      backgrounds to the $\nu_\mu$ and $\bar\nu_\mu$
      disappearance. Finally, the measurement of the composition,
      energy, and angle of the hadronic jet will provide a tight
      constraint on the overall hadronization models.

    \item {\bf Energy Scale:} The average density of about
      0.1\,g/cm$^3$ (close to that of liquid hydrogen) makes the STT
      mostly transparent to secondary particles --- the entire STT
      length is equivalent to $\sim 1.4 X_0$ including the various
      nuclear targets. As a result we will be able to measure
      accurately the 4-momenta of secondary particles as well as the
      missing-$P_T$ vector in the CC processes.  This redundant
      missing-$P_T$ vector measurement provides a most important
      constraint on the neutrino and antineutrino energy scales. The
      capabilities of this kinematic analysis have been demonstrated
      by the NOMAD experiment (with similar density and B field).  In
      addition, measurements of exclusive topologies like
      quasi-elastic, resonance and coherent meson production offer
      additional constraints on the nuclear effects affecting the
      neutrino energy scale (see Section 6.1.2 of~\cite{cdr-vol-2}).  %~\ref{sec-nd-sbp}).  
The requirements on the muon energy scale
      ($< 0.2\%$) and total hadron energy scale ($< 0.5\%$)
      uncertainties have been already achieved by the NOMAD
      experiment. The muon energy scale is calibrated with the mass
      constraint from the $K^0$ reconstructed in STT and was
      statistics limited in NOMAD. The total hadron energy scale is
      calibrated using the $y_{Bj}$ distribution in different energy
      bins. Compared to NOMAD, FGT will have $\times 100$ higher
      statistics and $\times 10$ higher granularity.
\end{itemize}

In summary the FGT will accurately quantify all four neutrino species and
predict the ratio FD/ND for them.  It will measure the 4-momenta of
the outgoing hadrons composing the hadronic jets in a variety of
nuclear targets, in essence proving a data-driven \textit{event generator}
which can be applied to the FD.

Since the FGT is based upon a different technology than the FD, it cannot
account for effects of LAr reconstruction inefficiencies in the
FD. The corresponding cancellation could be achieved only with an
identical ND, which to some extent is an ill-defined concept due to a
number of factors including size, beam profile and composition, rates, 
etc. However, given the detailed program to calibrate
LAr detectors in test beams and  multiple neutrino
experiments employing LAr detectors, by the time DUNE becomes
operational the reconstruction of particles in LAr will likely be well
understood. Finally, Section~\ref{sec:detectors-nd-alt} outlines
the enhancement of the ND complex via the placement of complementary
LAr detector(s) upstream of the FGT.

%\subsection{Matching ND Requirements for the Short Baseline Precision Measurements and Searches} 
\subsection{Short Baseline Precision Measurements and Searches} 
\label{cdrsec:detectors-nd-ref-fgt-req-sbp}

Sections 6.1.2
%~\ref{sec-nd-sbp} 
and 6.1.3 of~\cite{cdr-vol-2}
 %~\ref{sec-nd-np} 
summarize a rich physics program at the near site providing a
generational advance in precision measurements and sensitive searches.
This short-baseline physics program and the long-baseline oscillation
analyses share similar detector requirements and offer a deep synergy
and mutual feedback.  The reference FGT meets the requirements of the
short-baseline studies as briefly outlined below:
\begin{itemize}
    \item {\bf Resolution:} The FGT is designed to have an order of
      magnitude higher granularity than NOMAD, the most precise, high
      statistics neutrino experiment. The corresponding improvements
      include better tracking, electron/positron ID through TR, $dE/dx$
      measurement providing hadron-ID, 4$\pi$ calorimetry, 4$\pi$ muon
      coverage and a larger transverse area for event containment.
    \item {\bf Statistics:} The \MWadj{1.2} neutrino source at LBNF will
      offer a factor of 100 enhancement in statistics compared to
      NOMAD. The program of measuring $\nu$ and $\bar\nu$ interactions
      in a set of nuclear targets, including Ar and H, will enhance
      the physics potential of precision measurements and searches.
\end{itemize}
Sensitivity studies to the salient precision measurements and searches
can be found in~\cite{Adams:2013qkq, DPR}.

\subsection{Future Tasks to Quantify the Systematic Errors}
\label{cdrsec:detectors-nd-ref-fgt-req-future} 

%We need to undertake three outstanding tasks to quantify the
%systematic errors in oscillation studies and precision measurement
%program:
To quantify the systematic errors in oscillation studies and precision measurement
program, three tasks are still outstanding:\begin{itemize}
    \item {\bf Geant4 Simulation:} A Geant4 simulation of the FGT is
      needed to confirm and correct the projected systematic errors
      and the salient sensitivity studies.
     \item {\bf Event Reconstruction:} A program to reconstruct tracks
       in STT and to match the information from different
       subdetectors is needed to identify secondary particles.
     \item {\bf Translating ND-Measurements to FD:} A concerted effort
       needs to be mounted to transfer the precision measurements in
       ND to the far detector.
\end{itemize}    
The DUNE collaboration plans to pursue these issues with high priority
in the coming years before CD-2.

%% file: volume-detectors/nd-ref-sections/nd-ref-alts.tex
%%%%%%%%%%%%%%%%%%%%%%%%%%%%%%%% 
\section{Addition of a Liquid Argon Detector to the NND}
\label{sec:detectors-nd-alt}

The reference FGT ND design concept is not
identical to the far detector and it is not possible
to \textit{cancel} the event reconstruction errors exactly in a near-to-far
ratio as was done in MINOS.  The extent to which this lack of
cancellation will limit the ultimate precision of the experiment has
yet to be fully explored.  However, at the international Near Neutrino
Detector workshop held at Fermilab (July 2014) it was accepted that
the FGT offers a sound basis for moving forward and that a LArTPC
or a high-pressure gaseous-argon TPC placed upstream of the FGT would
enhance the near detector capability.

%We note that for LBNF/DUNE, an `identical near detector' concept
%is insufficient to fulfill the requirements for ND
%(see Chapter 6 of \volphys). 
The principal impediment to an
identical ND -- a LArTPC ND -- is the event-pileup problem
due to the high intensity of LBNF. Nevertheless, during the operation
of LBNF/DUNE there will be periods when the accelerator is running at
low intensity, for example during the initial ramp-up and during the
periodic shut-downs and accelerator upgrades. A $\sim$100-t LArTPC
stationed upstream of the FGT (the reference ND) would be able to
accumulate tens of thousands of neutrino interactions during the
low-intensity runs over the lifetime of the experiment.  The FGT will
act as a spectrometer for the emerging muons. In conjunction with the
measurements using nuclear targets in the FGT, including Ar gas, such a
LAr-ND would provide a means to accurately validate \textit{in situ} the
FGT predictions for a LAr detector before the extrapolation to the FD,
thus providing a valuable redundancy check.

Furthermore, for special neutrino interaction topologies, such as
neutrino-electron scattering, where there is a single electron or muon
in the final state, the combined LAr-ND and FGT configuration could
provide unique precision measurements.

Conceptual designs for a standalone LArTPC near detector and a
gaseous argon TPC near detector are under consideration and
could serve as starting points for the design of this addition to the
FGT.  Significant simulation and engineering studies are required to
understand whether a liquid or gas argon TPC is optimal for minimizing
systematic errors in the long-baseline measurements and to integrate
the additional detector system with the FGT design to make a coherent
near detector system.

%In the coming year, the DUNE near detector working group would explore the physics implication and cost estimates of such alternatives. 

%% file: volume-detectors/nd-ref-sections/nd-ref-blm.tex
%%%%%%%%%%%%%%%%%%%%%%%%%%%%%%%% 
\section{Beamline Measurements} 
\label{sec:detectors-nd-ref-blm}

This section outlines the DUNE strategy for measurements of secondary
beam particles downstream of  the beam absorber. 
These measurements are designed to provide constraints 
on the neutrino flux at the near and far
detectors, and monitor the pulse-to-pulse variation
 for beam diagnostic purposes. 
% A description of equipment for monitoring the proton beam's interaction with the proton target
% can be found in Volume 2: The Beamline at the Near Site. \fixme{The Physics Volume, Volume 2 does not have a ``Beamline at the near site chapter''}
The measurements and apparatus described here fall into
the category of equipment designed specifically for DUNE to
detect muons exiting the decay tunnel. 

\subsection{Design Considerations}
\label{subsec:detectors-nd-blm-design}

\begin{sloppypar}
The requirements for the beamline measurements,
as discussed in the NDC requirements documentation\cite{lbnfdune-cdr-req}, 
are related to how well the neutrino flux must be known.
%Given that DUNE does not have the luxury to construct identical 
%Near and Far Detectors, a near-far comparison is more complicated than it was in
%the MINOS experiment~\cite{gnumi-validation}, for example.   
While external hadron-production measurements can place 
constraints on the pion and kaon production in the target, they do not 
provide any constraint on the simulation of other key features, such 
as the horn focusing, secondary interactions, and the 
pion scattering and absorption in the air-filled decay volume. 

In addition to the external measurements, covered in
Section~\ref{sec:detectors-nd-ref-hadron}, that check the
simulation of the thick target, horn material, decay tunnel and
absorber, it is desirable to constrain the flux by making independent
measurements at the 4--5\% level of the muons that penetrate the
absorber. It would not be practical to do this for all penetrating
muons, but sufficient measurements at a few positions can be done in a
cost-effective way.
\end{sloppypar}
%
%\subsubsection{Muon Measurements}
%\label{subsubsec:detectors-nd-blm-muon-meas}

The primary physics goal of DUNE is to measure the transmutation
of $\nu_\mu$ to $\nu_e$ over the 1300\,km between Fermilab
and SURF. It is essential for DUNE to cross-check the estimate of 
beam $\nu_e$ using several methods.
There are two dominant sources of beam $\nu_e$: muon decays and kaon decays. 
The muon systems are designed to directly measure the 
muons that penetrate absorber with an energy 
threshold as low as possible, i.e., directly measure those muons 
whose decays are a major source of beam $\nu_e$. 
A measurement of the spectrum of those muons will translate
 directly into constraints on the spectrum of beam $\nu_e$.
That constraint has the advantage of being independent of poorly understood 
neutrino-nucleus cross sections.

Because muons and neutrinos come from the same parent pion and kaon
decays, a measurement of the absolute muon flux, in conjunction with the energy spectrum
seen in the muon monitors, can constrain the absolute neutrino flux.  
The goal for the DUNE muon monitors is to determine the absolute muon flux
to an accuracy of 5\% above a muon energy of 6\,GeV (which corresponds to
a neutrino energy of 1.6\,GeV) in the central part of the absorber.

It is essential to monitor the stability of the beam direction over
time. For example, above 6\,GeV, the ratio of the Far Detector flux over 
the Near Detector flux changes by 2\%.  
To keep the change in the neutrino beam less than 1\% in all energy bins,
the beam direction must be known to a precision of approximately 0.2\,mrad. 
Because the muon monitors will be located approximately 275\,m
from the beam target, this requires a measurement of the muons to an
accuracy of approximately 5\,cm.

%
%
%%%%%%%%%%%%%%%%%%%%%%%%%%%%%%%%%%%%%%%%%%%%%%%%%%%%%%%
\subsection{Muon-Measurement Facilities}
\label{subsec:detectors-nd-blm-muon-measurement-facilities}

The muon measurements are carried out in the region immediately
downstream of the hadron absorber at the end of the decay tunnel, below
the Absorber Service Building (LBNF 30).  A view of the absorber area
and the muon alcove is shown in Figure~\ref{fig:AbsorberElevationView}.  
\begin{cdrfigure}[Absorber conceptual design, elevation view]{AbsorberElevationView}
{Absorber conceptual design. The figure shows the elevation view of the 
absorber at the end of the decay tunnel. The beam axis is shown by
the blue line. The absorber is constructed of several different 
materials as shown: aluminum core in blue and grey, concrete 
(grey and tan), and steel (in brown and green).}
\includegraphics[width=4in,angle=0]{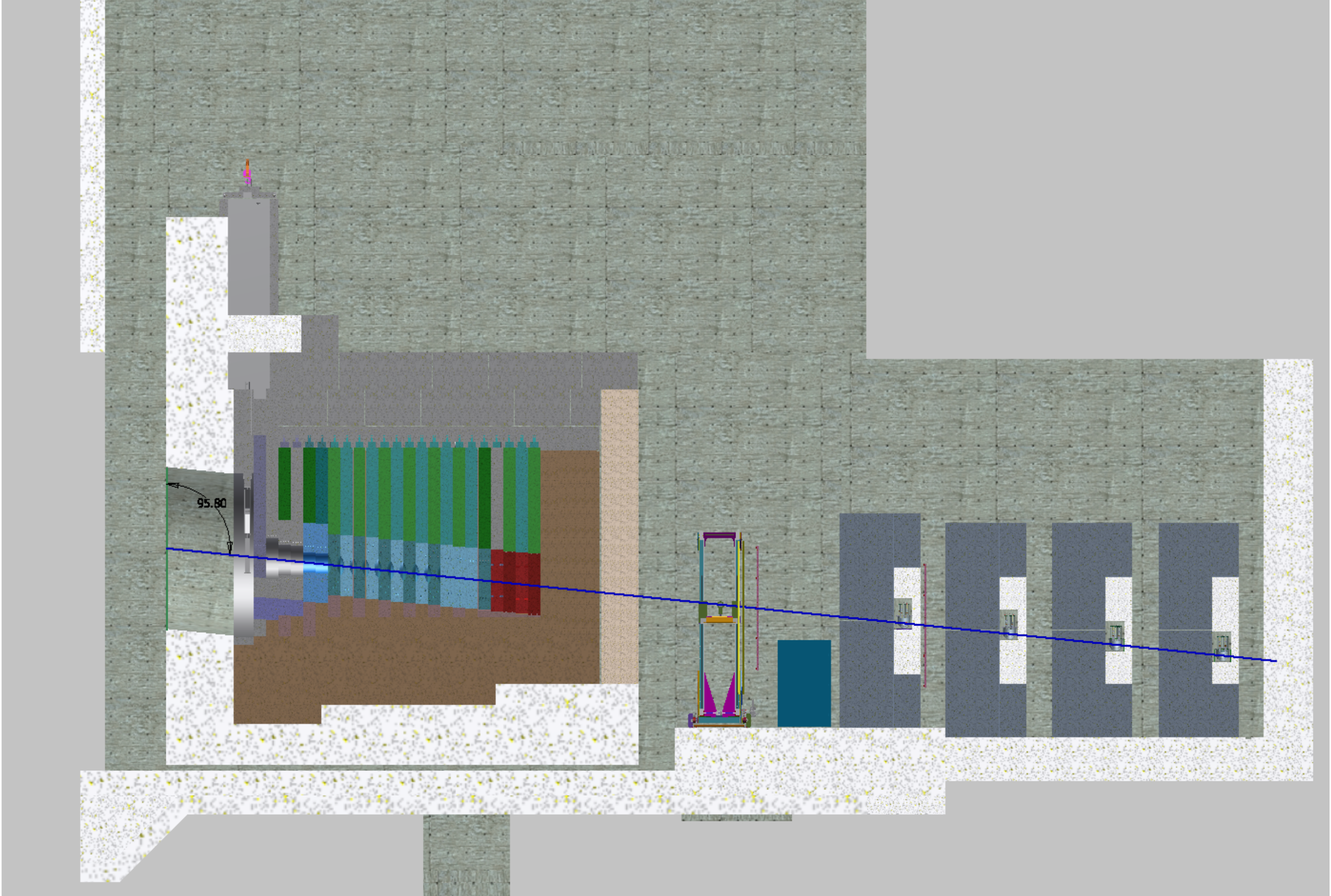}
\end{cdrfigure}
The axis of the decay pipe
cuts across the muon alcove at an angle and the size of the alcove is
largely determined by the requirement that it contain the shadow of
the four-meter-diameter decay pipe, projected through the alcove.
%, as shown in the elevation view of Figure~\ref{fig:AbsorberElevationView}.

Figure~\ref{fig:AbsorberElevationView} shows the downstream side of
the absorber and a conceptual layout of the muon systems described in
various sections of this chapter.  The absorber itself is encased in
concrete. The first set of muon-measurement devices, from left to
right, is a variable-pressure gas Cherenkov counters, which mounted on
a movable stand in order to scan across the rear surface of the
absorber.  Following that is an array of diamond ionization detectors
and finally a set of stopped-muon counters which are interspersed
between walls of steel ``blue blocks''.  The blue blocks are there to
provide several depths at which to monitor the stopped muons as they
range out in the material. A second array of ionization devices will
also be placed farther downstream within the blue blocks.

Figure~\ref{fig:AbsorberThickness} shows the energy lost by a
horizontal muon as it traverses the absorber, as a function of the
distance from the beam axis along a 45$^\circ$ line perpendicular to the beam axis. 
\begin{cdrfigure}[Energy loss in absorber]{AbsorberThickness}
{The energy loss a muon, parallel to the beam axis, experiences as it traverses 
the material in the absorber. The muon's energy loss is plotted versus the distance 
from the beam axis, along a 45$^\circ$ line perpendicular to the beam axis. Muons 
suffer between 4.7 and 9.3\,GeV of energy loss depending upon where they cross the absorber.}
\includegraphics[width=4in,angle=0]{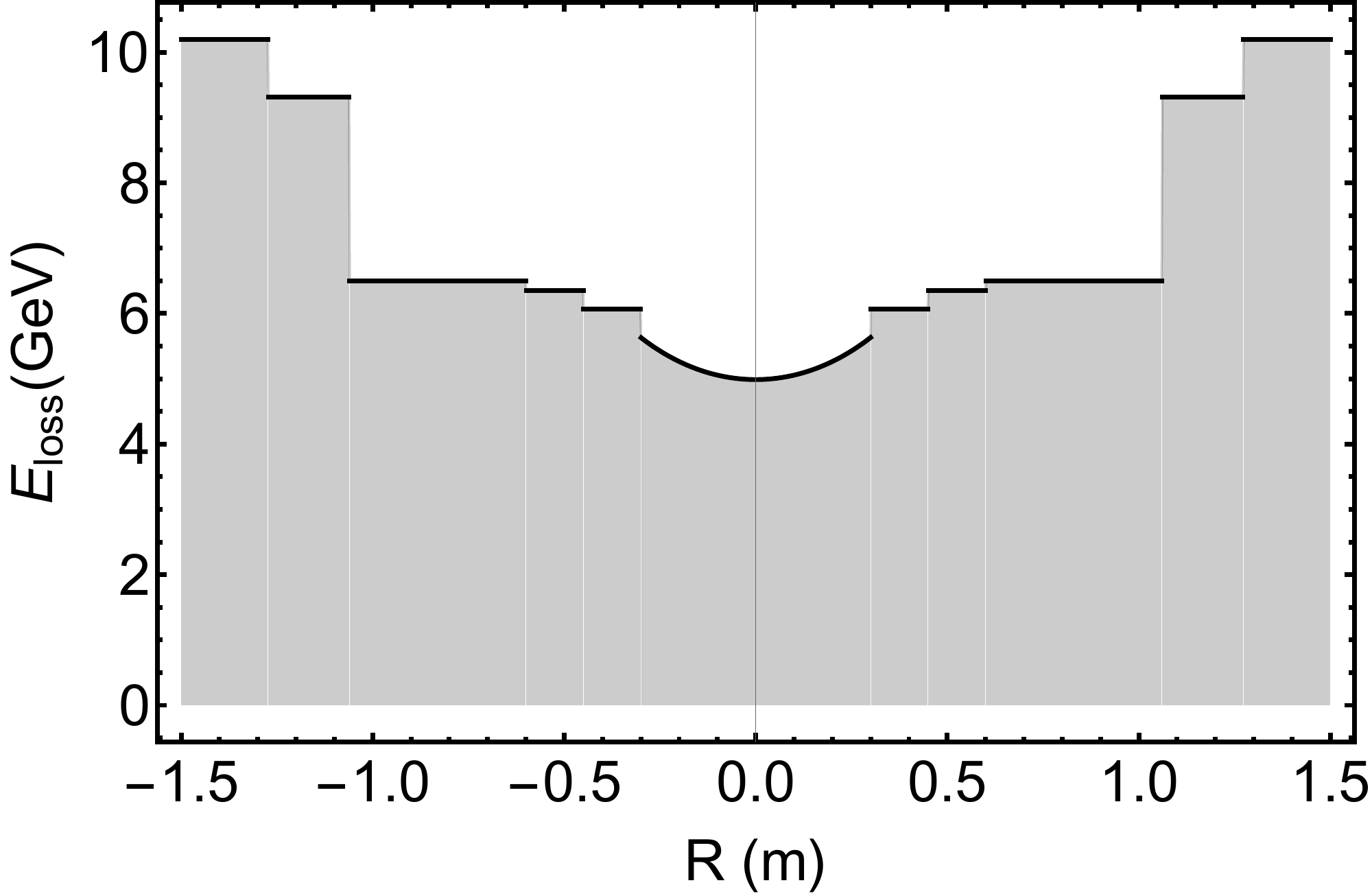}
\end{cdrfigure}
In the central region, roughly to a distance of 105\,cm, the muons lose
between 5.0 to 6.4\,GeV, so that the lowest-energy muons leaving the
absorber at that point correspond to neutrino energies of $\sim$1.5
to 2.0\,GeV. At a radius of roughly 105\,cm, the full thickness of steel
causes the muons 10\,GeV or more, corresponding to neutrino energies of
$\sim 2.6$\,GeV. From the perspective of the muon systems it will be
desirable to lower these thresholds if possible.

%
%  REmoved the individual muon component sections per Bill 4/23/15 (AH)
%%%%%%%%%%%%%%%%%%%%%%%%%%%%%%%%%%%%%%%%%%%%%%%%%%%%%%%

\subsection{Installation and Operation}

The muon system detectors, already fully calibrated in the NuMI beam,
will be installed as soon as the absorber hall (LBNF-30) is ready.
Because the system will be located in a radiation-controlled
environment that will not be accessible during beam operation, it is
essential that the electronics and gas handling system be both robust
and remotely operable.  The prototype system in use at the NuMI area
can be relocated for that purpose, or if desired a new system may be
constructed.  Periodic access will be required to the utilities area
to replace gas bottles.

The blue blocks associated with the muon systems will be installed
first.  The positioned stand for the Cherenkov detector system will be
installed next.  and then the various detectors will be installed.
The stopped-muon counters will be placed into the spaces between the
blue-block walls on support frames.  There will be access to the areas
between the shield blocks from the side, and the stopped-muon counters
will be designed so that they can be wheeled in from the side.  If
needed, they could then be moved around to measure the stopped-muon
rates across the muon beam.

The muon systems will be operated continuously in order to insure a
stable, high-quality neutrino beam.  The muon-monitor-system data will
be displayed in the control room on a spill-by-spill basis to monitor
the beam stability. Because the system will be located in a
radiation-controlled environment that will not be accessible during
the beam operation, it is essential that the electronics be designed
for remote operation.

\section{Hadron Production Measurements}
\label{sec:detectors-nd-ref-hadron}

\subsection{Introduction}

The technical components that would be needed to implement the
strategies described in this section are outside the scope of the DUNE
NDS conceptual design. The following information is included in this document
because it complements the conceptual design and expands the NDS
capabilities to more closely meet the mission need without increasing
the project cost.

\subsection{External Hadron-Production Measurements}
\label{sec:detectors-nd-external-hadron}

Uncertainties on hadron production will translate into uncertainties
in the neutrino fluxes in the DUNE far detector, since the neutrinos
are produced by hadrons decaying in the decay pipe. Precise
calculations of neutrino fluxes in high-energy accelerator beams are
limited at present by our knowledge of hadron production
cross-sections in hadron-nucleus collisions.  The modeling of
strong-interaction cascades and hadronic yields from ``thick'' targets
(up to a couple of interaction lengths) relies on detailed knowledge
of underlying physics and cross sections, which must be provided as a
starting point to simulations. The resulting prediction of the flux of
neutrinos, produced from decays of pions, kaons, and muons emerging
from a hadronic shower and beam line re-interactions, is an essential
part of simulations of most neutrino experiments.

Two-detector neutrino oscillation experiments predict the neutrino
flux at the far detector by using neutrino fluxes ``calibrated'' (or
appropriately scaled) by event energy spectra measured in the near
detector. However, these experiments rely on beam simulations since
the decay pipe (where most beam neutrinos are created) provides
different angular acceptance for the two detectors. In addition,
experiments using near and far detectors based on different detection
technologies further complicate the extrapolation. This chapter
outlines the DUNE strategy for augmenting the capabilities of the BLM
with external measurements of secondary-beam particles.
%\fixme{This was originally written when we just had 
%the BLM and no NND; no change needed here? Second point: This sentence should be in the intro paragraph of this chapter.}

\subsection{Background}

A complete knowledge of the momenta and decay points of the kaons, pions and
muons would be sufficient to  predict the un-oscillated flux of neutrinos
at the near and far detector locations. This would require knowledge of

\begin{itemize}
\item the phase-space distribution of the initial proton beam;
\item details of all materials present in the target, horn and decay pipe areas;
\item  the electromagnetic focusing characteristics of the magnetic horn;
\item the detailed development of the hadron cascade, spawned by the
initial proton, that passes through the target/horn/decay pipe; and
\item the meson-to-neutrino decay rates.
\end{itemize}

With careful engineering design and control of the materials in the target
area, these items can be simulated accurately, except for hadronic cascades in
the target, horn and decay pipe. The simulation of the hadronic cascade requires
accurate knowledge of the hadron scattering cross sections, for which there are no
first-principle calculations. These cross sections must therefore rely on models, which
in turn require hadron-production measurements that span particle type, particle
energy and the various materials found in the target, horn and decay pipe.

At present, a sufficient body of hadron-production
measurements does not exist to achieve DUNE's desired accuracy of
4--5\%, as determined by the irreducible error on the statistical
uncertainty for the appearance-measurement background, although this
is expected to improve over time. A program of hadroproduction
measurements has been approved as US-NA61 to run at CERN.

%% file: volume-detectors/nd-ref-sections/nd-ref-daq.tex
%%%%%%%%%%%%%%%%%%%%%%%%%%%%%%%% 
\section{The Data Acquisition System (DAQ) and Computing}
\label{cdrsec:detectors-nd-ref-daq-comp}

The scope of the Near Detector System DAQ (NDS-DAQ) and computing includes the
design, procurement, fabrication, testing, delivery and installation
of all the NDS-DAQ subsystems:
\begin{itemize}
\item NDS Master DAQ (NND-MDAQ),
\item Near Neutrino Detector DAQ (NND-DAQ),
\item Beamline Measurements DAQ (BLM-DAQ), and
\item NDS Computing.
\end{itemize}

\subsection{NDS DAQ}
\label{cdrsec:nd-gdaq-intro}

The Near Detector System (NDS) Data Acquisition system (NDS-DAQ)
collects raw data from each NDS individual DAQ, issues
triggers, adds precision timing data from a global positioning system
(GPS), and builds events.  The NDS-DAQ is made up of three parts, as
shown in the block diagram of Figure~\ref{fig:DAQ_Block}, a master DAQ
and one each for the near neutrino detector (NND, which is the FGT)
and the BLM systems. The names for these are, respectively, NDS-MDAQ,
NND-DAQ and BLM-DAQ.
\begin{cdrfigure}[Near Detector Systems DAQ block diagram]{DAQ_Block}
{Near Detector System DAQ block diagram: The NDS-DAQ consists 
of the NDS Master DAQ (green blocks), the Beamline Measurement DAQ (yellow summary 
block) and the Near Neutrino Detectors DAQ (orange summary block).  The 
NDS-DAQ connects to other portions of DUNE and LBNF, shown here in other colors (blue, 
light red, tan).}
\includegraphics[width=6in,angle=0]{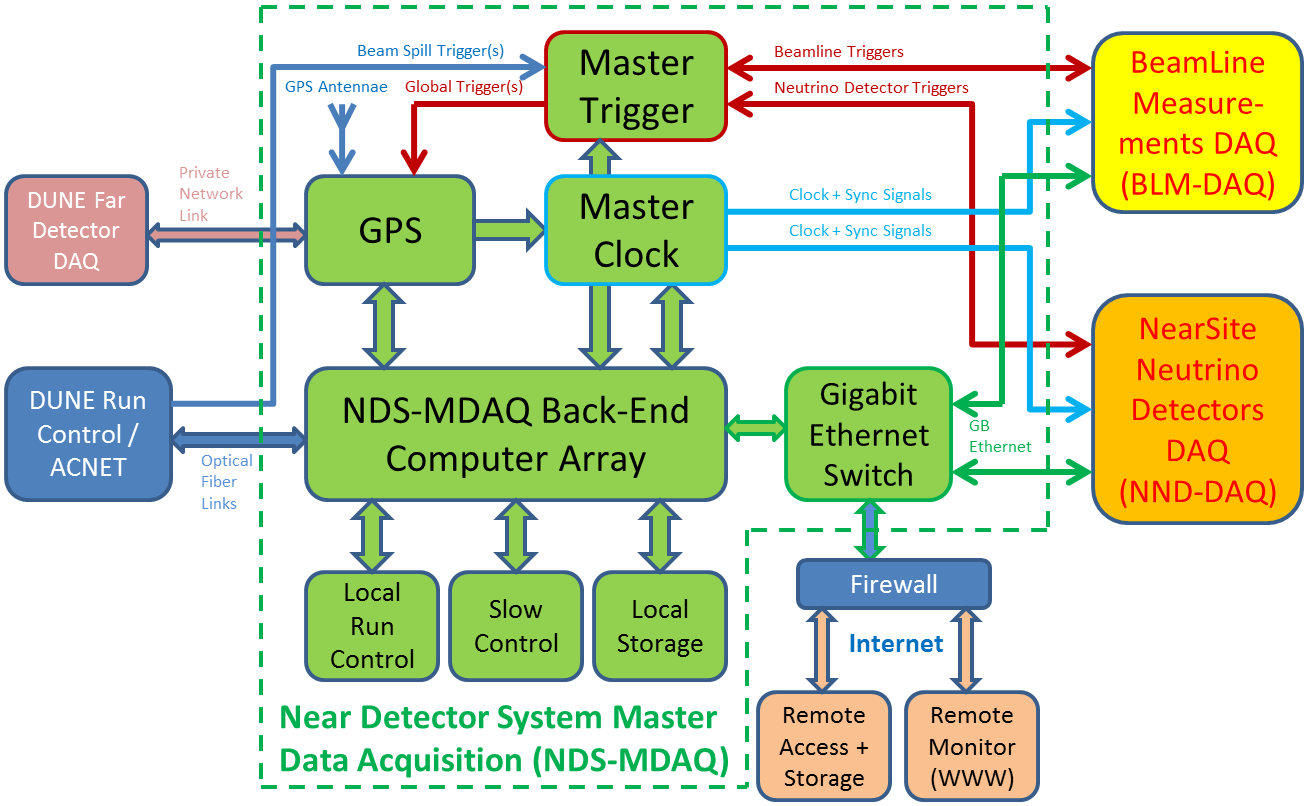}
\end{cdrfigure}

\subsubsection{NDS Master DAQ} 
\label{cdrsec:nd-master-daq}

The NDS Master DAQ (NDS-MDAQ) is designed to provide a high-level user
interface for local run control and data taking, as well as for secure
remote control and monitoring.  It will serve as the primary interface
to the NND-DAQ and BLM-DAQ and will include 
\begin{itemize}
\item the slow-control system,
\item online data and DAQ performance monitoring,  
\item raw data collection,
\item building of events, and
\item data storage.   
\end{itemize}
The NDS-MDAQ includes hardware two-way triggering for both the NND-DAQ
and BLM-DAQ, and GPS hardware for precision time-stamping and global
clock synchronization.  The design is currently based on a channel
count estimate of approximately 433,000 from the near neutrino
detector, plus $<1,000$ from the beamline detectors.  Custom
electronic components for the NDS-DAQ are based on existing custom
designs from other experiments, e.g., T2K and ATLAS, and implement
commercial components for the trigger modules, clock and timing
synchronization, GPS and environmental monitoring.

\subsubsection{Near Neutrino Detector DAQ (NND-DAQ)} 
\label{cdrsec:nd:nnd:daq}

The Near Neutrino Detector Data Acquisition system (NND-DAQ) collects
raw data from the DAQ in each NND subdetector and connects to the NDS
Master DAQ via Gigabit Ethernet. A block diagram of the NND-DAQ is
shown in Figure~\ref{fig:DAQ_NND}. 
\begin{cdrfigure}[Near Neutrino Detector DAQ block diagram]
{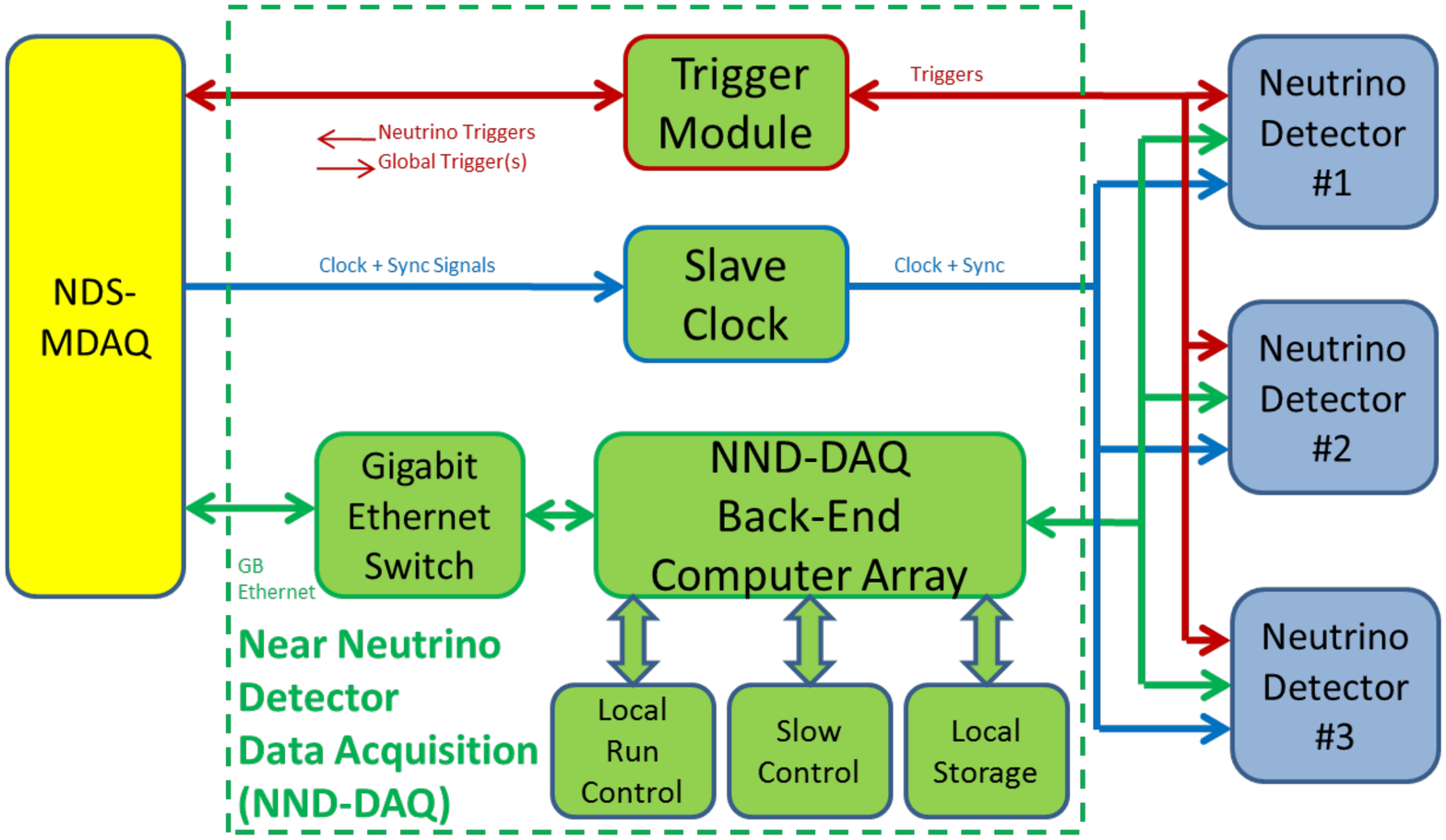}{A block diagram of the Near Neutrino Detector DAQ (NND-DAQ).}
\includegraphics[width=5in,angle=0]{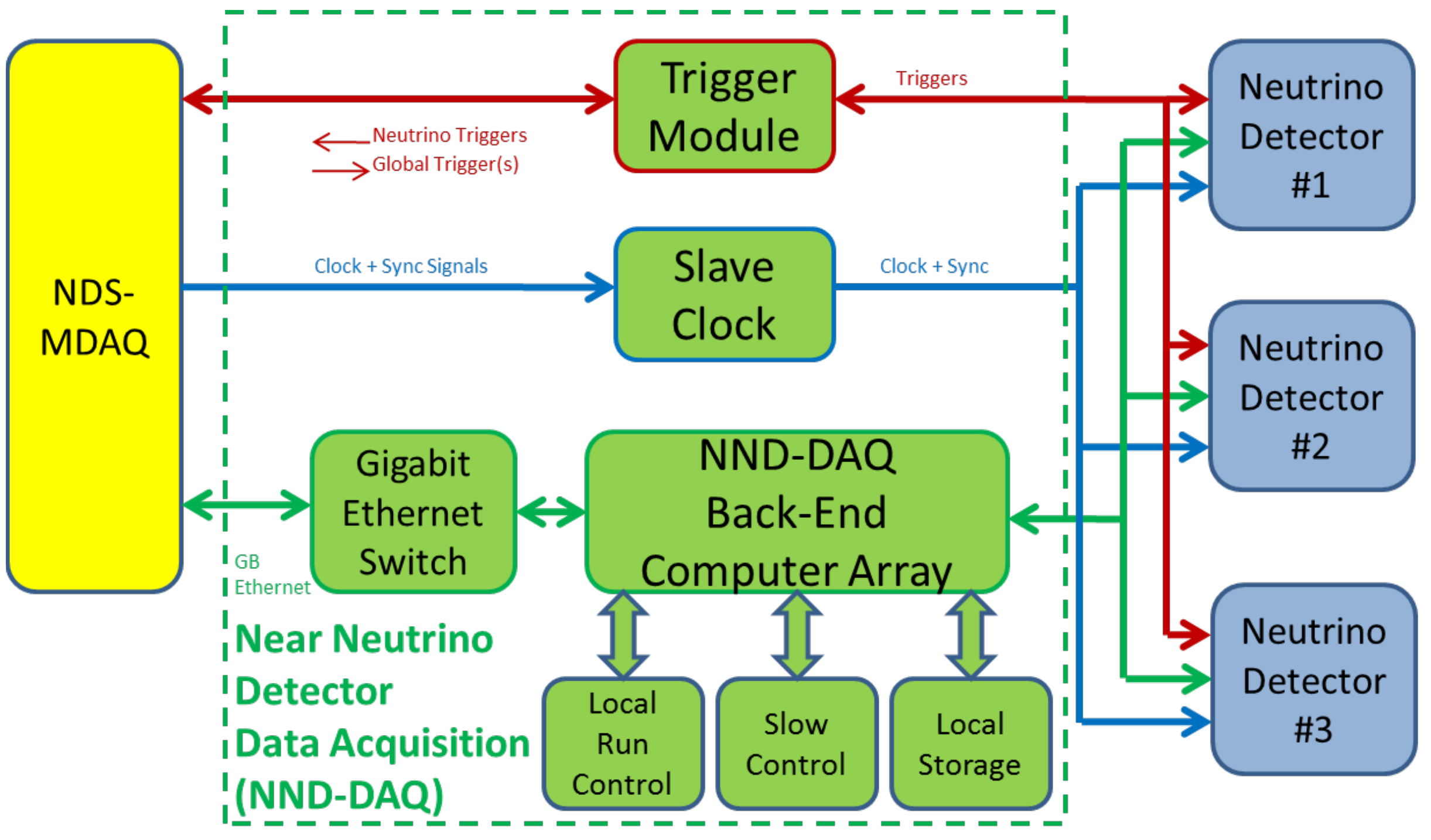}
\end{cdrfigure}
The NND-DAQ will mainly consist of a scalable back-end computer array,
interconnected to the individual subdetector DAQs via Gigabit
Ethernet and specialized electronics modules for trigger processing
and clock synchronization. It interfaces to the NDS-MDAQ for run
control and data collection. The NND-DAQ will also have its own local
run-control setup, consisting of a number of desktop workstations to
allow independent local runs that include NND subdetectors only; this
is useful during detector commissioning, calibration runs, stand-alone
cosmic runs, or other runs where the beam is stopped or not needed.

The quantity of computers required for the NND-DAQ back-end system is
highly dependent on the number of channels and expected data rates of
the individual neutrino detectors.  One back-end computer should be
able to handle approximately 3,000 channels for sustainable and
continuous runs. Assuming a total of 433,000 channels for all NND
subdetectors combined, about 150 back-end computers would be needed.

Trigger signals from each subdetector will be collected and
pre-processed by a trigger electronics module, similar in design to
the NDS trigger or master-clock modules of the NDS-MDAQ
design. Depending on the run mode, this module could feed local
trigger decisions to the detector DAQs for data collection, or it
could forward NDS triggers from the NDS-MDAQ or higher levels to the
NND subdetector DAQs.  A slave-clock electronics module, similar to
the master-clock module in the NDS-MDAQ, distributes clock- and
time-synchronization signals from the NDS-MDAQ to all NND
subdetectors.

\subsubsection{Beamline Measurements DAQ (BLM-DAQ)}

The BLM-DAQ will mainly consist of a scalable back-end computer array,
inter-connected to the individual beamline measurement detector DAQs
via Gigabit Ethernet and specialized electronics modules for trigger
processing and clock synchronization. It interfaces to the NDS-MDAQ
for run control and data collection. It will also have its own local
run-control setup, consisting of a number of desktop workstations to
allow independent local runs that include beamline measurement
detectors only; this is useful during detector commissioning,
calibration runs, stand-alone cosmic runs or other runs where the beam
is stopped or not needed.

\subsection{NDS Computing}
\label{cdrsec:nd-gdaq-global-computing}

The computing system encompasses two major activities: online
computing with required slow-control systems, and offline computing
for data analysis and event simulation.  The computing components are
based on currently available commercial computing and gigabit
networking technology, which is likely to improve over the next years
without driving costs up for the final design.

%% file: volume-detectors/chapter-sc.tex
\chapter{Software and Computing}
\label{ch:detectors-sc}

\section{Overview}

The three primary
components of the DUNE Software and Computing effort include the computing hardware, software
infrastructure and reconstruction. This chapter summarizes the much more detailed information presented in \anxrates~\cite{cdr-annex-rates}, which describes the data sources and rates for the far and near detector designs, and provides a reference to the parameters and assumptions used in estimating these characteristics, and \anxreco~\cite{cdr-annex-reco}. 

Simulated data samples provide the basis for detailed studies of detector performance, inform detector design choices, and enable the development of automated event reconstruction. Detailed Monte Carlo predictions of expected data distributions enable extraction of physics results from the DUNE experiment. Several important sources of systematic uncertainty come from detector modeling, and varying the assumptions incorporated in the simulations provides the mechanism by which these systematic uncertainties can be estimated. Simulations are also needed to extrapolate from auxiliary data samples (such as test-beam or in situ measurements, using events not passing signal-event selection requirements) to the signal selection samples. 

The Software and Computing effort is not part of the DUNE project; it 
is supported by non-project funds. It is coordinated by the
DUNE collaboration and across the LBNF/DUNE project. 

\section{Computing Infrastructure}
\label{sec:detectors-sc-infrastructure}

%\subsection{Overview}
There are many factors that influence the
data to be collected and processed in DUNE, e.g., rates, volume, etc.  Reference~\cite{cdr-annex-rates} 
contains information on both the reference and alternative
designs for the far detector.

\subsection{Raw Data Rates}
\label{sec:detectors-sc-infrastructure-data-rates}

\subsubsection{Types of data (using a supernova burst example)}
DUNE is a multipurpose apparatus and will pursue a variety of physics
goals.  %This will be reflected in different
%characteristics of data streams processed and collected in real time
%as well as offline and different strategies and algorithms for
%handling these streams.
This will result in data streams of differing
characteristics that will need to be processed and collected both in real time
and offline, as well as different strategies and algorithms for
handling these streams.

%As one example, consider the difference between neutrino oscillation
%physics with beam neutrinos on the one hand and the search for supernova
%neutrino bursts on the other.  Signals produced by ``beam events''
%will be characterized by total energy in the GeV range, so that
%various aspects of handling the signal and the data (e.g. thresholds
%for zero suppression etc) can be optimized for minimum-ionizing
%particles.

As one example, consider the difference between neutrino oscillation
physics with beam neutrinos and the search for supernova
neutrino bursts.  Signals produced by beam events
will be characterized by total energy in the GeV range, and therefore
various aspects of handling the signal and the data (e.g., thresholds
for zero suppression, etc.) can be optimized for minimum-ionizing
particles.
Since the energy scale of signals produced by supernova
neutrino burst interactions is in the range of tens of MeV, 
it is expected that lower thresholds will be needed
while processing these data in real time. This will result in
considerable additional contribution from radiological backgrounds, which 
leads to an expected data rate that is 
%The data rate that needs to be handled in the process of
%supernova neutrino bursts search can therefore be expected to be 
quite significant.

Another differentiating feature of supernova neutrino bursts is that
\textit{multiple neutrinos are expected to arrive and interact in the
  detector volume within seconds of each other}, as opposed to a
single vertex produced by a beam neutrino (or any other localized
interaction and/or decay). This presents an opportunity to apply the
DAQ architecture presented in Section~\ref{sec:detectors-fd-ref-daq} to make
these data ``self-triggering,'' i.e.,  to use the buffer memory in the
LArTPC detector readout to detect a corresponding signature in the
data stream and trigger recording of the potential supernova neutrino
burst event.

The characteristic time scale for such a supernova neutrino burst data
capture will be $\sim$\SI{10}{\second}.  Given the large amount of
data arriving within this time period (see~\cite{cdr-annex-rates}) and practical
limits on the bandwidth of the connection between the RCE data
processors and front-end computers in the DAQ, 
the buffer in the data processor will not have
sufficient capacity (due to design and cost considerations). It can
effectively buffer about \SI{0.4}{\second} of streaming data which is
likely enough for the trigger decision but not for the complete
supernova event capture.  Local storage attached to the
data processors will be necessary to record the data at full-stream
(no zero-suppression).  Preliminary estimates indicate that a
storage device such as a SSD (one or two per board) will have speed
sufficient for this purpose.  With the trigger properly tuned, the number
of times data are written to the SSD can be kept sufficiently low so
as to ensure their longevity.  Once captured in this manner, the data
can then be transmitted to the rest of the farm within the available
bandwidth.

The core elements of the DAQ system now exist as prototypes.  The
system as a whole, with the capabilities described in this section, is in the
conceptual design stage. % and information will be added to DUNE planning documents as it is developed.

%This section focuses on data streams present in oscillation physics studies with beam
%neutrinos, since these data will constitute the bulk of what's committed
%to mass storage, transmitted over networks, processed offline and in general have most significant
%infrastructure and cost implications.
%Issues and parameters related to other classes of data
%are covered in ``\anxrates'', and also in the DAQ and other sections.

\subsubsection{Assumptions}
\label{sec:detectors-sc-infrastructure-assumptions}
According to the present baseline design, the Far Detector will
consist of four identical modules of \tpcmodulemass each.  (For the
purposes of %estimating data characteristics in 
this document, %the issue of 
the effect of possible variations in the design of these modules on the 
estimation of data characteristics will not be
addressed.) A few basic assumptions have been made.
\begin{itemize}
\item Estimates correspond to the ``full detector,''
  i.e., they are effectively normalized to \dunedetectormass.
\item The accelerator spill cycle is \beamspillcycle with beam expected
  for \beamrunfraction of each calendar year.
\item Zero-suppression (ZS) thresholds will be set at levels that preserve
  signals from minimum-ionizing particles while effectively removing
  data due to electronics noise.
\item The DAQ will be able to trigger based on spill times and will be
  able to reject isolated $^{39}$Ar decays on at least a per-APA
  basis. (For DAQ details see Section~\ref{sec:detectors-fd-ref-daq}.)
\end{itemize}

\subsubsection{Far Detector LArTPC Parameters}
The basic parameters presented below are also listed in~\cite{cdr-annex-rates}.
\begin{itemize}
\item TPC channel count: \dunenumberchannels (i.e., four times the
  \daqchannelspermodule channel count for each \tpcmodulemass module)
\item Maximum drift Time: \tpcdrifttime
\item Number of drift time windows in one DAQ readout cycle: \daqdriftsperreadout
\item ADC clock frequency: $\sim$\daqsamplerate
\item ADC resolution (bits): 12
\end{itemize}

In addition to these parameters, there are other factors
affecting data rates and volumes, such as implementation of ZS in the DAQ RCE processors,
contribution from radiological and cosmological backgrounds, and DAQ
trigger configuration (e.g., the case of low-energy events). 

Non-ZS maximum event size (corresponding to a snapshot of the complete
TPC) can be calculated as a product of %the following numbers
\begin{itemize}
\item channel count,
\item number of ADC samples per total drift (collection) time,
\item drift time windows in one DAQ cycle, and
\item ADC resolution.
\end{itemize}

This results in a total of \dunefsreadoutsize %worth 
of TPC data in one readout (drift time). 

Zero suppression greatly reduces the event size.  An overly
conservative estimate (leaning to the higher end of the range of
values) based on a LArSoft Monte Carlo simulation of GeV-scale events
suggests a zero-suppressed and uncompressed event size of
$\sim$\beameventsize.  After compression this event size is expected
to reduce to $\sim$\beameventsizecompressed.  This particular simulation
employed a less-than-optimal schema for packing data and it is
expected that these sizes can be further reduced.

Some of the driving ZS and full-stream (FS) annual
data volumes are summarized in Table~\ref{tab:sc-zs-summary}.  Regarding the 
numbers in the row characterizing $^{39}$Ar, once DAQ-level rejection of isolated $^{39}$Ar decay 
events is invoked, a residual amount of data is accepted when the decay is accidentally
coincident with beam-$\nu$ activity. These numbers are therefore given for information only and do not 
represent the DUNE estimates of actual data to be committed to mass storage.
The data required to record this background is reduced to 3\% of the
``with-beam-$\nu$'' estimate of table~\ref{tab:sc-zs-summary} and is
thus negligible, being an order of magnitude smaller than the data
associated with the beam neutrino interactions themselves.

\input{volume-detectors/generated/zs-volume-summary}

\subsubsection{Far Detector Photon Detector (PD) Parameters}
There are variations in the basic parameters of the Photon Detector
currently in the R\&D stage. The numbers presented below should be
considered as ballpark values to be made more precise at a later time.
\begin{itemize}
\item Readout channel count: \num{24000} (i.e., four times the \num{6000} channel count for each 10-kt module)
\item Trigger rate is uncertain at this point due to ongoing
  investigation; one approach assumes one trigger per spill cycle
\item ADC resolution (bits): 12
\item ADC digitization frequency: \SI{150}{\MHz}
\end{itemize}
It is assumed that a few dozen samples will be recorded in each
channel, and that zero suppression of channels with signals
below a chosen threshold will be enforced, reducing the data volume by 
an order of magnitude.  This results in  360 kilobyte per spill
cycle, and as regards requirements on data handling, should be considered negligible compared to other data sources.

\subsubsection{Near Detector Data Rates and Parameters}
The near detector is subject to an intense R\&D effort and
its parameters are currently being optimized. The
most relevant parameters of the Fine-Grained Tracker (FGT) are listed below.
\begin{itemize}
\item Straw Tube Tracker (STT) readout channel count: \ndsstchannels
\item STT Drift Time: 120~ns
\item STT ADC clock frequency and resolution (bits): \SI{3}{\ns} intervals, 10 bit
\item ECAL channel count: \ndecalchannels
\item Muon Detector Resistive Plane Chambers (RPC) channel count: \ndmuidchannels
\item Average expected event rate per spill: \textasciitilde 1.5
\end{itemize}
Since there are large uncertainties in estimates of the detector
occupancy levels per event, broad assumptions must be made in order to
estimate the data rate. The current estimate (as quoted in the Near
Detector section of~\cite{cdr-annex-rates}) is
\textasciitilde \nddatarate, which translates into \textasciitilde
\SI{30}{\tera\byte\per\year}.

% Assuming that 10\% occupancy in the STT and 40 samples
%per trigger, one arrives to \textasciitilde 1MB of data per event. Under same assumption, ECAL will contribute \textasciitilde 0.25MB
%and the Muon Detector \textasciitilde  0.75MB.
%Based on these parameters, the upper limit of the ND data rate can be estimated as 1.5MB/s. This translates into \textasciitilde 45TB/year. 

\subsection{Processed Data}
\label{sec:detectors-sc-infrastructure-processed-data}
For the purposes of this document, \textit{processed data} is defined as most of the
data that is not considered ``raw,'' i.e., it is data \textit{derived} from raw 
(including possibly multiple stages of calibration and reconstruction)
as well as data produced as a result of Monte Carlo studies.

There are uncertainties in anticipated quantities of all of these
types of data. Table \ref{tab:sc-zs-summary} contains a range of
numbers reflecting limiting cases such as ZS vs FS.  Depending on the
%exact 
optimum readout strategy, an annual raw data volume of
\SI{1}{\tera\byte} to \SI{1}{\peta\byte} may be collected.  Assuming
that the data undergoes a few processing stages, DUNE %one 
can expect %the need 
to handle as much as \textasciitilde \SI{2}{\peta\byte} of data
annually for reconstruction and a lesser volume for final analysis
purposes.

For Monte Carlo, at the time of writing, the typical annual volume of data
produced has been of the order of a few tens of terabytes.  Initial
expectations are that the MC sample size for beam events will need to
be 10--100 times that of the data.  With collaboration growth
and more detailed studies (e.g., of systematics) undertaken, 
%expectation is that 
this estimate is likely to increase.

\subsection{Computing Model}
\label{sec:detectors-sc-infrastructure-computing-model}

\subsubsection{Distributed Computing}

Given that the collaboration is large and widely dispersed
geographically, a fully distributed approach to computing is planned,
based on the experience of the LHC
experiments. This includes not only ``traditional'' grid technologies
as deployed during the first decade of this
century, but also more recent technologies such as cloud computing and the Big
Data methodology. This combination will allow the collaboration to better
leverage resources and expertise from many of its member institutions
and improve the overall long-term scalability of its computing
platform.

DUNE will operate a distributed network of federated resources, for
both CPU power and storage capability. This will allow for streamlined
incorporation of computing facilities as they become available at
member institutions and %thus is particularly amenable to accommodate
for staged construction and commissioning of the detector subsystems. A
modern Workload Management System will be deployed on top of grid and
cloud resources to provide computing power to DUNE researchers.

\subsubsection{Raw Data Transmission and Storage Strategy}
Fermilab will be the principal data storage center for the DUNE experiment. It
will serve as a hub where the data from both the facility (e.g., the beam,
target and cryogenics) and the %various 
far and near detector systems are collected, catalogued and committed to mass
storage. This will obviously require transmission of data over
considerable distances (certainly for the far detector). %In addition,
The far detector DAQ systems %are being designed to 
will be located underground in the vicinity of the far detector modules at SURF, adding the %which results in an additional 
step of transmitting the data from 4850L to the surface.

Raw data to be collected from the DUNE detectors are considered
``precious'' due to the high cost of operating both the facility (LBNF) and the DUNE detectors. % that are part of DUNE. 
This leads to three
basic design elements in the data transmission and storage chain:
\begin{itemize}
\item Buffering:
\begin{itemize}
\item %Adequate 
Buffers will be provided for the DAQ systems to
  mitigate possible downtime of the network connection between 4850L
  and the surface.
\item Buffers will be provided at the surface facility to mitigate
  downtime of the network connection between the far site and Fermilab.
\end{itemize}
\item Robust transmission: data transfer needs to be instrumented with
  redundant checks (such as checksum calculation), monitoring, error
  correction and retry logic.
\item Redundant replicas: it is a common practice in industry and
  research (e.g., the LHC experiments) to have a total of three geographically distributed copies
  of ``precious'' data.  %Such geographical 
  Distribution of the replicas may include countries
  other than the United States, where the data will be collected.
  This provides protection against catastrophic events (such as
  natural disasters) at any given participating data center, %participating in thisscheme, 
  and facilitates rebuilding (``healing'') lost data should
  such an event happen.
\end{itemize}

\subsubsection{Data Management}
\label{sec:detectors-sc-infrastructure-computing-model-data-mgt}

Data will be placed into mass storage at Fermilab. As described
above, additional copies (replicas) will be distributed to other
properly equipped computing centers. For example, consideration is
given to both Brookhaven National Laboratory and NERSC as candidates
for the placement of extra replicas. A given replica does not need to
reside in its entirety at a single data center; the replicas can be
``striped'' across multiple data centers if that becomes optimal at
the time the Computing Model is implemented.

Recent progress in network and storage technologies have made possible
\textit{federation of storage} across multiple member-institution data centers. % located at member institutions. 
In this approach, data can be effectively shared
and utilized via the network (``data in the grid''). 

%One example of an advanced federated system of this type is XRootD.
For data distribution, DUNE will use a combination of managed data movement between
sites (such as ``dataset subscription,'' primarily for managed
production), and a network of XRootD servers (XRootD is an example of an advanced federated system) for caching processed data
and for analysis.  A file catalog and a metadata system
will be required for efficient data management at scale. Efforts
will be made to leverage experience of member institutions in this
area, and to reuse existing systems or design ideas whenever
possible.

\subsection{Computing Implications of the Dual-Phase Far Detector Design}
\label{sec:detectors-sc-alternate}
Parameters of the alternative Far Detector design (based on the
dual-phase technology) are listed in Chapter 2 of~\cite{cdr-annex-rates}. The readout channel count is 614,400 (i.e., four times the
          153,600 channel count for each \ktadj{10} detector module) and the drift time is 7.5~ms.
The Photon Detector readout channel count is 720 (i.e., four times the 180
          channel count for each 10-kt module).

According to some estimates listed in~\cite{cdr-annex-rates}, the ``Full Stream''
readout will produce 16.09~GB of data for each candidate event. This is
about 65\% of the data volume in one readout cycle of the reference
design.  Although signal processing strategies may be implemented
differently in the dual-phase design, it can be argued that the total
data rate will be of the same order of magnitude or less than that of the
reference design.

\section{Near Detector Physics Software}
\label{sec:detectors-sc-nd-physics-software}

This section summarizes the %A longer description of the 
current status of the near detector
simulation and reconstruction, described in detail in~\cite{cdr-annex-reco}. % with an abbreviated summary here.

Two approaches are being  pursued for the simulation of
the DUNE near detector.  The first is a fast Monte Carlo based on
parameterized detector responses. The GENIE\cite{GENIE} generator is
used to model the interactions of neutrinos with nuclei in the
detector, and a parameterization of the achieved NOMAD reconstruction
performance is used to model the detector response.  
The second
approach, under development, is a full GEANT4-based simulation. 
 
 The fast Monte Carlo tool is based on work done for the
far detector (see Appendix A.3 of~\cite{Adams:2013qkq}) and is capable of
rapidly evaluating the sensitivity of the detector design to a broad
variety of analyses targeting specific final states.  The full
GEANT4-based simulation and subsequent reconstruction chains will be
used to inform the parameterized responses of the fast Monte Carlo, as
well as being indispensable tools for simulating and extracting
results from the near detector.
Figure~\ref{fig:ndeventdisplaychaptersc} shows the trajectory of a
negatively charged muon with an initial momentum of 1~GeV propagating
in the straw tube tracker, as simulated using GEANT4.
\begin{cdrfigure}[Trajectory of a 1-GeV $\mu^-$ simulated in the near detector.]{ndeventdisplaychaptersc}
{The trajectory of a 1 GeV $\mu^-$ produced by the GEANT4 simulation of the near detector's straw-tube tracker.}
\includegraphics[width=0.7\textwidth]{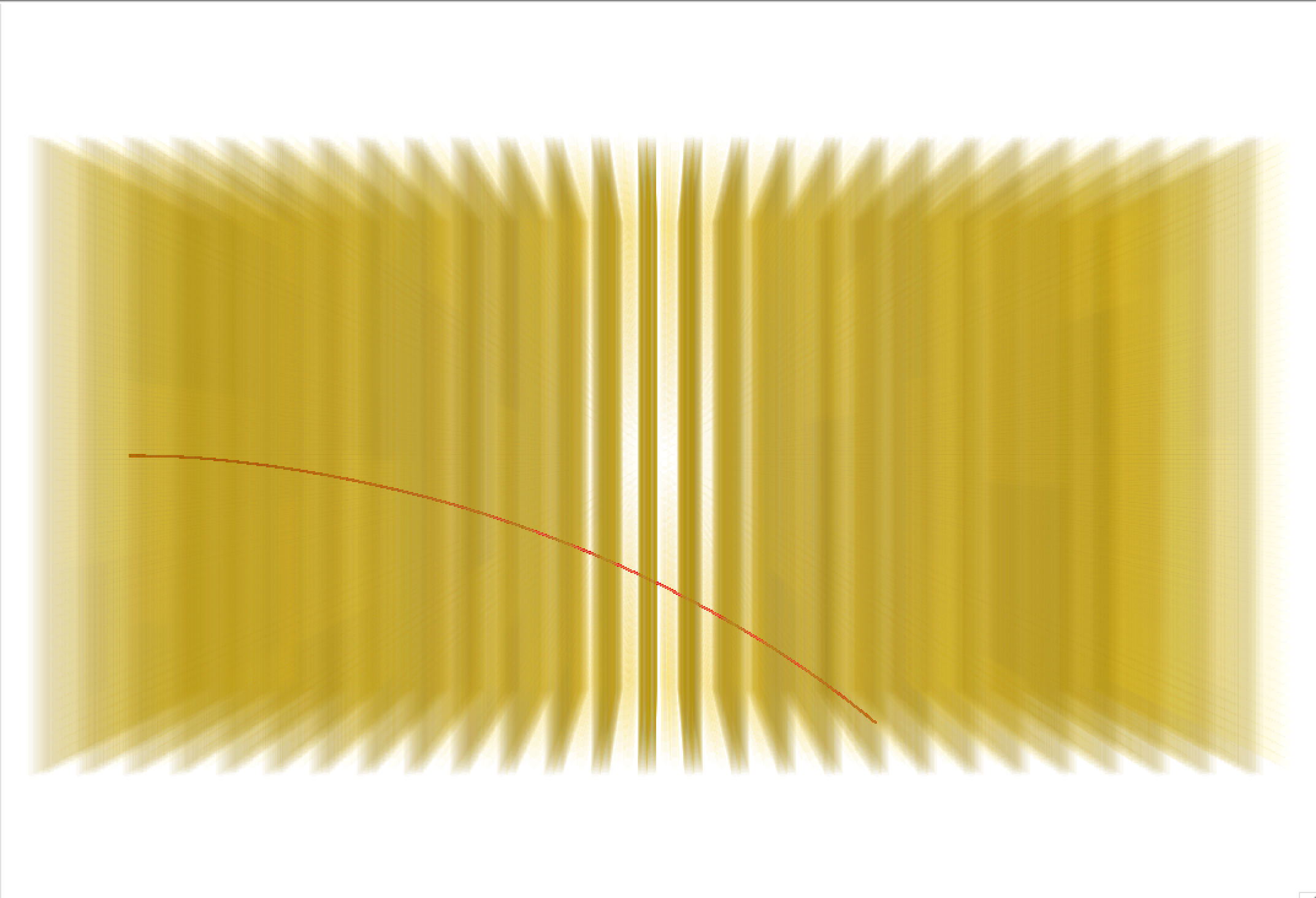}
\end{cdrfigure}

\section{Far Detector Physics Software}
\label{sec:detectors-sc-physics-software}

This section summarizes %Longer descriptions of 
the single-phase and dual-phase far detector
simulations and reconstruction, described in detail in~\cite{cdr-annex-reco}. % with an abbreviated summary here.

%\subsection{Simulation}
%\label{sec:detectors-sc-physics-software-simulation}
% Beam simulation in the Beam Requirements chapter
%\subsubsection{Beam Simulation}
%\label{sec:detectors-sc-physics-software-simulation-beam}
% ND simulation and reconstruction in the ND chapter
%

\subsection{Far Detector Simulation}
\label{sec:detectors-sc-physics-software-simulation-fd}

Detailed GEANT4-based~\cite{GEANT4:NIM,GEANT4} Monte Carlo simulations
have been developed for the single-phase and dual-phase far detector
designs, incorporating both the TPC modules and the photon detection
systems. These simulations provide a basis for detailed studies of
detector design and performance, and also enable the development of
automated event-reconstruction algorithms.

The single-phase detector simulation is implemented in
LArSoft\cite{Church:2013hea}, which provides a common simulation
framework for LArTPC experiments.  LArSoft is based on the \textit{art}
framework\cite{Green:2012gv}, and is supported by the Fermilab
Scientific Computing Division.  The comparison of data from
ArgoNeuT\cite{Anderson:2012vc,Anderson:2012mra} with LArSoft
simulations gives confidence in the reliability of the detector
simulation.  Future data from
LArIAT\cite{Adamson:2013/02/28tla,Cavanna:2014iqa},
MicroBooNE\cite{Chen:2007ae,Jones:2011ci,microboonecdr}, and the
35-ton prototype (Section~\ref{sec:proto-35t}) will allow further tuning
of the LArSoft simulation as experience is gained.  The dual-phase
detector simulation and hit-level reconstruction are based on the
Qscan\cite{lussi:thesis} package, which has been developed over the
past decade, and is currently being used for technical design and
physics studies for the \cerndualproto{} program.

Events are generated using either the GENIE\cite{GENIE} simulation of
neutrino-nucleus interactions, the
CRY\cite{Cosmic-CRY,Cosmic-CRY-protons,CRY-url} cosmic-ray generator,
a radiological decay simulator written specifically for LArSoft using
the decay spectra in~\cite{docdb-8797}, a particle gun or one of
several text-file-based particle input sources. GEANT4 
simulates the trajectories of particles and their energy
deposition.  Custom algorithms have been developed to propagate the
drifting charge and scintillation photons through the detector and to
simulate the response characteristics of the TPC wires, readout
electronics and photon detectors.
Figure~\ref{fig:larsofteventdisplays} shows some examples of simulated
accelerator neutrino interactions in the MicroBooNE detector.
\begin{cdrfigure}[Simulated neutrino interactions in MicroBooNE]{larsofteventdisplays}
{Examples of accelerator neutrino interactions, simulated by LArSoft
  in the MicroBooNE detector. The panels show 2D projections of
  different event types.  The top panel shows a $\nu_{\mu}$
  charged-current interaction with a stopped muon followed by a decay
  Michel electron; the middle panel shows a $\nu_{e}$ charged-current
  quasi-elastic interaction with a single electron and proton in the
  final state; the bottom panel shows a neutral-current interaction
  with a $\pi^{0}$ in the final state that decayed into two photons
  with separate conversion vertices.}
\includegraphics[width=\textwidth]{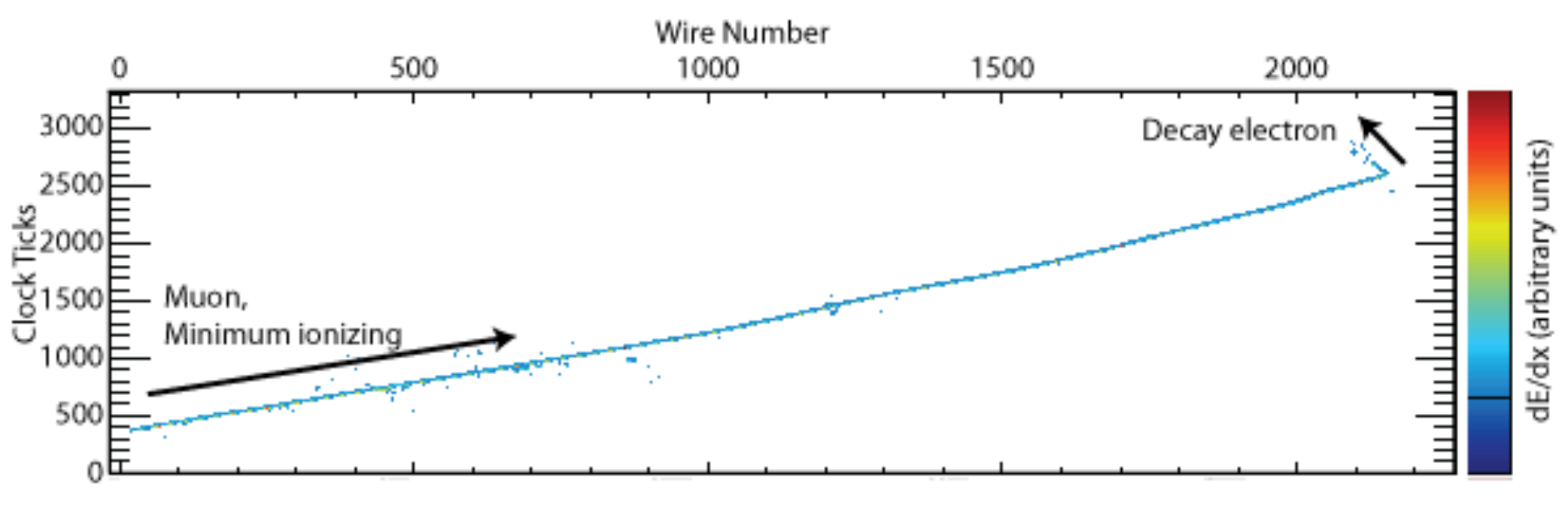}
\includegraphics[width=\textwidth]{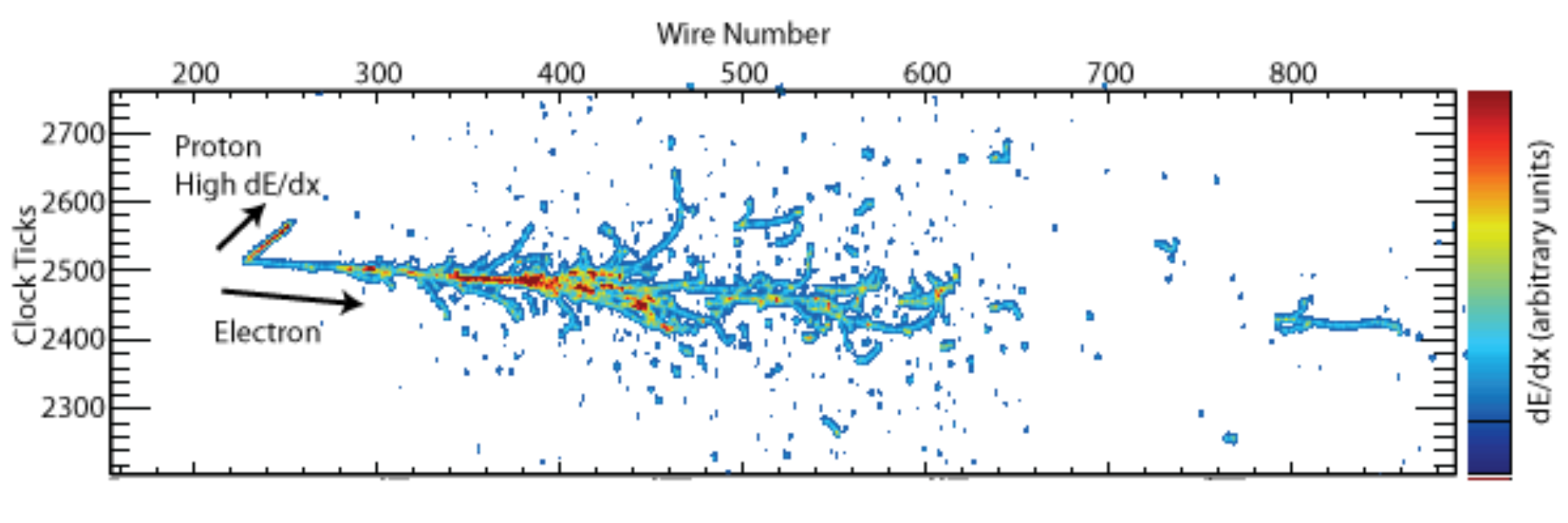}
\includegraphics[width=\textwidth]{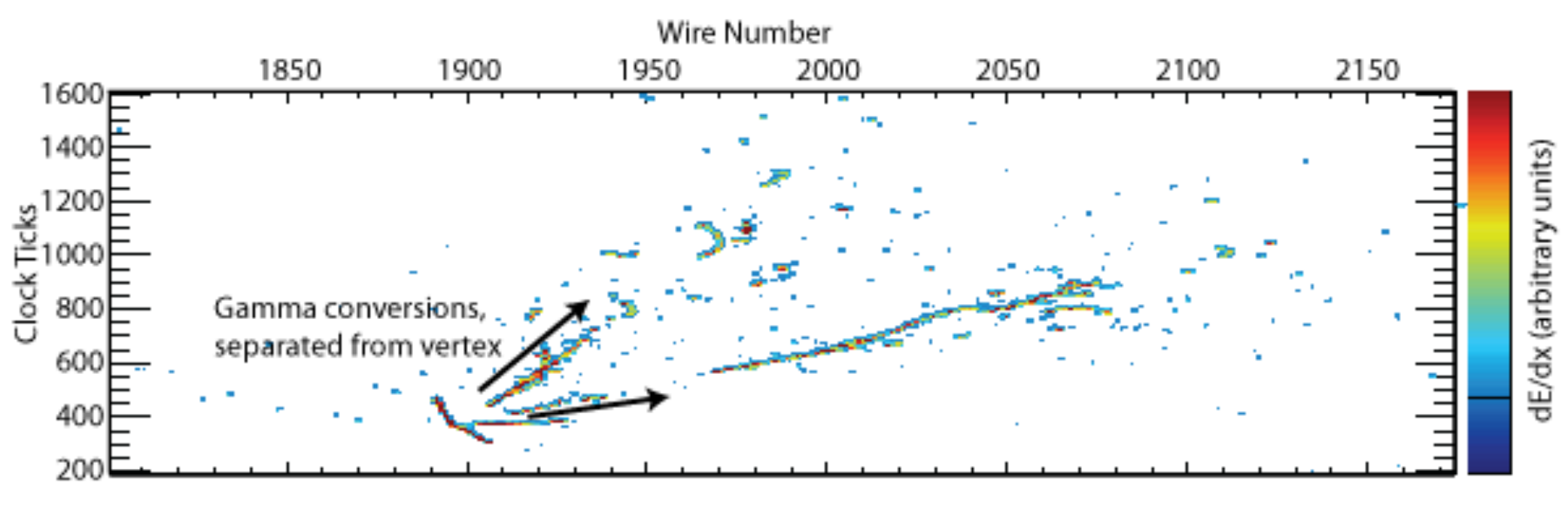}
\end{cdrfigure}

\subsection{Far Detector Reconstruction}
\label{sec:detectors-sc-physics-software-reconstruction-fd}

The reconstruction of particle interactions in LArTPC detectors is an
active area of research that has advanced significantly in recent
years.  In particular, the analysis of the data from the
ICARUS\cite{Amerio:2004ze,icarus-url,ICARUS-pizero,Antonello:2012hu}
and ArgoNEUT
experiments\cite{Adamson:2013/02/28tla,argoneut-url,Acciarri:2013met}
required the development of a variety of new reconstruction
techniques, forming the basis for precision neutrino physics
measurements.  Accurate reconstruction is needed not only of neutrino
scattering events from the beam, but also atmospheric neutrino events,
supernova burst neutrino interactions and nucleon decay events, each
with its own requirements.  With the advance of both single-phase and
dual-phase technologies and expansion of the experimental program to
include MicroBooNE\cite{Chen:2007ae,microboone-url}, the 35-ton
prototype and the CERN test experiments, the reconstruction tools have
grown in both volume and sophistication, supported by powerful
software frameworks such as LArSoft and Qscan.

Fully automated chains of event-reconstruction algorithms are being
developed for for both the single-phase and
dual-phase detector designs.  The first stage of reconstruction involves the
processing of the ADC wire signals and the identification of pulses,
or ``hits'' in the 2D space of wire number and charge
arrival time.  These hits provide the input for a series of
pattern-recognition algorithms, which form 2D and 3D clusters,
representing individual particle tracks and showers.  A set of
high-level algorithms is used to reconstruct the 3D vertex and
trajectory of each particle, identify the type of particle and
determine the four-momentum.  While each stage of the reconstruction
chain has been implemented, the algorithms -- in particular those
addressing the higher-level aspects of reconstruction such as particle
identification -- are rather preliminary and are in active
development.

\subsubsection{TPC Signal Processing, Hit Finding, and Disambiguation}

The signal-processing steps in the single-phase and dual-phase
detector designs are similar but are accomplished with separate software.
Both proceed first by decompressing the raw data and filtering the
noise using a frequency-based filter.  The single-phase
signal-processing algorithm also deconvolves the detector and
electronics responses at this step.  Both the single-phase and
dual-phase hit-finding algorithms then subtract the baselines and fit
pulse shapes to the filtered raw data.  The hit-finding algorithms are
able to fit multiple overlapping hits.  The main parameters of the
hits are the arrival time, the integrated charge, and the width.  A
raw ADC sum is also retained in the description of a hit, which often
carries a better measurement of the total charge.  The current
algorithms are found to perform well in ArgoNeuT
analyses\cite{Anderson:2012vc} for the single-phase software and
during several phases of R$\&$D and prototyping on small-scale
dual-phase LAr-LEM-TPC
setups\cite{Badertscher:2008rf,Badertscher:2012dq}.
Figure~\ref{fig:lbnoeventdisplay} shows example event displays of the
reconstructed hits in both real and simulated data.

\begin{cdrfigure}[Dual-phase LArTPC-reconstructed events for data and MC]{lbnoeventdisplay}
{
%Badertscher:2012dq
Dual-phase LArTPC-reconstructed events for data and MC.  Top: Cosmic
ray event displays for a hadronic shower candidate.  Bottom:
Reconstructed hits for a MC simulation of a 5-GeV $\nu_{\mu}$
interaction.  The secondary particles produced in the two interactions
are distinguished by different colors, based on the MC truth
information (blue=muon, green=electron, red=proton, cyan=pion).  From
Ref.\cite{Badertscher:2012dq}.  }
\includegraphics[width=0.99\textwidth]{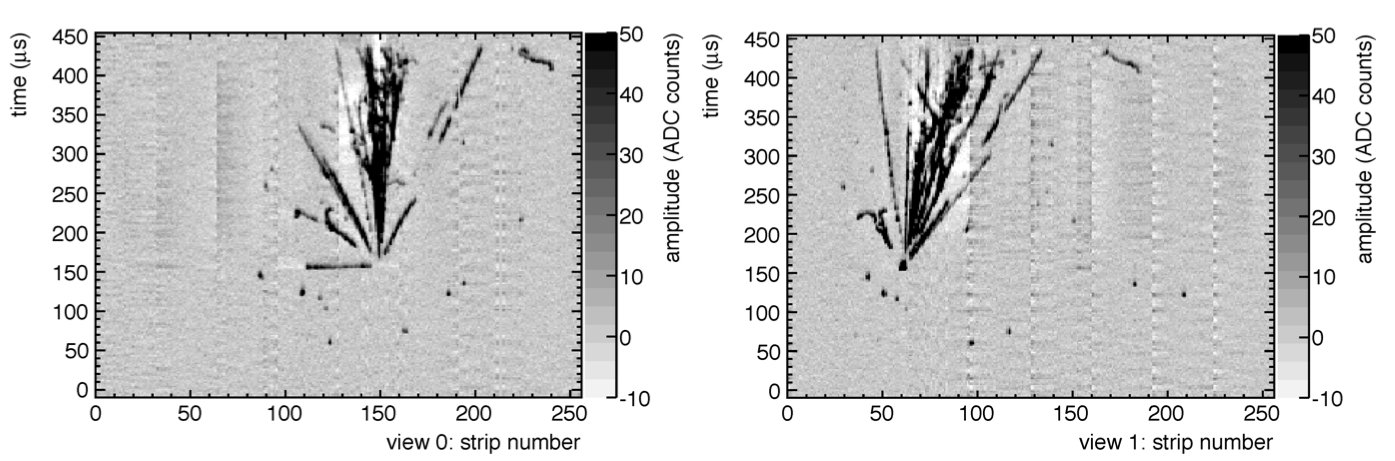}
\includegraphics[width=0.99\textwidth]{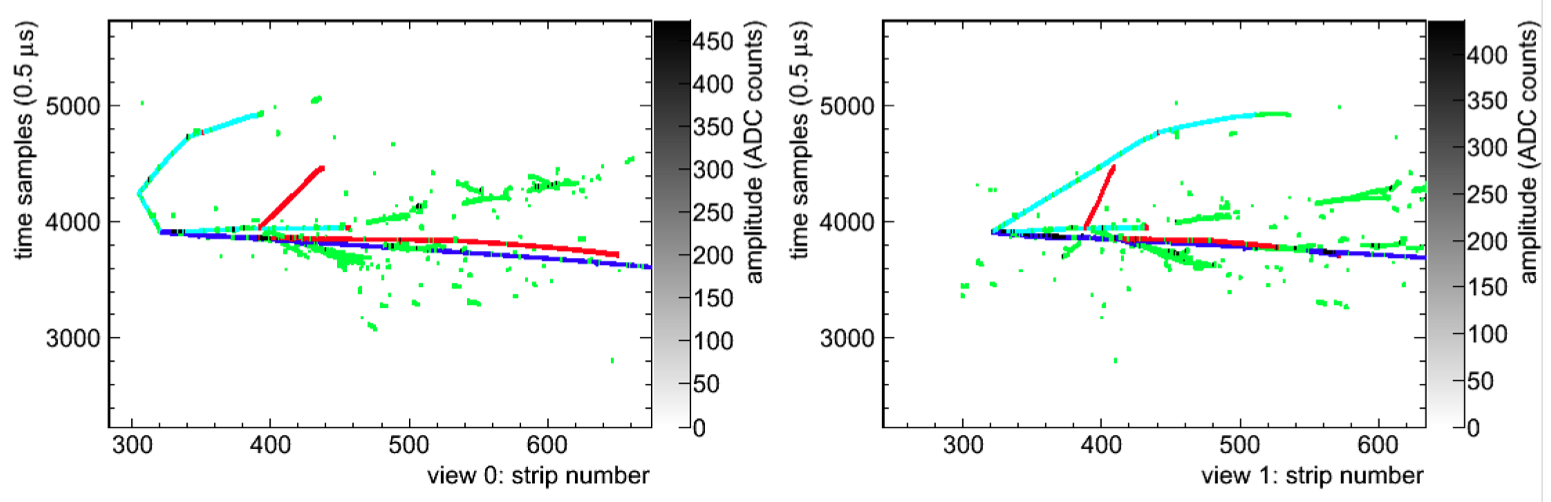}
\end{cdrfigure}

The wrapping of induction-plane wires in the single-phase APA design
introduces an additional discrete ambiguity in the data by connecting
multiple wire segments to each DAQ channel. A ``disambiguation''
algorithm is used to break the ambiguity and determine which wire
segment generated the charge on each hit.  The algorithm forms
associations between the collection and induction views, identifying
``triplets'' of hits that have intersecting wire segments and
consistent arrival times. In most events, the majority of hits are
associated with a single wire segment, and can be trivially
disambiguated.  The remaining hits are then disambiguated by
clustering them with trivially disambiguated hits.

\subsubsection{Photon Detector Signal Reconstruction}

Photon detector signals are processed in similar ways to those on the
TPC wires.  Noise is filtered out, and hits are identified as pulses
above the pedestal.  Hits are grouped together into clusters in time,
called ``flashes,'' for subsequent association with clusters in the
TPC.  Each flash has a time, a total integrated charge, and a position
estimate.  The time of an interaction is important in order to help
reject cosmic-ray events and also to determine the absolute position
of an event along the drift direction.  This position is important in
order to correct for finite electron lifetime effects for proper
charge measurement, which is important for particle identification and
extraction of physics results.  Signal events which can be out of time
from the beam include atmospheric neutrinos, supernova burst
neutrinos, and proton decay interactions.

\subsubsection{TPC Pattern Recognition}

The reconstruction of particles in 3D can be accomplished either by
forming 2D clusters and associating them between views, or by first
associating 2D hits between views and then clustering the resulting 3D
hits.  The clustering of hits in LArTPC detectors is a challenging
task due to the variety and complexity of event topologies.  However,
several automated 2D and 3D pattern-recognition algorithms have been
implemented using a range of techniques.

One promising suite of reconstruction tools is the PANDORA software
development kit\cite{Marshall:2013bda,Marshall:2012hh}, which provides
fully automated pattern recognition for both single-phase and
dual-phase technologies.  PANDORA implements a modular approach to
pattern recognition, in which events are reconstructed using a large
chain of algorithms.  Several 2D pattern-recognition algorithms are
first applied that cluster together nearby hits based on event
topology.  The resulting 2D clusters are then associated between views
and built into 3D tracks and showers, modifying the 2D clustering as
needed to improve the 3D consistency of the event.  Vertex-finding
algorithms are also applied, and neutrino events are reconstructed by
associating the 3D particles to the primary interaction vertex.

Figure~\ref{fig:pandoraefficiency} shows the current efficiency for
reconstructing the leading final-state lepton as a function of its
momentum for 5-GeV $\nu_{e}$ and $\nu_{\mu}$ charged-current
interactions simulated in the MicroBooNE detector; the DUNE
single-phase detector is expected to perform similarly, although the
multiple TPC geometry with wrapped wires requires additional software
effort.
%In both samples, the reconstruction efficiency increases rapidly with momentum,
%rising above 90\% at 500\,MeV and reaching approximately 100\% at 2\,GeV.
Figure~\ref{fig:recoannexpandoravertexresolution} shows the spatial
resolution for reconstructing the primary interaction vertex in these
5-GeV event samples, projected onto the $x$, $y$ and $z$ axes. An
estimate of the overall vertex resolution is obtained by taking the
68\% quantile of 3D vertex residuals, which yields 2.2~cm (2.5~cm)
for $\nu_{\mu}$CC ($\nu_{e}$CC) events.
\begin{cdrfigure}[PANDORA reconstruction efficiency]{pandoraefficiency}
{Reconstruction efficiency of Pandora pattern recognition algorithms
 for the leading final-state lepton in 5-GeV $\nu_{\mu}$ CC (left) and
 $\nu_{e}$ CC (right) neutrino interactions, plotted as a function of
 the lepton momentum. The reconstruction performance is evaluated
 using the MicroBooNE detector geometry. }
\includegraphics[width=0.49\textwidth]{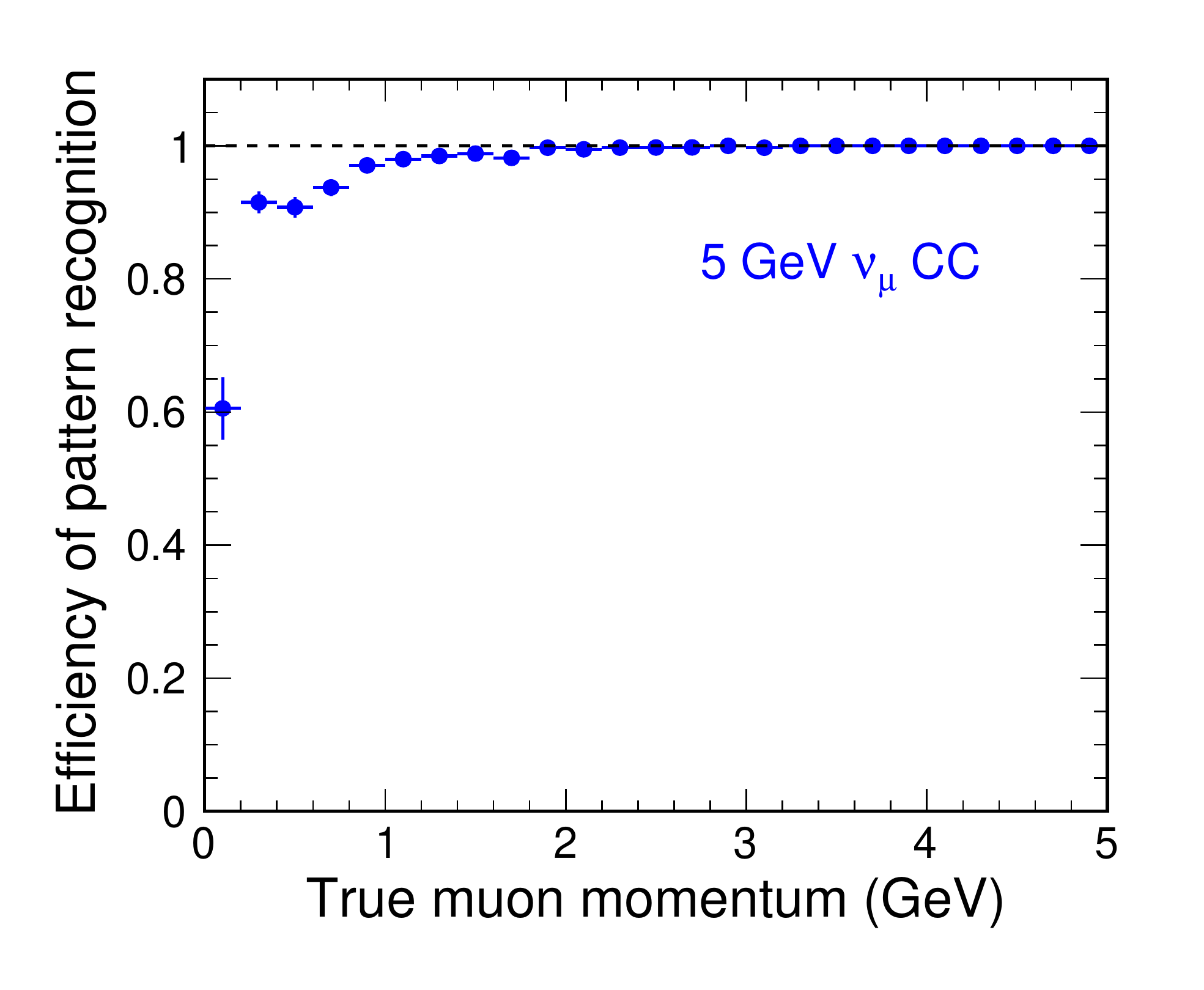}
\includegraphics[width=0.49\textwidth]{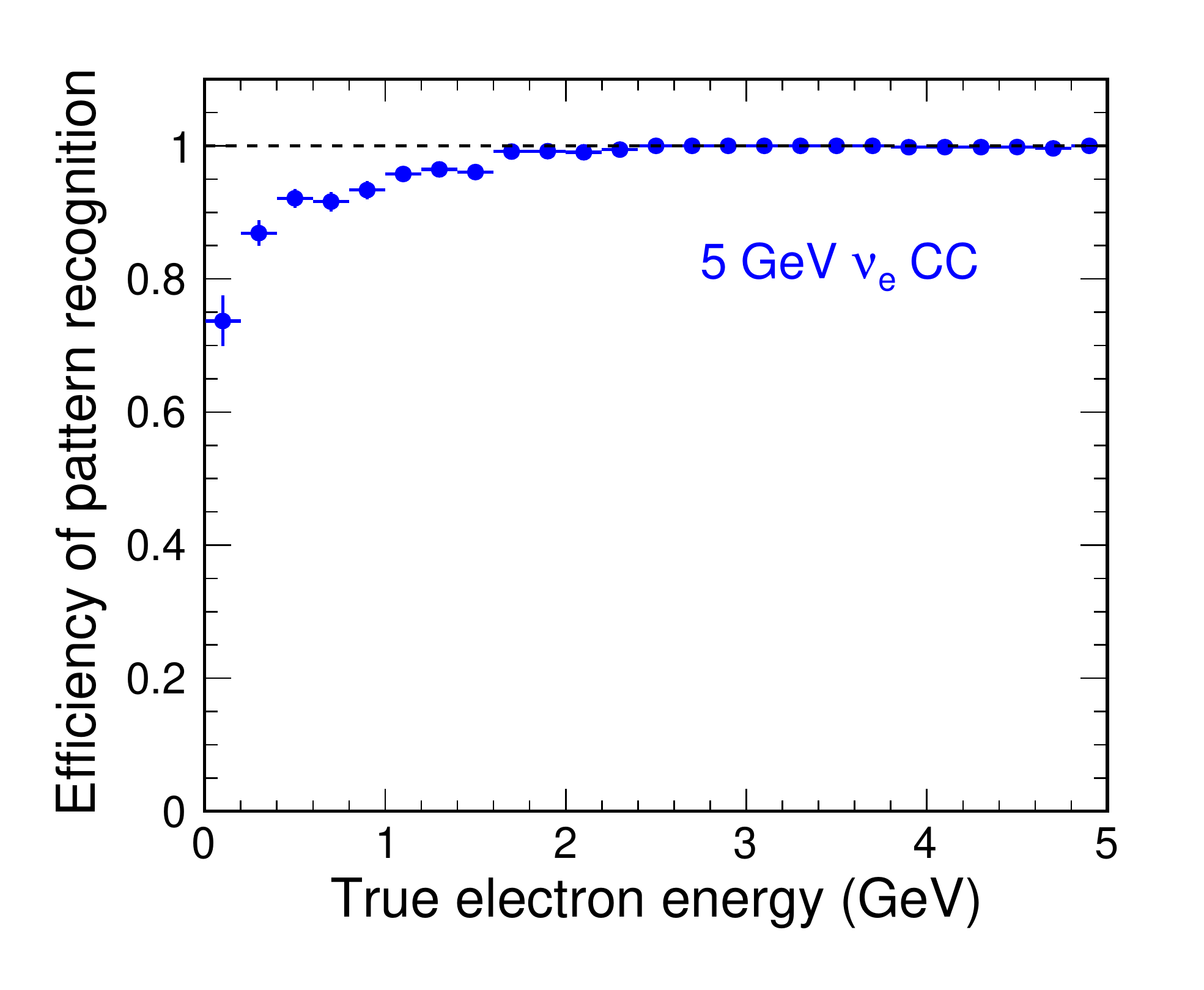}
\end{cdrfigure}
\begin{cdrfigure}[PANDORA vertex resolution]{recoannexpandoravertexresolution}
{Distribution of 2D residuals between reconstructed and simulated interaction
 vertex for 5-GeV $\nu_{\mu}$ CC (left) and $\nu_{e}$ CC (right) interactions in the MicroBooNE detector.
 The $x$ axis is oriented along the drift field, the $y$ axis runs parallel 
 to the collection plane wires, and the $z$ axis points along the beam direction.}
\includegraphics[width=0.49\textwidth]{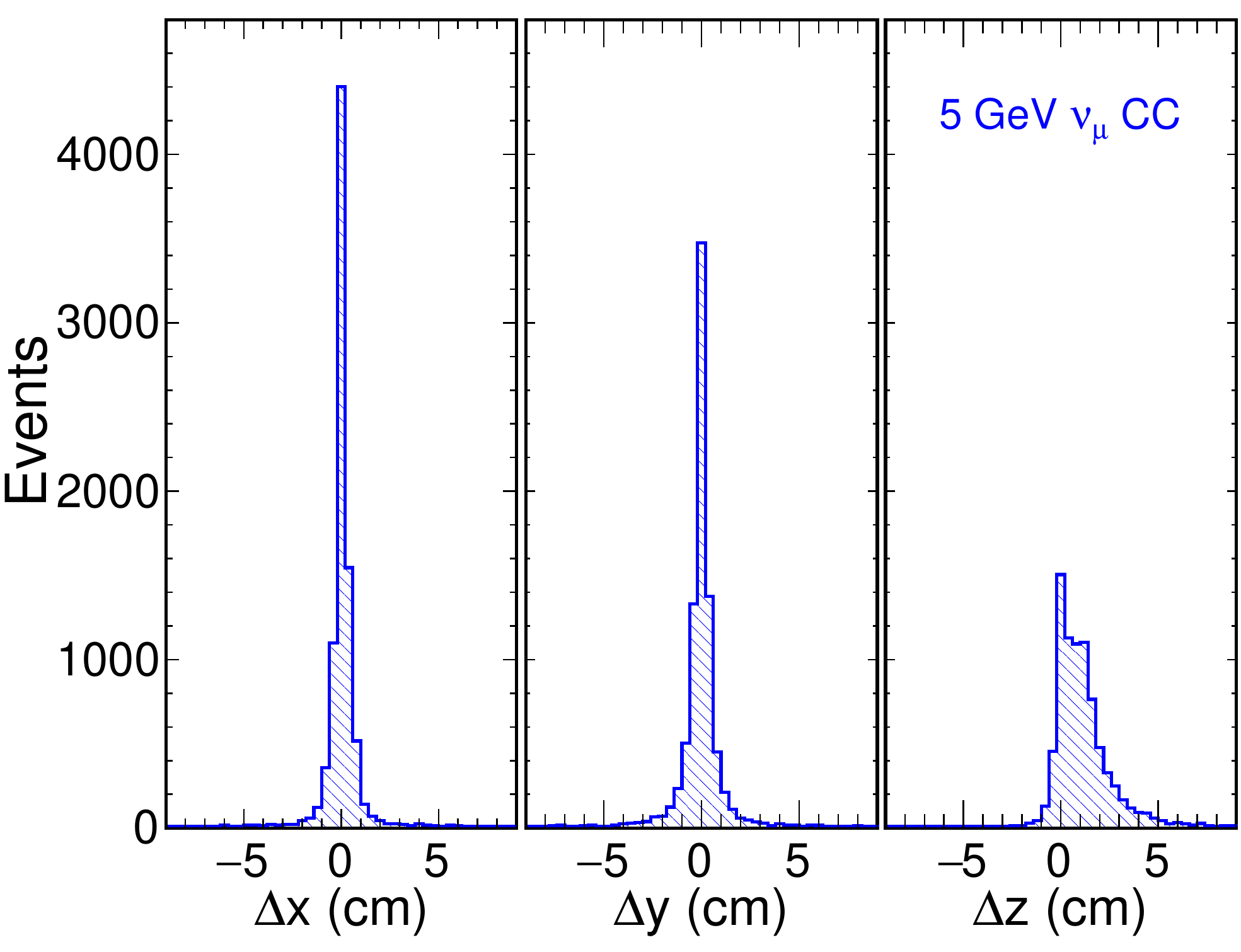}
\includegraphics[width=0.49\textwidth]{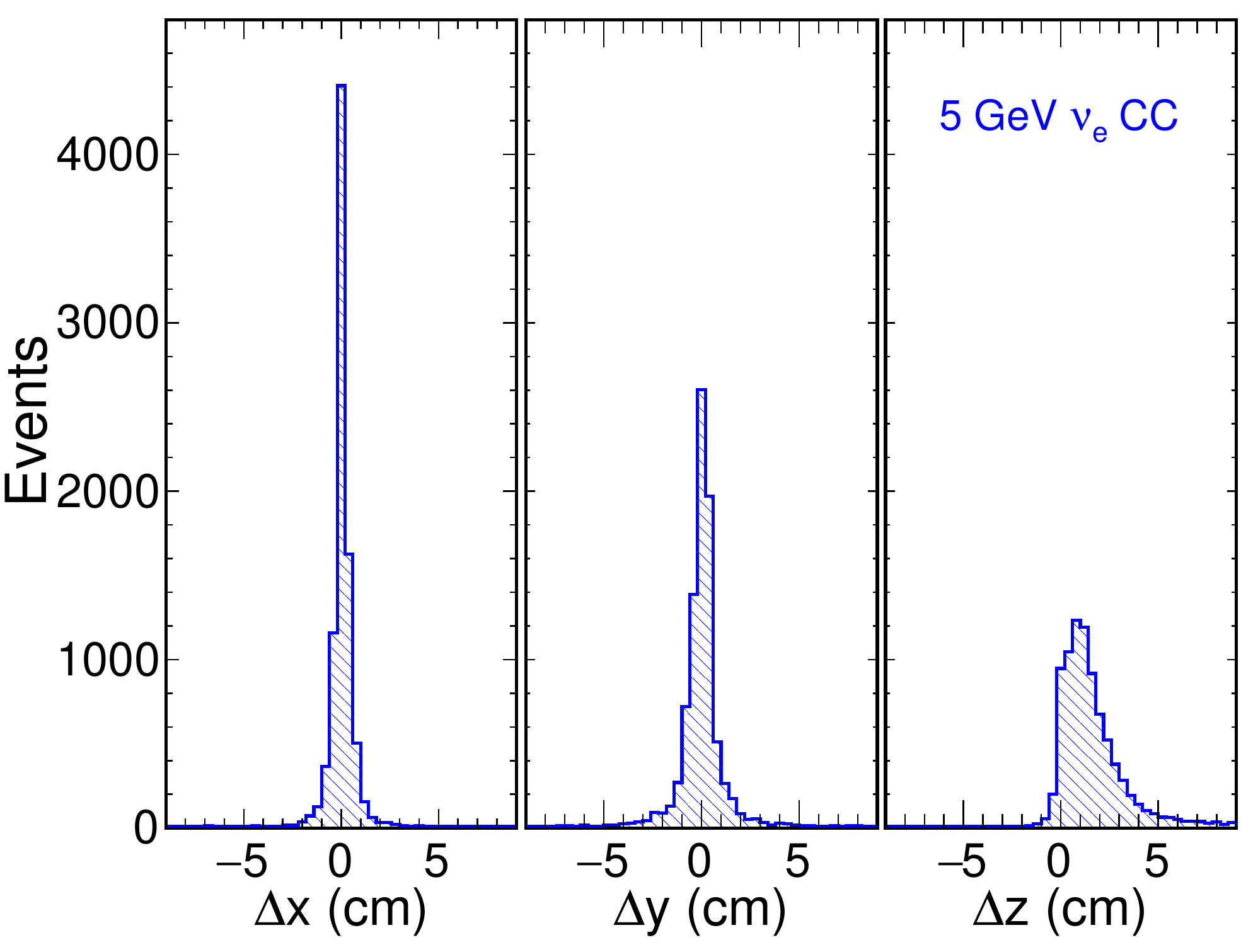}
\end{cdrfigure}

\subsubsection{Track Fitting and Shower Measurement}

After the pattern recognition stage, a series of high-level
reconstruction algorithms is applied to the 2D and 3D clusters; these algorithms
fit the trajectories of particle tracks and measure the spatial and
calorimetric properties of electromagnetic and hadronic showers.
Several high-level techniques have been demonstrated for use in LArTPC
detectors using both real and simulated data.

The Kalman filter technique\cite{kalman} is well-established in
high-energy physics, and has been applied to 3D track reconstruction
in liquid argon by ICARUS\cite{Ankowski:2006ts}.  The technique
incorporates the effects of multiple Coulomb scattering, enabling a
scattering-based measurement of the track momentum, which is shown by
ICARUS to have a resolution as good as $\Delta p/p \approx 10\%$ for
the most favorable track lengths.  The data from ICARUS have also
been used to develop a precise track-reconstruction algorithm, which
builds a 3D trajectory for each track by simultaneously optimizing its
2D projections to match the observed data\cite{Antonello:2012hu}.
Another promising track reconstruction technique, based on the local
principal curve algorithm, has been implemented for the dual-phase
detector, and is shown to provide a precise reconstruction of two-body
final states~\cite{Back:2013cva,LAGUNA-LBNO-deliv}.

A full 3D reconstruction of electromagnetic showers is currently in
development.  In the present scheme, the first stage is an examination
of clusters in terms of their 2D parameters, and a selection of
shower-like clusters for further analysis. The 3D start position,
principal axis, and shower-shape variables are then reconstructed by
matching up 2D hits between views.  These 3D parameters, combined with
calorimetric information, enable a measurement of the total shower
energy as well discrimination between electrons and converted photons,
based on the ionization energy in the initial part of the shower. The
kinematic reconstruction of final-state neutral pions from their
$\pi^{0} \rightarrow \gamma\gamma$ decays can be performed by
combining together associated pairs of photons.

\subsubsection{Calorimetry and Particle Identification}

The reconstructed energy of hits follows from the measured charge
after corrections are made for sources of charge loss.  The energy of
physics objects can then be reconstructed by summing the energy of the
associated hits and, when this is combined with a reconstructed
trajectory, a measurement of the ionization density $dE/dx$ can be
made, which is an important input to particle identification.  In
order to reconstruct this information, the measured charge on each hit
is first obtained from fits to the pulse shapes.  The charge loss due
to the finite electron lifetime is corrected based on the time of the
event, and the path length corresponding to each hit is calculated
based on the event trajectory.  The effects of recombination, known as
``charge quenching,'' are corrected using a modified Box
model\cite{Thomas:1987zz} or Birks' Law\cite{Birks:1964zz}.  The
identity of a particle track that ranges out in the active detector
volume may be ascertained by analyzing the ionization density $dE/dx$
as a function of the range from the end of the track, and comparing
with the predictions for different particle species.

In a LArTPC, electromagnetic showers may be classified as
having been initiated by an electron or a photon using the $dE/dx$ of
the initial $\sim$2.5~cm of the shower.  Electron-initiated showers
are expected to have $dE/dx$ of one MIP in the initial part, while
photon-initiated showers are expected to have twice that.  Current
algorithms achieve a performance of 80\% electron efficiency with 90\%
photon rejection, and %with 
a higher efficiency for fully-reconstructed
showers.

%Do we have any calorimetry plots?  
% not for the main CDR.  Perhaps the annex

\subsubsection{Neutrino Event Reconstruction and Classification}

Once the visible particles in an event have been reconstructed
individually, the combined information is used to reconstruct and
classify the overall event.  The identification of neutrino event
types is based on a multivariate
analysis\cite{Back:2013cva,WA105_TDR,LAGUNA-LBNO-deliv,LAGUNA-LBNO-EOI},
which constructs a number of characteristic topological and
calorimetric variables, based on the reconstructed final-state
particles. In the present scheme, a Boosted Decision Tree algorithm is
used to calculate signal and background likelihoods for particular
event hypotheses. The current performance has been evaluated using
fully reconstructed $\nu_{e}$ and $\nu_{\mu}$ charged-current
interactions with two-body final states, simulated in the dual-phase
far detector\cite{LAGUNA-LBNO-deliv}.  The correct hypothesis is
chosen for 92\% (79\%) of $\nu_{\mu}$ ($\nu_{e}$) quasi-elastic
interactions with a lepton and proton in the final state, and 79\%
(71\%) of $\nu_{\mu}$ ($\nu_{e}$) resonance interactions with a lepton
and charged pion in the final state.  For selected events, the
neutrino energy is estimated kinematically for quasi-elastic
interactions using a two-body approximation; otherwise a
calorimetric energy measurement is applied.  The calorimetric
reconstruction takes into account the quenching factors of the
different particles, assuming that all hits not associated with the
primary lepton are due to hadronic activity.
Figure~\ref{fig:recoenergynue} shows the resulting energy
reconstruction for $\nu_e$ CCQE and CC1$\pi^{+}$ events.
\begin{cdrfigure}[Reconstruction of electron neutrino energy]{recoenergynue}
{Performance of neutrino energy measurement, evaluated using the dual-phase far detector simulation. 
Distributions of reconstructed versus true neutrino are shown for $\nu_{e}$ CCQE events (left),
assuming two-body kinematics, and $\nu_{e}$  CC1$\pi^{+}$ events (right),
using a calorimetric energy estimation.}
\includegraphics[width=0.49\textwidth]{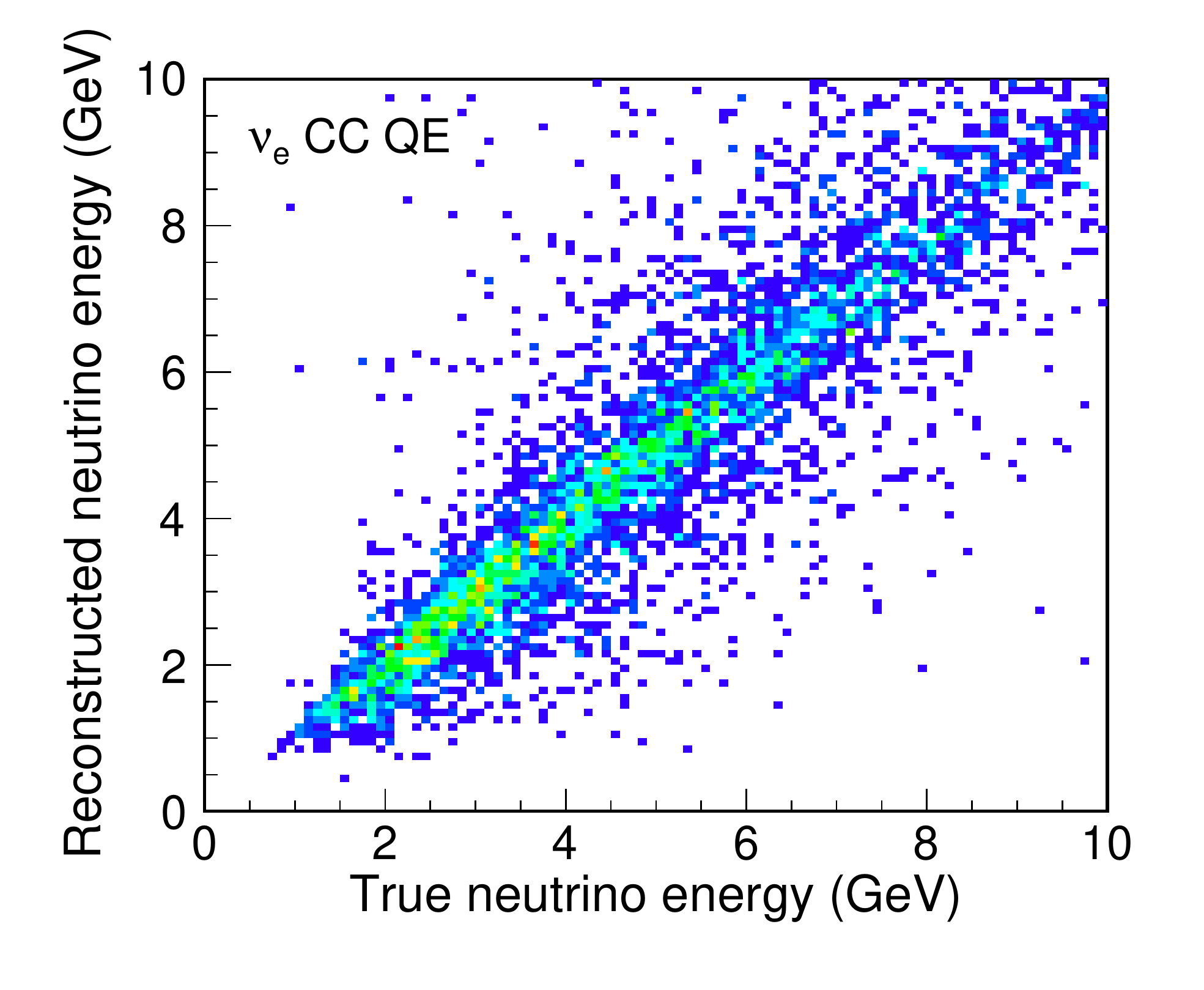}
\includegraphics[width=0.49\textwidth]{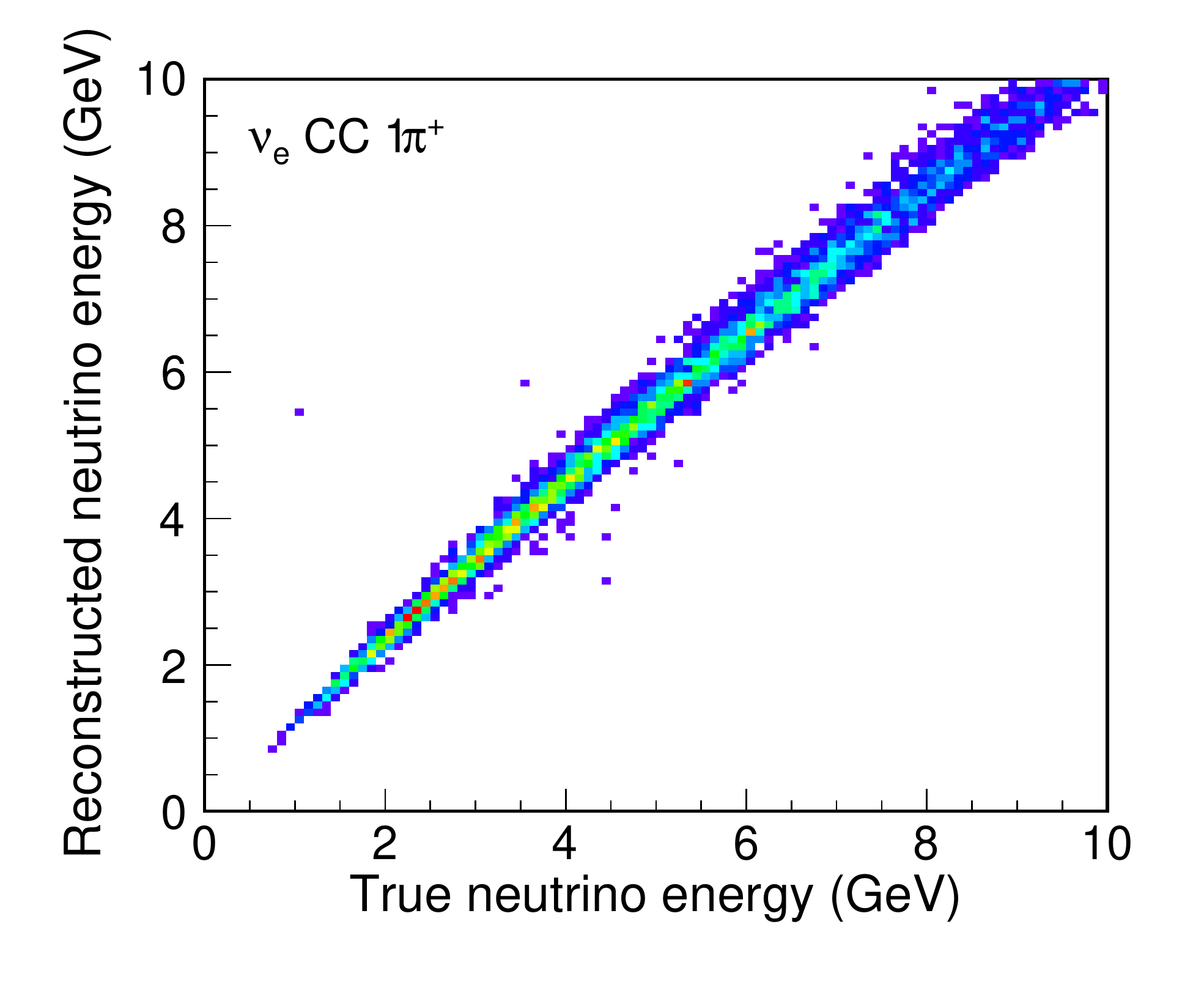}
\end{cdrfigure}

%%%% Do we need a summary?
% not really -- 5 pages is a little tight; this is the summary

%% file: volume-detectors/generated/zs-volume-summary.tex
% do not edit, this is generated by dune-params

\begin{cdrtable}[Annual data volume estimations for zero-suppressed (ZS) data from various sources.]{rrrrr}{sc-zs-summary}{Annual data volume estimations for zero-suppressed (ZS) data from various sources. An additional full-stream (FS) data estimation is given for supernova burst (SNB).}
Source & Event Rate & Event Size & Data Rate & Annual Data Volume \\ \toprowrule
$^{39}Ar$ (ZS) & \SI[round-mode=places,round-precision=1]{11.16}{\mega\hertz} & \SI[round-mode=places,round-precision=0]{150.0}{\byte} & \SI[round-mode=places,round-precision=1]{1.674}{\giga\byte\per\second} &  \SI[round-mode=places,round-precision=0]{52.8262940816}{\peta\byte}\\
all in-spill & & & & \SI[round-mode=places,round-precision=0]{158.558121686}{\tera\byte} \\
with-beam-$\nu$ & & & & \SI[round-mode=places,round-precision=0]{79.279060843}{\giga\byte} \\
\colhline
cosmic-$\mu$ (ZS) & \SI[round-mode=places,round-precision=3]{0.258947264}{\hertz} &
\SI[round-mode=places,round-precision=1]{2.5}{\mega\byte} & \SI[round-mode=places,round-precision=1]{647.36816}{\kilo\byte\per\second} &
\SI[round-mode=places,round-precision=0]{20.4289491035}{\tera\byte} \\
\colhline
beam-$\nu$ (ZS) & \SI[round-mode=places,round-precision=0]{8770.19567714}{\per\year} & \SI[round-mode=places,round-precision=1]{2.5}{\mega\byte} &
\SI[round-mode=places,round-precision=2]{0.69479167}{\kilo\byte\per\second} & \SI[round-mode=places,round-precision=0]{21.9254891928}{\giga\byte} \\
beam-$\nu$ (FS) & \SI[round-mode=places,round-precision=0]{8770.19567714}{\per\year} & \SI[round-mode=places,round-precision=1]{24.8832}{\giga\byte} &
\SI[round-mode=places,round-precision=0]{6.915456}{\mega\byte\per\second} & \SI[round-mode=places,round-precision=0]{218.230533073}{\tera\byte} \\
\colhline
SNB cand. (ZS) & \SI[round-mode=places,round-precision=0]{12.0}{\per\year} & \SI[round-mode=places,round-precision=1]{16.74}{\giga\byte} &
\SI[round-mode=places,round-precision=0]{6365.63904105}{\byte\per\second} & \SI[round-mode=places,round-precision=0]{200.88}{\giga\byte} \\
SNB cand. (FS) & \SI[round-mode=places,round-precision=0]{12.0}{\per\year} & \SI[round-mode=places,round-precision=1]{46.08}{\tera\byte} &
\SI[round-mode=places,round-precision=1]{17.522620}{\mega\byte\per\second} & \SI[round-mode=places,round-precision=0]{552.96}{\tera\byte} \\
\end{cdrtable}

%% file: volume-detectors/chapter-proto.tex
\chapter{Prototyping Strategy}
\label{ch:detectors-proto}

%%%%%%%%%%%%%%%%%%%%%%%%%%%%%%%%
\input{volume-detectors/proto-sections/proto-overview}

\input{volume-detectors/proto-sections/proto-35t}
\input{volume-detectors/proto-sections/proto-cern-single}
\input{volume-detectors/proto-sections/proto-wa105}

\input{volume-detectors/proto-sections/proto-nd}
\input{volume-detectors/proto-sections/proto-sbn-connect}

%% file: volume-detectors/proto-sections/proto-overview.tex
%%%%%%%%%%%%%%%%%%%%%%%%%%%%%%%%
\section{Overview}
\label{sec:proto-overview}

This chapter describes the prototyping strategy for the DUNE  far and near detectors and the efforts that are underway or being planned.  These include:
\begin{itemize}
\item the 35-ton prototype at Fermilab,
\item the single-phase DUNE prototype at CERN,
\item the dual-phase LArTPC prototype at CERN known as WA105, and
\item prototypes for the Near Neutrino Detector (NND) and Beamline Measurement (BLM) systems
\item the LArTPC-based short-baseline neutrino physics program (SBN) at Fermilab.
%\item The near detector Beamline Measurements prototype (for the detector described in \ref{sec:detectors-nd-ref-blm})
\end{itemize}

The single-phase LArTPC prototyping efforts, i.e., the 35-t at
Fermilab and the DUNE prototypes at CERN, have evolved from the plans
made initially for LBNE, as summarized in Chapter 7 of \anxlbnefd~\cite{cdr-annex-lbne-design} 
and fully described in the \textit{Integrated Plan for LArTPC Neutrino
  Detectors in the U.S}, a report which was submitted to the DOE in
2009.  The 2009 report outlined an R\&D program with the goal of
demonstrating a scalable LArTPC far detector design for a
long-baseline neutrino oscillation experiment.  The following list of
the detector development components is taken from the Executive
Summary of the 2009 report (and edited to remove out-of-date
information).
\begin{itemize}
   \item Materials Test Stand (MTS) program
   to address questions pertaining to maintenance of argon purity;
    \item electronics test stands at Fermilab and BNL;
    \item the Liquid Argon Purity Demonstrator (LAPD) at Fermilab;
    \item the ArgoNeuT prototype LArTPC;
    \item the MicroBooNE experiment, a physics experiment that will
      advance our understanding of LArTPC technology;
    \item a software development effort that is well integrated across
      present and planned LArTPC detectors;
    \item a membrane-cryostat mechanical prototype to evaluate this technology;
    \item an installation and integration prototype for LBNE, to
      understand issues pertaining to detector assembly, particularly
      in an underground environment;
    \item $\sim$5\%-scale electronics systems test to understand
      system-wide issues as well as individual component reliability; and
    \item a calibration test stand that would consist of a small TPC
      to be exposed to a test beam for calibration studies, relevant
      for evaluation of physics sensitivities.
\end{itemize}

This detector development plan has largely been enacted.  The MTS is a
standard tool used to assess any materials planned for use in LArTPCs.
LAPD operated successfully, showing the viability of a ``piston purge'' in
place of evacuation to attain the desired argon purity.  The membrane
cryostat mechanical prototype became the DUNE 35-t prototype and
repeated the demonstration of the piston purge in its Phase-1
operation.  ArgoNeuT collected quality data in the NuMI beam and
helped spur the integrated reconstruction software development effort
now known as LArSoft.  The proposed calibration test stand has become
the LArIAT facility at Fermilab.  The development of cold electronics
for LArTPCs continues, with a goal of incorporating most of the signal
processing and formatting into analog and digital ASICs.  The
installation and integration prototype for LBNE has evolved into the
35-t Phase-2 and single-phase CERN prototypes for DUNE, described
in this Chapter.
 
At the time the LBNE document was written, the 35-t membrane cryostat
had not yet been fabricated; Section~\ref{sec:proto-35t} provides a
summary of the recent and near-future 35-t operations.  The 1-kt
prototype described in the LBNE document is no longer planned and has
been replaced by the DUNE single-phase LArTPC prototype at CERN, which
is summarized in Section~\ref{sec:proto-cern-single} and fully
described in \anxcernproto~\cite{cdr-annex-singleph-proto}.  The prototyping plans made for LBNE were
in turn a part of a larger LArTPC detector development program at
Fermilab, which is also linked to the short-baseline program at
Fermilab; these connections are described in
Section~\ref{sec:sbn_connect}.

The dual-phase WA105 prototype detector has evolved from the
European-based design studies performed for LAGUNA-LBNO; the
development of these is summarized in
Section~\ref{sec:detectors-fd-ref-ov}.  The WA105 detector provides a
large-scale implementation of a dual-phase LArTPC as summarized 
in Section~\ref{sec:proto-cern-double} and fully described in
\anxdualtdr~\cite{WA105_TDR}.  A 20-t engineering prototype of WA105, also described in
Section~\ref{sec:detectors-fd-ref-ov}, is planned for the immediate
future.

Both the single-phase and dual-phase prototypes at CERN will be
operated in a test beam and calibrated with charged particles (pions,
protons, muons, electrons). The data from the CERN prototypes will be
combined with data from other operating LArTPCs, such as those in the
Fermilab short-baseline and test beam programs and used to refine the
simulations of the DUNE far detectors.

The near detector prototyping plans described in
Section~\ref{sec:proto-nd} are under development and will
utilize particle test beams at CERN and Fermilab.  Prototypes
of certain short-baseline systems have already been operated in the
NuMI beam at Fermilab.

The prototyping efforts enable the Collaboration
to acquire the knowledge and skills necessary for successful
construction and implementation of the DUNE detectors. They provide
guidance in areas such as procurement, construction and installation
techniques, in addition to refinement of technical designs and validation of
cost and schedule estimates.  Operation of the prototypes will provide
opportunities to test data-taking and data-handling assumptions and
to enhance the development of data analysis tools.  Finally, the
Fermilab short-baseline program offers experience with large neutrino
event samples in LArTPC detectors as well as an opportunity to make
detailed measurements of neutrino-argon interactions that are
important for DUNE physics.

%% file: volume-detectors/proto-sections/proto-35t.tex
%%%%%%%%%%%%%%%%%%%%%%%%%%%%%%%%
\section{The 35-t Prototype}
\label{sec:proto-35t}

The 35-t (metric ton) LAr prototype was designed to demonstrate that a
non-evacuable membrane cryostat could satisfy the DUNE far detector
requirement that oxygen contamination of the LAr be less than 100
parts per trillion (ppt) and stably maintain that level.  In addition,
construction and operation of the 35-t cryostat has also served to
identify requirements for procurement of materials and services, to
inform procedures for construction and to inform best practices for
safe operation.  Construction and operation of the 35-t membrane
cryostat alone (without detector elements), now called ``Phase-1,''
was successfully completed in 2014.

The second phase of 35-t prototype operations (Phase-2) includes
installation and operation of a small-scale, single-phase LArTPC and
photon detector in this cryostat, focusing on the performance
of active detector elements within the LAr volume.  Phase-2 is
currently under construction and plans to take data in winter 2015-2016.

\subsection{35-t Phase-1}

The 35-t membrane cryostat was built by the Japanese company IHI,
contracted by the LBNE project.  
Table~\ref{tab:35Tdimensions} lists the construction materials and
dimensions; more information can be found in~\cite{bib:membcryo1573}.

\begin{cdrtable}[35-t prototype materials and dimensions]{ll}{35Tdimensions}
{35-t prototype materials and dimensions}
Parameter & Value \\ \toprowrule
Cryostat Volume	&      29.16 m3\\ \colhline
Liquid argon total mass	 &     38.6 metric tons\\ \colhline
Inner dimensions	&      4.0 m (L) x 2.7 m (W) x 2.7 m (H)\\ \colhline
Outer dimensions        &      5.4 m (L) x 4.1 m (W) x 4.1 m (H)\\ \colhline
Membrane		&      2.0 mm thick corrugated 304 SS\\ \colhline
Insulation		&      0.4 m polyurethane foam\\ \colhline
Secondary barrier system	   &   0.1 mm thick fiberglass\\ \colhline
Vapor barrier	Normal	  &    1.2 mm thick carbon steel\\ \colhline
Steel reinforced concrete	    &  0.3 m thick layer\\
\end{cdrtable}

The prototype was built in Fermilab's PC-4 facility near the Liquid
Argon Purity Demonstrator (LAPD)~\cite{bib:lapdP07005} 
allowing re-use of a large portion of the LAPD
cryogenics-process equipment.  The proximity and
size (30~t) of the LAPD vessel offers the possibility using
LAPD as a partial storage vessel for LAr if it ever becomes necessary
to empty the 35-t cryostat.

The 35-t system employs a submersible pump to move LAr from the
cryostat through the filters.  Figure~\ref{fig:35cutaway} shows a
cutaway view of the cryostat and a photograph of the interior of the
completed cryostat.
\begin{cdrfigure}[35-t prototype cryostat, cutaway view]{35cutaway}{(left) Cutaway view of the 35-t prototype cryostat. (right) Interior
photograph of the completed cryostat.}
\includegraphics[width=0.60\textwidth]{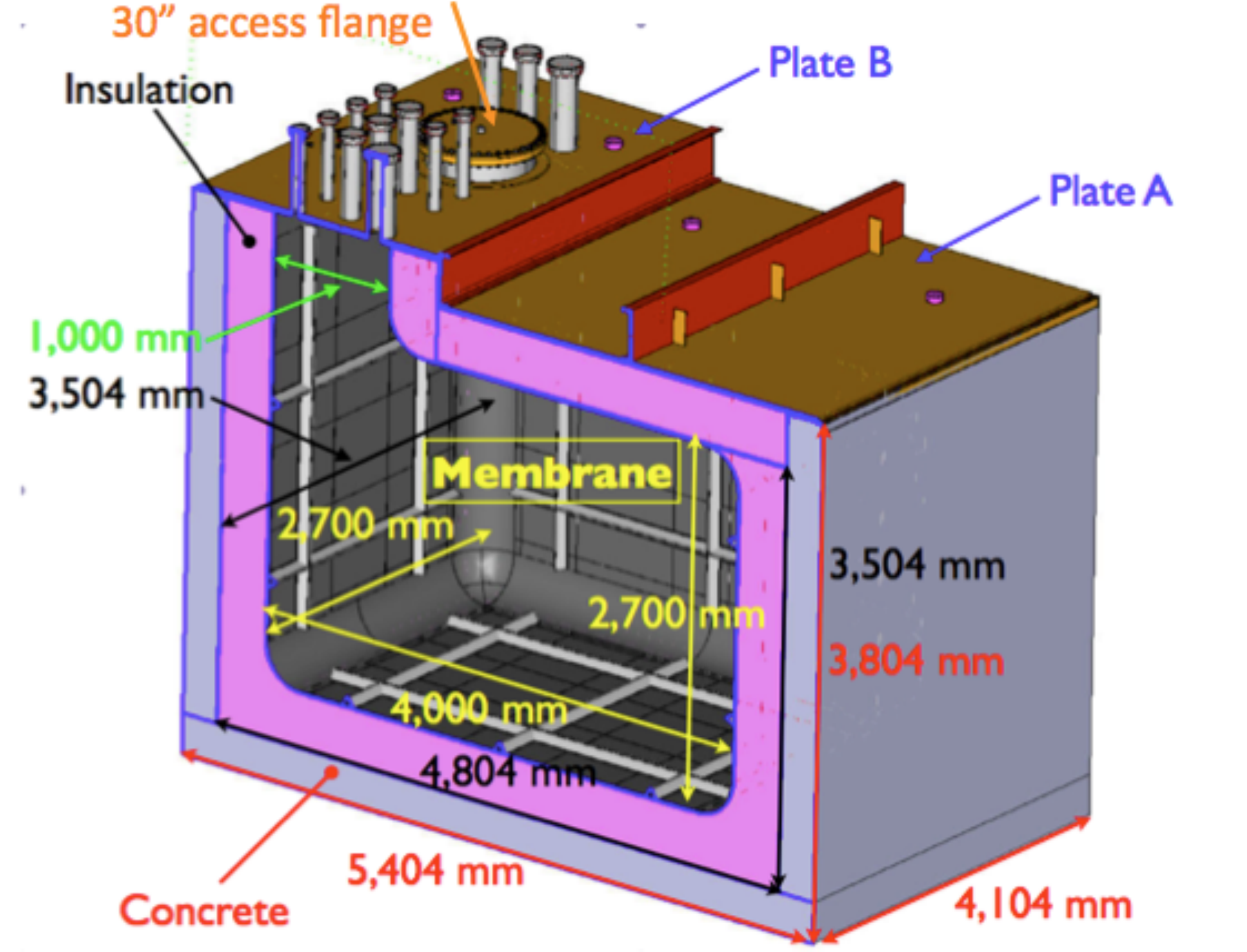}
\includegraphics[width=0.35\textwidth]{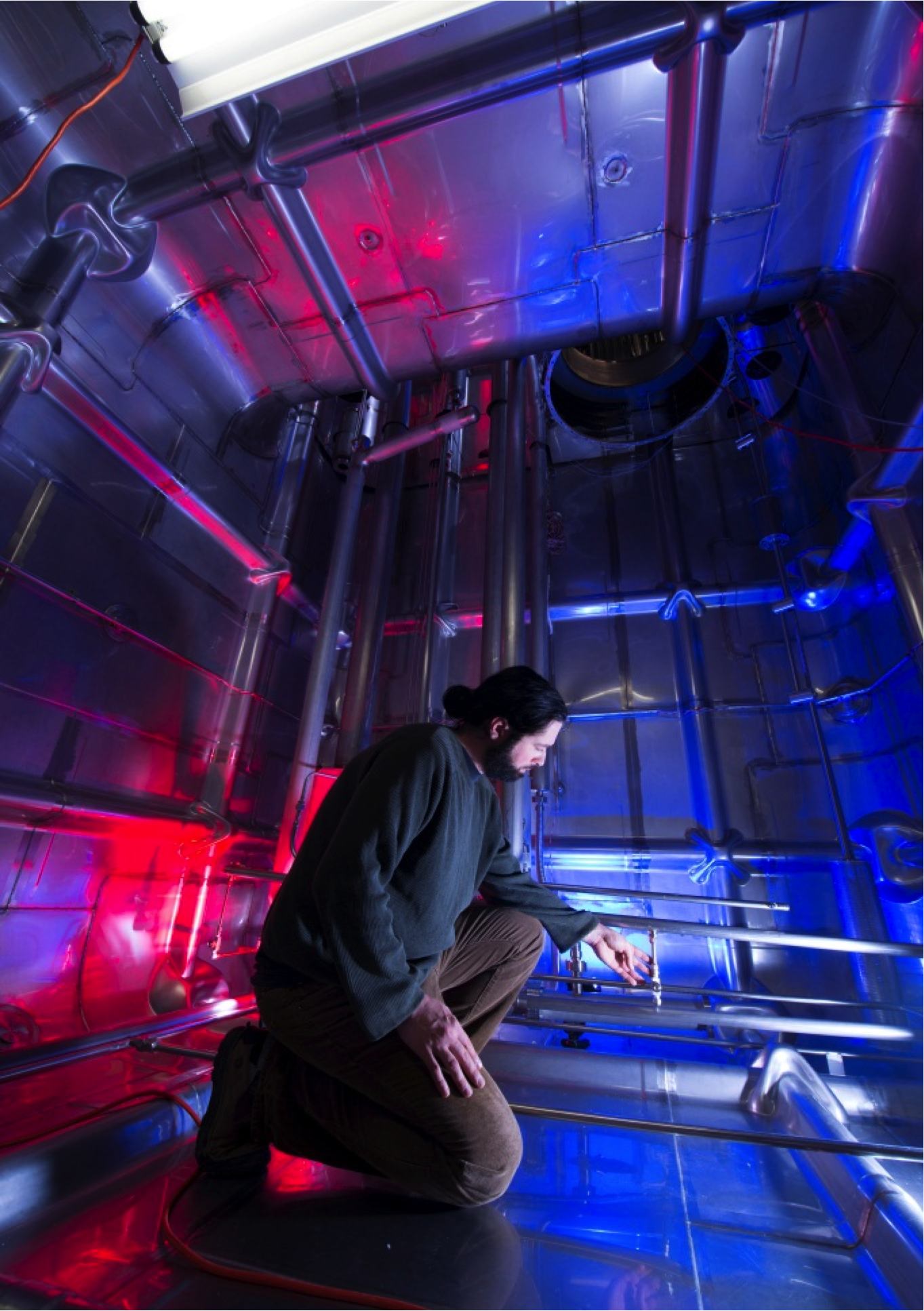}
\end{cdrfigure}

The techniques of membrane cryostat construction were demonstrated to
be suitable for high-purity LArTPC operations.  In particular, welding of
corrugated panels, removal of leak-checking dye penetrant and
ammonia-activated leak-detecting paints and post-construction-cleaning
methods were tested and found to be suitable.

As was demonstrated by LAPD, initial removal of impurities within the
cryostat can be achieved by purging with gaseous argon. Accordingly,
this was adopted for the 35-t.  Figure~\ref{fig:35TPurge} graphically
shows the first step of the purification process, removal of the
ambient air.  The initial state, $t=0$, reflects the initial values
for oxygen, water and nitrogen in the ``dry air'' state.
\begin{cdrfigure}[Gas argon purge and recirculation in 35-t cryostat]{35TPurge}{Progress of the gas argon purge as it removes impurities  from the 35-t. The shown quantities are measured by various gas analyzers. The first stage of the purification is a process called the ``Piston Purge''.  The second stage is ``Recirculation with Filtering.'' The gap between the two steps was due to troubleshooting a leak.}
\includegraphics[width=0.8\textwidth]{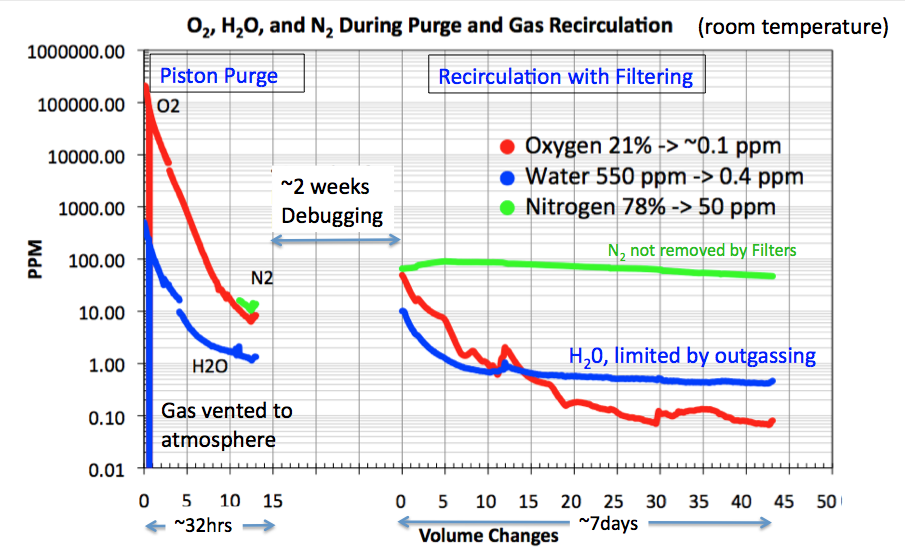}
\end{cdrfigure}

Once the room-temperature, gas purge ceased to improve purity,
the cooldown and LAr-fill stage began.  A gas/liquid spray method was
used to cool the cryostat.  This generated a turbulent mixing of cold
gas in the cryostat and cooled the entire surface.  The cooldown rate
was maintained within the limits specified by the cryostat
manufacturer.  Upon completion of the cooldown, LAr was transferred
into the cryostat and purification via recirculation loop started.

During recirculation and purification, dedicated purity monitors were
used to measure electron lifetime, which can be translated into
equivalent oxygen contamination levels.
Figure~\ref{fig:35TElectronLifetime} shows the electron lifetime from
the start of the LAr pump operation until the end of the Phase-1 run.
In general, the electron lifetime improved as a function of pump
on-time; despite several events that spoiled the lifetime, 
electron lifetimes of $>$2~ms were routinely achieved.
\begin{cdrfigure}[LAr electron lifetimes, 35-t cryostat]{35TElectronLifetime}{LAr electron lifetmes as measured by
Cryostat Purity Monitors. Significant events are annotated on the plot. Major divisions on horizontal axis
are one week periods. Equivalent purity levels are shown as dashed horizontal lines.}
\includegraphics[width=0.7\textwidth]{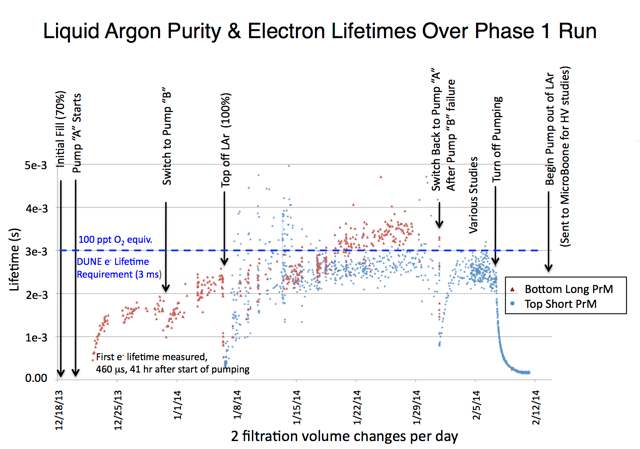}
\end{cdrfigure}

The 35-t Phase-1 run successfully demonstrated that there is nothing
inherent to membrane cryostat technology that would preclude achieving
the stated goals of the DUNE far detector. In addition, experience
gained in operating the 35-t system will inform future design
decisions, e.g., developing plans for replacing pumps in a way that
prevents loss of purity.  Future system designs could avoid the
coupling of acoustical vibrations into the cryostat by locating the
pumps externally; this would have the added benefit of facilitating
maintenance and repair.

\subsection{35-t Phase-2}
\label{sec:proto-35t-ph2}

Phase-2 of the 35-t prototype includes a fully operational TPC and
photon detector in the cryostat.  Commissioning of the TPC is expected
in August 2015 after TPC installation, purge, cooldown and LAr fill. \fixme{check}
Phase-2 operation is planned for several months of cosmic ray running.
External plastic scintillator paddles placed around the cryostat will
provide both the trigger and rough position measurements of the
incoming cosmic rays.  Figure~\ref{fig:35TTPC} shows the trial
assembly of the TPC outside of the cryostat along with a model of the
TPC inside the cryostat.
\begin{cdrfigure}[35-t cryostat with TPC]{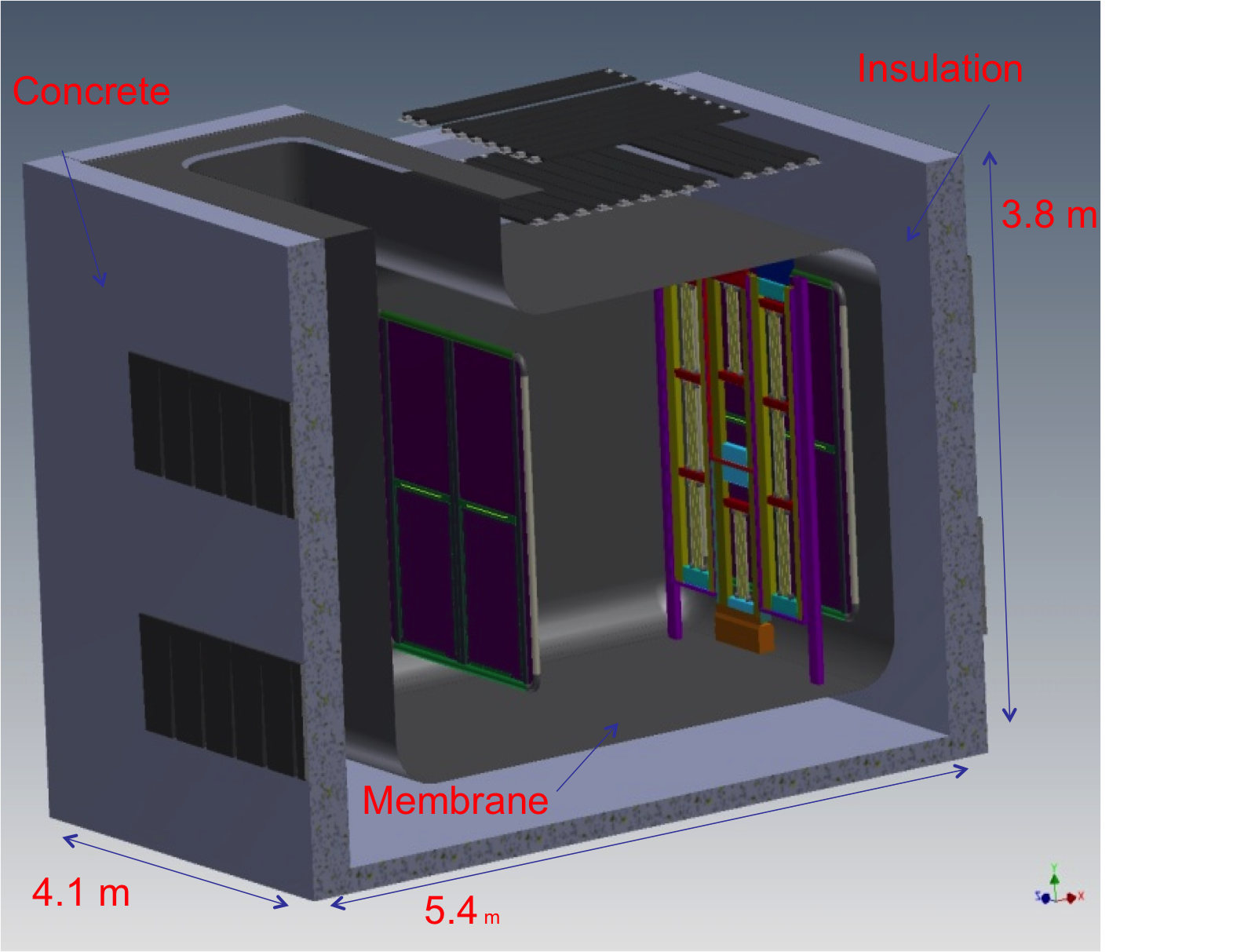}{(left) Rendering of the
35-t cryostat with TPC and photon detectors installed.  Note the
position of the APA, which is asymmetrically located between the CPAs
(purple) and splits the volume into one large and one very small %two
separate drift regions.  The length of the longer one of these is
close to what is proposed for the far detector.  The other has a
shorter drift length due to lack of space.  (right) A trial assembly
of the TPC.  } \includegraphics[width=0.55\textwidth]{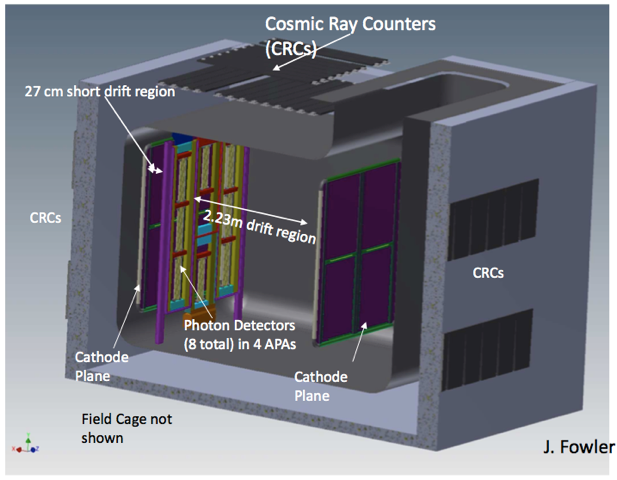}
  \includegraphics[width=0.35\textwidth]{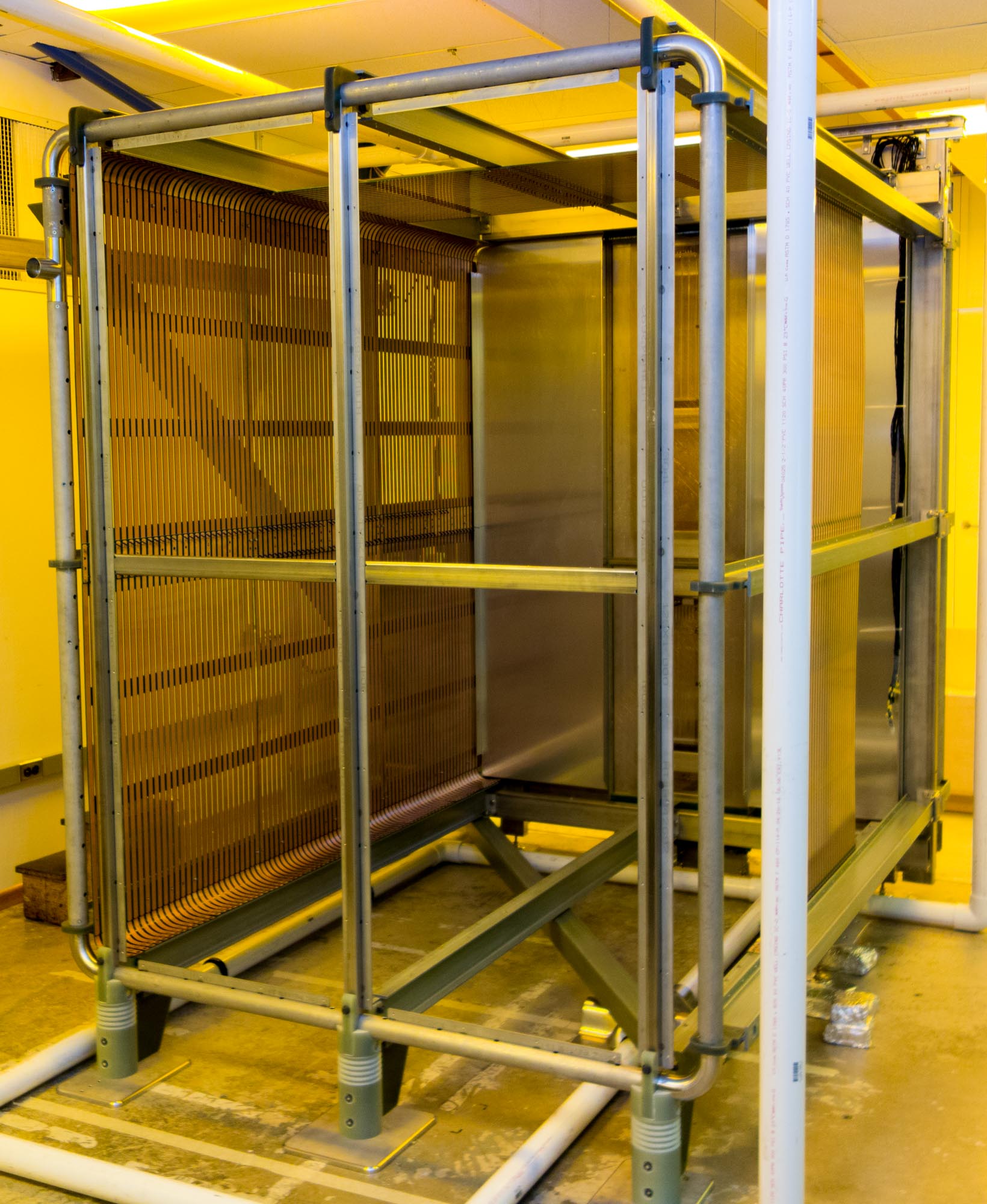}
\end{cdrfigure}

The Phase-2 prototype incorporates many novel DUNE single-phase far
detector design elements as described in previous sections of this
document and allows these to be tested in an operational TPC.  Some of
the more important aspects are collected in Table~\ref{tab:35TDesign}.
\begin{cdrtable}[35-t Prototype design elements]{lcl}{35TDesign}{35-t Design Elements}
 Design Aspect& Section & How Tested\\ \toprowrule
Modular APAs with wrapped wires & \ref{subsec:fd-ref-apa}&Build small-scale APA Modules with FD design\\
\colhline
Vertical Gaps between APAs &\ref{sec:detectors-fd-ref-tpc}& Assemble APAs side-by-side.\\
&&Study reco'd tracks that cross the gaps.\\ \colhline
Horizontal Gaps between APAs &\ref{sec:detectors-fd-ref-tpc}& Build two shorter APAs and stack vertically\\
&&Study reco'd tracks that cross the gaps\\ \colhline
Field cage constructed of &\ref{sec:detectors-fd-ref-tpc}&Operate at HV
and measure field uniformity\\
FR4 Printed Circuit Board \\ \colhline
APAs immersed in active volume &\ref{sec:detectors-fd-ref-tpc}& Study reconstructed tracks that cross APAs\\ \colhline
Cold Digital Electronics & \ref{sec:detectors-fd-ref-ce} & Measure noise performance etc. {\it in situ}\\ \colhline
Waveguide-style Photon Detector& \ref{sec:detectors-fd-ref-pd}&Install in APAs. Measure lightyield\\ \colhline
Triggerless-capable DAQ & \ref{sec:detectors-fd-ref-daq} & Take data using multiple DAQ modes\\
\end{cdrtable}
As can be seen from Table~\ref{tab:35TDesign}, successful tests of many of the new
design features will require simulation, reconstruction and analysis of 35-t data.
This will be performed using the LArSoft package, which is also used to simulate and
reconstruct data from ArgoNeuT, MicroBoone and LArIAT.
Reuse of software developed for those experiments will greatly facilitate 35-t developments;
however, the novel hardware features of the 35-t prototype necessitate new software developments, including:
\begin{itemize}
\item{code to divide the wrapped wires into as many as five individual linear segments.
A hit on a single electronic channel can, in principle, be related to a
signal on any of these segments.}
\item{``disambiguation'' code to identify which of the possible wire segments was actually responsible
for the observed hit.}
\item{code for determining the start time of the event ($t_0$). Since the 35-t prototype DAQ can
run ``triggerless,'' methods are needed for finding the $t_0$ in data. Information from the external
scintillator paddles as well as the internal photon detectors can be used.}
\item{Code for ``stitching'' together track segments observed in different tracking volumes.
Since hits can come from either side of the four APAs, there are
effectively eight separate tracking volumes,
which are treated as separate TPCs.}
\end{itemize}

With these simulation and reconstruction tools in hand, ``physics''
analysis of the data can be undertaken in the areas of validation of
new detector design elements and analysis of basic LArTPC performance.
Among the highest-priority analysis tasks are:
\begin{itemize}
\item{basic detector performance: signal/noise, purity measured with tracks, track direction resolution,
photon detector light yield,}
\item{measurement of distortions due to space charge and field non-uniformity, and}
\item{identification of different types of particles: muons, protons, neutrons, pions.}
\end{itemize}

The results obtained by operating and analyzing data from the 35-t
Phase-2 prototype are expected to be very valuable in refining the
CERN single-phase prototype design, in preparation for the first
\ktadj{10} DUNE far detector module.

%% file: volume-detectors/proto-sections/proto-cern-single.tex
%%%%%%%%%%%%%%%%%%%%%%%%%%%%%%%%
\section{The CERN Single-Phase Prototype}
\label{sec:proto-cern-single}

A CERN single-phase prototype detector and accompanying beam-test
program is in preparation. As an \textit{engineering} prototype, it is
intended to validate the construction of the components planned for the
first DUNE \ktadj{10} detector module at scale and thereby mitigate
risks associated with extrapolating small-scale versions of the
single-phase LArTPC technology to a full 10-kt detector module.  It is
intended to benchmark the operation of full-scale detector
elements and perform measurements in a well characterized
charged-particle beam --- an essential step.

The prototype will incorporate components with the same
dimensions and features as those for the first 10-kt DUNE far detector
module.

\subsection{Program of Tests and Measurements}

Besides validating the performance of the detector components,
planning and constructing the CERN prototype will establish and
commission production sites and test the installation procedure.
Further, before the beam test, many basic detector-performance
parameters can be established with cosmic-ray muons.  These data will
aid in identification of potentially problematic components, leading
to future improvements and optimizations of the detector design.  Once
it is exposed to a test beam of charged particles of different types
and energies it will collect data that can be combined with results
from LArIAT and the short-baseline program at Fermilab.  Together
these measurements will be used to validate MC simulations, and they
will serve as data input to DUNE sensitivity studies and allow
validation and tuning of tools for event reconstruction and particle
identification.  The following detector performance measurements are
anticipated:
 \begin{itemize}
 \item characterize performance of a full-scale TPC module,
 \item study performance of the photon detection system,
 \item test and evaluate the performance of detector calibration tools (e.g., the laser system),
  \item verify functionality of cold TPC electronics under LAr cryogenic conditions,
  \item perform full-scale structural test under LAr cryogenic conditions,
  \item verify argon contamination levels and associated mitigation procedures,
  \item develop and test installation procedures for full-scale detector components, and
  \item identify flaws and inefficiencies in the manufacturing process.
\end{itemize}

The physics sensitivity of the DUNE experiment has so far been
estimated based on detector performance characteristics published in
the literature, simulation-based estimates and a variety of
assumptions about the anticipated performance of the future detector
and event reconstruction and particle-identification algorithms.  This
engineering prototype and the test beam measurements aim to replace
these assumptions with measurements to use for full-scale DUNE
detector components and the algorithms and thereby enhance the
accuracy and reliability of the DUNE physics-sensitivity projections.
The collection of beam measurements will serve both as a calibration
data set for tuning the MC simulations and as a reference data set for
the future DUNE detector.
In order to make precise measurements, the DUNE 
detector will
need to accurately identify and measure the energy of the particles
produced in the neutrino interaction with argon, which will range from
hundreds of MeV to several GeV.

More specifically, the goals of the prototype detector beam-test measurements include
the use of a charged-particle beam to
\begin{itemize}
\item measure the detector calorimetric response for
\begin{itemize}
	\item hadronic showers and
	\item electromagnetic showers;
\end{itemize}
\item study e/$\gamma$-separation capabilities;
\item measure event reconstruction efficiencies as function of energy
  and particle type;
\item measure performance of particle identification algorithms as
  function of energy for realistic detector conditions;
\item assess single-particle track calibration and reconstruction;
\item validate accuracy of MC simulations for relevant particle energy and orientation; 
\item study other topics with the collected data sets:
 \begin{itemize}
    \item pion interaction kinematics and cross sections, 
    \item kaon interaction cross section to characterize proton decay backgrounds, and
    \item muon capture for charge identification.
 \end{itemize}
\end{itemize}
A detailed enumeration of the desired minimum integrated particle
counts as a function of charged-particle species and momentum is
nearing completion. This has led to development of a run plan based on
realistic beam composition, particle energies and efficiency
information.

An invited technical proposal for the CERN single-phase detector and
beam-test program
%\cite{CERN_single-phase_proposal} 
was submitted
to the CERN SPSC in September 2015. This proposal is \anxcernproto~\cite{cdr-annex-singleph-proto}. The plan includes a first beam run in
2018 before the long shutdown of the LHC. Experience gained from
construction, installation and commissioning of this prototype, as
well as performance tests with cosmic-ray data are expected to lead to
an optimization of corresponding phases of the DUNE single-phase far
detector module(s).

\subsection{Detector Configuration and Components}

As mentioned above, the prototype detector components have 
the same dimensions and features as those of the far detector reference design described in
Chapter~\ref{ch:detectors-fd-ref}. This includes the TPC and photon
detector components, as well as their positioning and spacing within
the cryostat.

\subsubsection{TPC Configuration}

The size of the prototype is in large part determined by the
requirement to fully contain hadronic showers of up to several GeV in
energy.  The particle containment of hadronic showers initiated by
charged pions or protons is a critical feature for calorimetric
measurements. Simulation studies indicate that showers initiated by
\GeVadj{10} primary pions and protons are contained within a volume
measuring 6~m in the longitudinal and 5 $\times$ 5~m$^2$ in the
transverse directions. With the basic APA unit measuring
6 $\times$ 2.3~m$^2$, the arrangement identified as satisfying the
requirement consists of two times three APAs side-by-side, a central
cathode and two drift volumes each with 3.6~m drift
length. Figure~\ref{fig:CERN_single_TPC} shows a view of the CERN
single-phase TPC along with the field cage and a view of the TPC
within the cryostat.
\begin{cdrfigure}[Cutaway view  of the CERN single-phase prototype TPC]{CERN_single_TPC}{View of the CERN single-phase detector TPC (left) and inserted in the cryostat (right). }
\includegraphics[width=0.40\textwidth]{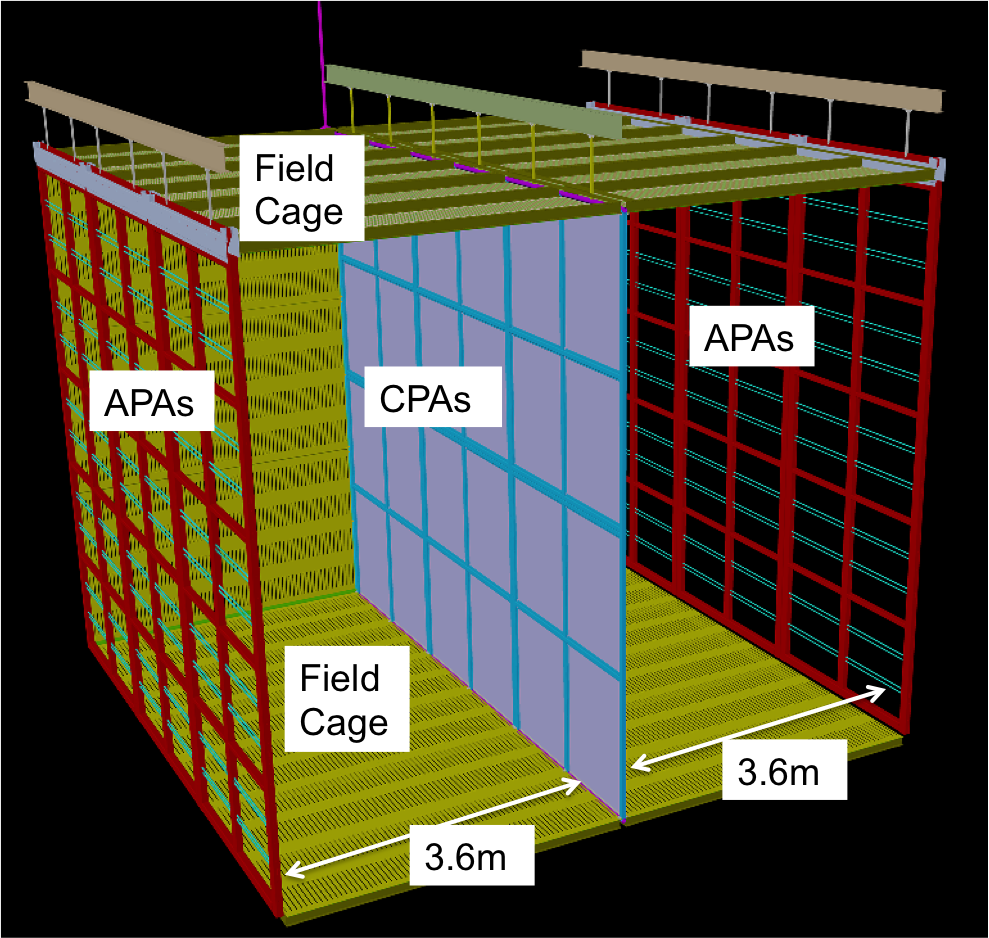}
\includegraphics[width=0.59\textwidth]{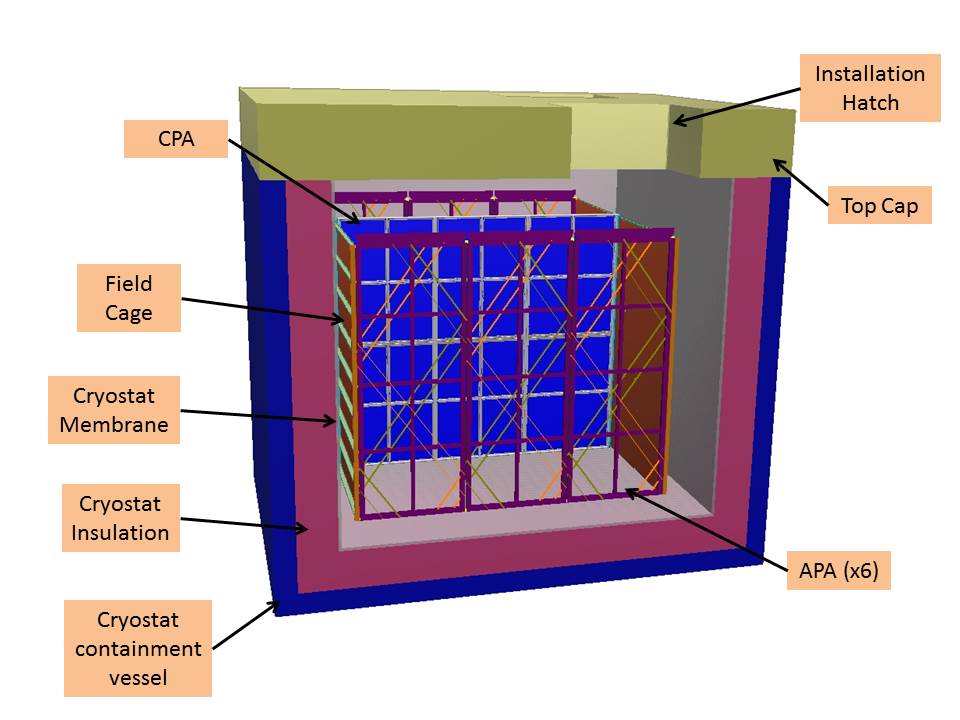}
\end{cdrfigure}
The TPC readout, photon-detection system, DAQ, slow control and
monitoring, as well as the key issues of the installation procedure
are described in corresponding sections of
Chapter~\ref{ch:detectors-fd-ref}.

\subsubsection{Cryostat}

The CERN prototype uses a membrane tank technology with internal
dimensions of 7.8~m (tranverse) $\times$ 8.9~m (parallel) $\times$ 8.1~m
(height).  It contains 725 tons of LAr, equivalent to about
520~m$^3$ (where the remaining volume contains the gas ullage). 
The active (fiducial) detector mass of LAr amounts to
400~tons (300~tons).  The external cryostat dimensions are 10.6~m
(tranverse) $\times$ 11.7~m (parallel) $\times$ 10.9~m (height).

The cryostat design is scaled up from the 35-t prototype
cryostat\cite{bib:membcryo1573}, described in
Section~\ref{sec:proto-35t}.  Unlike the 35-t cryostat, it uses a
steel outer supporting structure with an inside metal liner.  It is
similar to the WA105 dual-phase prototype detector cryostat and to
that for the Fermilab Short-Baseline Near Detector
(SBND)\cite{bib:SBND}.  The support structure rests on I-beams to
allow for air circulation underneath the cryostat; this maintains the
temperature within the allowable limits.  A stainless-steel membrane
contains the LAr within the cryostat. The pressure loading of the
cryogenic liquid is transmitted through rigid foam insulation to the
surrounding outer support structure. The membrane is corrugated to
provide strain relief resulting from temperature-related expansion and
contraction. The cryostat top cap consists of the same layers as the
cryostat walls.  From the inside out, the layers include the
stainless-steel primary membrane, intermediate insulation layers and
vapor barrier; they all continue across the top of the detector
providing a leak-tight seal.  The cryostat roof is a removable steel
truss structure to which stiffened steel plates are welded from the
underside. They form a flat vapor-barrier surface onto which the roof
insulation attaches directly.

The truss structure rests on the top of the supporting structure where
a positive structural connection between the two is made in order to
resist the upward force caused by the slightly pressurized argon in
the ullage space. The hydrostatic load of the LAr in the cryostat is
carried by the floor and the sidewalls. In order to meet the maximum
deflection of 3~mm between APA and CPA and to decouple the detector
form possible sources of vibrations, the TPCs are connected to an
external bridge over the top of the plate supported on the floor of
the building. Everything else within the cryostat (electronics,
sensors, cryogenic and gas plumbing connections) is supported by the
steel plates under the truss structure.

All piping and electrical penetrations into the interior of the
cryostat are made through the top plate.  Penetrations are clustered
in one region.  The top cap has two large openings for TPC
installation, and a manhole to allow entry into the tank after the
hatches have been closed.

\subsubsection{Cryogenics System}

The main goals of the cryogenics system are to purge the cryostat
prior to the start of the operations (with argon gas in open and
closed loops), cool the cryostat and fill it with LAr.  The LAr is
continuously purified and the boil-off argon gas is captured,
recondensed and purified.  The design calls for a 10-ms electron
lifetime (30 ppt O$_2$ equivalent), a quantity that is measured by the
detector.

The LAr-receiving facility includes a storage dewar and ambient
vaporizer to deliver LAr and gaseous Ar to the cryostat. The LAr goes through
the LAr handling and purification system, whereas the gaseous Ar goes through
%the gaseous Ar 
its own purification system before entering the cryostat.  Studies are
ongoing to standardize the filtration scheme and select the optimal
filter medium for both the prototype and future detectors.

During operation, an external LAr pump circulates the bulk of the
cryogen through the LAr purification system. The nominal LAr
purification flow rate completes one full volume exchange in 5.5~days.
The boil-off gas is recondensed and sent to the LAr
purification system before re-entering the vessel.

The proposed LAr cryogenics system is based on that of the 35-t
prototype, MicroBooNE and SBND, %Short-Baseline Near Detector,
and the current plans for the DUNE single-phase far detector module.

%% file: volume-detectors/proto-sections/proto-wa105.tex
%%%%%%%%%%%%%%%%%%%%%%%%%%%%%%%%
\section{The WA105 Dual-Phase Demonstrator}
\label{sec:proto-cern-double}

In recent years, two consecutive FP7 Design Studies
(LAGUNA/LAGUNA-LBNO) have led to the development of a conceptual
design (fully engineered and costed) for a 20-kt/50-kt GLACIER-type
underground neutrino detector. In these studies, an underground
implementation has been assumed {\textit{ab initio} 
and such constraints have been important and taken into account in
design choices. The LAGUNA-LBNO design study, completed in August
2014, has produced many technnical developments focused on the
construction of large and affordable liquid argon underground
detectors addressing the complete investigation of three-flavor
neutrino oscillations and the determination of their still unknown
parameters.
These detectors will be very powerful for non-beam studies as well,
such as proton decay, atmospheric neutrinos and supernova neutrinos.

The WA105 experiment, approved in 2013, is designed to provide a full-scale demonstration
of these technological developments. It will be exposed to a beam of
charged hadrons/electrons/muons of 0.5--20~GeV/c to characterize the
detector response to hadronic and electromagnetic showers.  A detailed
description of the experiment is available in the Technical Design
Report of 2014, \anxdualtdr~\cite{WA105_TDR},  
and an up-to-date picture of
technical developments can be found in the Status Report\cite{WA105_SREP}
submitted to the SPSC CERN committee in March
2015. These developments form the basis of the
alternative far detector design, described in
Chapter~\ref{ch:detectors-fd-alt}.

The WA105 demonstrator is a dual-phase LArTPC with an active volume of
6 $\times$ 6 $\times$ 6~m$^3$.

These dimensions are motivated by the 4 $\times$ 4~m$^2$ Charge Readout
Plane (CRP) units that are the basic readout components of the
large-scale LAGUNA/LBNO 20--50-kt detectors.
The 6 $\times$ 6 $\times$ 6~m$^3$ active volume is consistent with a fiducial
volume that accommodates the CRP size and provides full containment of
hadronic showers.
Surface operation prohibits drift lengths above 6~m. The footprint of
the active volume corresponds to 1:20 of the surface of the LBNO \ktadj{20}
detector. The active volume contains about 300 tonnes of LAr. The
important parameters of the detector are presented in
Table~\ref{tab:demo_para} and Figures~\ref{fig:6by6_open},~\ref{fig:6by6_plan}
and~\ref{fig:6by6_vert} provide a 3D drawing and two cut views.
\begin{cdrtable}[Parameters for the WA105 demonstrator]{lcc}{demo_para}{Parameters for the WA105 demonstrator}
Parameter & Units & Value \\ \toprowrule
Liquid argon density & t/m$^3$& 1.38 \\ \colhline
Liquid argon volume height & m& 7.6 \\ \colhline
Active liquid argon height& m  & 5.99 \\ \colhline
Hydrostatic pressure at the bottom& bar & 1.03 \\ \colhline
Inner vessel size (WxLxH) &m$^3$ & 8.3 $\times$ 8.3 $\times$ 8.1\\ \colhline
Inner vessel base surface& m$^2$& 67.6 \\ \colhline
Total liquid argon volume& m$^3$ & 509.6 \\ \colhline
Total liquid argon mass & t & 705 \\ \colhline
Active LAr area & m$^2$& 36 \\ \colhline
Charge readout module (0.5 x 0.5 m$^2$) & & 36\\ \colhline
N of signal feedthrough & & 12 \\ \colhline
N of readout channels & & 7680\\ \colhline
N of PMT & & 36 \\
\end{cdrtable}
\begin{cdrfigure}[Illustration of the WA105  6 $\times$ 6 $\times$ 6~m$^3$ demonstrator with inner detector]{6by6_open}{Illustration of the  6 $\times$ 6 $\times$ 6~m$^3$  demonstrator with the
detector inside the cryostat}
\includegraphics[width=0.8\linewidth]{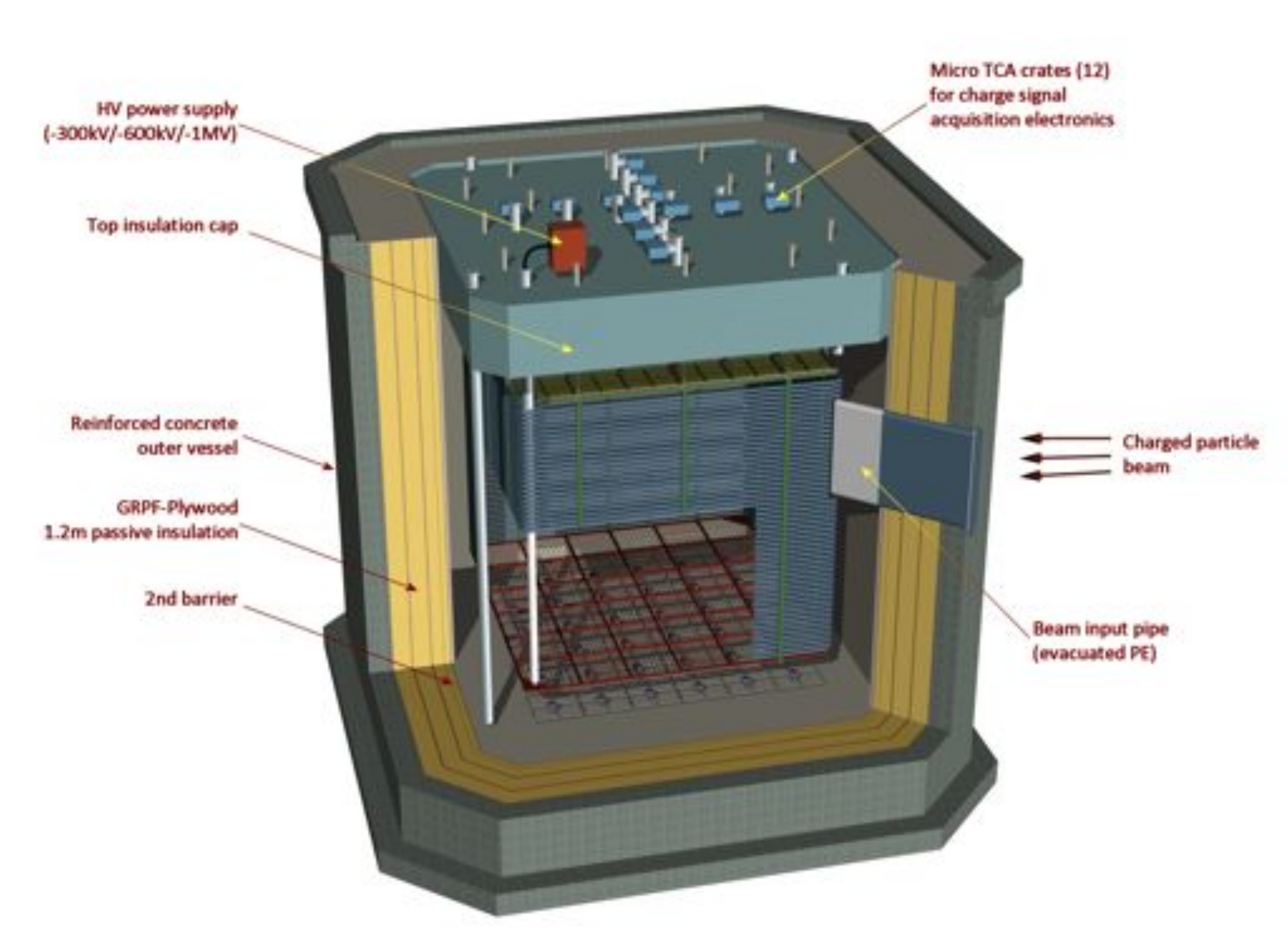}
\end{cdrfigure}
\begin{cdrfigure}[Plan view section of the WA105 6 $\times$ 6 $\times$ 6~m$^3$ demonstrator]{6by6_plan}{\small Plan view section of the $6\times 6\times 6 $~$m^3$ demonstrator}
\includegraphics[width=0.7\linewidth]{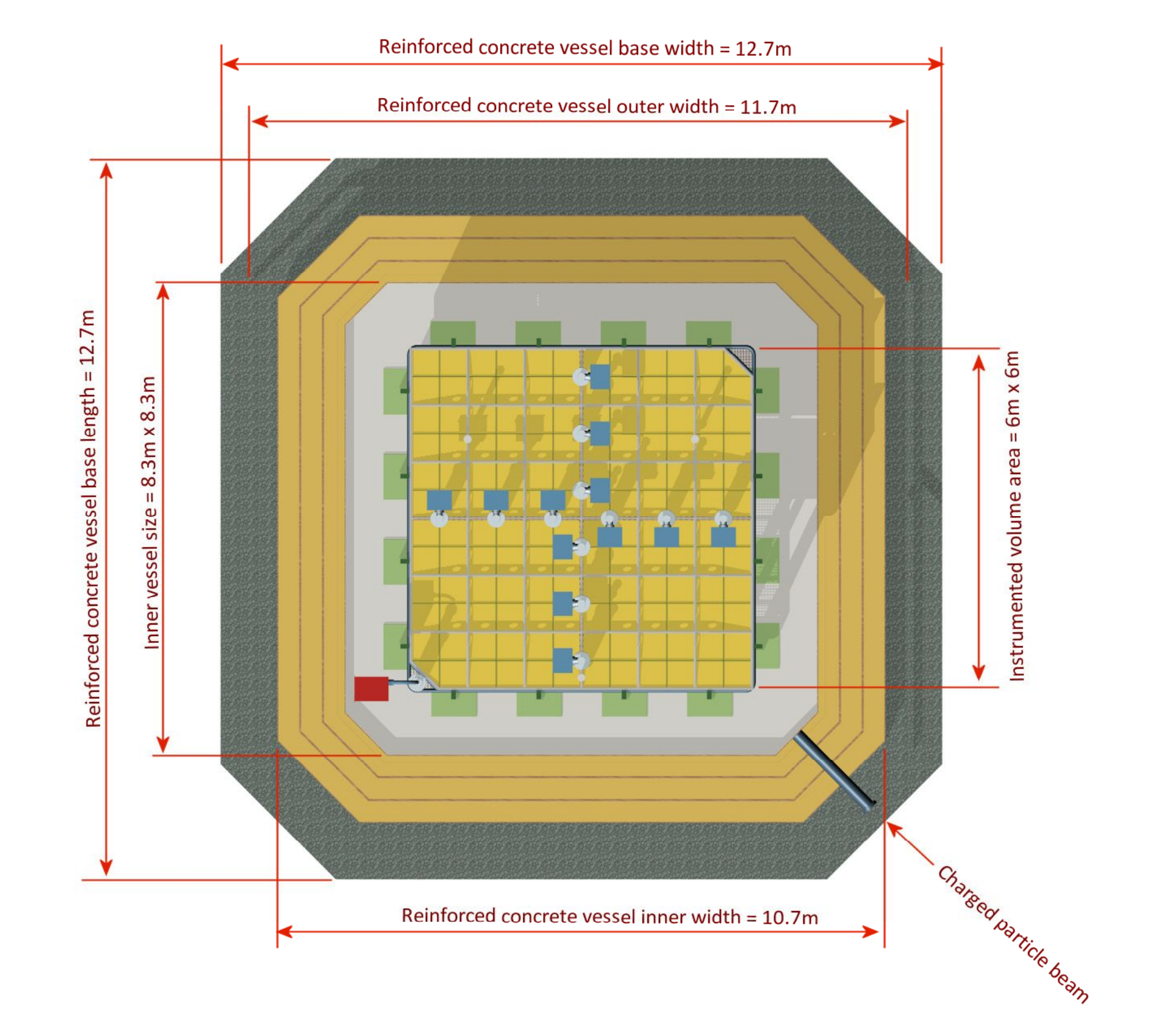}
\end{cdrfigure}
\begin{cdrfigure}[\small Vertical cross section of the  WA105 6 $\times$ 6 $\times$ 6~m$^3$ demonstrator]{6by6_vert}{\small Vertical cross section of the 6 $\times$ 6 $\times$ 6~m$^3$ demonstrator}
\includegraphics[width=0.7\linewidth]{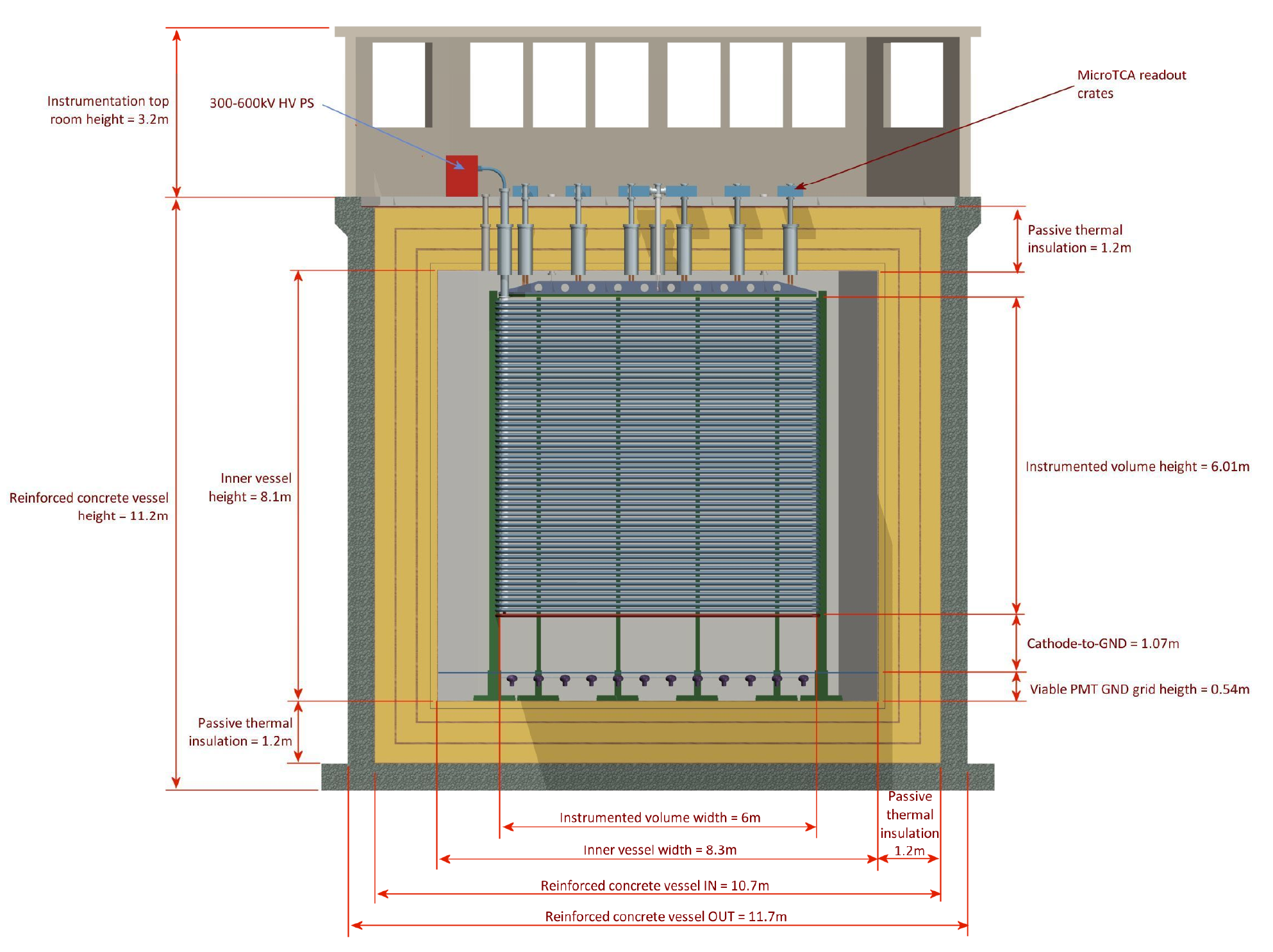}
\end{cdrfigure}

The dual-phase LArTPC design drifts ionization electrons
vertically through the LAr in a uniform electric field up to the
liquid-vapor interface, where they are extracted from the liquid into
the gas phase.

In the gas, the Charge Readout Plane (CRP) described in
Section~\ref{sec:detectors-fd-alt-chg-readout}, 
extracts, multiplies and collects the charge. 

The drift path in the WA105 demonstrator reaches 6~m. The detector
will operate with drift fields of 0.5~kV/cm and 1~kV/cm, corresponding
to cathode voltages of $-$300--600~kV. 600~kV is the voltage also
needed for the operation of the DUNE dual-phase detector module with 12~m drift at
0.5~kV/cm. The CRP has an active surface of 36~m$^2$, with its anode subdivided into
strips of 3.125-mm pitch and 3-m length, for a total of \num{7680}
readout channels.

The WA105 demonstrator will establish the techniques developed for
the 20/50-kt LBNO detectors, in particular:
\begin{itemize}
\item{tank construction technique based on the LNG industry with non-evacuated vessel,}
\item{purification system,}
\item{long drift,}
\item{HV system 300--600~kV, large hanging field cage,}
\item{large area double-phase charge readout,}
\item{accessible cryogenic front-end electronics and cheap DAQ electronics, and}
\item{long-term stability of UV light readout.}
\end{itemize}

Furthermore, the 6 $\times$ 6 $\times$ 6~m$^3$ demonstrator exposed to the
test beam promises a rich physics program to:
\begin{itemize}
\item{assess detector performance in reconstructing hadronic showers; the most demanding task in neutrino interactions;}
\item{measure hadronic and electromagnetic calorimetry and PID performance;}
\item{full-scale software development, simulation and reconstruction;}
\item{collect high-statistic hadronic interaction samples with unprecedented granularity and resolution for the study of hadronic interactions and nuclear effects;}
\item{assess physics capabilities of the dual-phase versus
  single-phase performance, in particular: high S/N, 3-mm pitch,
  absence of materials in long drift space, two collection views, no
  ambiguities; and}
\item{study systematics related to
  the reconstruction of the hadronic system (resolution and energy
  scale), electron-identification efficiencies and $\pi^0$ rejection and particle $dE/dx$ identification for proton decay.}
\end{itemize}

The 6 $\times$ 6 $\times$ 6~m$^3$ WA105 detector is expected to start 
taking data in 2018 in the EHN1 Hall extension currently under
construction at CERN.  The detector components are in an advanced
state of design/prototyping, or preproduction. 
Completion of the WA105 detector design and the
preparation of its construction have been progressing very quickly.
Many technical design details are benefitting from the
implementation of a 20-t (3 $\times$ 3 $\times$ 1~m$^3$) prototype (see Figure~\ref{fig:3by1}),
which corresponds to the readout cell size for the DUNE dual-phase detector module. 
\begin{cdrfigure}[Exploded view of the WA105 3 $\times$ 3 $\times$ 1~m$^3$  prototype]{3by1}{Exploded view of the 3$\times$3$\times$1~m$^3$  prototype}
\includegraphics[width=0.95\linewidth]{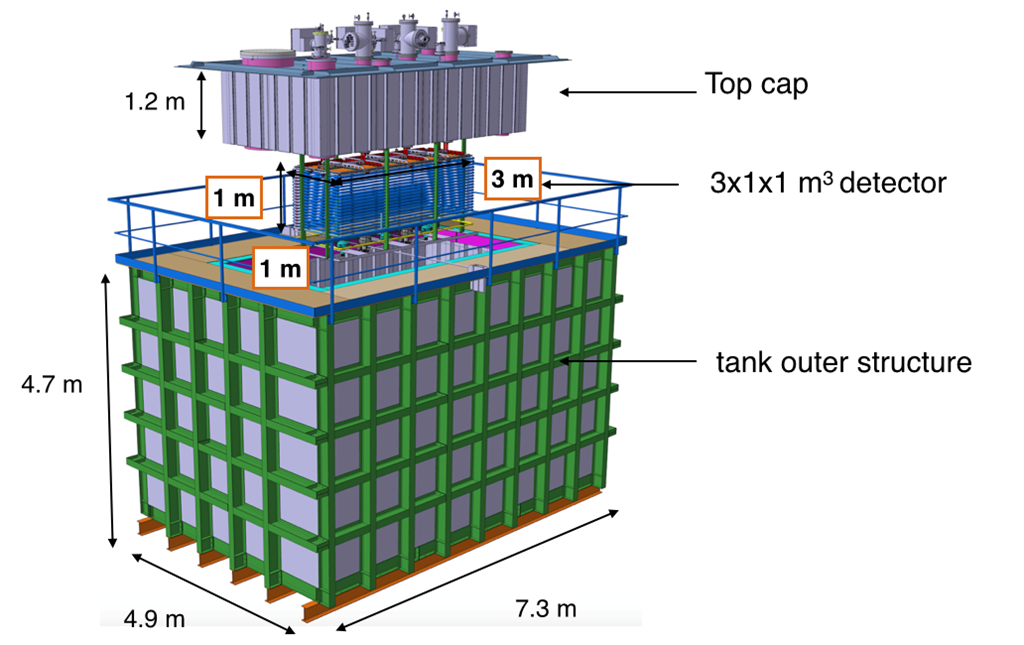}
\end{cdrfigure} 
Development of this 20-t prototype has verified
the complete system integration: 
production of fully engineered prototype versions of many detector
parts, including installation details and ancillary services;
establishment of the Quality Assessment (QA) procedures for the
construction, installation and commissioning chains; establishment of
the procurement processes for the major detector components; and
validation of the cost and schedule estimates for WA105.  The
3$\times$3$\times$1~m$^3$ 20-t prototype represents a technical test
bed and integration exercise to accelerate the design, procurement, QA
and commissioning activities needed for the WA105 demonstrator.
In particular, a complete procedure for construction of
GTT-licensed corrugated membrane cryostats 
been established at CERN and a full chain for the procurement,
processing, assembly and commissioning of the LEM detectors and 
anodes has been implemented.

%% file: volume-detectors/proto-sections/proto-nd.tex
%%%%%%%%%%%%%%%%%%%%%%%%%%%%%%%%

\section{Near Detector Prototypes}
\label{sec:proto-nd}
Near detector prototypes are planned for both the neutrino detector and beamline measurements systems.

\subsection{Near Neutrino Detector Prototypes}
\label{sec:proto-nd-nnd}
The prototyping plan for the Fine-Grained Tracker (FGT) near neutrino
detector, as described in Section~\ref{cdrsec:detectors-nd-ref},
involves the following major steps:
\begin{itemize}
\item Straw-Tube Tracking detector (STT) prototyping,
\item ECAL prototyping,
\item MuID -- RPC development, and
\item Dipole magnet studies.
\end{itemize}

A schematic of the FGT is given in Figure~\ref{fig:STT_schematic}. The
prototyping activity for the reference design will be developed
jointly by the participating collaborators in India, with some
contributions from institutions in the U.S. or other countries.  The
prototyping work is spread over a duration of three years. The plan is
detailed below.

\subsubsection{Straw-Tube Tracking Detector}

The proposed Straw-Tube Tracking (STT) detector design
provides the central active tracking of the FGT and use straws of
1-cm diameter fabricated from an inner carbon-loaded Kapton (XC) wall
and a second aluminum-coated outer Kapton (HN) wall. The details of the
STT design are available in
Section~\ref{cdrsec:detectors-nd-ref-fgt-stt}. The prototype design
has two layers of 60 straws each.  The straws will have the
same dimensions as listed in
Section~\ref{cdrsec:detectors-nd-ref-fgt-stt}, but half the nominal
length, i.e., 1.8~m. The major milestones in the STT prototyping
are highlighted below.

%\textbf{\textit{Design and Fabrication of STT Prototype}} \\
The three-year STT R\&D and prototyping phase will start with the 3D
design of a prototype module.  This will include optimization of
parameters for the prototype assembly and validation of the mechanical
structure using Finite Element Method (FEM) techniques. This process
will be a self-feeding system with input from GEANT detector
simulations.  Figure~\ref{fig:STT_SimulationG4} shows the STT
simulation.  
The STT prototype will be fabricated and will undergo
extensive tests both in the laboratory and 
in particle beams at CERN.
\begin{cdrfigure}[GEANT4 simulation of 1-cm straws for the STT prototype]
{STT_SimulationG4}{GEANT4 simulation of 1~cm straws for the STT prototype}
\includegraphics[width=0.6\textwidth]{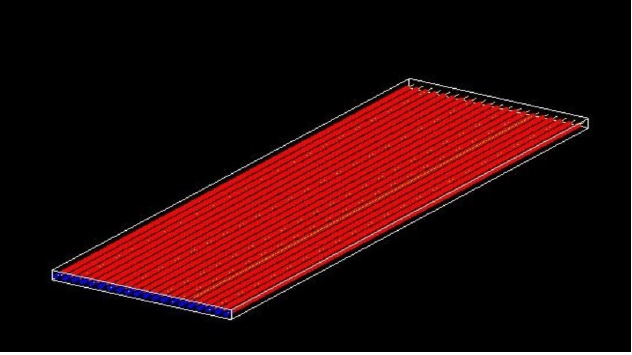}
\end{cdrfigure}

%\textbf{ \textit{Design and prototyping of radiators and nuclear targets}} \\
As described in Section~\ref{cdrsec:detectors-nd-ref-fgt-stt}, a key
feature of the STT is the capability to integrate a series of
nuclear targets for (anti)neutrino interactions.  The main
target is provided by the radiators that are made of thin polypropylene foils
(Figure~\ref{fig:STT_Detail}).  The design of the radiator targets has
been optimized with simulations of the Transition Radiation (TR)
with emphasis on integration into the mechanical structure of
the STT modules.  The production and design of the plastic foils was
discussed with vendors and a half-scale
(1.8~m$\times$1.8~m) prototype of the radiator targets will be produced to demonstrate
assembly, mechanical properties and overall
performance.
A preliminary design has been developed for the
pressurized Ar gas target (Figure~\ref{fig:STT_ArTargets}), based on
the use of 0.5-in diameter stainless steel tubes.
Prototypes of the pressurized Ar gas tubes will be built to 
optimize the design of the tubes, including the tube diameter
and the possibility of using carbon-composite tubes. 
Construction of small-scale prototypes of the Ca and C targets is also planned.
\begin{cdrfigure}[Design of the radiator target plane with pressurized Ar gas for STT]
{STT_ArTargets}{Design of the radiator target plane with pressurized Ar gas for STT}
\includegraphics[width=0.6\textwidth]{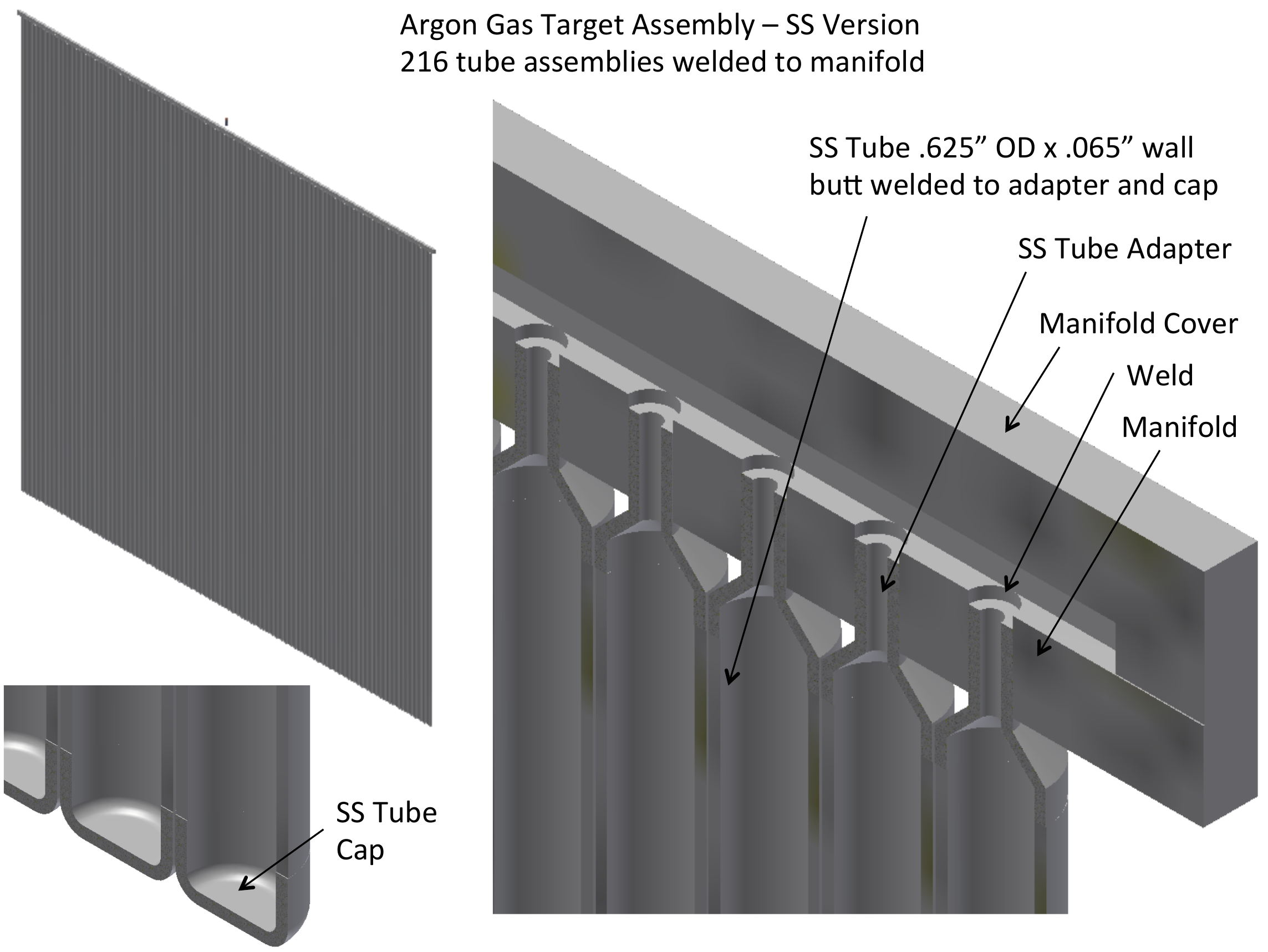}
\end{cdrfigure}

%\textbf{ \textit{Anode Wire Studies}} \\
The straw tube sense wires were initially proposed as 30-$\mu$m
diameter gold-plated tungsten, similar to the COMPASS design. In order
to minimize the material budget of the mechanical frames used for the
STT modules, it is important to reduce the wire tension. To this end,
the prototyping includes a detailed study of the possibility of using
20-$\mu$m wires instead of the default 30-$\mu$m. The tensile strength
of these wires inside the straw tubes could affect the signal
generation over a long period due to sagging; a detailed study is in
progress.  Figure~\ref{fig:STT_WireTensionTests} shows the tension
measurement results for 20--30~$\mu$m wires using the
induced-resonance method.  The proposed tension limit on the sense
wires is 70~g.
\begin{cdrfigure}[STT wire-tension measurement studies (20~$\mu$m and 30~$\mu$m)]
{STT_WireTensionTests}{STT wire-tension measurement studies for 20~$\mu$m and 30~$\mu$m.}
\includegraphics[width=1.0\textwidth]{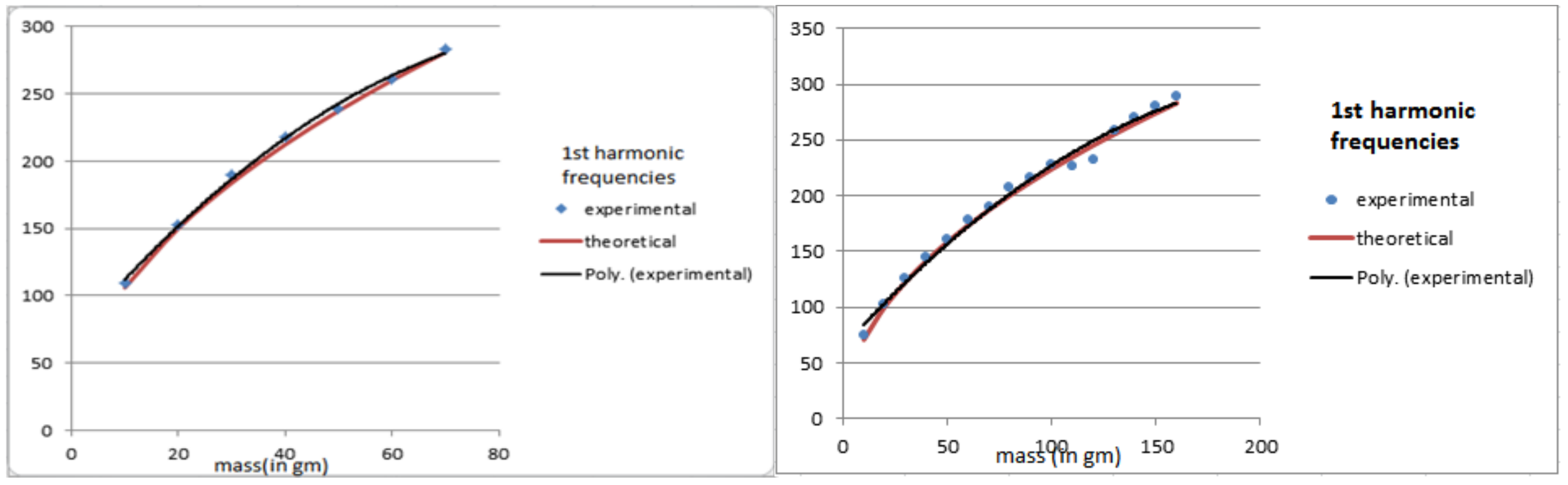}
\end{cdrfigure}

%\textbf{ \textit{Test Straw Chamber}} \\
A test chamber with 48 straws of the same dimensions as those for the
FGT but with 1-m length has been built and is available for
operational studies aimed at understanding the gas flow rates and
finalizing the preamplifier selection parameters.
Figure~\ref{fig:STT_TestStrawChamber} shows a signal pulse with
Ar+CO$_2$ (80:20) gas taken with cosmic rays. The voltage versus
amplitude for one of the straws is also shown in
Figure~\ref{fig:STT_TestStrawChamber} to establish the QA/QC procedure
for the fabricated straws.
\begin{cdrfigure}[Pulse and voltage-amplitude for the test STT chamber]
{STT_TestStrawChamber}{Example pulse from the test straw chamber (left) and
measurement of voltage vs. amplitude for one of the straws in the test chamber (right).}
\includegraphics[width=1.0\textwidth]{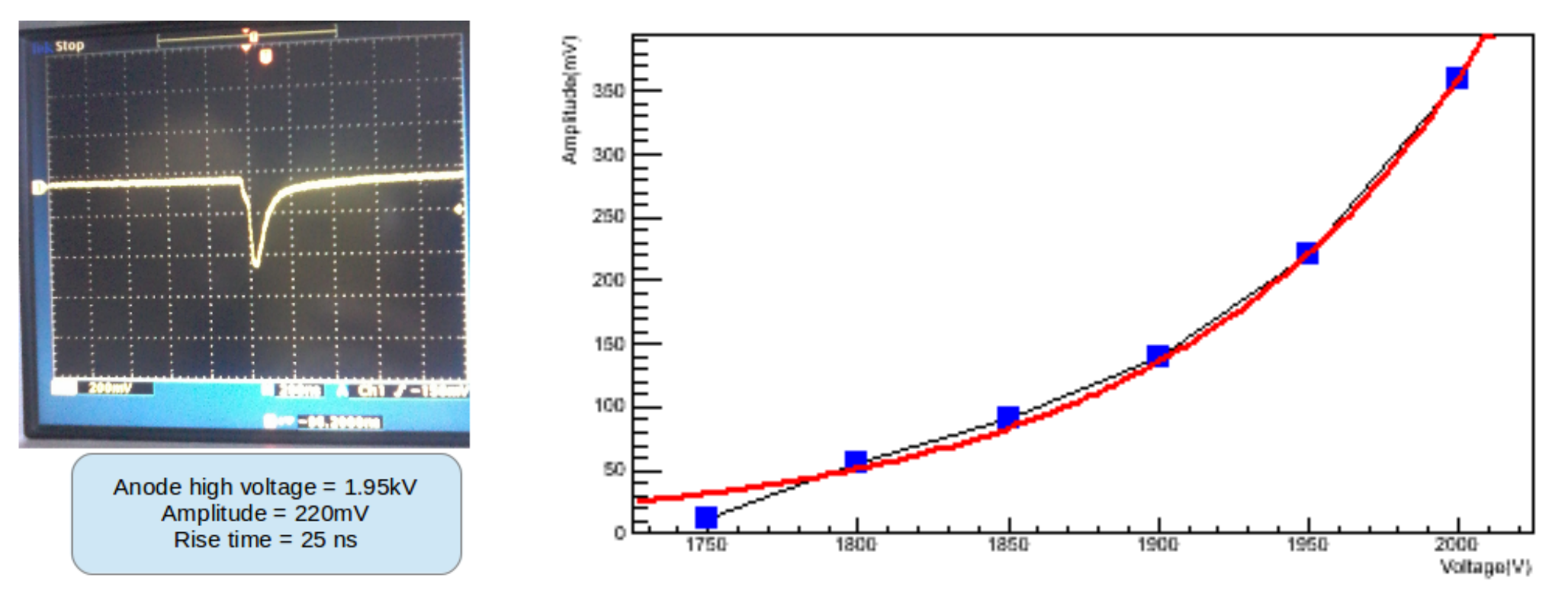}
\end{cdrfigure}

%\textbf{ \textit{Front-end Electronics}} \\
We are performing tests of prototype electronics for the signal
readout.  A four-channel preamplifier has been tested with the test
chamber using a radioactive source and the signal has been recorded as
shown in Figure~\ref{fig:STT_SignalPreAmp}.
\begin{cdrfigure}[Signal from single straw (STT) using the BARC preamp and source]
{STT_SignalPreAmp}{Signal from single straw using the BARC (Bhabha Atomic Research Center) preamp and source.}
\includegraphics[width=0.6\textwidth]{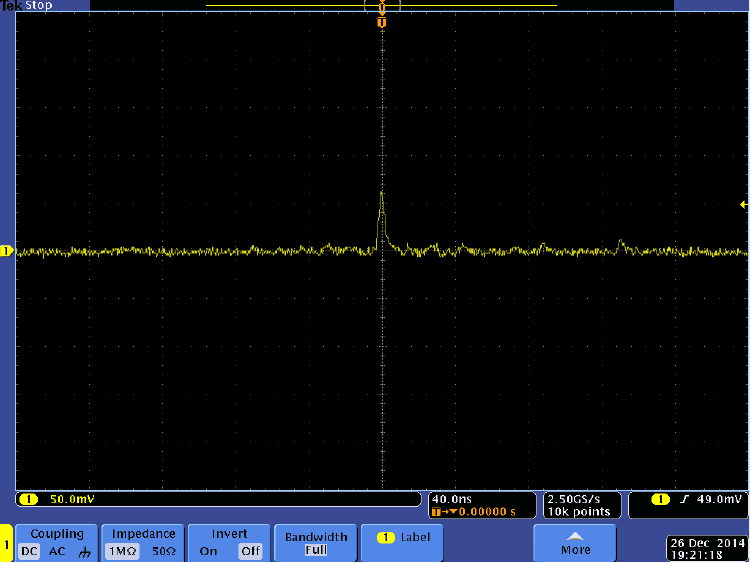}
\end{cdrfigure}
The back-end DAQ is still being worked out and would follow the
description in Section~\ref{cdrsec:detectors-nd-ref-fgt-instrum}. At
present both CAMAC- and VME-based DAQ are available. In addition, a
$\mu$TCA-based fast DAQ has also been setup.

%\textbf{ \textit{Other Activities}} \\
Other activities are in progress. As part of the prototyping, 50 straws of 1.8~m from Lamina
Dielectrics Ltd. and 1~km of 30~$\mu$m anode wire from Luma
has been procured. The optical bench for the fabrication of the
straws has been setup.  Two pre-mixed gas bottles of Ar+CO$_2$ have been
procured. The operational gas mixture of
Xe+CO$_2$ will be added
soon. Local industry and vendors have been identified for
manufacturing nozzles, end-plugs, wire-spacers and steel
balls. Local workshops are available to fabricate the mechanical
structure to hold the straws in the prototype design and also to
fabricate a test stand for studies of efficiency and characteristics with
a radioactive source. Wire stringing, straw gluing and other tooling
setups are still to be established.

%\textbf{ \textit{Design of the Full-scale STT Modules}} \\
%We will optimize t
The final design of the STT modules will be optimized based on the
results obtained from the STT prototype and the related prototyping
activities listed above. This task includes a detailed FEM analysis to
assess the mechanical structure and the choice of final
materials. 
%The final design is intended to be ready for the fabrication of the full-scale STT detector. 

\subsubsection{ECAL Detector}

In the FGT, the ECAL detector will have $4\pi$ coverage outside
the STT.  The detailed description of this
detector is given in Section~\ref{cdrsec:detectors-nd-ref-fgt-ecal}.
The ECAL prototype will be a 2 $\times$2-m$^2$ module similar to the
downstream-ECAL design.  The half-scale downstream ECAL prototype
construction, which uses Pb as the absorber and extruded scintillator
with embedded fiber as the active detector system, will involve the
following steps:
\begin{itemize}
\item procure materials (plastic scintillator bars, WLS fibers,
  SiPM, Pb sheets, etc.;)
\item set up mechanism to ensure the quality of the scintillator bars,
  fibers and Pb sheets;
\item set up tools for the characterization of SiPMs;
\item assemble scintillator bars in an aluminum frame for a
  prototype layer formation;
\item undertake R\&D for the coupling of the fiber to the SiPMs as well
  as the inserting of fiber in the scintillator;
\item develop readout electronics for the prototype and set up a cosmic
  test stand with full DAQ; and
\item complete ECAL mechanical design. 
\end{itemize}

The ECAL readout system is centered on a highly sensitive/high-gain
SiPM.  During the R\&D phase, SiPMs from Hamamatru, AdvanSiD and SiPM
developed in India by SCL will be compared.  Discussions have been
started with all the vendors.

Optimization of the ECAL detector geometry with GEANT4 simulations has
been initiated. The geometry in the current GEANT4 simulation includes
58 layers of alternating horizontal and vertical scintillator layers
per 1.75~mm Pb along the $z$-direction. In the present configuration
each scintillator layer is made of plastic scintillator bars of
dimensions 4~m $\times$ 2.5~cm $\times$ 1~cm, resulting in 160 bars per
layer, and \num{9280} scintillator bars for the downstream ECAL .
Figure~\ref{fig:ECAL_SimulationG4} shows the longitudinal view of the
electromagnetic shower in the downstream ECAL by 2-GeV
photons. Figure~\ref{fig:ECAL_detail} shows the design of the
Pb-scintillator assembly configuration for the ECAL.
\begin{cdrfigure}[Longitudinal view of the EM shower in
the downstream ECAL by 2-GeV photons]
{ECAL_SimulationG4}{Longitudinal view of the electromagnetic shower in
the downstream ECAL by 2 GeV photons.}
\includegraphics[width=0.6\textwidth]{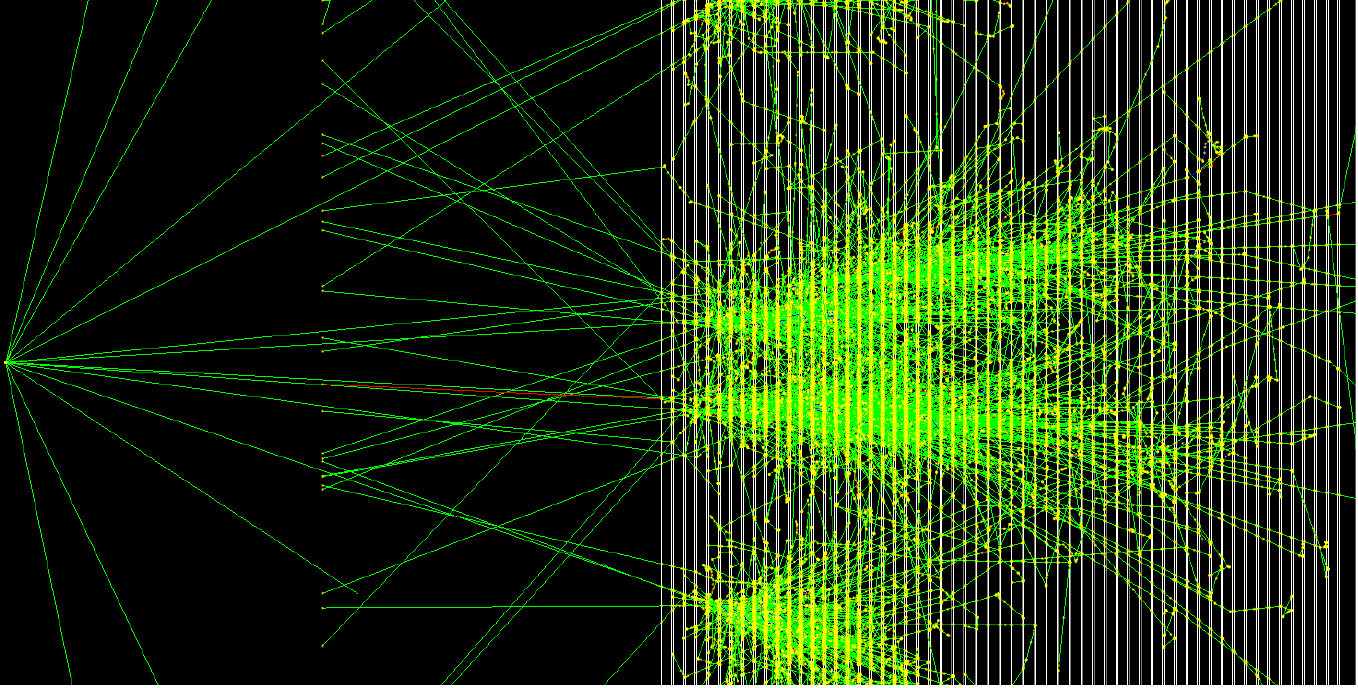}
\end{cdrfigure}

For the construction of the prototype and for the assembly of the
actual detector a space of dimension 32~m $\times$ 12~m has been
identified. Construction of a class 10,000 clean room covering a
laboratory space of 12~m $\times$ 12~m is under consideration.
Figure~\ref{fig:ECAL_LabIITG} shows the schematic diagram of the
laboratory refurbishment plan for the ECAL R\&D and fabrication work.
\begin{cdrfigure}[Laboratory refurbishment plan for ECAL R\&D and assembly]
{ECAL_LabIITG}{Laboratory refurbishment plan for the ECAL R\&D and assembly work.}
\includegraphics[width=0.8\textwidth]{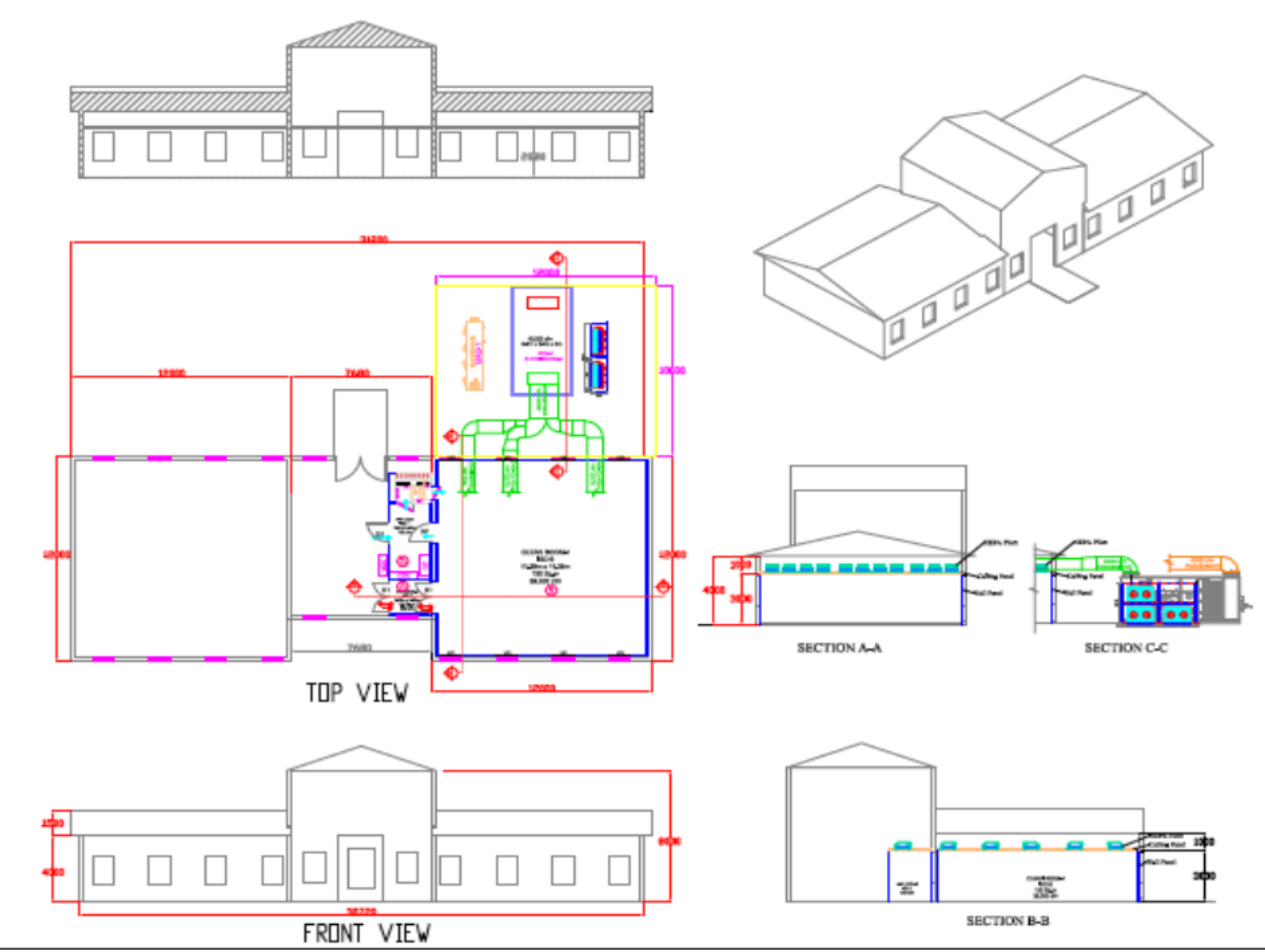}
\end{cdrfigure}

\subsubsection{Dipole Magnet Development}

The massive dipole magnet (see Section~\ref{cdrsec:detectors-nd-ref-fgt-magnet}) for the FGT 
is critical for the particle-momentum measurements, will 
provide space for the MuID--RPC installation in the magnet steel
and will provide structural support for the FGT. 
The planned magnet prototype
includes the engineering development of the tooling and
infrastructure 
that will be used to produce one C out of the total eight Cs of the
8.0-m long dipole.  The same C will be utilized in the final magnet
assembly. In a similar way, one of the four coils
will be assembled to establish the coil winding procedure and measure the
operating characteristics.  Field simulation work is very
advanced (see Figure~\ref{fig:Magnet_Bfield}) and the mechanical
designs are being produced.  (Steel dimensions are being optimized to house the muon
identification detectors.) Since it will be a
closed system, access to the inner detector systems is under extensive
study.

\subsubsection{MuID--RPC Detector}

Muon identification is accomplished via Bakelite Resistive Plate
Chamber (RPC) detectors.  The RPC mounting structure will be provided
by the magnet steel (on sides and ends).  Extensive R\&D for RPCs is
being applied to the prototyping of the muon identifiers. The size of
the FGT RPCs makes it challenging to procure the raw material from
industry; however, an Indian company has been identified and a
large 2.4-m $\times$ 1.2~m RPC prototype has been assembled (see
Section~\ref{cdrsec:detectors-nd-ref-fgt-muonid}). The I-V
characteristics obtained for these RPCs are very encouraging. More
RPCs will be fabricated during the prototyping phase and will be
tested for sustained efficiency with variation in ambient parameters,
as the Bakelite is sensitive to such changes. Some of the measured
quantities are shown in Figure~\ref{fig:RPC_PrototypeTests}.  
\begin{cdrfigure}[RPC characteristics measured during the prototype development]
{RPC_PrototypeTests}{RPC characteristics measured during the prototype development.}
\includegraphics[width=1.0\textwidth]{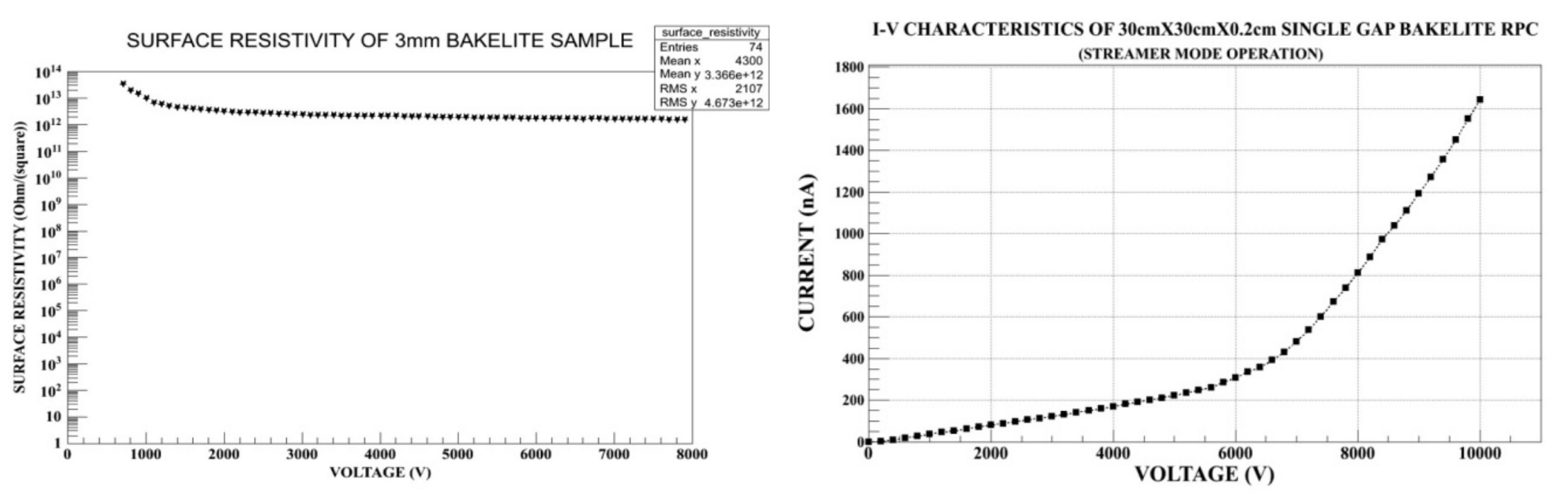}
\end{cdrfigure}
The readout electronics are being developed with input from INO-ICAL
detector~\cite{1748-0221-7-10-P10003} R\&D. The standard RPC gases
will need to be replaced with safer ones for underground operation.
An initiative in this direction will be taken up during the
prototyping phase.

\subsection{Beamline Measurement Detectors Prototyping Plan}
\label{sec:proto-nd-blm}
This Section describes recent and ongoing prototyping efforts for the detectors described in Section~\ref{sec:detectors-nd-ref-blm}.

\subsubsection{Prototype Development for the Cherenkov and Ionization Detectors}
\label{subsec:proto-blm-muon-cherenkov-proto}

A prototype Cherenkov counter, along with associated fully automated
gas systems, HV systems, and a data acquisition system has been
constructed and is undergoing testing in the NuMI neutrino beam's Muon
Alcove 2. In addition, three diamond detectors~\cite{ref:CERNdiamond} (from CERN)
for ionization measurements have  been installed in the alcove. 
Figure~\ref{fig:Alcove2Cherenkov} shows the prototype detectors in the
NuMI Alcove 2.
\begin{cdrfigure}[Muon gas Cherenkov counter]{Alcove2Cherenkov}
{A prototype muon gas Cherenkov detector for DUNE.
Muons travel through an L-shaped 4-inch Conflat pipe filled with a
pressurized gas. A flat mirror mirrors directs the optical photons
to a photo multiplier. The lower right inset shows the 20~bar MKS
pressure reading achieved by the Cherenkov gas system, and the inset
on the upper right shows the CERN/Cividec diamond detectors mounted to the Cherenkov housing.}
\includegraphics[width=6.in]{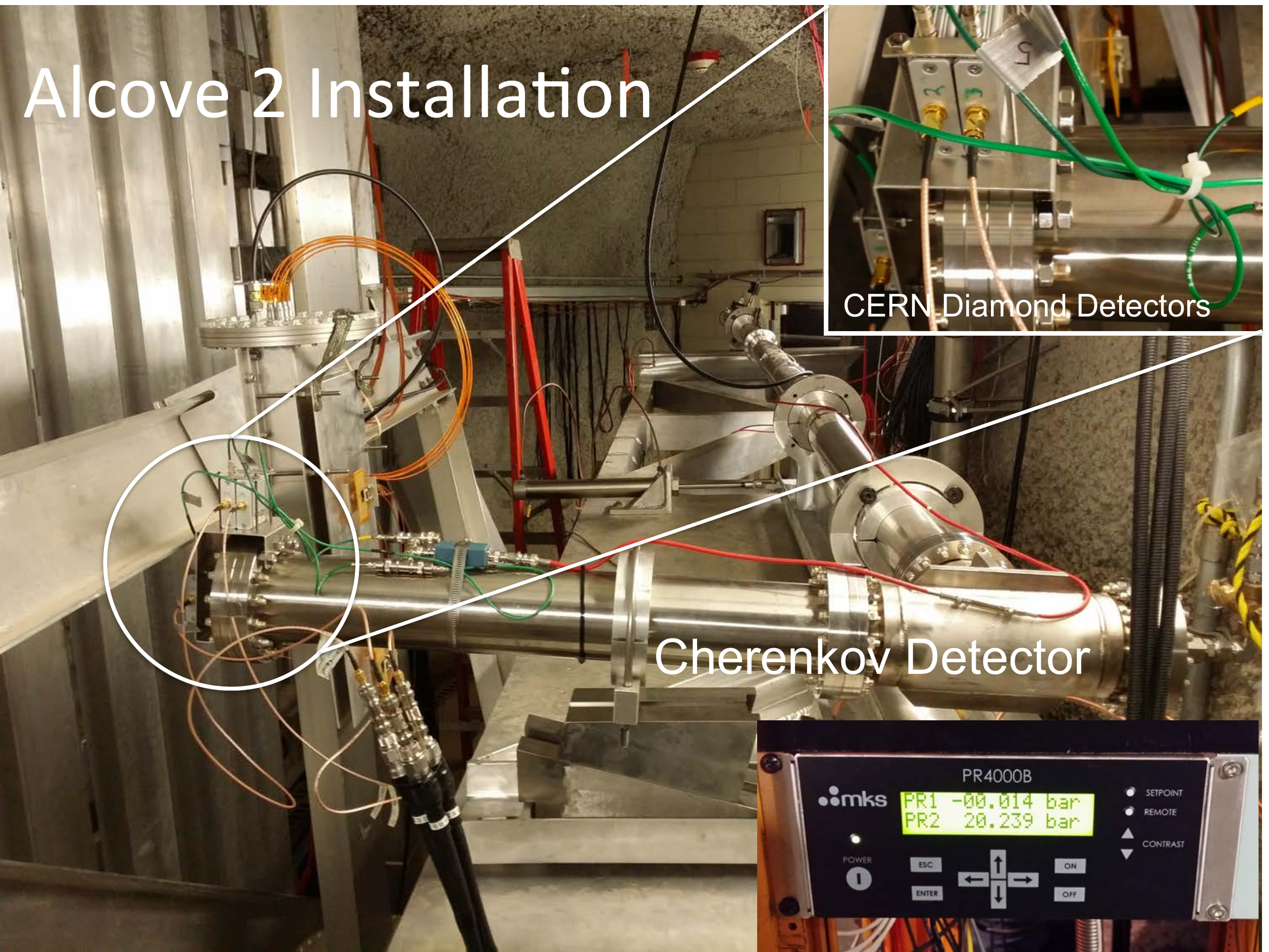}
\end{cdrfigure}

The counter has an automated gas system with an adjustable % settable 
pressure that
ranges from vacuum to 20~atm, corresponding to muon Cherenkov
thresholds of 200~GeV/c and 1~GeV/c, respectively. When operated at
vacuum, a photomultiplier tube (PMT) registers all background light unrelated to the gas,
e.g., transition radiation and light from particles hitting the window and
PMT glass.  These contributions are observed to be very small relative
to the coherent, directional Cherenkov light.

The counter is constructed with a 1-m long radiator section as shown
in Figure~\ref{fig:CherenkovCounterDetail} . A 20-foot extension
allows the reflected Cherenkov light to travel to a sapphire pressure
window viewed by a PMT.
\begin{cdrfigure}[Muon gas Cherenkov counter detail]{CherenkovCounterDetail}
{A prototype muon gas Cherenkov detector for DUNE.  }
\includegraphics[width=6.in]{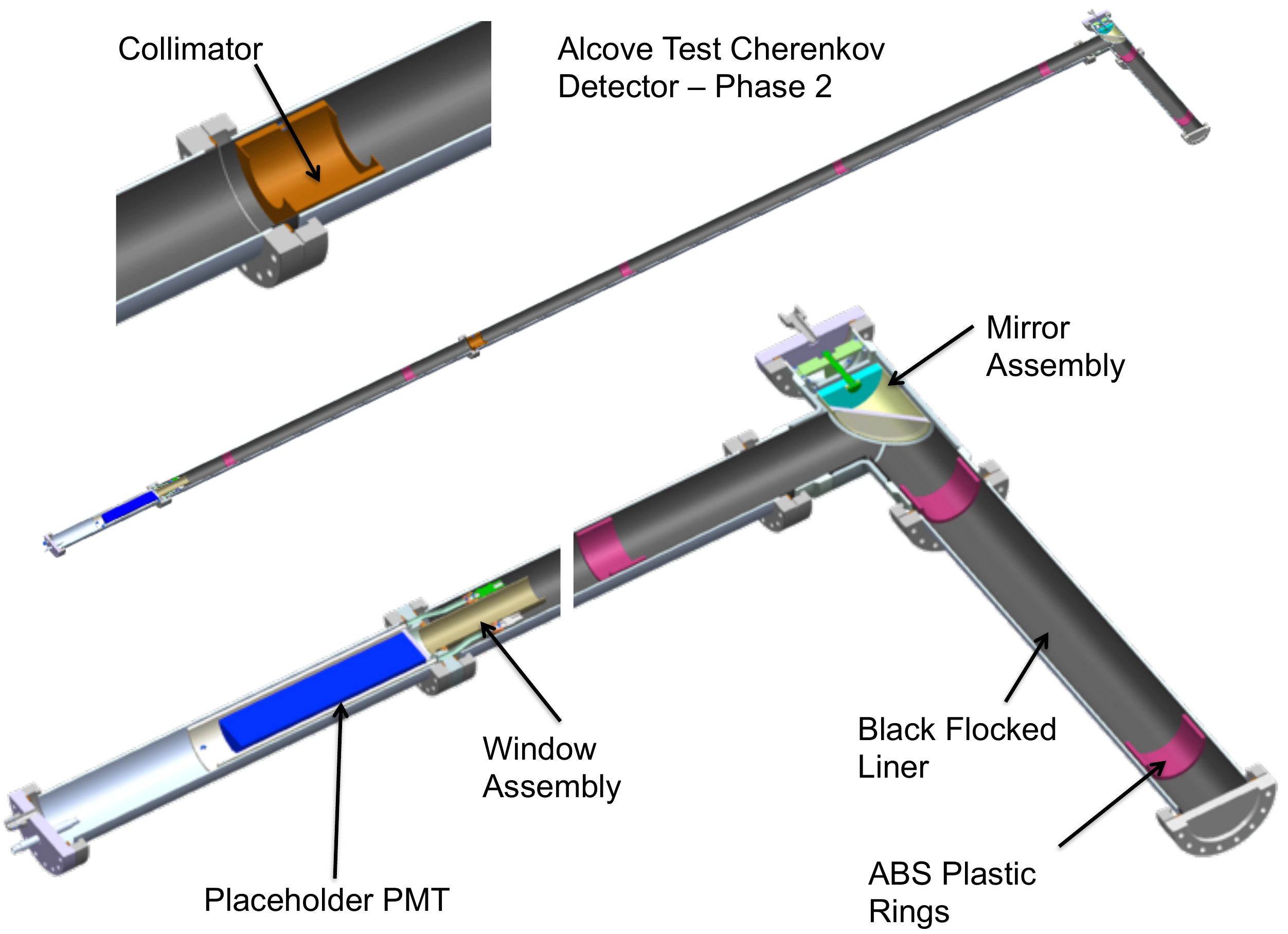}
\end{cdrfigure}

The prototype has been fully integrated into NuMI operations and
real-time waveforms can be viewed online as shown in
Figure~\ref{fig:MuonDetectorWaveforms}.
\begin{cdrfigure}[Muon detector waveforms]{MuonDetectorWaveforms}
{The real-time display of the muon detector prototypes in operation
on the NuMI beam line. The top two panels are the Cherenkov counter
and CERN diamond detector. The signals are
transmitted through low-loss heliax cable, then the waveform
is digitized at 2.5~GHz with a 12-bit dynamic range, and is
recorded onto disk storage for analysis. The signal from the
muons is contained in the short beam pulse ``buckets'' created
by the accelerator RF structure. The fast timing allows the
prompt muon signal to be easily separated from potential backgrounds
such as stopped-muon decays, beta decays, and neutrons.}
\includegraphics[width=5in]{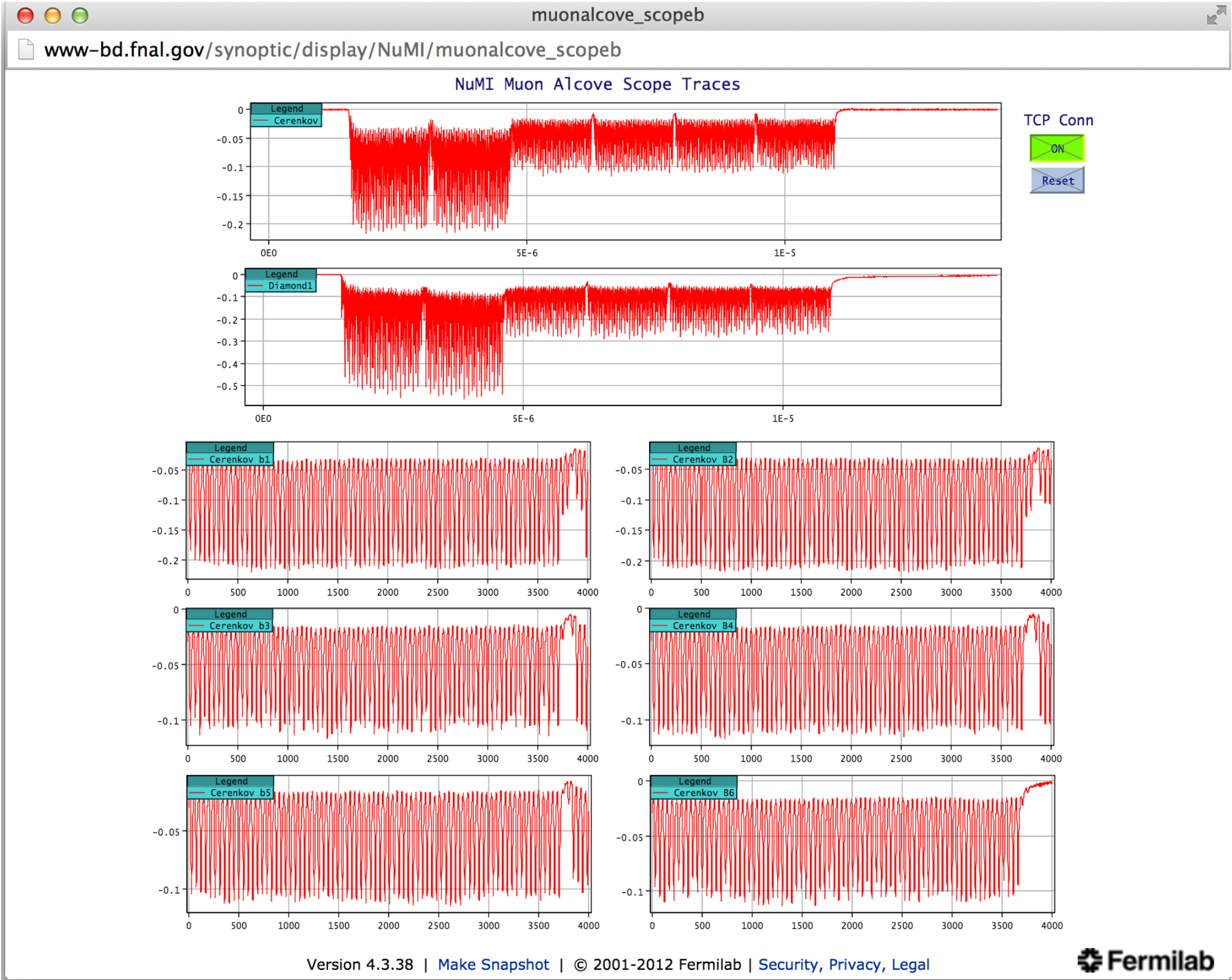}
\end{cdrfigure}
The top panel shows the waveform from the Cherenkov counter at 2~atm
gas pressure, which corresponds to a muon momentum threshold of
3~GeV/c. The second panel shows the waveform from a 9~mm $\times$ 9~mm
diamond detector mounted to the front flange of the Cherenkov radiator
section, as shown in the inset of Figure~\ref{fig:Alcove2Cherenkov}.

The extracted NuMI proton beam produces a signal in the Resistive Wall
Monitor (RWM) that is recorded with an identical digitizer. This allows a direct,
bucket-by-bucket (individual proton pulses) comparison of the proton
current onto the NuMI primary proton target. The muons are measured
after the absorber with a 400-ps time resolution.

\subsubsection{Prototype Development of the Stopped-Muon Counters}

Prototype development activity for the Michel-electron detectors will
be divided into studies of (1) the rate of particles and the radiation environment where
the detectors will be located, and (2) development of the counters
themselves.

The radiation environment will be studied with Monte Carlo
simulation and measurement from initial prototype detectors
in the NuMI muon alcoves~\cite{ref:NuMIBeamMonitors}.
The prototypes will be installed in the alcoves in 2016 and 2017.
Studies will be performed to determine if the photon sensors
can survive the radiation environment at the location of the Michel
detector. If the sensors can survive, they can be attached directly to
the Cherenkov medium; if not, optical guides will have to bring the
light to a lower-radiation area, to the side of the beam. Potential
radiation damage to the Cherenkov radiator itself will also be
studied.

The detector design will focus on selecting radiator and shielding
material, photon-detection technology and control/readout
hardware. Possible radiators include ones that use aerogel (these may be designed to
be replaced periodically) and ones that use flowing liquids such as H$_2$O or
mineral oil. Long-timescale saturation from the very high-rate
environment of the beam spill could affect the photon-counting
devices~\cite{ref:HighRateCounting}. Thus, it will likely be necessary
to design fast-switching, high-voltage circuits that turn on the
photon counters in the first few microseconds after the spill is
over. A similar system was developed in the 1990s for the Brookhaven
Muon (g-2) Experiment~\cite{ref:G2} .

%\textbf{ \textit{Current Prototyping Activities}} \\
A second set of muon detectors, the final 
%\fixme{really want to say `final'?i - yes} 
DUNE design, are being
constructed at this time (2015). They are being installed directly
behind the NuMI proton beam dump (Muon Alcove 1), mounted on a movable stand which has undergone an engineering
review at Fermilab. 
The entire setup, detectors and stand, will be
suitable for use in the %DUNE 
LBNF beam. The higher-radiation environment of Alcove 1 is more representative %similar 
than Alcove 2 of the conditions in the eventual DUNE installation, and will 
allow a more accurate calibration in the NuMI
beam. The setup will be eventually transferred to the DUNE Absorber Hall. 

%% file: volume-detectors/proto-sections/proto-sbn-connect.tex
%%%%%%%%%%%%%%%%%%%%%%%%%%%%%%%%
\section{Connections to the Short-Baseline Program at Fermilab}
\label{sec:sbn_connect}

% Top part moved to overview of chapter. Anne 5/22
 
%Not all prototypes must be executed by DUNE in order for DUNE to benefit.  
DUNE will benefit from a range of past and ongoing efforts at
Fermilab. Some have been evolving in tandem with the former LBNE and
present DUNE efforts. The strategy behind MicroBooNE --- to
incorporate some detector development aspects in an experiment with
goals to investigate short-baseline neutrino physics --- expanded to
include detectors upstream and downstream.  The Fermilab
Short-Baseline Program on the Booster Neutrino Beamline now consists
of the Short-Baseline Near Detector (SBND), MicroBooNE, and the ICARUS
T-600; the program is fully described in a recent
proposal~\cite{Antonello:2015lea}.  There is significant overlap in
the collaboration membership of DUNE and the three short-baseline
detectors.
 
Each of the short-baseline detectors shares some technical elements with each other and/or with the DUNE far detector prototypes e.g.,   
 cryogenic system design,
 argon purification techniques and
 cold electronics.

In other aspects, e.g., the design details of the anode wire planes, %however, 
the detectors are very different.  The SBND is most similar to the DUNE single-phase detector design, having adopted the 35-t APA-CPA-field cage design, while the MicroBooNE TPC field cage follows the ICARUS design.  The cold electronics installed on the MicroBooNE TPC represent an initial step in an ongoing program; the 35-t and the SBND 
implement subsequent outgrowths of
%utilize subsequent steps in the 
cold electronics development.  While commissioning its cryogenics system, MicroBooNE conducted investigations of the voltage breakdown in high-purity argon; the results prompted some design adjustments to the field cage adopted by the 35-t Phase-2 and the SBND, demonstrating the sharing and feedback of technical developments.  

Coordinated development of reconstruction software for LArTPC detectors is a major outcome of the 2009 \textit{Integrated Plan}.  LArSoft is fully supported by the Fermilab Scientific Computing Division and has contributors from all of the operating and planned LArTPC experiments at Fermilab.  Track and shower reconstruction methods, and particle identification techniques , are already shared between ArgoNeuT, MicroBooNE, LArIAT and the 35-t.  Real data from these detectors is assisting DUNE simulation efforts.  The Short-Baseline experiments, starting with MicroBooNE, will develop neutrino interaction classification techniques based on the details revealed by their fine-grained tracking capabilities, and are likely to exert a strong influence on DUNE oscillation analyses.

%% file: volume-detectors/chapter-summary.tex
\chapter{Summary of DUNE Detectors}
\label{ch:detectors-summary}

The DUNE experiment is a world-leading, international physics
experiment, bringing together a global neutrino community as well as
leading experts in nucleon decay and particle astrophysics to explore
key questions at the forefront of particle physics and
astrophysics. The massive, high-reolution near and far detectors will
enable an extensive suite of new physics measurements that are
expected to result in groundbreaking discoveries.

The far detector will be located deep underground at the 4850L of
SURF.  Its 40-kt fiducial mass of LAr will enable sensitive studies of
long-baseline oscillations with a 1,300~km baseline, as well as a rich
program in astroparticle physics and nucleon decay searches.  The far
detector configuration consists of four LArTPCs.  They provide
excellent tracking and calorimetry performance, high signal efficiency
and effective background discrimination, all of which converge to
provide an overall excellent capability to precisely measure neutrino
events and reconstruct kinematical properties with high
resolution. The full imaging of events will enable study of neutrino
interactions and other rare events to unprecedented levels.

The magnetized, high-resolution near detector will measure the
spectrum and flavor composition of the neutrino beam extremely
precisely. It is able to discriminate neutrino flavor through
particle identification and separate neutrino and antineutrino fluxes
through charge discrimination of electrons and muons produced in the
neutrino charged current-interactions. These capabilities  enable DUNE to
reach unprecedented sensitivity in long-baseline neutrino oscillation
studies.  This is the primary role of the near detector, however, its
exposure to an intense flux of neutrinos will provide an opportunity to
collect unprecedentedly high neutrino interaction statistics, making
possible a wealth of fundamental neutrino interaction measurements, an
important component of the DUNE Collaboration's ancillary scientific
goals.

%% file: common/final.tex
% this is added just after end of document

\cleardoublepage
\renewcommand{\bibname}{References}
\bibliographystyle{ieeetr}
\bibliography{common/citedb}